\documentclass[article,11pt]{emulateapj}
\usepackage{graphicx}
\usepackage{times}
\usepackage{natbib}
\usepackage{amsmath}
\usepackage{amssymb}
\def\jgr{J. Geophys. Res. }

\def\aj{Astron. J.}
\def\apj{Astrophys. J.}

\def\grl{Geophys. Res. Let.}

\def\planss{Planet. \& Space Sci.}

\def\asht{AsH$_3$ }
\def\ashtx{AsH$_3$}
\def\hto{H$_2$O }
\def\htox{H$_2$O}

\def\icarus{Icarus}

\def\deg{$^\circ$ }
\def\degx{$^\circ$}

\def\mum{$\mu$m }
\def\mumx{$\mu$m}
\def\chf{CH$_4$ }
\def\chfx{CH$_4$}
\def\chtd{CH$_3$D }

\def\chisq{$\chi^2$ }
\def\chisqx{$\chi^2$}

\def\hto{H$_2$O }
\def\htox{H$_2$O}
\def\pht{PH$_3$ }
\def\phtx{PH$_3$}
\def\pthf{P$_2$H$_4$ }
\def\pthfx{P$_2$H$_4$}
\def\nht{NH$_3$ }
\def\nhtx{NH$_3$}
\def\nhfsh{NH$_4$SH }
\def\nhfshx{NH$_4$SH}

\begin{document}

\title{Evolution of Saturn's north polar color and cloud structure 
  between 2012 and 2017 inferred from Cassini VIMS and ISS observations}
\author{L.A. Sromovsky$^1$, K. H. Baines$^1$, and P.M. Fry$^1$}
\affil{$^1$Space Science and Engineering Center, University of Wisconsin-Madison,
1225 West Dayton Street, Madison, WI 53706, USA}
\slugcomment{Journal reference: L.A. Sromovsky, K.H. Baines and P.M. Fry, 
Icarus, 362 (2021) 114409}

\begin{abstract}

Cassini/ISS imagery and Cassini/VIMS spectral imaging observations
from 0.35 to 5.12 \mum show that Saturn's north polar region (70\degx
-- 90\degx N) evolved significantly between 2012 and 2017, with the
region poleward of the hexagon changing from dark blue/green to a
moderately brighter gold color, except for the inner eye region
(88.2\degx -- 90\degx N), which remained relatively unchanged.  These
and even more dramatic near-IR changes can be reproduced by an aerosol
model of four compact layers consisting of a stratospheric haze at an
effective pressure near 50 mbar, a deeper haze of putative diphosphine
particles typically near 300 mbar, an ammonia cloud layer with a base pressure
between 0.4 bar and 1.3 bar, and a deeper cloud of a possible mix of
\nhfsh and water ice particles within the 2.7 to 4.5 bar region.
Between the eye and the hexagon boundary near 75\degx N were many
small discrete bright cloud features that VIMS spectra indicate have
increased opacity in the ammonia cloud layer.  Our analysis of the
background clouds between the discrete features shows that between
2013 and 2016 the effective pressures of most layers changed
very little, except for the ammonia ice layer, which decreased from
about 1 bar to 0.4 bar near the edge of the eye, but increased to 1
bar inside the eye. Inside the hexagon there were large increases in
optical depth, by up to a factor of 10 near the eye for the putative diphosphine
layer and by a factor of four over most of the hexagon
interior.  Inside the eye, aerosol optical depths were very low,
suggesting downwelling motions.  The high contrast between eye and
surroundings in 2016 was due to substantial
increases in optical depths outside the eye. The color change from
blue/green to gold inside most of the hexagon region can be explained
in our model almost entirely by changes in the stratospheric haze, which
 increased between 2013 and 2016 by a factor of four in optical
depth and by almost a factor of three in the short-wavelength peak of
its wavelength-dependent imaginary index. A plausible mechanism for
increasing aerosol opacity with time is the action of photochemistry
as the north polar region became increasingly exposed to solar UV
radiation.  For 2013 we found an ammonia mixing ratio of about 50$\times 10^{-6}$
in the depleted region between 4 bars and the \nht condensation level
($\sim$ 1 bar), but the \nht results for 2016 are unclear due to very high
retrieval uncertainties associated with increased aerosol opacity.  We
retrieved a deep abundance of about 5$\times 10^{-6}$ for \pht
and a pressure breakpoint (where the \pht abundance begins to decline
with altitude) that coincided with the main cloud top near 300 mbar,
except when that cloud opacity was very low, at which point the \pht
breakpoint pressure generally increased substantially, consistent with
prior suggestions that the cloud layers shield \pht from destruction
by UV radiation above the clouds.  We found an average \asht mixing
ratio of 2$\times 10^{-9}$  with some evidence for a decline
with altitude above the main cloud layer.
\end{abstract}

\maketitle
\shortauthors{Sromovsky et al.}
\shorttitle{Saturn's north polar cloud evolution.}

\newpage

\section{Introduction} 

\begin{figure*}[!htb]\centering
\includegraphics[width=6in]{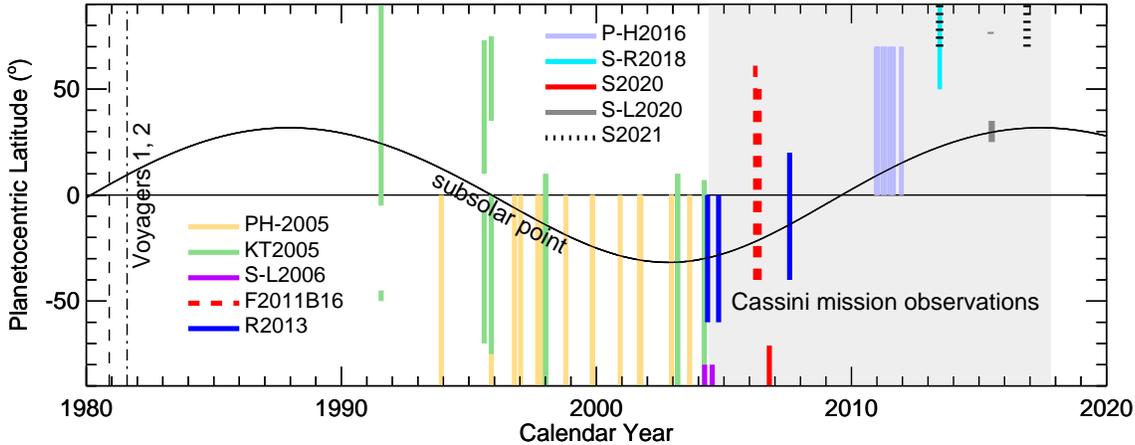}
\caption{Saturn's subsolar latitude versus time marks its seasonal
  progress as it approached northern summer solstice in May 2017.
  Other summer solstices over this Saturn year were in December 1987 (northern) and October
  2002 (southern).  Our selected VIMS observation times, in 2013 and
  late 2016, and latitudinal coverage are indicated by thick dotted
  lines at the upper right. The approximate times and latitude ranges
  analyzed in selected prior publications are shown by vertical bars
  with color and lines style indicated in the legend, with reference
  codes formed from author initials and publication date for the
  following references: \cite{Perez-Hoyos2005}, \cite{Kark2005},
  \cite{Sanchez-Lavega2006}, \cite{Fletcher2011vims} and
  \cite{Barstow2016} (combined as by F2011B16), \cite{Roman2013},
  \cite{Perez-Hoyos2016}, \cite{Sanz-Requena2018},
  \cite{Sro2020spole}, \cite{Sanchez-Lavega2020hazes}, and this work
  (labeled as S2021). Year labels are centered on tick marks placed at
  the beginning of the labeled year. These do not include
  storm-focused publications covering small local regions.}
\label{Fig:seasons}
\end{figure*}

The Cassini observations of Saturn's atmosphere extended from 2004
until atmospheric entry  on 15 September 2017.
As illustrated in the plot of subsolar latitude versus time
(Fig. \ref{Fig:seasons}), the mission extended over nearly a complete
seasonal half-cycle, beginning just a few years after the October 2002
 southern summer solstice and ending a few months after the May 2017
northern summer solstice. Both polar regions were found to be
encircled by strong cyclonic vortex circulations
\citep{Orton2005Sci,Fletcher2008,Baines2009cyclone,Antunano2015,Sayanagi2017,
Antunano2018,Fletcher2018NatCo,Achterberg2018}, with a visual
similarity to hurricanes on earth, including dark eyes
 centered on the poles and, in the case of the southern
vortex, what appeared to be eyewall clouds casting long shadows
\citep{Dyudina2008Sci,Dyudina2009}.  However, a detailed radiative
transfer analysis of the southern region did not find any deep
convective wall clouds, but instead a region of generally low aerosol optical
depth, with shadows cast by spatially sharp but small optical depth
transitions in relatively translucent layers
\citep{Sro2020spole,Sro2020shadows}.  Also found in the south polar
region were discrete bright features with 3-\mum spectral signatures,
characteristic of ammonia ice clouds, which are generally not visible
on Saturn except in association with convective storm systems such as
those in Storm Alley \citep{Baines2009stormclouds,Sro2018dark} or in
the Great Storm of 2010-2011 \citep{Sro2013gws}.  The corresponding
spectra of these south polar features were well modeled by locally
increased optical depths in the ammonia cloud layer, which were
visible in the polar region due to the small optical depth of the
overlying haze layers \citep{Sro2020spole}.

The north polar region is remarkable not only for the existence of a
hexagonal weather pattern associated with a meandering zonal jet
\citep{Sayanagi2019book}, but also for the remarkable color
transformation that occurred within the hexagon as the planet
approached northern summer solstice. As shown in press-release images
reproduced in Fig.\ \ref{Fig:issevol}, the natural color of the region
transformed from a blue/green color to an orange/gold color, a
presumed result of solar UV photochemistry producing a haze of
particles in the upper troposphere that began to resemble in color the
top haze layer seen over most of the rest of Saturn. In spite of this
upper haze layer, underlying discrete ammonia cloud features were also
seen inside the hexagon, even when the haze was its thickest, near the
end of the Cassini mission. This was demonstrated by an initial
analysis of the 2017 high spatial resolution Grand Finale VIMS
observations of the north polar region close to and within its eye
\citep{Baines2018GeoRL}. Besides finding evidence for local convection
associated with discrete ammonia clouds, they also found generally
very low optical depths, especially within the eye region, which did
not undergo the color transformation seen in the rest of the polar
region, and where aerosols were almost completely absent.  The
relatively unchanged appearance of the eye and appearance of discrete
cloud features within the hexagon can also be seen in the June 2013
and April 2017 mosaics in Fig.\ \ref{Fig:issevol}.

Here we present a more extensive analysis of the north polar region,
covering a wider range of latitudes from inside the eye to just
outside the hexagon, allowing us to define the characteristics of the
polar cloud bands and their temporal evolution during Saturn's
approach to its May 2017 northern summer solstice.  Our analysis is based primarily on
observations of the Cassini Visual and Infrared Mapping Spectrometer
(VIMS) and to a lesser degree on the bandpass filter observations of
the Cassini Imaging Science Subsystem (ISS). VIMS is unique in covering
both visual and near-IR spectral bands and providing constraints from
both reflected sunlight and thermal emission by Saturn.  We first
describe the observations we used, followed by a description of our
vertical structure and composition model, and next a description of
our radiation transfer model.  We finally describe the results of
fitting models to the spectral observations and their implications for
the evolution of aerosols and color in Saturn's north polar region, as
well as changes in the vertical distribution of phosphine.

\section{ISS Observations}

\subsection{ISS instrument characteristics and data reduction}

The Cassini Imaging Science Subsystem (ISS) is described by
\cite{Porco2004SSR}.  It has a narrow angle camera (NAC) with a field
of view (FOV) 0.35\deg across, and a wide angle camera (WAC) with a FOV of
3.5\degx, both using 1024-pixel square CCD arrays with pixel scales of
1.24 and 12.4 arcseconds/pixel respectively (in the unbinned imaging
mode).  For our analysis we used WAC images (identified in
Table\ \ref{Tbl:obslist1}) to provide the best combination of 
spatial resolution and polar latitude coverage as well
as samples of short and long wavelength continuum filters,
as well as an intermediate methane band.  The image files were retrieved from the
NASA Planetary Data System’s Imaging Node and processed with the USGS
ISIS 3 cisscal application, which is derived from the IDL cisscal
application developed by the Cassini Imaging Central Laboratory for
Operations (CICLOPS). This cisscal application produces images in I/F
units (the ratio of target brightness relative to that of a
unit-albedo Lambertian reflector illuminated and viewed normally at
the same distance from sun). Ephemeris and pointing data allowing
transformations between image and planet coordinates are disseminated
by NASA's Navigation and Ancillary Information Facility
\citep{Acton1996}.
 
\begin{figure*}[!t]\centering
\includegraphics[width=6in]{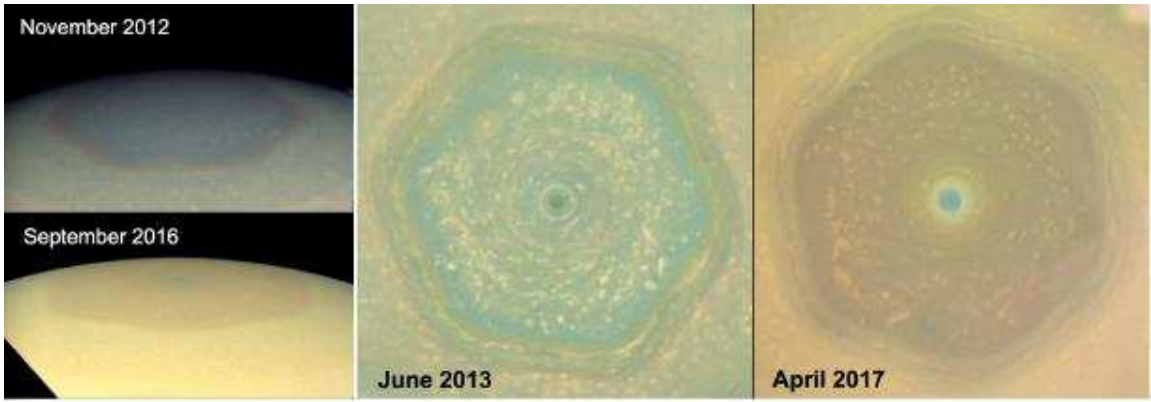}
\caption{In the left panel, taken from NASA press release PIA21049, are color 
composite images of the north polar region from 2012 and 2016, during which 
the color inside the hexagon appeared to change from blue to orange.  
A remapped polar stereographic view of the transition, taken from NASA press 
release PIA21611, combined wide-angle camera images in red, green, and
blue filters to produce these approximations of natural color views. 
The original images have a resolution of about 80 km/pixel (2013) and
 80-14 km/pixel (2017), while the remapped images both
have a scale of 25 km/pixel. }
\label{Fig:issevol}
\end{figure*}


\begin{table*}[!htb]\centering
\caption{Observing conditions for ISS images we used.}
\vspace{0.1in}
\setlength\tabcolsep{3pt}
\begin{tabular}{ l c c c c c r}
               &                   &   UT Date    &          &  Pixel & Phase & Fig.\\
ISS image ID   &  ISS Filter       &  YYYY-DOY  &  Time    &  size (km)  & angle (\degx) & ref.    \\
\hline
  W1749823714 &  VIO (440 nm)    &  2013-164   &  13:13:05.936  &  68.6   &  94.7  &\ref{Fig:isstriple}A \\
  W1749823766 &  MT2 (727 nm)    &  2013-164   &  13:13:57.563  &  68.6   &  94.7  &\ref{Fig:isstriple}B \\
  W1749823807 &  CB2 (752 nm)    &  2013-164   &  13:14:39.035  &  68.6   &  94.7  &\ref{Fig:isstriple}C \\
  W1859479945 &  VIO (440 nm)    &  2016-338   &  17:05:17.086  & 152.9   &  96.2  &\ref{Fig:isstriple}D \\
  W1859479996 &  MT2 (727 nm)    &  2016-338   &  17:06:08.150  & 152.9   &  96.1  &\ref{Fig:isstriple}E \\
  W1859480030 &  CB2 (752 nm)    &  2016-338   &  17:06:42.169  &  76.4  &   96.1  &\ref{Fig:isstriple}F \\
\hline
\end{tabular}\label{Tbl:obslist1}
\parbox{5in}{$^1$Observation IDs are ISS\_192SA\_NPOLMOV001\_VIMS (2013) and ISS\_251SA\_NPOLMOV001\_VIMS (2016). \par
}
\end{table*}

\begin{figure}[!ht]\centering
\includegraphics[width=3in]{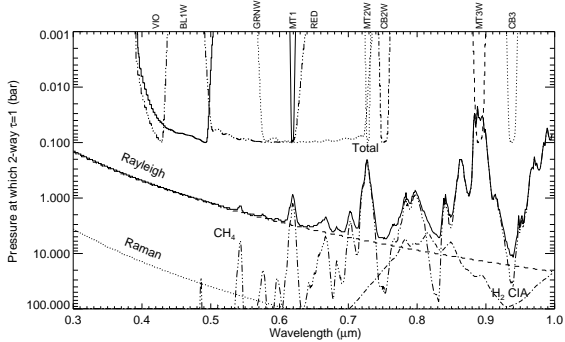}
\caption{Vertical penetration depths in a clear atmosphere as a function
of wavelength for Raman scattering, \chf absorption, and collision
induced absorption (CIA) individually and combined (solid curve),
 with selected ISS filter transmissions normalized to 0.1 at
their peaks.}
\label{Fig:ccdpendepth}
\end{figure}

The penetration depths of selected ISS filters we used are illustrated
in Fig.\ \ref{Fig:ccdpendepth}, where normalized filter transmissions
are overlaid on a plot of the pressure at which the 2-way vertical optical depth
in a clear atmosphere reaches unity.  The violet (440 nm) filtered image is
unaffected by methane absorption, and in a clear atmosphere would record
high I/F values produced by Rayleigh scattering over large optical
depths. But in Saturn's atmosphere a UV-absorbing haze decreases
shortwave reflectivity and produces the tan color characterizing most
of Saturn. The MT2 (727 nm) and MT3 (890 nm) filters sample
intermediate and strong methane bands, and are mainly sensitive to
scattering by upper tropospheric and stratospheric hazes respectively.  The CB2
filter samples a continuum wavelength which is sensitive to aerosol scattering
originating at the several bar level, is much less sensitive to
Rayleigh scattering, and only weakly sensitive to absorption by methane
and hydrogen collision-induced absorption (CIA).  At the incident
solar angles applicable to polar observations, these penetration
depths reach pressures that are often a factor of two or more smaller.
For the quantitative analysis we describe in later sections, the
viewing and illumination geometry is fully accounted for.  Here we
make use of Fig.\ \ref{Fig:ccdpendepth} for qualitative
interpretations described below.

\subsection{ISS characterization of Saturn's north polar region. }

The basic morphologies of Saturn's north polar cloud features as it
approached its northern summer solstice are displayed in Fig.\ \ref{Fig:isstriple}, 
using ISS wide angle camera images taken on 13 June 2013 (panels A-C) and
on 3 December 2016 (D-F).  The main features in 2013 are the eye region (bright in A, and
dark in B and C), which is about 1.2\deg in
radius, an intermediate region interior to the hexagon (generally dark
in B-C and E-F, but bright in A and D), which extends
out to about 75\degx N in planetocentric latitude, and a middle band (bright in B) that extends from
the hexagon out to 63.5\degx N.  A similar morphology is present in 2016 images, but
there are significant differences with respect to wavelength, latitude, and time
that provide important clues regarding the nature of the cloud
structure within this region.  These are interpreted with
the help of penetration depth calculations displayed in
Fig.\ \ref{Fig:ccdpendepth}.   

\subsubsection{Qualitative interpretation of 2013 ISS images.}
\begin{figure*}[!ht]
\includegraphics[width=6.2in]{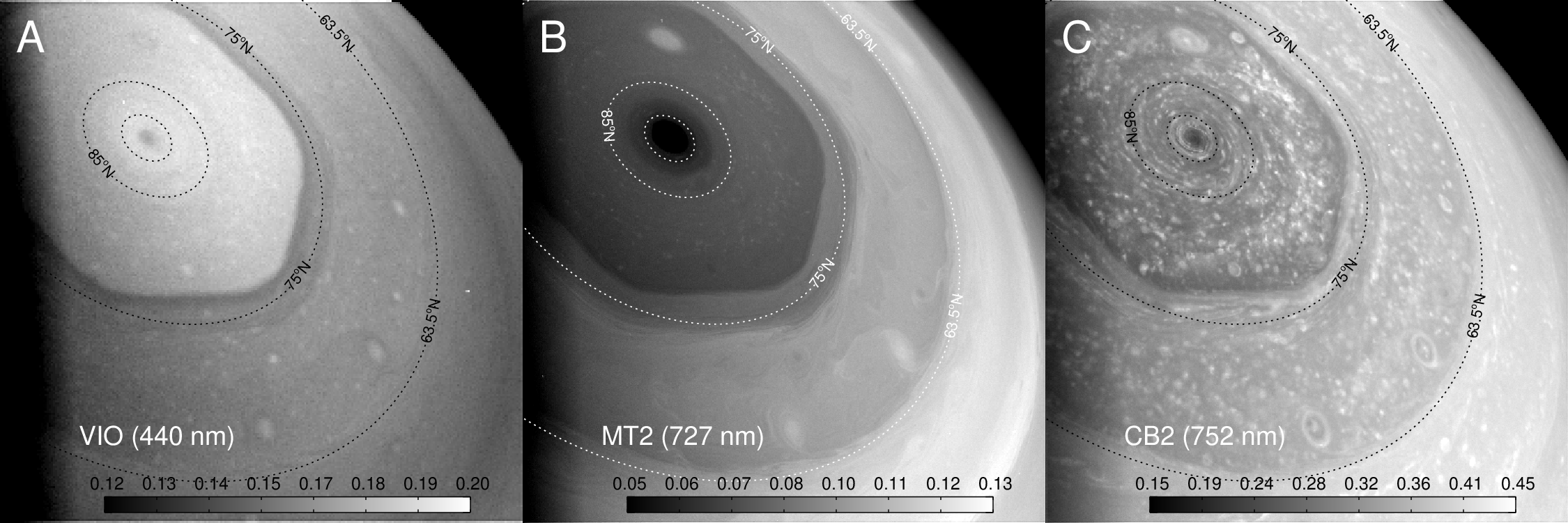}
\includegraphics[width=6.2in]{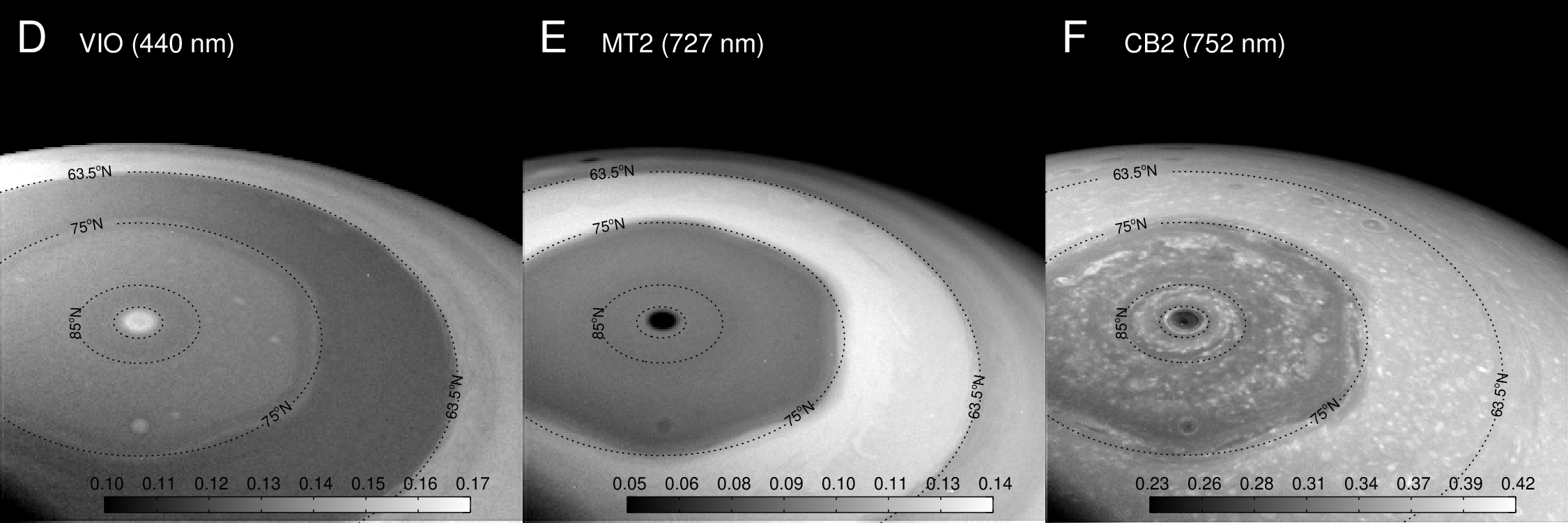}
\caption{Cassini/ISS wide-angle images taken on 13 June 2013, with an
  image scale of 69 km/pixel (A-C) and on 3 December 2016 (D-F), with
  image scales of 153 km (D, E) and 76.4 km (F).  Images D-F were also
  used in NASA press release PIA21053, which misidentified two of the
  filters and the date. We removed much of the brightness variationa
  across the images using a Minnaert correction in which the I/F at
  the pole (at image coordinates $x_p, y_p$) is unchanged, but away
  from the pole is corrected to $I/F(x,y)\times (\mu
  (x_p,y_p)/\mu(x,y))^{k-1}(\mu_0(x_p,y_p)/\mu_0(x,y))^k$, where $\mu$
  and $\mu_0$ are cosines of the observer and solar zenith angles
  respectively and $k$ chosen to be 0.82,
  0.9, and 0.85 for UV, MT2, and CB2 images respectively. Latitudes
  are planetocentric and the unlabeled contour is at 88\deg
  N.}
\label{Fig:isstriple}
\end{figure*}

In the violet image of Fig.\ \ref{Fig:isstriple}A the interior of
the hexagon is much brighter than the surrounding clouds, which will later
be shown to be a result of reduced aerosol absorption and greater visibility
of Rayleigh scattering. A strikingly different appearance is provided
by the 727-nm image in panel B, where the interior of the hexagon is seen to
be darker than surrounding clouds.  Here methane absorption provides a 
dark background that is not brightened much by the smaller amounts of high altitude aerosols 
available to scatter light above that absorption.  The central eye
region is especially dark due to even smaller amounts of aerosol scattering.  
Another strikingly different appearance is seen in the 752-nm image in panel C.  
Because methane absorption is
relatively weak at this wavelength, deeper discrete bright cloud features
can be seen through the relatively translucent overlying aerosol layers. Near-IR
observations show that these discrete features are associated
with enhanced scattering in the ammonia ice layer, as seen for similar
features found in the south polar region \citep{Sro2020spole}.
 The significant depth of these features is
evident from their nearly complete absence from the images with
significant Rayleigh scattering (panel A) or significant methane absorption
(panel B).  
Note that there seem to be two hexagonal boundaries in the 2013 images (most
visible in panel B): one just poleward of the 75\deg N contour and
a second just outside that contour.  The inner boundary is the
most sharply defined for all three wavelengths.    

\subsubsection{Qualitative interpretation of 2016 ISS images. } 
Panels D-F of Fig.\ \ref{Fig:isstriple} provide views of the same region
at the same wavelengths but almost 3.5 years after the views in panels A-C.  
The violet image in panel D
shows that a dramatic change has occurred in the upper aerosol layers during that interval. 
 The bright region inside
the hexagon has darkened dramatically due to the increased abundance
of absorbing aerosols, with the exception of the small eye region, which has
apparently not seen much of an increase in aerosol absorption. (Although
the eye seems brighter in D than in A, this is due to a change in
the image enhancement.) Nor has
the eye seen an increase in aerosol scattering, which is evident from the
fact that it remained dark in the 727-nm image in panel E.  
Another region that has seen an increased amount of upper level
aerosols is the region between the hexagon boundary (around 75\deg N planetocentric)
and the 63.5\deg N contour, which has gotten darker in violet (due to more
aerosol absorption at short wavelengths) and brighter in the 727-nm image due to more aerosol scattering
at longer wavelengths where chromophore absorption is not important.

In panel D of Fig.\ \ref{Fig:isstriple}, the polar region from
63.5\degx N to 75\degx N is relatively dark compared to its
surroundings, with an increase in I/F at the edge of the hexagon near
75\degx N, and another increase at 63.5\degx N. Hazes inside
the hexagon boundary appear to be more absorbing than those further from the
pole.  The exception is the bright eye of the north polar vortex.  The
high brightness of the eye is not due to increased aerosol scattering because it
appears relatively dark in panel E, which is sensitive to upper
tropospheric aerosols (down to a few 100 mbars), and also in panel F, which can sense scattering
by much deeper aerosols (down to the several bar level). This indicates that there is very little
aerosol scattering within the eye to significant depths. The eye is
also dark in the strong methane band at 890 nm (not shown), indicating
that it is also relatively free of stratospheric aerosols.  
There are also several bright spots in panel D in addition to the
bright eye.  Though dimmer than the eye, they appear to be of similar origin,
i.e. increased Rayleigh scattering due to reduced aerosol absorption.
At least one of these, located directly below the eye and just inside the hexagon,
also appears darker than its surroundings in panels E and F, confirming
that interpretation.

The 2013 image in the MT2 (727 nm) filter (panel E) looks similar to the
corresponding image from late 2016 (panel B), except that there is a contrast reversal between clouds
south of 63.5\degx N and those north of it.  In 2013 the middle band
was darker than the outer band, but in 2016 the middle band was
brighter.  In the middle band the increase in UV-absorbing haze
between 2013 and 2016 is a plausible source of increased brightness
because the absorption of that haze is not significant at longer
wavelengths. The polar region in the CB2 filter looks
much the same during the two observing times, with the exception that
the contrast between the middle band and the outer cloud band, which
was significant in 2013, has virtually disappeared by the end of 2016.
This might also be due to the increased optical thickness of polar
haze by the latter date.  On the other hand, the numbers and
distributions of discrete bright features seem very similar in the CB2
images, a result that seems initially inconsistent with the impression
given by the color composite images shown in Fig.\ \ref{Fig:issevol}.
This is a result of the CB2 filter (752 nm) being much less sensitive to scattering
by small haze particles that dominate the appearance at shorter wavelengths.

\subsubsection{Changes in the hexagon boundary}

An odd result of the 2013-2016 comparison in Fig.\ \ref{Fig:isstriple}
is that the hexagon seems to have increased in size between 2013 and
late 2016, with its outer boundary seemingly moving from a position well inside the 75\degx N latitude
circle in 2013 to essentially straddling it at the later date,
most easily seen in comparing the CB2 filter images (Fig.\ \ref{Fig:isstriple}C and F).
The situation is more complicated for the other two filters, both of which
seemed to have an extra band of darker brightness (panel D) or intermediate brightness
 (panel E) with an inner boundary inside the 75 \degx N circle and an outer
boundary outside the circle. It appears that by late 2016 both boundaries
moved closer to the 75\degx N circle.

\subsubsection{Color changes}

The northern polar region's transition
from winter to summer included the formation of what \cite{Sayanagi2016DPS}
called a bright polar hood (see Fig.\ \ref{Fig:issevol}), which
we will see is also connected to the
previously noted color change inside the hexagon, from
blue/green to a more golden color.  A much smaller effect was seen in the
small eye region, where the color remained blue/green over the entire
period. Two factors that might contribute to the different behavior of
the eye are (1) the region closest to the pole is the least exposed to
sunlight and thus haze production would be lower and (2) as suggested
by the sharp boundary of the eye region, there may be a dynamical effect
similar to what is seen at the eye of an earthly hurricane, namely a
downwelling that keeps the region relatively free of upper level
aerosols. Such a downwelling has also been invoked, e.g. by \cite{Fletcher2008}, to explain
warm temperatures and depleted \pht gas.  A more accurate description of the color changes is
provided by VIMS spectral observations.

\section{VIMS Observations}

\subsection{VIMS instrument characteristics and data reduction}

The Visual and Infrared Mapping Spectrometer (VIMS) includes two
separate spectrometers. As described by \cite{Brown2004}, the so called
visual spectrometer covers the 0.35-1.05 \mum spectral range using
 96 bands with a 7.3-nm spacing, and the
near-IR spectrometer covers the range of 0.85-5.12 \mum using 256
contiguous channels sampling the spectrum at intervals of
approximately 0.016 \mumx. The instantaneous field of view of each
pixel pair, as combined in the observations we used, covers 0.5 $\times$
0.5 milliradians, and a typical frame has dimensions of 64 pixels by
64 pixels.  Due to effects of order-sorting filter joints, we avoided
comparisons between models and observations at the most strongly
affected regions near 1.64 \mum and 3.85 \mumx, but found that the
effects of the 2.98-\mum joint produced only a relatively small local
 depression of about 10\%, which was not a significant issue \citep{Sro2013gws}.

The VIMS data sets we selected (identified in
Table\ \ref{Tbl:obslist2}) provide an efficient combination of
spatial coverage from the north pole out to just beyond the hexagon boundary,
and at the same time sufficient spatial resolution to resolve key
aerosol features. These data were reduced using the USGS ISIS3
\citep{Anderson2004} vimscal program, which was derived from the
software provided by the VIMS team (and is available on PDS archive
volumes). We used the RC19 calibration as described by
\cite{Clark2018cal}, with the exception that we had to correct serious
errors in how the RC19 visual spectrometer calibration was implemented
in the ISIS3 software. These corrections are described later in
Sec.\ \ref{Sec:caldiff}.  Conversion to I/F (reflectivity relative to
a normally illuminated Lambertian reflector) used the solar spectrum
packaged with the ISIS3 and PDS-supplied software, which is now based
on a more modern solar reference described by \cite{Thompson2015sun}.
Computation of planet coordinates and illumination parameters for each
pixel of a VIMS cube utilized kernels supplied by JPL NAIF system and
SPICELIB software \citep{Acton1996}.

\begin{table*}[!htb]\centering
\caption{Observing conditions for VIMS VIS and IR data cubes we used to constrain models.}
\setlength\tabcolsep{3pt}
\begin{tabular}{ l c c c c c r}
                             &                   &     UT Date  &        Start/End &   Pixel &    Phase & Fig. \\[-0in]
         Observation ID$^1$ &     Cube Version &   {\footnotesize YYYY-MM-DD} &   Time & size &
   angle & Ref. \\
\hline\\[-0.05in]
{\small 192SA\_NPOLVORT001} & {\small V1749893170\_1} & 2013-06-14 & \parbox{0.9in}{08:30:42.07 08:39:22.16}  & 348 km & 46.06\deg & \ref{Fig:vimsvis2013}, \ref{Fig:vimsnir2013}\\[0.1in]
{\small 249SA\_NPOLMOV001} & {\small V1858161491\_1} & 2016-11-18 & \parbox{0.9in}{10:51:12.07 10:58:25.56}  & 312 km & 78.95\deg &\ref{Fig:vimsvis2016}, \ref{Fig:vimsnir2016}\\[0.1in]
\hline
\end{tabular}\label{Tbl:obslist2}
\parbox{5in}{$^1$The full observation ID has a prefix VIMS\_ and suffix \_PRIME for row 2.
Cube IDs have suffixes \_1\_ir and \_1\_vis to distinguish near-IR and visual cubes.  \par
}
\end{table*}

\subsection{VIMS visual channel observations}

\begin{figure*}[!hbt]
\includegraphics[width=6.2in]{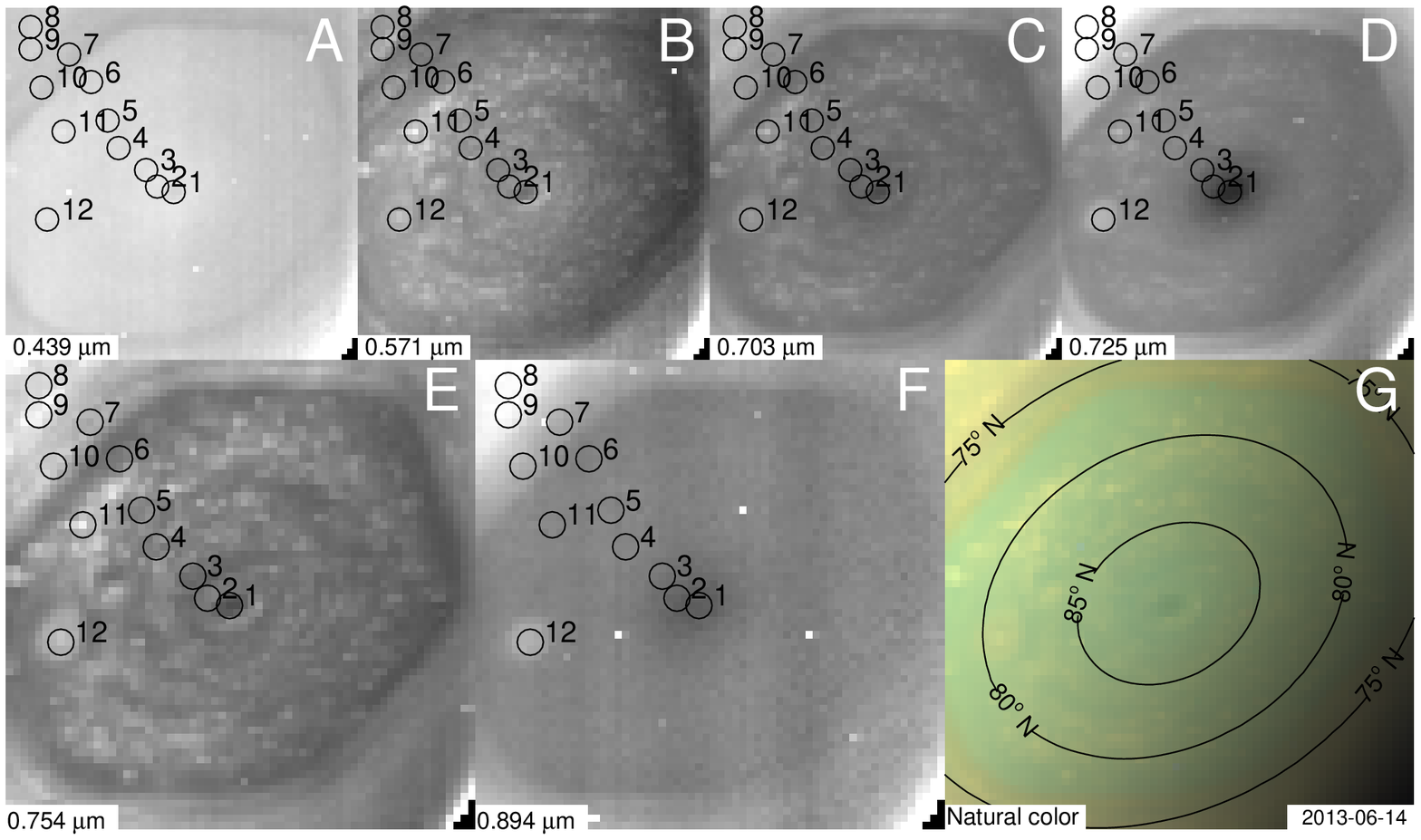}
\includegraphics[width=6.2in]{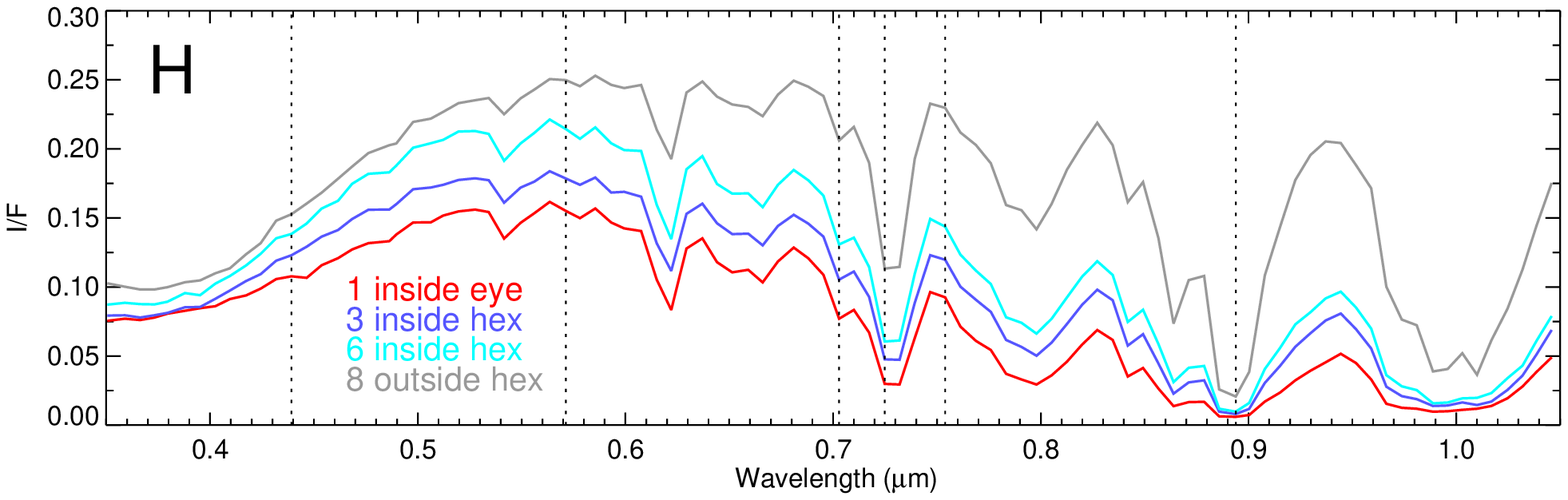}
\caption{VIMS 14 June 2013 visual channel images (A-F), a natural
  color image (G) computed from the spectrum at each pixel, and
  selected visual spectra (H) from locations encircled and numbered in
  images A-F.  The spectral observations were destriped using smoothed
  7-column averages to reduce column-to-column offset changes, as
  described by \cite{Adriani2007}, except for $\lambda < 0.46$ \mum
  for which we used a median instead of an average. The vertical
  dashed lines in H indicate wavelengths of images in panels
  A-F. Numbered circles mark locations from which modeled spectra were
  extracted.}
\label{Fig:vimsvis2013}
\end{figure*}

\begin{figure*}[!ht]
\includegraphics[width=6.2in]{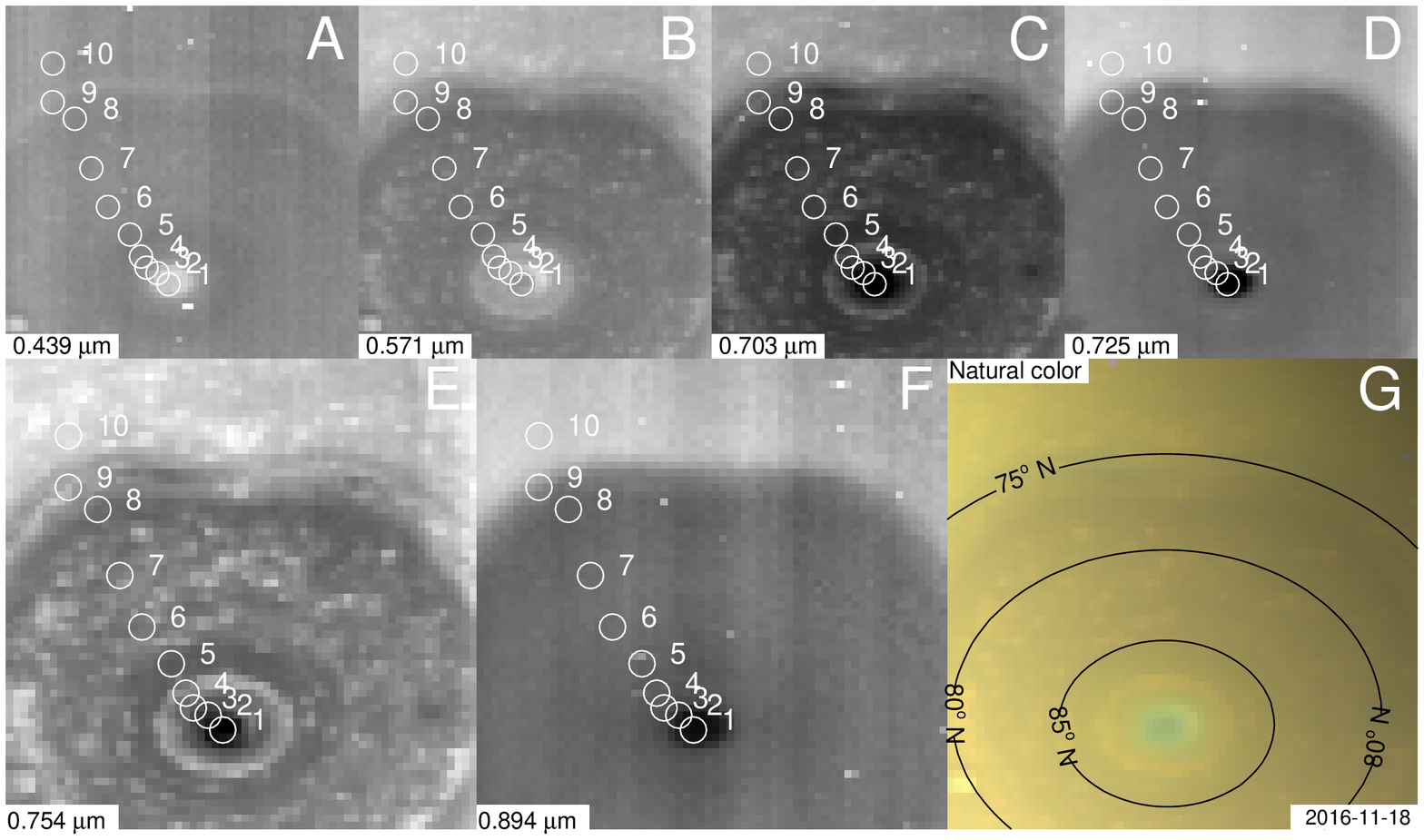}
\includegraphics[width=6.2in]{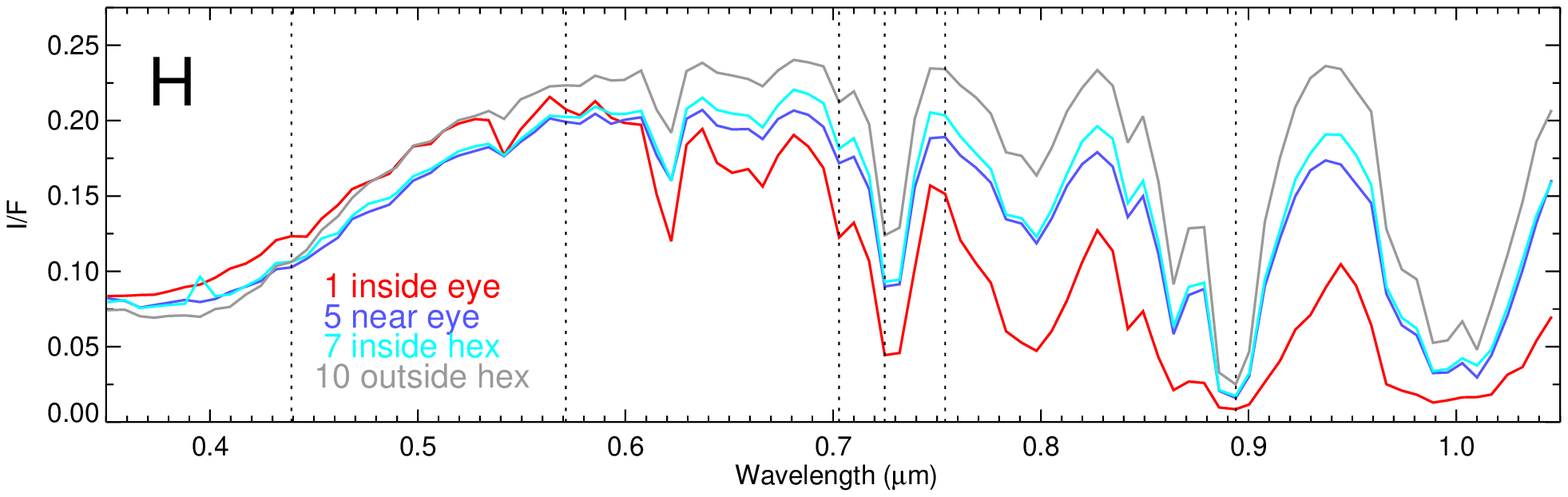}
\caption{As in Fig.\ \ref{Fig:vimsvis2013} except for 2016 November 18
  VIMS visual channel observations. From the spectra we see that the
continuum reflectivities of the region inside the hexagon do not decline
much with wavelength (except in the eye) compared to the 2013 results
in Fig.\ \ref{Fig:vimsvis2013}. This is due to the presence of significant
scattering by aerosol particles.}
\label{Fig:vimsvis2016}
\end{figure*}

VIMS visual channel observations of Saturn's north polar region in
June 2013 and November 2016 are shown in Figs.\ \ref{Fig:vimsvis2013}
and \ref{Fig:vimsvis2016} respectively. These display spatial structure
and spectral variations consistent with the
ISS bandpass filter images displayed in Fig.\ \ref{Fig:issevol} and
Fig.\ \ref{Fig:isstriple}. For both VIMS observations we used the full
spectrum to compute tristimulus values (the RGB primary multipliers) using CIE1931 color matching
functions \citep{CIE1932} and the natural color using the IEC61966-2-1 standard sRGB
color space with a D65 white reference \citep{IEC99}. This provides greater
accuracy than is possible with bandpass filters, although color
differences between our results and the press release results in Fig.\ \ref{Fig:issevol} are not
very significant.  Our VIMS-based color images are
shown in panel G of both figures, using a linear stretch from zero,
without attempting to correct for illumination changes across the
images.  These confirm that the region
transformed from a blue/green color in 2013 to a gold color near the
end of 2016, except for the eye, which remained relatively
unchanged in color and somewhat more blue than the rest of the
hexagon interior in 2013.  

All the spectral samples show a significant drop
in I/F at wavelengths below 570 nm, by a factor of two or more by 400
nm.  That decline is due to particulate absorption, probably in a
photochemical haze, and gives Saturn its overall tan color.  What
compound is responsible for that absorption is unknown. Possible
candidates are discussed by \cite{West2009satbook} and refer
to the same possible candidates discussed for Jupiter's red chromophore
by \cite{West1986}. We will later show that the red chromophore needed
for Saturn is likely significantly different from that which works
well on Jupiter.

The 2013 spectral samples taken inside the hexagon (but outside the eye) have a similar shape
characterized by a strong downward gradient from green to red.  This is
qualitatively consistent with scattering by relatively small particles (a combination
of Rayleigh scattering and small-particle haze scattering).  Their declining
peak I/F as the pole is approached is mainly due to declining illumination as
the terminator is approached (near the lower right).  In panel A, for which we used
a Minnaert function to approximately correct for changes in viewing and illumination
across the image, we see that the entire region inside the hexagon is of similar brightness
at short wavelengths, as we saw in the ISS image centered at a similar wavelength
in panel A of Fig.\ \ref{Fig:isstriple}.  

The spectral samples taken from the 2016 VIMS visual observations
(Fig.\ \ref{Fig:vimsvis2016}) present a different picture in which the
spectra inside the hexagon look more like the spectrum taken just
outside the hexagon, except for the eye spectrum, which has a strong
downward gradient from green to red that accounts for the blue/green
color of the eye in panel G. Due to relatively constant illumination
for the chosen spectral samples, those spectra interior to the hexagon
are much more similar than the corresponding samples from 2013. This
is qualitatively consistent with increased aerosol scattering by
relatively larger particles, presumably generated by photochemistry,
although the lack of any obvious latitudinal gradient contradicts the
idea that haze production might be proportional to the local solar
incident UV flux. It is consistent with haze production depending on the daily
mean actinic flux, which is less dependent on the solar zenith angle.

The discrete bright features seen at continuum wavelengths are
relatively deep given their low contrast at 725 nm and their
disappearance at 894 nm. We will later show that they are indeed
ammonia ice clouds seen underneath the translucent overlying haze.
The main exception is the bright spot at the left edge of the 2013
images, which has a spectrum (not shown) that is similar to that of
the clouds outside the hexagon, and extends high enough to be seen at
894 nm.  This feature is not as apparent in color images, especially
without a hard stretch.  It can be seen in the 2013 mosaic of
Fig.\ \ref{Fig:issevol}, about 80\% of the distance from the pole to
the hexagon boundary along a line 45\deg CCW from vertical.  The
feature does not appear to be present in 2016 images.

\subsection{VIMS near-IR north polar observations}

\subsubsection{Near-IR overview}

The VIMS near-IR views of Saturn's north polar region in late 2013 and
in 2016 are displayed in Figs.\ \ref{Fig:vimsnir2013} and
\ref{Fig:vimsnir2016} respectively, where we show images at six sample
wavelengths and a color composite in which ammonia ice clouds attain a
magenta color (the composite assigns R and B to continuum wavelengths
and G to a wavelength at which ammonia ice is a strong absorber, so
that where ammonia ice is present there is a lack of green, which
creates magenta as the sum of red and blue).  Panels A, C, and E show
continuum wavelength 1.58 \mum and pseudo continuum wavelengths 2.73
\mum and 4.08 \mumx, respectively.  As evident from the penetration
depth profile displayed in Fig.\ \ref{Fig:nirpendepth}, these are at
wavelengths of local minima in gas absorption, allowing the deeper
views of aerosol scattering contributions.  A deep penetration is also
available at 5.05 \mum (shown in panel F), but Saturn's thermal
emission at this wavelength greatly exceeds scattered sunlight, and is
controlled by the blocking effect of deep aerosol layers. The bright
regions in panel F are where this blocking effect is greatly reduced
by a lack of deep absorbing aerosols.  Perhaps surprisingly, the eye
is not a region of unusually high emission in either 2013 or 2016, or
in 2008 \citep{Baines2009cyclone}, indicating that, at least at these
times, it is not clear of aerosols down to the 5 bar level. Thus, if
descending motion is responsible for the relative clarity of the upper
troposphere within the eye, that descent does not continue much below
the several bar level.

\subsubsection{5-\mum emission features}

\begin{figure*}[!ht]
\includegraphics[width=6.2in]{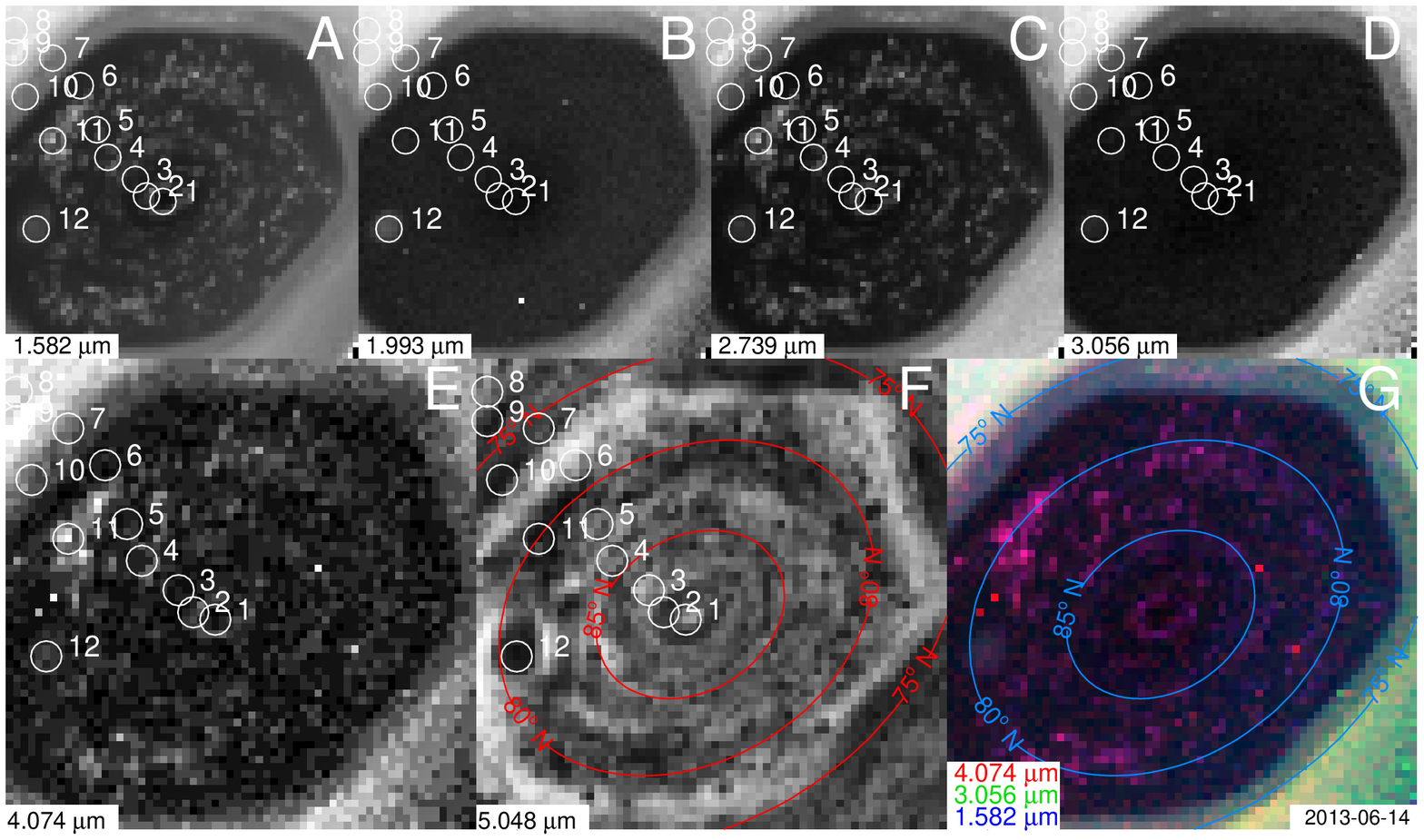}
\includegraphics[width=6.2in]{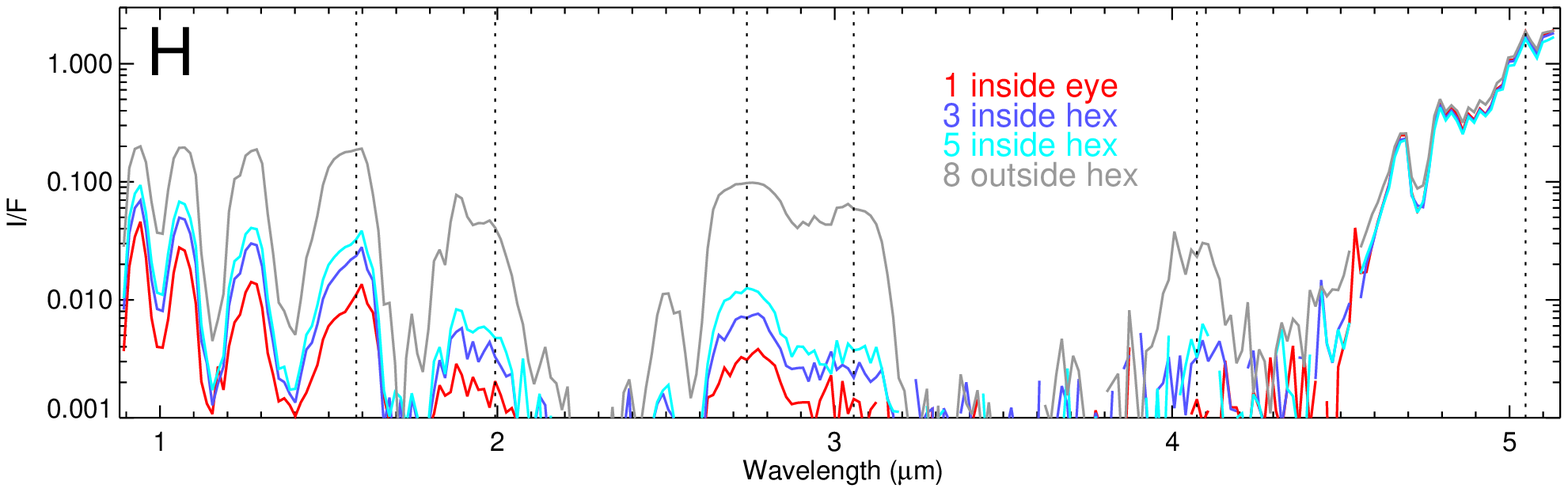}
\caption{VIMS 14 June 2013 near-IR channel images (A-F), a pseudo
  color composite image (G), and plots of selected spectra (H) from
  locations circled and numbered in panels A-F. The color composite is
  constructed so that discrete clouds with ammonia ice absorption at
  3.05 \mum appear with a magenta color. Panel F is an image dominated
  by Saturn's thermal emission at 5.1 \mumx, which accounts for what
  might appear to be an impossibly large I/F.  Wavelengths at which
  images are displayed are indicated in the spectral plot by vertical
  dotted lines as well as labels within the images.  Note the very low
  reflectivity inside the hexagon boundary, compared to what was seen
  in late 2016 (Fig.\ \ref{Fig:vimsnir2016}).}
\label{Fig:vimsnir2013}
\end{figure*}

\begin{figure*}[!hbt]
\includegraphics[width=6.2in]{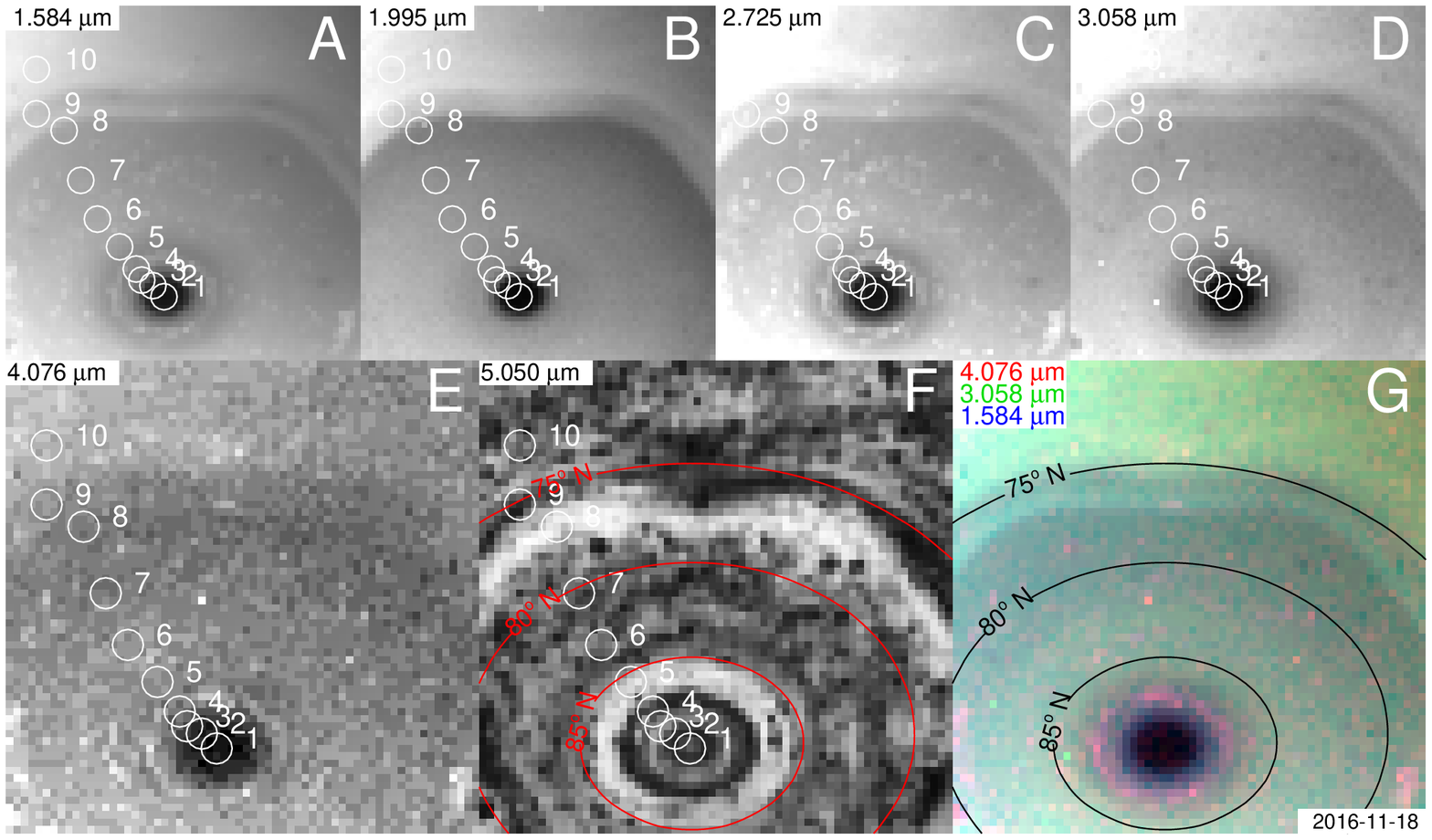}
\includegraphics[width=6.2in]{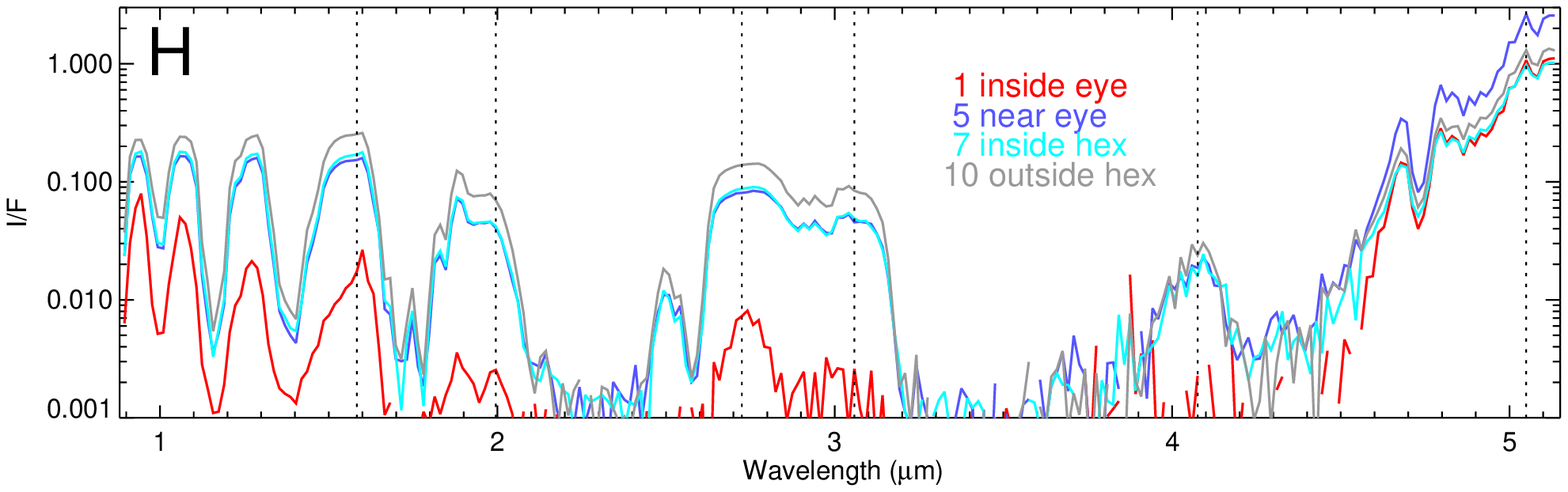}
\caption{As in Fig.\ \ref{Fig:vimsnir2013}, except for 2016 November 18
observations by VIMS.}
\label{Fig:vimsnir2016}
\end{figure*}

\begin{figure*}[!ht]
\includegraphics[width=6.2in]{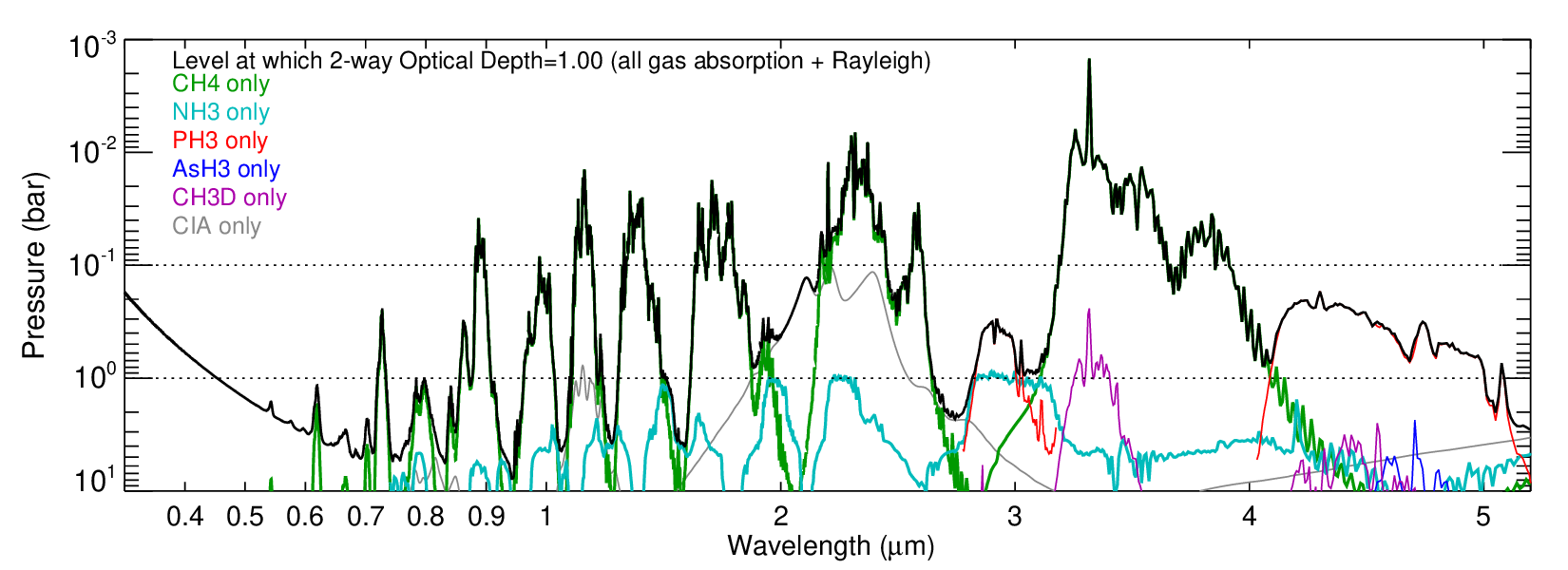}
\caption{Vertical penetration depth in a clear atmosphere for various single gases (colored)
and combined gases (black), computed for gas mixing ratio profiles displayed in Fig.\ \ref{Fig:gasmix}.}
\label{Fig:nirpendepth}
\end{figure*}

There are also strong patterns in the 5-\mum emission that are either
very muted or undetectable in the upper tropospheric clouds seen in
panels A, C, and E of Figs. \ref{Fig:vimsnir2013} and
\ref{Fig:vimsnir2016}.  Perhaps the most prominent of these is the
dramatic bright ring of emission seen in 2016
(Fig.\ \ref{Fig:vimsnir2016}F), which intersects numbered circle 5
(near 86\degx N).  This region is only slightly darker in panels A and
C, and unremarkable in panel E.  A similar example can be found by
comparing locations 7 and 8.  These locations appear equally bright in images A-E,
but show a large difference in 5-\mum emission levels in panel F.  On
the other hand, there are cases where there is a strong correlation
between features at these different wavelengths.  A prominent example
is between the dark ring of low emission in F (near 87\degx N,
intersecting location 4) and the bright ring of clouds in panels A, C,
and E, which appear as magenta (ammonia ice-signature) clouds in G,
both of which are indicators of increased cloud optical depth.
However, the brightness modulations at 5 \mum appear to be occurring
in a deeper layer than those in the upper troposphere, as also
inferred from a quantitative analysis of Saturn's south polar cloud
structure \citep{Sro2020spole}. This also seems to be true across Saturn
\citep{Baines2005,Baines2009cyclone,Fletcher2011Sci}.

\subsubsection{Near-IR changes between 2013 and 2016}

The near-IR views of the north polar region provide a much
different and more dramatic picture of the aerosol changes that occurred
between 2013 and 2016 (Figs. \ref{Fig:vimsnir2013} and \ref{Fig:vimsnir2016}).
At near-IR wavelengths the inside of the hexagon was much darker in 2013 than in 2016.  The
lack of aerosol scattering in 2013 is much more dramatic in the
near-IR than in the visible because in the latter case the lack of aerosol
contributions is seen against a bright background of largely conservative
Rayleigh scattering, while in the near-IR the background is much
darker due to gaseous absorption, primarily by methane. In 2013 the
near-IR spectrum of the eye showed a falloff with increasing
wavelength that is similar to that seen in 2016, but even a bit
steeper.  The greatest spectral difference between these two times
is in spectra from the interior of the hexagon.
In 2013 the mid-hexagon spectra are much more similar to the eye
spectrum than to the spectrum outside the hexagon.  They also show a
steep falloff of I/F with increasing wavelength, though not quite as
steep as for the eye spectrum. 

Different results are seen in 2016 (Fig.\ \ref{Fig:vimsnir2016}). In this case spectral samples
 between the eye and 
the outer hexagon boundary (e.g. 5 and 7) are very similar to each other (ignoring the
thermal emission difference beyond 4.6 \mum), and in spectral shape
very similar to the spectral sample (10) from just outside the hexagon. They are just slightly
lower in I/F, partly a result of reduced illumination. Their lack of a strong decline in
continuum I/F with wavelength is indicative of relatively large aerosol particles compared to
what is normally seen in stratospheric hazes. The spectrum from the eye is dramatically
different: continuum I/F values decline rapidly with wavelength, disappearing into
the noise at 4.08 \mumx. What little aerosol scattering is produced inside most of the eye
appears to arise from small particles, and there the optical depth of aerosol contributions
must be very low. However, analysis of close-up observations of the eye by \cite{Baines2018GeoRL}
have found exceptionally large particles in a bright discrete ammonia-signature
feature in this region.

There are some other noteworthy differences in the morphology and
discrete features between 2013 and 2016. The 14 June 2013 observations
captured an unusual discrete feature that is bright in the 1.993-\mum
image in panel B, but quite dark in the 5.1-\mum image in panel F. It
is encircled and labeled as feature 12 here and in Fig.\ \ref{Fig:vimsvis2013}, where
it also appears as a bright feature in panels C-F.
It is located at 80.7\degx N planetocentric and 82.4\deg E.
 There is no similar feature observed on
18 November 2016.  Note that the 5-\mum image from 2013 does not
contain the bright emission ring at 86\degx N that is seen in the 2016
image.

To understand these interesting spectral changes in the north polar region in terms
of the evolution of cloud structure and composition in the north polar region, we need to
carry out radiation transfer modeling.  But a prelude to that is to find a way to
combine both visual and near-IR spectral bands into a single co-located spectrum
that can provide the maximum possible constraints on the vertical structure of
aerosols and gases.

\subsection{Combining visual and near-IR spectral observations}

We initially tried to constrain the vertical cloud structure of the
polar regions from just the near-IR VIMS observations, but there are
significant advantages in carrying out a combined fit to both spectral
ranges simultaneously.  While the near-IR region offers the advantages
of a variety of methane band strengths,  accurate line-by-line correlated-k
models covering much of the range, sensitivity to greater depths and to
thermal emission in the 5-\mum region, sensitivity to trace gases, and
a wavelength range insensitive to Rayleigh scattering in favor of aerosol scattering,
there are also significant advantages to the visual spectral range.
These include much greater sensitivity to very small particles, sensitivity
to the chromophore that provides most of Saturn's color, and to its
vertical location, as well as sensitivity to Rayleigh scattering at
wavelengths where methane absorption is negligible.  However, there
are two problems in combining visual and near-IR spectral observations:
differences in spatial sampling and potential differences in calibration.

\subsubsection{Accounting for spatial sampling differences}
 The first problem
with combined analysis is that the two spectrometers have slightly
different fields of view and are not precisely bore-sighted. According
to \cite{Brown2004} the near-IR FOV is 0.495$\pm$0.003 mrad, while
the visual FOV is 0.506$\pm$0.003 mrad. The bore-sight misalignment
pre-launch was less than 0.3 pixels, but in 1999 (in flight) the IR-VIS offset was found to be
be -1 pixels in the X direction and +2 pixels in Y. And in 2001 it was
measured to be -1 and 0 respectively \citep{Brown2004}.  Using the spectral region
where the VIMS-VIS and VIMS-IR bands overlap, we estimate, for the 2013 and 2016 cubes
we analyzed, that the VIS - IR offset
is closer to 2 pixels in X and 1.5 in Y. 

\begin{figure*}[!htb]\centering
\includegraphics[width=6.2in]{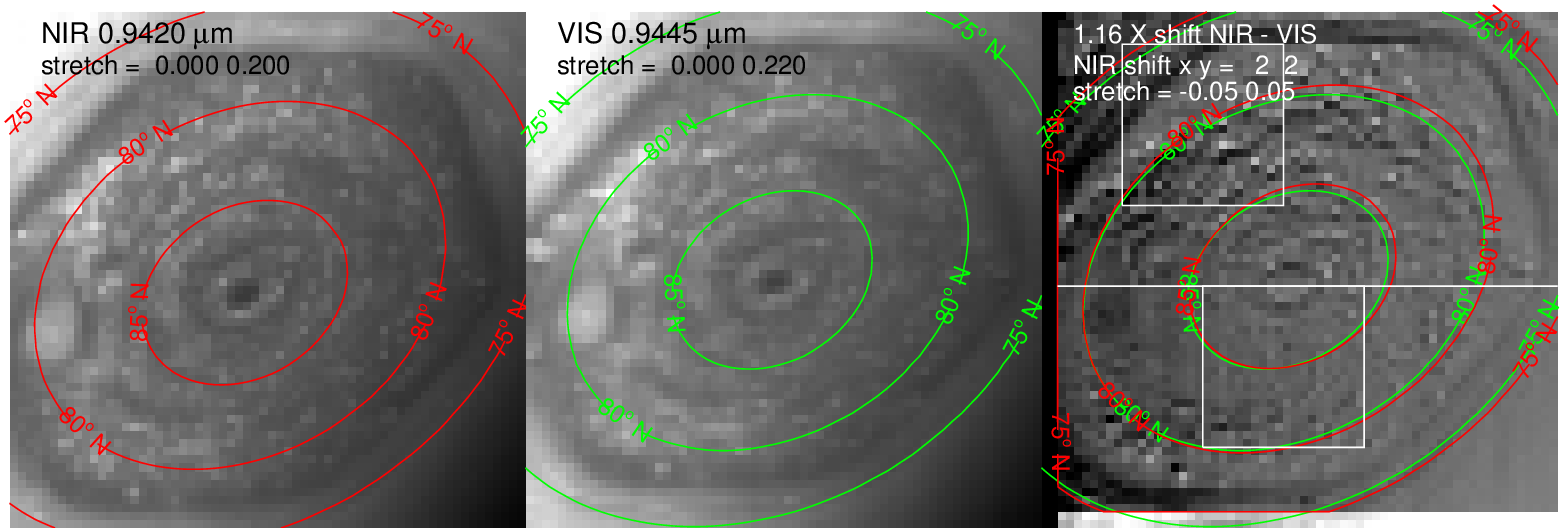}
\includegraphics[width=6.2in]{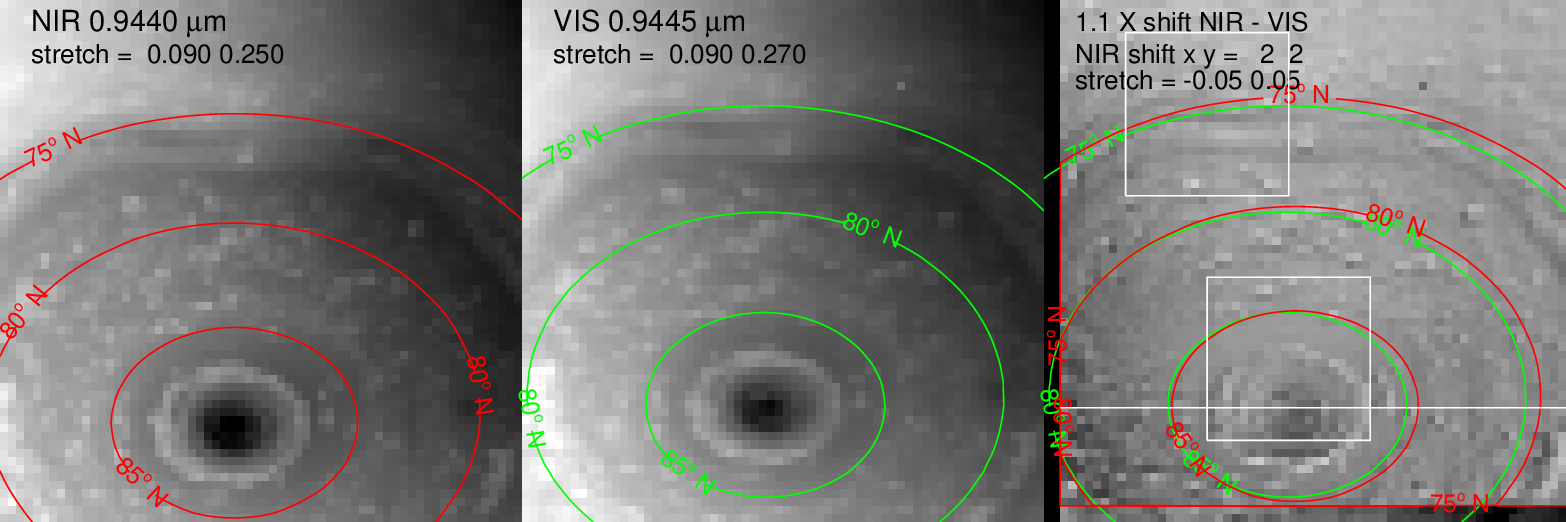}
\caption{Near-IR and Visual images from June 2013 (top) and November 2016 (bottom), showing differences
  in scale and bore-sight offset, with the help of the difference image
  at the right, where the near IR image is shifted 2 pixels to the
  right and 2 pixels upward and multiplied by 1.16 (top) or 1.1 (bottom) before subtracting
  the visual image, the multiplication factor accounting for slight
  differences in wavelength and/or calibration. The latitude contours in red
  (from the near-IR image navigation) are also shifted to illustrate scale
  differences and to compare with the unshifted visible navigation in green.  Note that in regions where
  navigations align (with this offset), the difference image is spatially
smooth and near zero (gray).  The white boxes are averaging regions referred to
in Fig. 11}
\label{Fig:visnirdiff}
\end{figure*}

Fig.\ \ref{Fig:visnirdiff} illustrates the effects of different image
scales and bore-sighting offsets.  However, it also shows that when
sampled at the same planet locations, the I/F differences
are small.  They are also small in regions of spatial homogeneity.
Thus our approach to spatial sampling is to take pixels that are
closest to the same position on the planet and avoid regions of large
pixel-to-pixel variations.  The pixel closest to the north pole in
2013 visual images of Fig.\ \ref{Fig:vimsvis2013} is at x = 29 and y =
13, which is centered at 89.62\degx N and 322.43\degx E, while in the
near-IR images of Fig.\ \ref{Fig:vimsnir2013} the pixel closest to the
pole is at x = 28 and y = 11, centered at 89.72\degx N and 318.39\degx
E.  Near the pole, a displacement of 1 pixel away from the pole
corresponds to a difference in latitude of 0.37\degx.  This makes it
difficult to obtain accurate combined spectra for small discrete
features, but less of a problem for regions that are locally
homogeneous or slowly varying from pixel to pixel, which are the
regions we try to sample.

\subsubsection{Accounting for calibration differences}\label{Sec:caldiff}

Potential differences in radiometric calibrations were investigated by
comparing visual and near-IR spectra in the spectral overlap region
between 0.88 \mum and 1.05 \mumx.  However, it is first necessary to
correct the latest RC19 visual channel calibration because the ISIS3
VIMSCAL implementation failed to include a multiplier that is
essentially the inverse responsivity of the visual detector array.
The multiplier that was included for both visual and near-IR
instruments is displayed in Fig. 7 of the final VIMS calibration
report \citep{Clark2018cal}. The IR portion includes the inverse
responsivity factor, but the visual portion (the first 96 values) did
not. Instead, it includes the ratio of RC19 to RC17 inverse
throughputs (RC17 and RC19 refer to earlier and later radiometric
calibrations of the VIMS instrument), which appears to be mainly just
the ratio of the new solar reference \citep{Thompson2015sun} to the
prior solar reference \citep{Thekaekara1974} that was used in the RC17
calibration. Fortunately, the correct multiplier accounting for
detector responsivity was available.  Inserting that multiplier into
the ISIS3 calibration code yields visual I/F spectra that are very
similar to those obtained from the RC17 calibration, except somewhat
smoother and smaller by a factor of about 0.95 over most of the
spectral band.  The radiance (termed specific energy) cubes were
changed more significantly than the I/F cubes, roughly in accord with
the difference between the old and new solar references.  The new
solar reference was also used in the near-IR VIMS spectrometer
calibration.  Although there is a substantial difference in the
near-IR between the two solar references, this did not lead to a large
change in I/F spectra because of other changes in the calibration.
However, it did lead to 10-20\% decreases in the radiances near 1 \mum
and near 3 \mum respectively.

\begin{figure*}[!htb]\centering
\includegraphics[width=3.in]{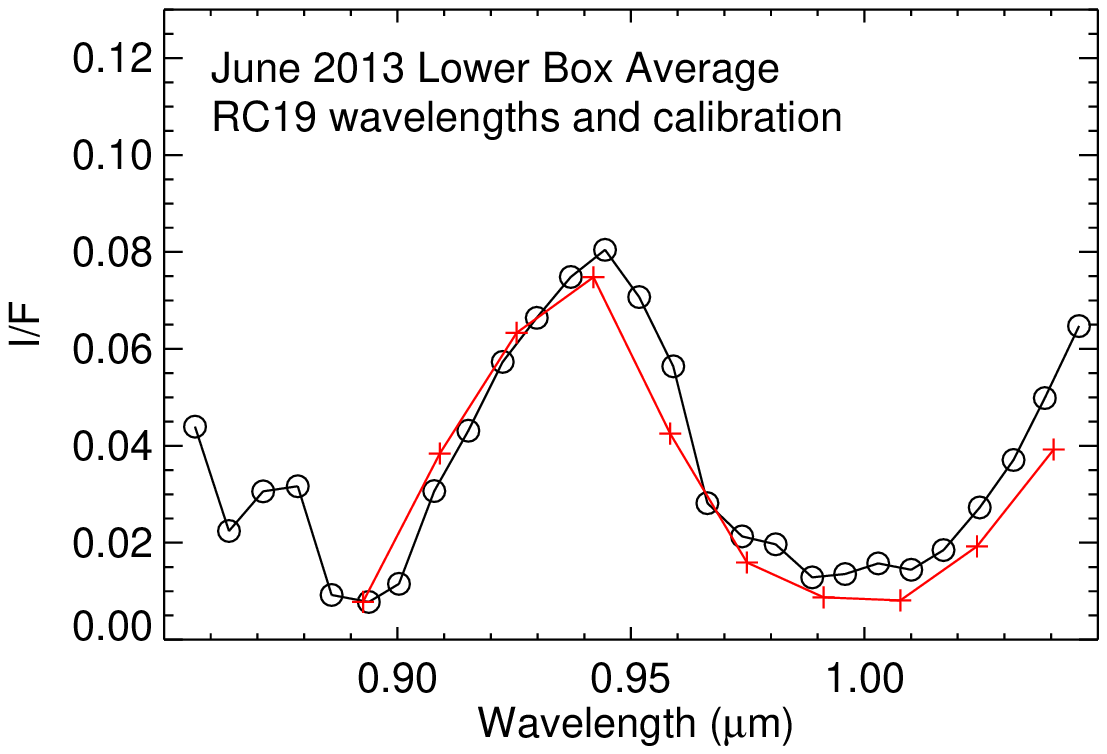}
\includegraphics[width=3.in]{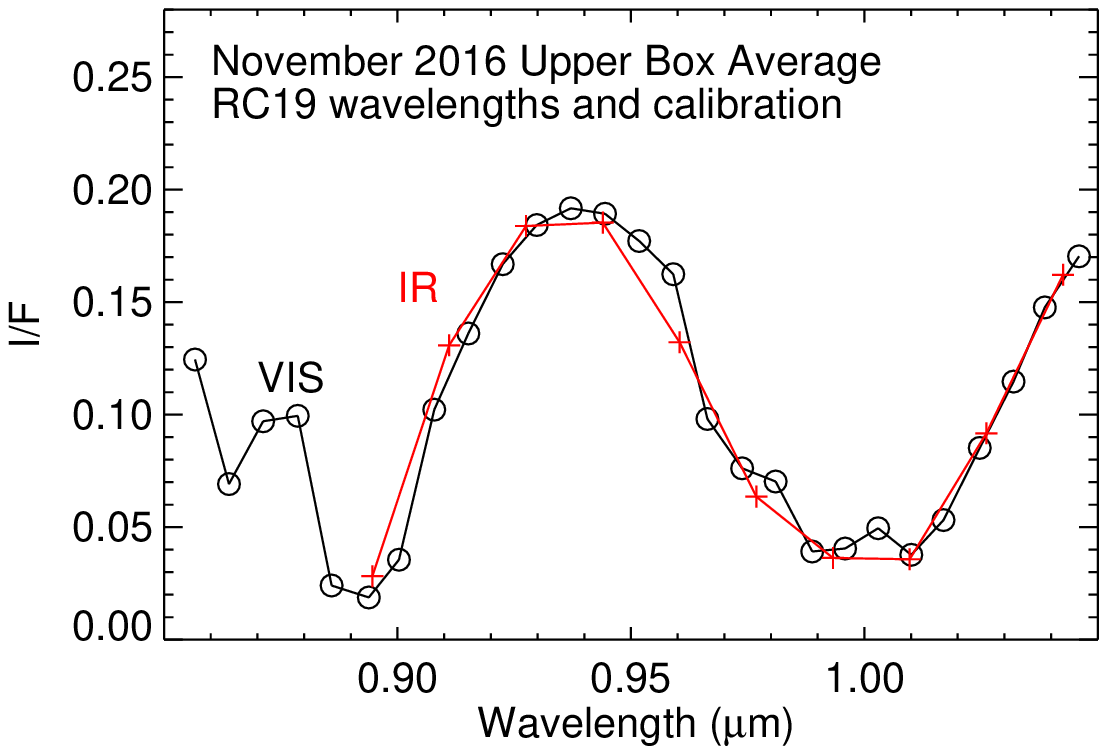}
\caption{Comparison of VIMS near-IR (red, +) and visual (black, o) 
spectra in their overlap region for spectra in 2013 (averaged over  the lower box in the upper right
panel of Fig.\ \ref{Fig:visnirdiff}) and in 2016 (averaged over the upper box
in the lower right panel of Fig.\ \ref{Fig:visnirdiff}).}
\label{Fig:visnirspec}
\end{figure*}

With the RC19 visual channel calibration corrected, we can then
 make meaningful comparisons of visual and near-IR I/F spectra in
the overlap region from 980 nm to 1050 nm.
Two comparisons are shown in Fig.\ \ref{Fig:visnirspec}, in the left
panel for a 2013 average (over the lower box in the upper right panel
of Fig.\ \ref{Fig:visnirdiff}) and in the right panel for a 2016
average (over the upper box in the lower right panel of
Fig.\ \ref{Fig:visnirdiff}).  Because of response falloff at the edges
of their respective spectral ranges, the near IR results (red, +) are least
reliable at the shortest wavelength and the visual results (open
circles) are least reliable at the longest wavelengths. Some of the
difference between the two spectra are due to lower resolution of the
near-IR spectrometer. Also note that the near-IR wavelengths have
shifted relative to the visual channels between 2013 and 2016.  The
agreement between the 2016 spectra is remarkably good, and provides
little evidence for a discrepancy.  The 2013 comparison is less
comforting but does not provide sufficient justification for changing
either of the calibrations.  To create a single spectrum that joins
the visual and near-IR into a single spectrum, we chose to join the
spectra in the middle of the overlap region, using the visual spectrum
for wavelengths less than 970 nm and the IR spectrum for wavelengths
longer than 970 nm.  After completion of model fits, we found that
a joint at 940 nm instead of 970 nm would have improved fit quality somewhat.

\subsubsection{Noise and offset estimates}

\begin{figure}[!htb]\centering
\includegraphics[width=3.2in]{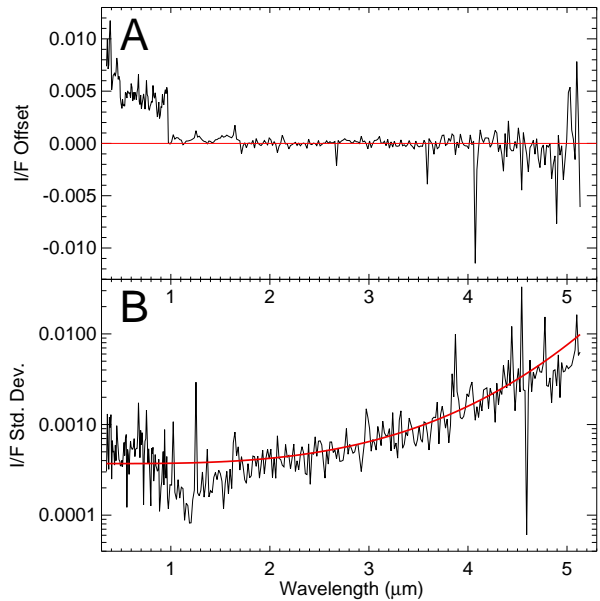}
\caption{Mean (A) and standard deviation (B) of
nine VIMS space view spectra.  The red curve in the bottom
panel displays the adopted noise model defined in the text.}\label{Fig:space}
\end{figure}

\cite{Sro2010vims} investigated near-IR VIMS noise levels by comparing
observations of Jupiter taken just 2 minutes apart, and found standard
deviations of measurements of the same location on Jupiter were only
of the order of 0.1-0.2\% for moderate to high signal levels.  That
noise level is well below other sources of uncertainty.  Their crude
I/F noise estimate for very low signal levels was $\sim$5$\times
10^{-4}$, which is worth accounting for.  We tried to make new noise
level estimates in two ways, first by measuring the standard deviation of adjacent
measurements in relatively spatially smooth regions. Because spatial
smoothness is also a function of wavelength, this was not effective at
all wavelengths.  The most reliable results were obtained at
wavelengths for which the measured I/F was not significantly larger
than the noise level.  Better results were obtained from cubes
containing a view of space.  The standard deviation of nine space
measurements (also in I/F units) roughly followed an analytical model for which SD(space) = $\exp(a
+ c\times(\lambda -\lambda_0)^3)$, with $a=-7.6$, and $c = 0.03$, for
$\lambda_0 = 0.35$ \mumx.  This model is displayed in comparison with
the space observations in Fig.\ \ref{Fig:space}.  This is also shown
as the dashed component in Fig. 13.  Although there are
large discrepancies at wavelengths less than 1.3 \mumx, these are not
significant compared to the signal levels in this part of the
spectrum.

\begin{figure*}[!ht]\centering
\includegraphics[width=5.5in]{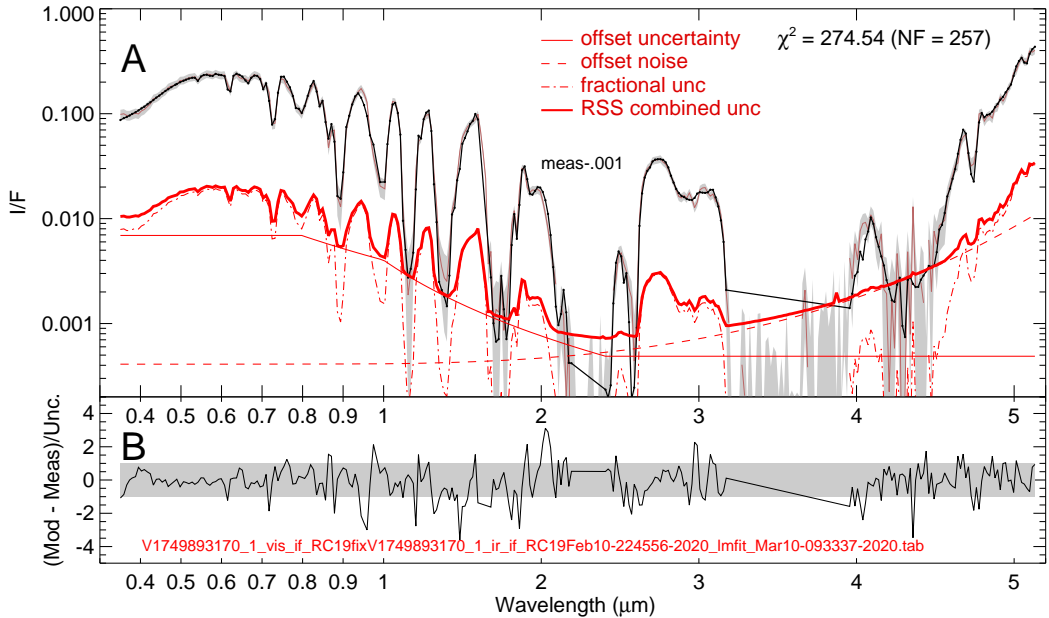}
\caption{{\bf A:} The components of our error model in comparison with
  a VIMS combined visual and near-IR spectrum of Saturn (gray shaded
  band) and a model spectrum (solid black).  See text for
explanation of these components. {\bf B:} the model -
  measured difference divided by the model uncertainty estimate, with
  the gray shaded region indicating the one-$\sigma$ uncertainty band.}
\label{Fig:unc}
\end{figure*}

The I/F offset displayed in the top panel of Fig.\ \ref{Fig:space} is
apparently not valid for views of the planetary disk, suggesting that
scattered light within the instrument might be a factor.  Subtracting
this offset from the spectra did not result in improved fit quality,
and it is apparent from other considerations that there are offsets
present in VIMS planetary spectra.  \cite{Sro2010vims} compared VIMS
near-IR spectra of Jupiter to groundbased observations and HST NICMOS
observations that indicated offsets of the order of 0.001 in I/F units
at wavelengths near 2.4 \mumx.  We also found indications of similar
offsets, indicated by comparing deep minima in model spectra at 2.58
\mum with VIMS measurements of much reduced absorption depths that
could not be matched by any model we tried that was consistent with
other parts of the spectrum.  Subtracting an offset of near 0.001 in I/F
from the measured VIMS spectra substantially eliminated this
discrepancy, as well as improved fits in other low I/F regions of the
spectrum.  However, the exact level of this offset is uncertain, and
that uncertainty grows at shorter wavelengths, especially at visual channel
wavelengths, where striping effects, even though largely corrected,
do leave a residual uncertainty.  The region between 0.85 and 2.5 \mum
is a region that produces significant fitting errors that contribute
a large fraction to \chisq when wavelength independent error models
are assumed, and much of that is in the region of I/F minima, where
offset uncertainties might be a significant factor.  It is also possible
that stray light  and errors in modeling methane absorption
are factors. To include all these effects in a way that balances the
\chisq contributions more equally in wavelength, we used a
wavelength dependent offset uncertainty model with a maximum of SDmax
for $\lambda < 0.8$ \mum that is 6\% of the mean I/F of the two
spectral peaks between 1.2 and 1.4 \mumx. For  $\lambda > 0.8$ \mumx,
it declines as $(\lambda/0.8\ \mu\mathrm{m})^{-2.4}$ up to 
 $\lambda = 2.4$ \mumx, beyond which it remains constant at that
value. This second component (offset uncertainty) is shown
by a thin solid line in Fig. 13.  To these two components we added a fractional uncertainty
of 0.08 times the observed spectral I/F. This third component
is a crude estimate of the combined effects of calibration and modeling
errors.  These multiple uncertainty contributions
are illustrated in Fig.\ \ref{Fig:unc}.

\subsection{VIMS combined visual and IR spectra}

\begin{figure*}[!htb]\centering
\includegraphics[width=2.05in]{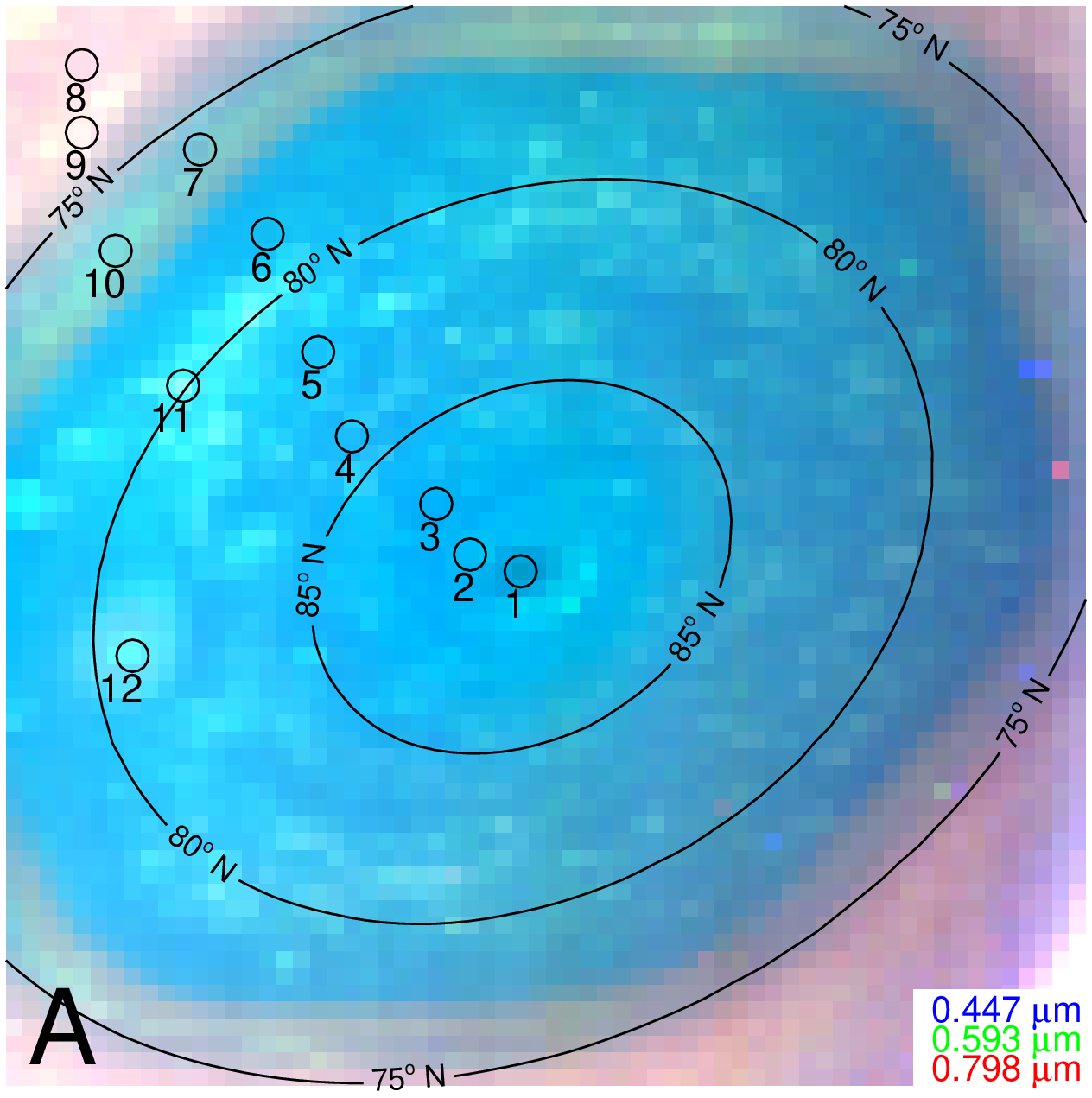}  
\includegraphics[width=2.05in]{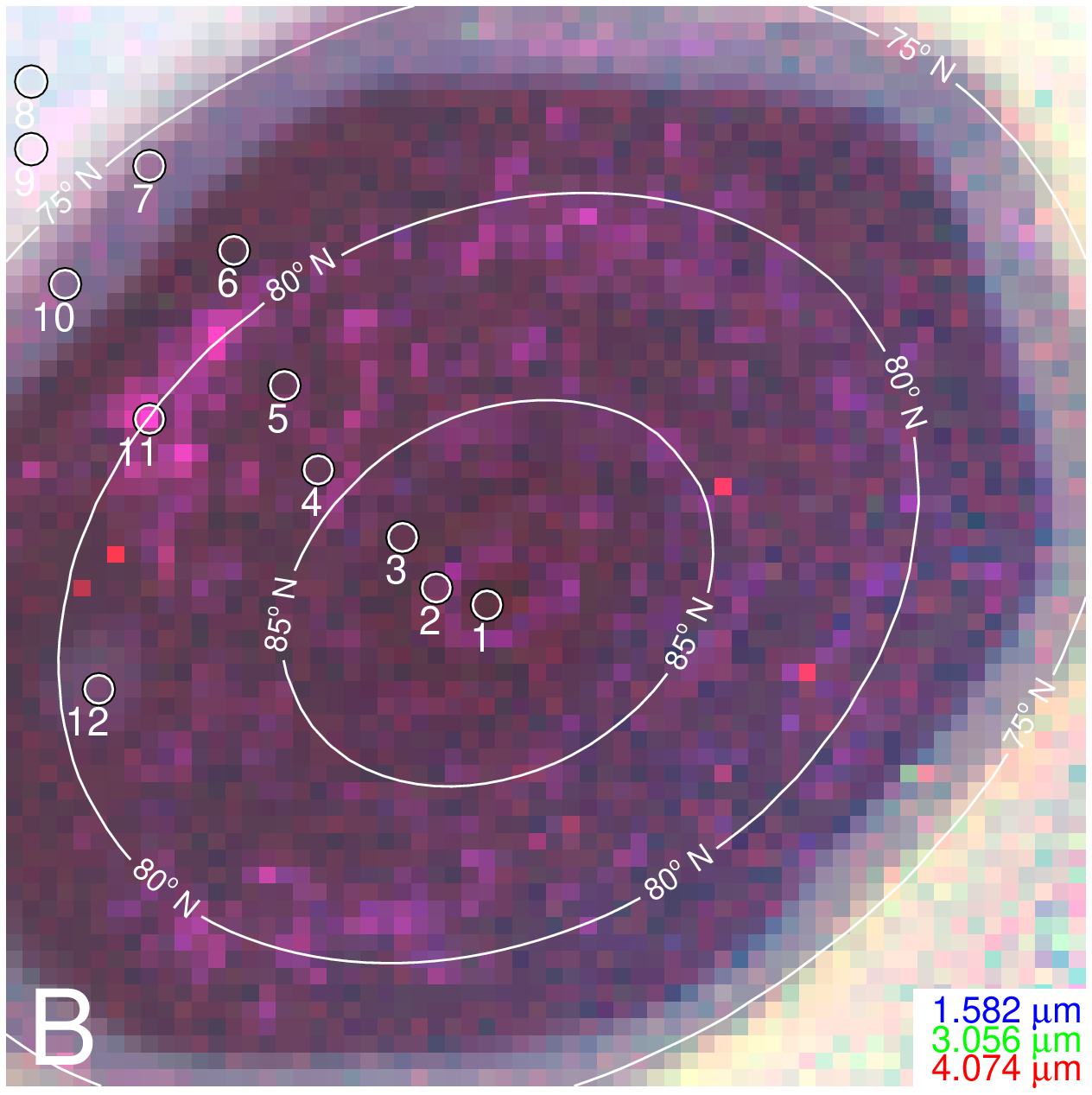} 
\includegraphics[width=2.05in]{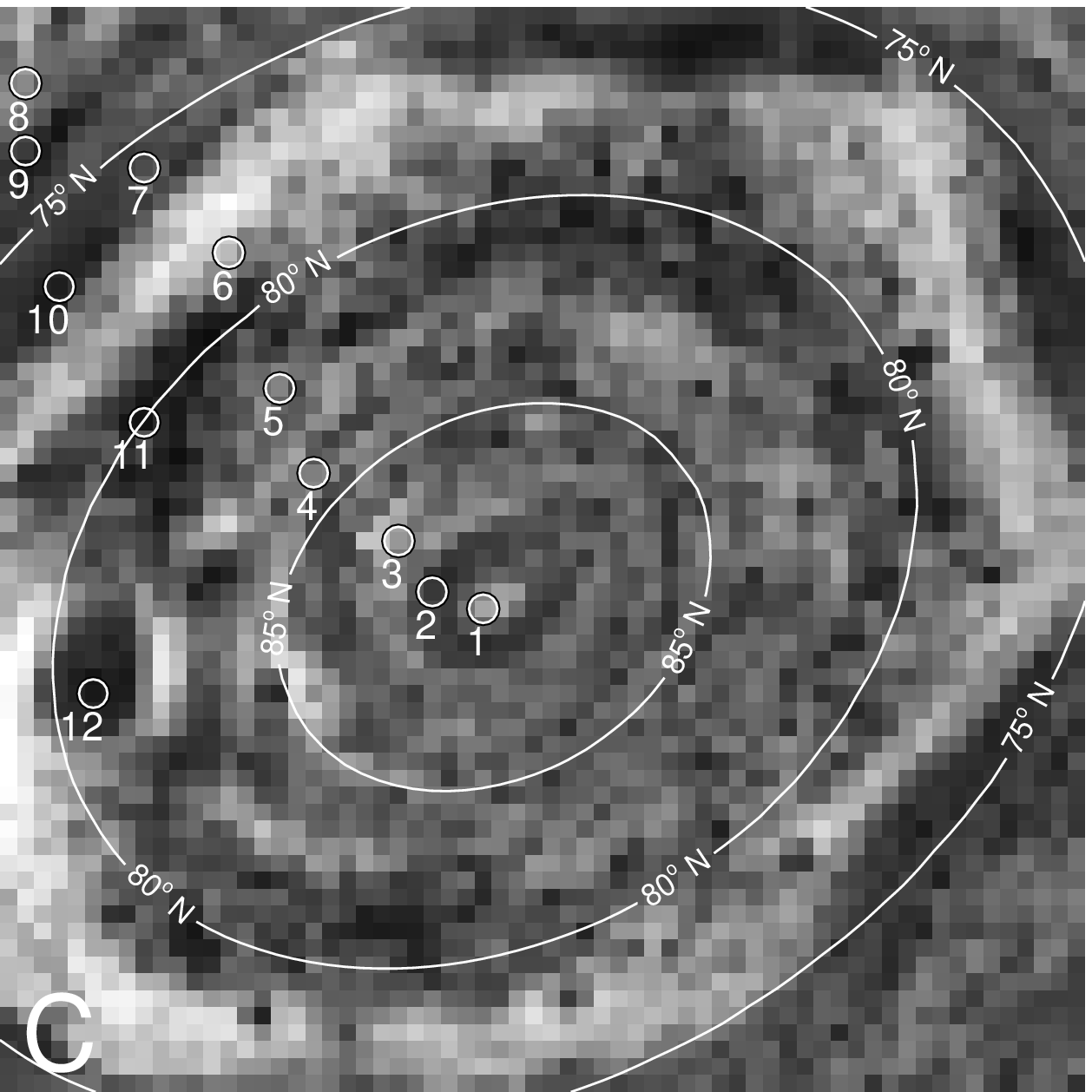} 
\includegraphics[width=6.2in]{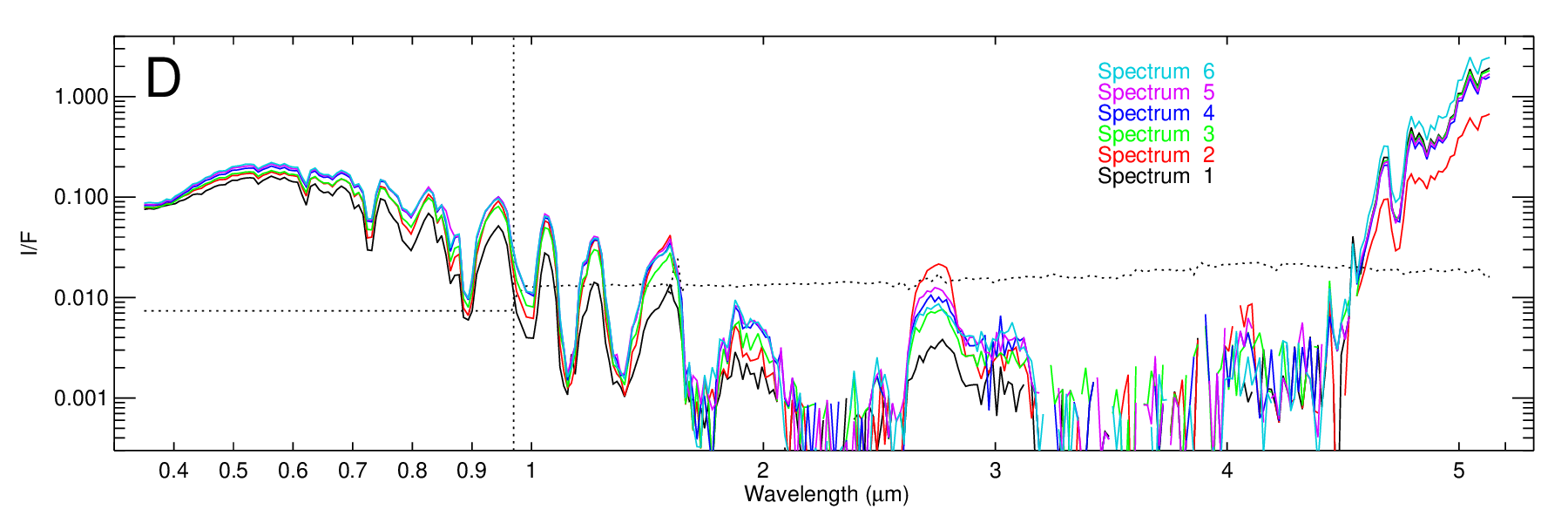}\\  
\includegraphics[width=6.2in]{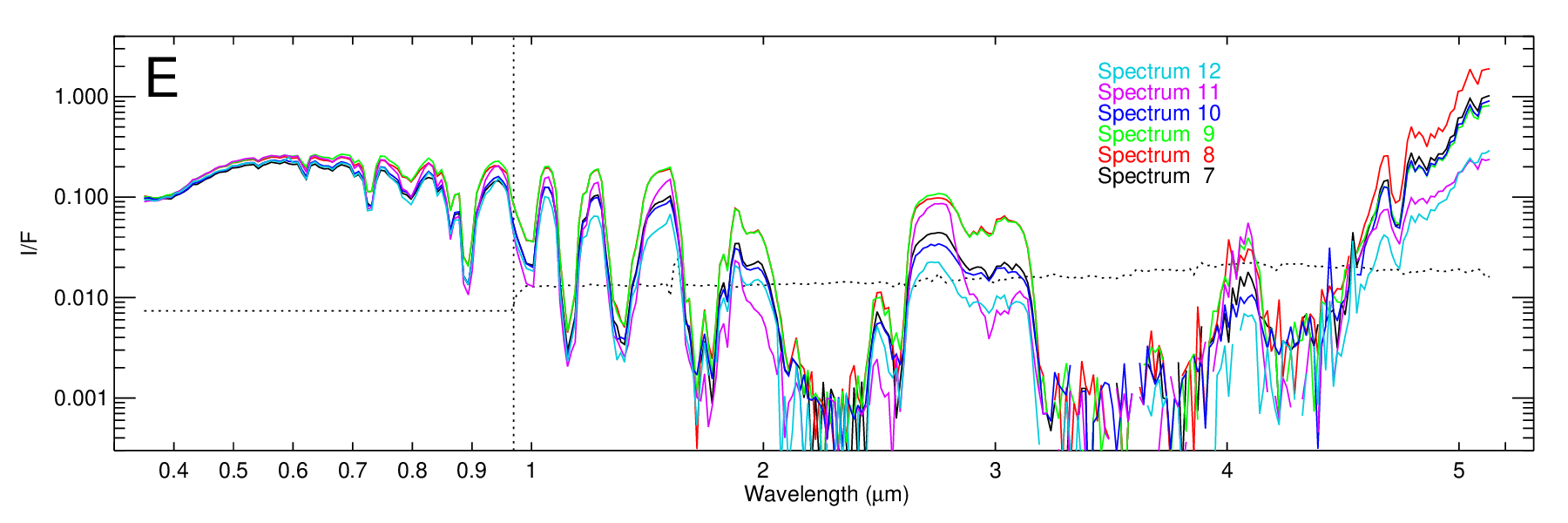}\\ 
\caption{VIMS 2013 observations and selected spectral sampling
  locations indicated by numbered circles in reflected-light visual
  (A) and near-IR (B) composites, and also at a thermal emission
  wavelength of 5.04 \mum (C). Spectra from these locations are shown
  in panels D and E. Note the nearly factor of ten drop between 2.7
  \mum and 3 \mum for spectrum 11 (E), and its sharp drop between
1.9 and 2 \mumx,  which are signatures of absorption by \nht
  ice particles. Similar, but more attenuated features can be seen
in spectrum 2 (D). The vertical dotted lines in D and E mark the
  wavelength at which visual and near-IR VIMS spectra are joined.  The
  nearly horizontal dotted line is the FWHM (in \mumx) as a function of
  wavelength. \label{Fig:2013combined}}
\end{figure*}

\begin{figure*}[!htb]\centering
\includegraphics[width=2.05in]{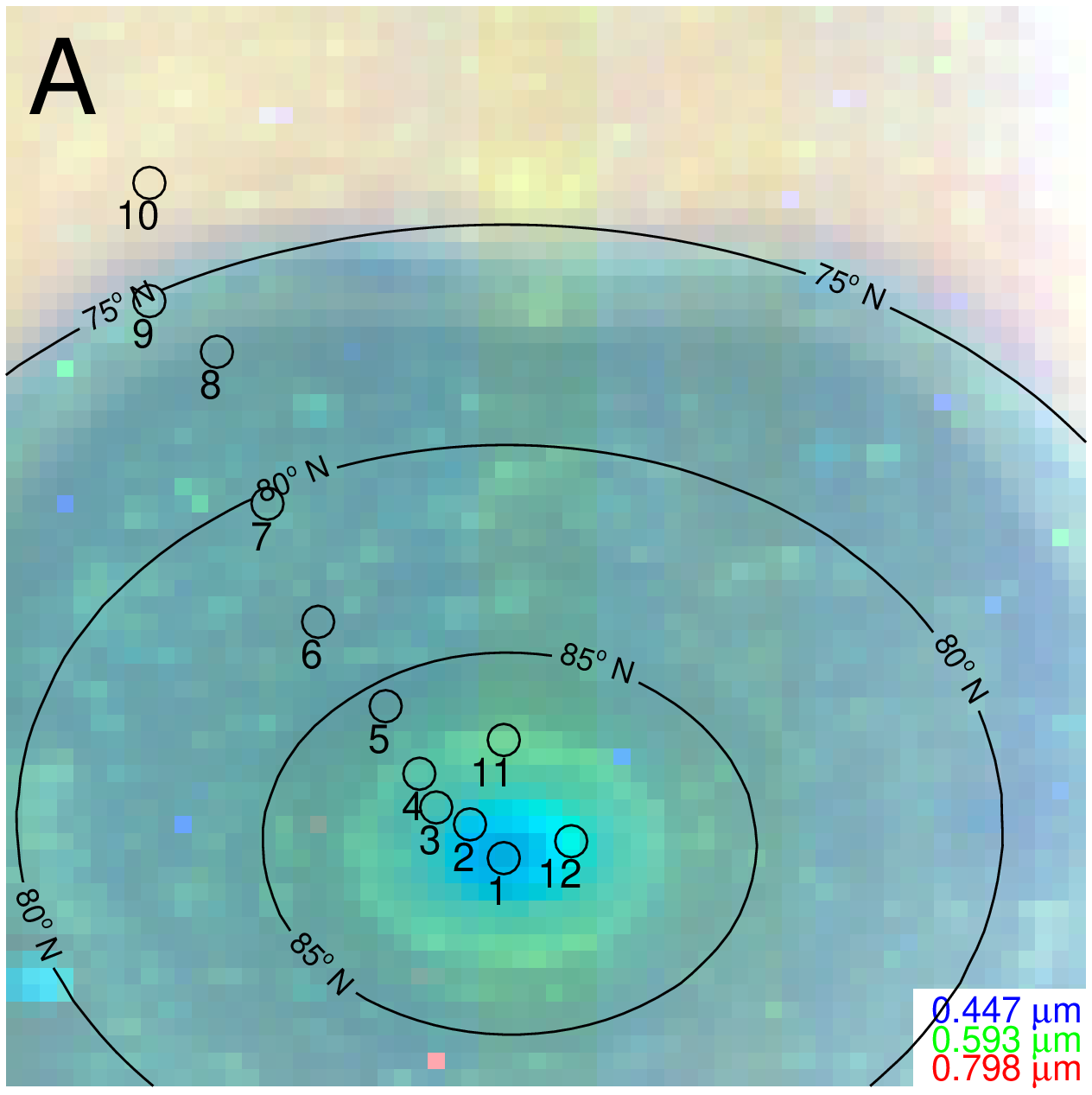}  
\includegraphics[width=2.05in]{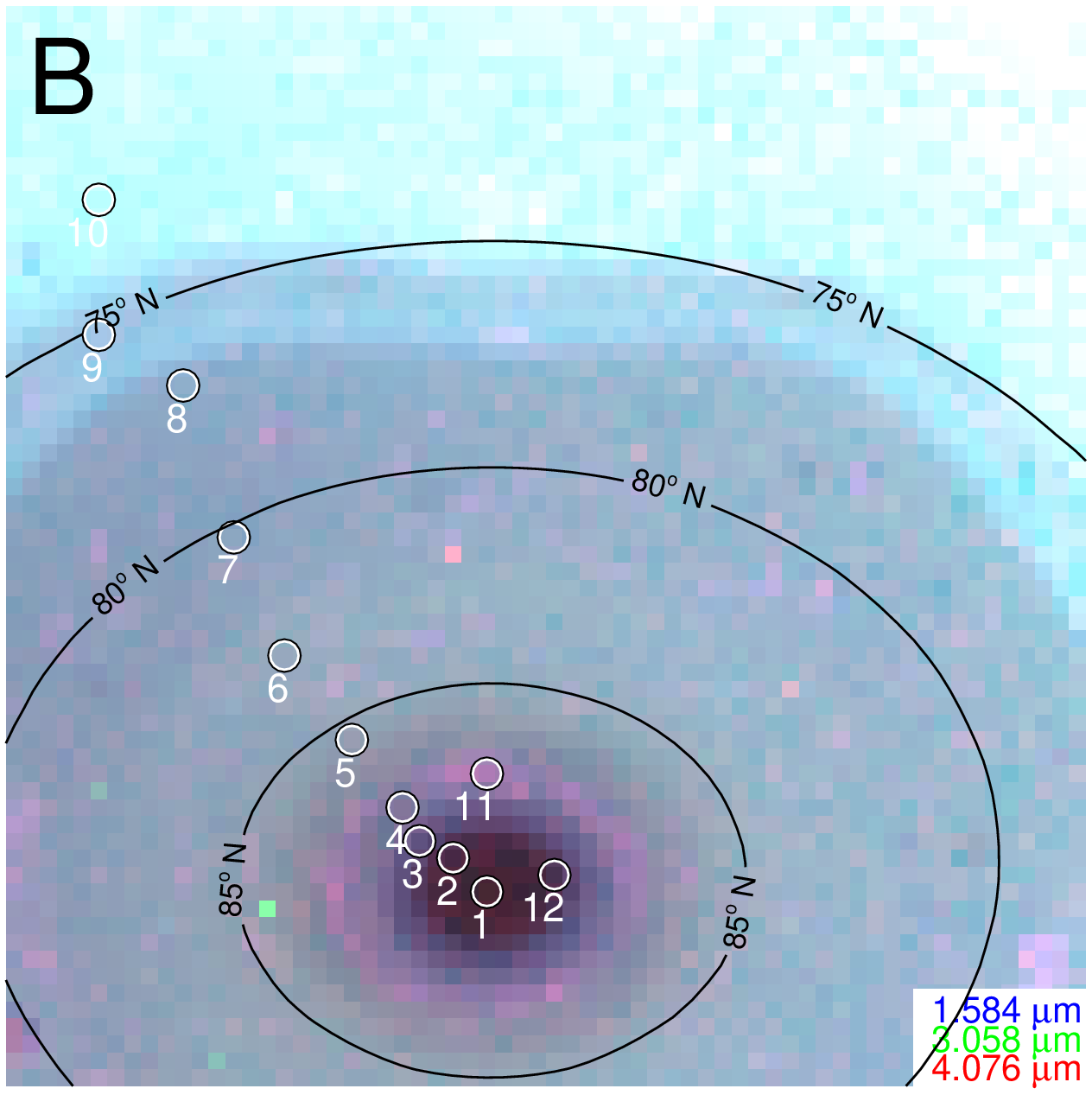} 
\includegraphics[width=2.05in]{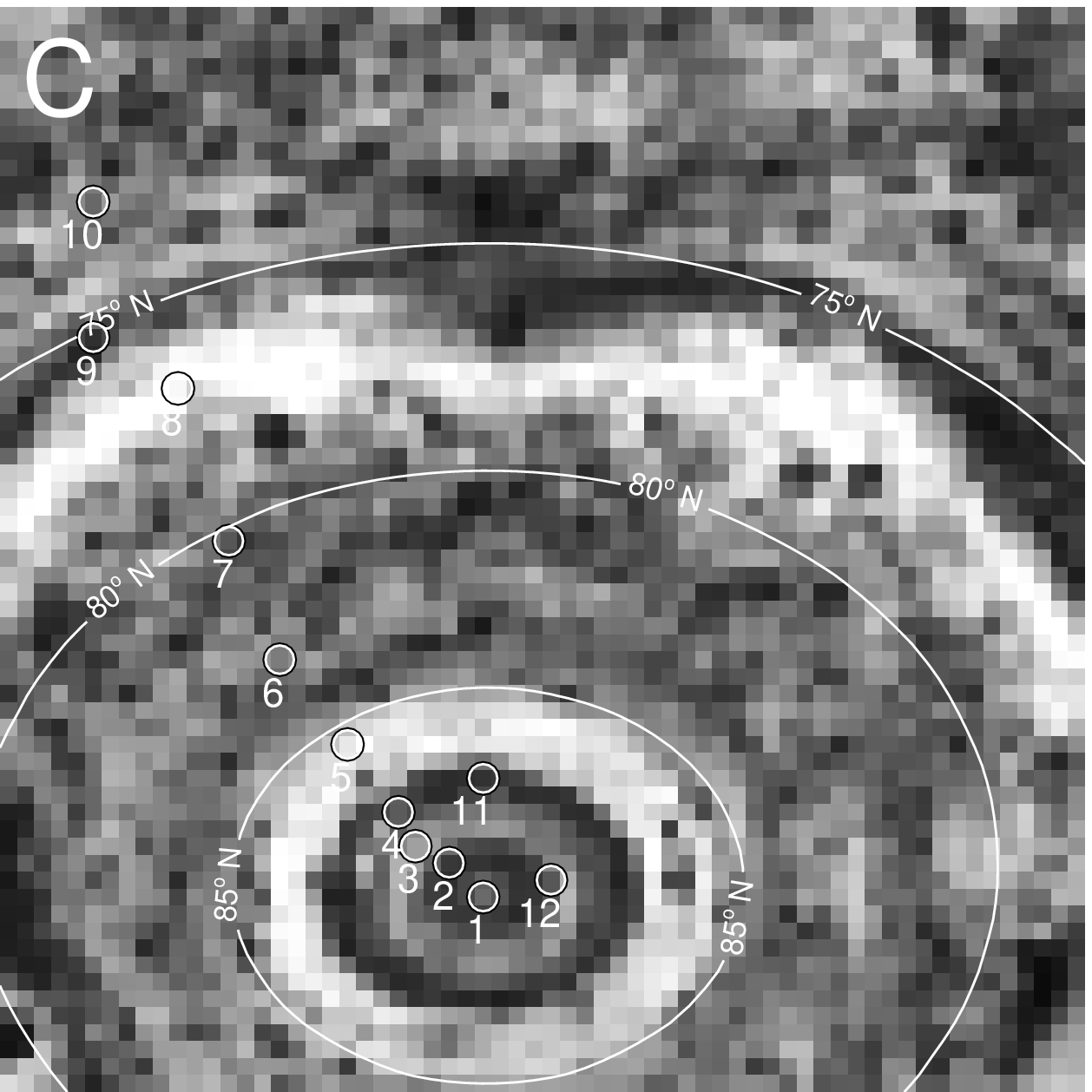} 
\includegraphics[width=6.2in]{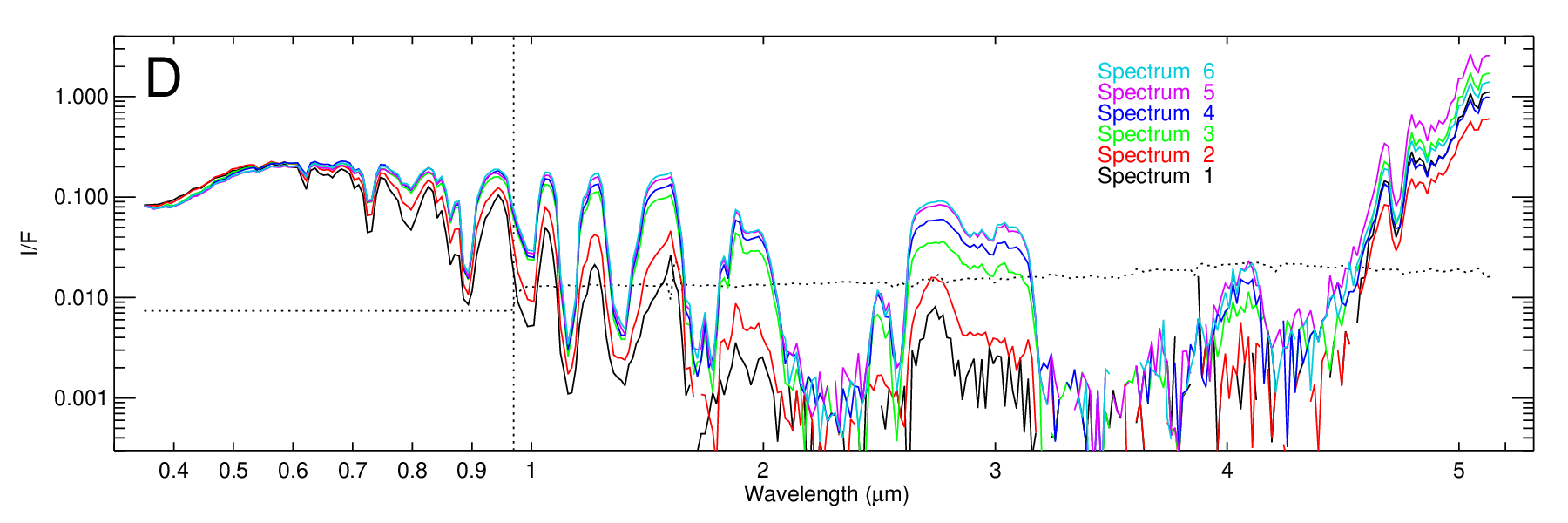}\\  
\includegraphics[width=6.2in]{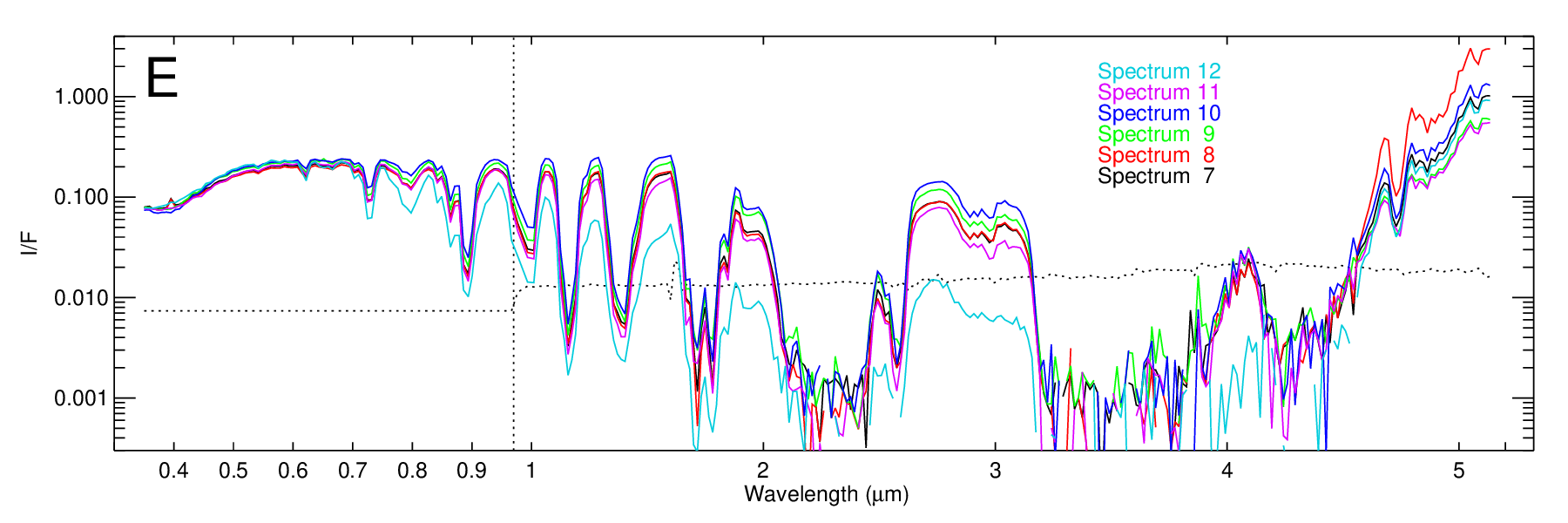}\\ 
\caption{As in Fig.\ \ref{Fig:2013combined}, except for 2016 observations
and target selections.  Note the nearly spectrally flat continuum
values for spectra inside the hexagon but outside the eye (spectra
5-9) compared to those from similar locations in 2013
 (spectra 3-7 in Fig.\ \ref{Fig:2013combined}). \label{Fig:2016combined}}
\end{figure*}

Our combined visual and near-IR spectra are illustrated in Figs.\ \ref{Fig:2013combined} 
and \ref{Fig:2016combined} for 2013 and 2016 observations respectively.  In each figure
twelve composite spectra are plotted for image locations illustrated
in three images: a visual composite, a near-IR composite, and a thermal emission at 5.2 \mum
monochromatic image.  The first 10 of these spectra are from the same locations
identified in previous figures.  For the 2013 observations, the spectra within the hexagon are seen
to be relatively dark, with the main exceptions being for a discrete ammonia cloud (spectrum
11).  For the 2016 observations, the spectra within the hexagon are generally significantly
brighter, except for the inner eye region, which is the only truly dark region.

\section{Radiative transfer modeling}

Our parameterization of atmospheric composition and the vertical
distribution and scattering properties of aerosols, our modeling of
gas absorption, and our treatment of multiple scattering and thermal
emission generally follow \cite{Sro2020spole}, the main exception
being that we here analyze combined visual and near-IR spectra.  
Our model parameters
for this study are listed and described in Table\ \ref{Tbl:paramlist}.
Model parameters were constrained in a two-step process. First, we
used trial and error calculations guided by sensitivity studies to
form a crude initial fit.  These initial guesses were dramatically
refined using a form of the Levenberg-Marquardt non-linear regression
algorithm described by \cite{Press1992}, which also provides
uncertainty estimates for the model parameters based on spectral
measurement uncertainties.  In fitting model spectra to observed
VIMS spectra we omitted the 1.61--1.68 \mum region to avoid the effects of
an order-sorting filter joint, and also the 2.2--2.4 \mum and
3.18--3.95 \mum regions due to their very poor signal to noise ratios.
In the following subsections we describe our assumed atmospheric
structure and composition, gas absorption models, parameterization of
cloud structure, parameterization of chromophore absorption,
constraining deep cloud composition, constraining of chromophore
location, and estimation of sensitivity to model parameters and to
initial guesses.

\begin{table*}[!htb]\centering
\caption{Cloud and gas model parameters used in spectral
  calculations.}
\begin{tabular}{r l l }
Param. (unit) & Description & Value\\
\hline
$p_1$ (bar) & stratospheric haze base pressure & adjustable\\
$r_1$ (\mumx) & effective radius of stratospheric particles & adjustable\\
$n_1(\lambda)$ & refractive index of stratospheric particles & $n_1=1.4 +i\times n_{1i}(\lambda)$\\
$\tau_1$ & stratospheric haze optical depth at 1 \mum & adjustable\\
\hline
$p_{2}$ (bar) & base of main visible cloud layer (P$_4$ or \pthf ?)  & adjustable\\
$r_2$ (\mumx) & effective radius of main cloud particles & adjustable\\
$n_2(\lambda)$ & refractive index of main cloud & $n_2=1.74$\\
$\tau_2$ & optical depth of upper cloud at 1 \mum & adjustable\\
\hline
$p_3$ (bar) & base pressure of putative \nht cloud & adjustable\\
$r_3$ (\mumx) & effective radius of \nht cloud particles & adjustable\\
$n_3(\lambda)$ & refractive index of \nht cloud particles & \cite{Martonchik1984} \\
$\tau_3$ & optical depth of \nht cloud at 1 \mum & adjustable\\
\hline
$p_{4}$ (bar) & base of deep cloud (\nhfsh + \hto?) & adjustable\\
$r_4$ (\mumx) & effective radius of deep cloud particles & adjustable\\
$n_4(\lambda)$ & refractive index of deep cloud particles &1.6 + $i\times n_{4i}(\lambda)$  \\
$\tau_4$  & optical depth of deep cloud at 1 \mum & adjustable\\
\hline
NH$_3v1$ & \nht VMR for Pcond $<$ P $<$ 4 bars & 12 ppm or adjustable \\
$p_b$ (bar) & \pht break-point pressure & adjustable\\ 
$\alpha_0$ & \pht volume mixing ratio for $p > p_b$ & adjustable\\
$f$  & \pht to H$_2$ scale height ratio for $p < p_b$  & fixed at 0.1\\ 
AsH$_3$v & AsH$_3$ volume mixing ratio & adjustable\\
\hline\\
\end{tabular}
\parbox{4.5in}{NOTE: These compact cloud layers are assumed to have a top pressure
  that is 0.9 times the bottom pressure and a particle to gas scale
  height ratio of 1.0; aerosol particles are assumed to scatter like
  spheres with a gamma size distribution with variance parameter
  $b=0.05$ for layers 1-3 and 0.1 for layer 4, with distribution
  function $n(r) = \mbox{constant}\times r^{(1-3b)/b} e^{-r/ab}$,
  where with $a = r_\mathrm{i}$, for $i=1,..,4$, and $b=$
  dimensionless variance, following \cite{Hansen1974}. Optical depths
  are given for a wavelength of 1 \mumx. $n_{1i}(\lambda)$ and
  $n_{2i}(\lambda)$ are either set equal to zero or adjusted using 2-3
  free parameters in a model based on Eq.\ \ref{Eq:chromod}. The
  imaginary index $n_{4i}(\lambda)$ is an empirical function discussed
  in Sec.\ \ref{Sec:deep}. We also tried refractive indexes of \nhfsh and \hto for layer 4.}
\label{Tbl:paramlist}
\end{table*}

\subsection{Atmospheric structure and composition}
\begin{figure}[!bt]\centering
\includegraphics[width=3.1in]{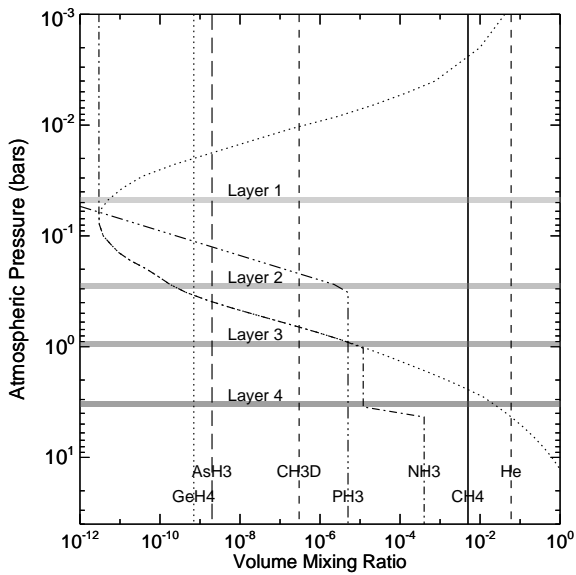}
\caption{Volume mixing ratios (VMR) of spectroscopically important gases
in the atmosphere of Saturn.  These values were either assumed
or derived from fitting the spectrum from the background cloud at
location 2. The VMR for germane, to which VIMS is not sensitive,
is from \cite{Bjoraker1986}. See text for other VMR references. The dotted curve traces the
saturation vapor pressure profile of \nhtx. Horizontal gray bars show typical vertical
locations of our four compact aerosol layers of distinctly different
compositions. }
\label{Fig:gasmix}
\end{figure}

We used \cite{Lindal1985} to define the temperature structure between
0.2 mbar and 1.3 bars, and approximated the structure at deeper levels
using a dry adiabatic extrapolation. We also assumed their value of
0.0638 for the He/H$_2$ number density ratio, which is within the
0.04-0.075 range recently derived from Cassini Composite Infrared
Spectrometer observations by \cite{Achterberg2020}.  Our assumed
nominal composition of the atmosphere as a function of pressure is
displayed in Fig.\ \ref{Fig:gasmix}.  For methane we chose the
\cite{Fletcher2009ch4saturn} volume mixing ratio (VMR) of
(4.7$\pm$0.2)$\times 10^{-3}$, which corresponds to a \chfx/H$_2$
ratio of 5.3$\times 10^{-3}$.  For \chtd we also used the
\cite{Fletcher2009ch4saturn} VMR value of 3$\times 10^{-7}$.  We
assumed an ammonia vapor profile with a deep uniformly mixed region
for $p >$ 4 bars with a VMR of 400 PPM, another uniformly mixed region
with a lower VMR between 4 bars and the ammonia condensation pressure,
above which we assumed that the ammonia VMR followed a saturated vapor
pressure profile up to the tropopause, and above that we again assumed
a uniform mixing ratio.  This profile is illustrated by the dot-dash
line in Fig.\ \ref{Fig:gasmix}.  It was inspired by the results of
\cite{Briggs1989} who found a deep mixing ratio of 480$\pm$100 ppm and
a depleted mixing ratio of 70-110 ppm near 2 bars.  We chose a
somewhat lower deep value that is roughly in the middle of the
\cite{Laraia2013} limits of 360-480 ppm.  However, the only \nht
profile parameter that our VIMS observations are sensitive to is the
mixing ratio in the depleted region, which is presumably produced by
the formation of an \nhfsh cloud near 4 bars.  Our initial attempts to
fit that parameter resulted in very low values with very high
uncertainties.  We thus assumed in our first round of latitude
dependent fits a fixed value of 12 ppm, which we chose to produce an
ammonia condensation pressure near 1 bar because that is close to the
upper limit of the base pressures we found for the ammonia cloud layer
(layer 3 in our model).  We subsequently learned that it was mainly
the 2016 observations that presented fitting problems. It appears that
the VIMS sensitivity to the ammonia mixing ratio decreases
substantially wherever upper level aerosol opacity is high, leading to
at least poor uncertainty estimates, and perhaps also erroneously low
VMR values. For the 2013 observations, for which the upper
tropospheric aerosol opacity is low, we were able to obtain reasonably
well constrained fits, and we thus included our \nht fits in our final
results.

Other spectroscopically important gases that were constrained by our fits
to VIMS observations included arsine (\ashtx) and phosphine (\phtx).
 We fit an adjustable vertically uniform
mixing ratio for Arsine and, following \cite{Fletcher2009ph3}, we parameterized
 the \pht VMR profile
using three parameters: a deep mixing ratio  $\alpha_0$, a break point pressure $p_b$,
and a ratio $f$ of the \pht scale height to the pressure scale height. 
 At pressures less than $p_b$, the \pht mixing ratio can be written
as \begin{eqnarray} \alpha(p) = \alpha_0 (p/p_b)^{(1 - f)/f} \quad
  \mathrm{for} \quad p<p_b.\label{Eq:prof1}
\end{eqnarray}

\noindent
 Because we found a high correlation between spectral effects of $p_b$
 and $f$, only two of our three parameters could be well constrained by
 the VIMS observations. Because our attempts to fit all three
parameters usually resulted in low (but poorly constrained) values of $f$,
we chose to fix $f$ at 0.1, and decided fit the other two parameters.
   Our fixed value of $f$ is at the low end of the range
of values found by \cite{Fletcher2011vims} when they assumed a fixed
breakpoint pressure.

\subsection{Gas absorption models}
We used gas absorption models 
 described by \cite{Sro2018dark} and \cite{Sro2013gws}  and references cited therein.  For
methane we used correlated-k models
based on line-by-line calculations down to 1.268 \mum
\citep{Sro2012LBL}, but for shorter wavelengths used correlated-k fits
to band models of \cite{Kark2010ch4} (we used P. Irwin's fits
available at http://users.ox.ac.uk/$\sim$atmp0035/ktables/ in files
ch4\_karkoschka\_IR.par.gz and ch4\_karkoschka\_vis.par.gz).  The \nht
absorption model fits are from \cite{Sro2010iso}, which are based primarily on band models of
\cite{Bowles2008}.
Absorption models for phosphine (\phtx) and arsine
(AsH$_3$) are the same as described by \cite{Sro2013gws}.
Collision-induced absorption (CIA) for H$_2$ and H$_2$-He was
calculated using programs downloaded from the Atmospheres Node of the
Planetary Data System, which are documented by \cite{Borysow1991h2h2f,
  Borysow1993errat} for the H$_2$-H$_2$ fundamental band,
\cite{Zheng1995h2h2o1} for the first H$_2$-H$_2$ overtone band, and by
\cite{Borysow1992h2he} for H$_2$-He bands.

\subsection{Parameterization of cloud structure}\label{Sec:parcloud}

Our parameterization of cloud structure diverges from the 
 vertically diffuse layer structures of the type commonly
used in prior publications.  Instead, we here use a
structure with four compact layers with fitted vertical positions and
distinct compositions.  In limited trials we found
that somewhat thicker layers could improve fits to some spectra (an example is given
in a later section), but others were best fit with compact layers.
Compact layers also provided the best fits for \cite{Fletcher2011vims}
in modeling night-side 5-\mum spectra, which were, however,
insensitive to the stratospheric haze.    But even in the polar
stratosphere, there is some evidence for compact layers from limb
observations \citep{Sanchez-Lavega2020hazes}.  While we think more
effort in constraining layer thicknesses with VIMS spectra is likely
to be productive, especially with improved signal/noise ratios from
spatial averaging and/or center-to-limb analysis, this is a fairly
complex topic that is left for future work.  Using compact structures
in this study avoids having to constrain both an upper boundary as
well as a scale height for each layer, yet still allowed us to make
accurate fits to the observations and leads to more stable solutions.
Because we have many methane bands of varying strengths, we avoid
ambiguities between pressure and optical depth, and also have
such great sensitivity to vertical structure that our retrievals
do not suffer much from the lack of center-to-limb constraints.

\begin{figure*}[!t]\centering
\hspace{-0.15in}\includegraphics[width=3.15in]{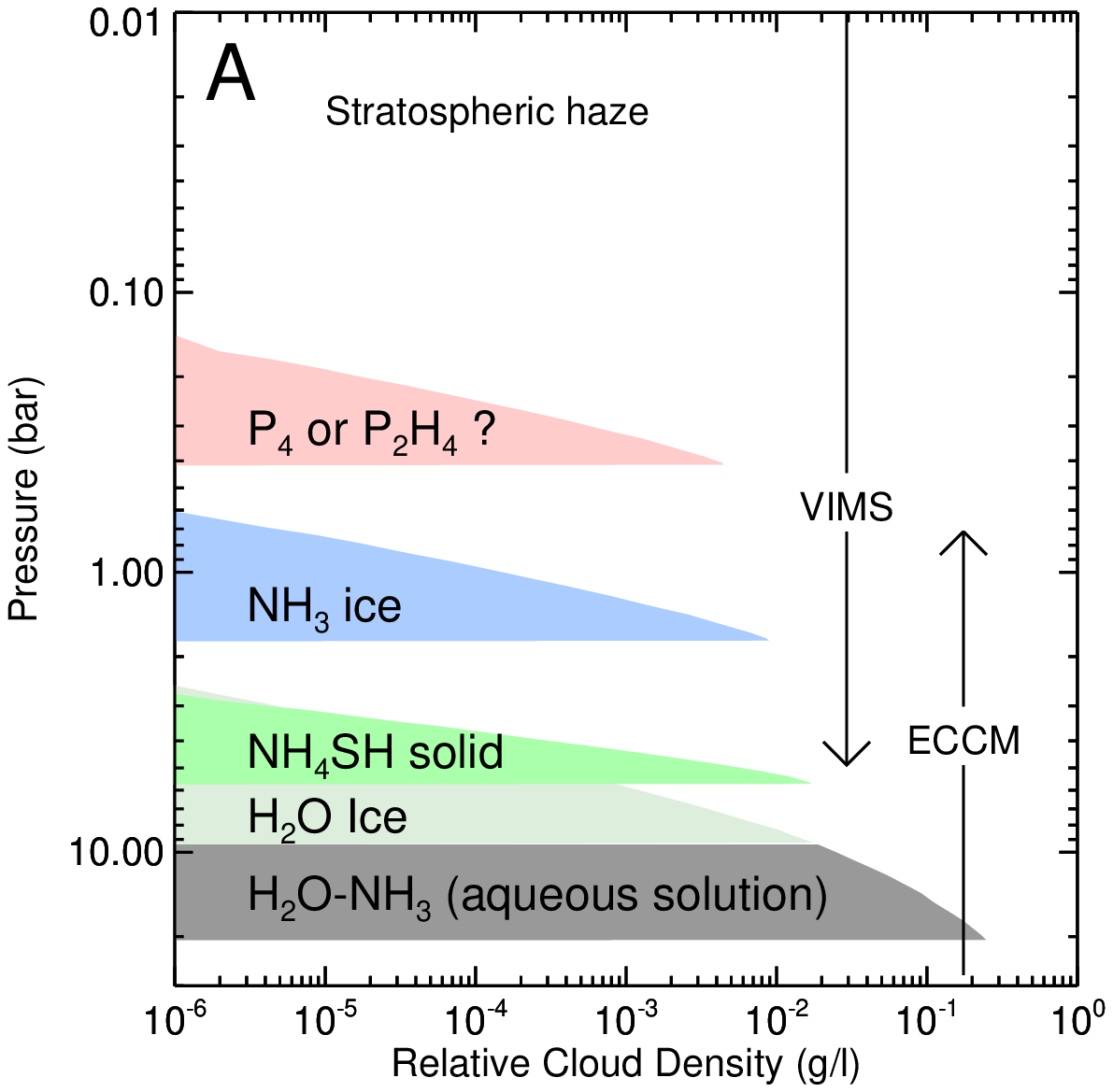}\hspace{-0.15in}
\includegraphics[width=3.15in]{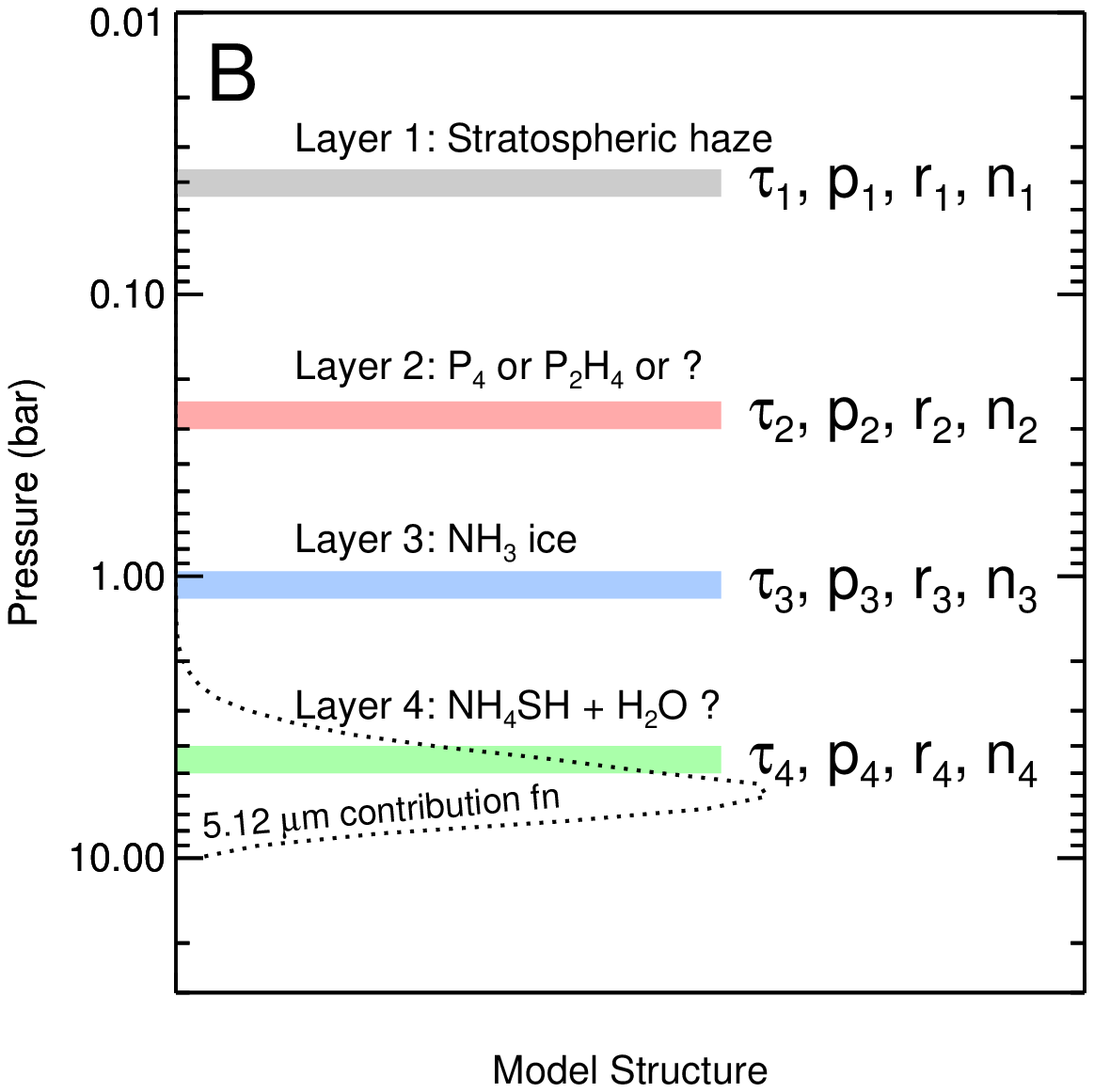}
\caption{{\bf A:} Equilibrium cloud condensation model (ECCM) of the
  composition and vertical distribution of clouds on Saturn for P $>$
  500 mbar according to \cite{Atreya2005SSR}.  All the ECCM clouds
  have densities expected to greatly exceed actual amounts. For
  P $<$ 500 mb, a cloud layer of unknown composition (possibly P$_4$
  or \pthfx) has been inferred from radiation transfer modeling (see
  text).  The density profile of that cloud is chosen for illustrative
  purposes.  The downward arrow indicates the pressure depth to
  which VIMS observations might be sensitive to cloud
  composition. {\bf B:} Our model structure using four compact
  cloud layers with compositions corresponding to expectations shown in
  A.  Also
  shown is the clear-atmosphere scaled contribution function for
  emission at 5.12 \mum from \cite{Sro2016}.}\label{Fig:eccm}\label{Fig:cloudmodel}
\end{figure*}

Our chosen layer compositions are partly based on expectations from
equilibrium cloud condensation (ECCM) modeling
\citep{Weidenschilling1973,Atreya2005SSR}, as illustrated in
Fig.\ \ref{Fig:eccm}.  The lower aerosol optical depth in the polar
regions and the wide spectral range of the VIMS visual and near-IR
observations provide a unique capability to probe deeper layers as
well as constrain composition to some degree. However, because the
ECCM model does not provide a well defined composition for the
stratospheric haze (which we generally found at an effective
pressure of about 50 mbar) or the main upper cloud layer on Saturn (which we
found to have an effective pressure at the 200-300 mbar level in the polar region), we
needed to make some assumptions.

At least one of the top two layers is likely composed of a complex
hydrocarbon that provides the short wavelength absorption responsible
for Saturn's generally tan color, which we will refer to as Saturn's
chromophore.   We considered chromophores in either layer
and both layers, but concluded from an analysis presented
in Sec.\ \ref{Sec:vertloc} that the stratospheric haze alone was a
viable location for the chromophore and provided the most convenient
modeling behavior.  We decided to model the stratospheric layer
as a haze of spherical particles with an adjustable 
pressure, an adjustable optical depth, an adjustable radius, and a
real and imaginary index parameterized to function as the chromophore,
 as described in Section\ \ref{Sec:chrompar}.

The second layer, which is optically much thicker than
the stratospheric haze, and is the main visible cloud layer in most
regions on Saturn, appears to have no discernible ammonia
ice spectral signature \citep{Kerola1997} or any other near-IR spectral absorption
features that could be used to identify it \citep{Sro2013gws}. A possible composition
suggested by \cite{Fouchet2009} on photochemical grounds is
diphosphine (\pthfx), the optical properties of which are not known
well enough to test this possibility.  We also treated this layer as
comprised of spherical particles with a real refractive index of 1.74,
which is the value for \pthf at 195 K according to
\cite{Wohlfarth2008}.   We also found from test fits at a variety of locations that it generally
 produced slightly better fits than a real index of 1.4.   Although we considered using an imaginary index
with strong short-wavelength absorption (serving as the chromophore), our
final fits treated this layer as non-absorbing over the VIMS spectral range.

The best justified composition is for the third layer, which is
composed of \nht ice, a conclusion based on detection of ammonia ice
absorption features at 3-3.1 \mum and near 2 \mum and the fact that the
fitted pressure of this layer are at a level where ammonia condensation is plausible
for the fitted ammonia mixing ratio. An ammonia ice layer also fits
well in models of south polar VIMS spectra \citep{Sro2020spole}. The
ammonia spectral signature can be readily seen in the north polar 2013 spectrum
from location 2 (Fig.\ \ref{Fig:2013combined}D) and much more
dramatically in the spectrum from location 11
(Fig.\ \ref{Fig:2013combined}E).  Both of these provide
further confirmation of our identification of \nht ice as the
composition of layer-3 particles, as discussed later in Section\ \ref{Sec:nh3ice}.
  We also modeled the scattering of this layer using a
distribution of spherical particles with an adjustable effective
radius and optical depth.  

Because VIMS observations do not penetrate much deeper than the 5-bar
level, our ability to constrain aerosols at the water cloud level is
poor.  Based on the work described in Sec.\ \ref{Sec:deep}, which did
not find acceptable fits using water ice or \nhfshx, we selected a
real index of 1.6, which is a somewhat arbitrary value between that of
water ice and \nhfshx, and a wavelength dependent imaginary index that
crudely varied as some mix of these candidates, but provided generally
more absorption, and seemed to provide a decent overall fit to the
observations. Instead of following \cite{Sro2020spole} in assuming a
fixed particle size, optical depth, and base pressure, with only an
adjustable top pressure for the bottom layer, we here assumed
spherical particles with an adjustable radius.  We assumed that all
layers were vertically thin.  To avoid having to constrain four
additional parameters, we used an adjustable base pressure and a top
pressure 90\% of the base pressure, which allows for layer overlap if
necessary (conversion of our results to the case of more extended
layers is discussed in Sec.\ \ref{Sec:convert}).  For all layers we
assumed a gamma size distribution \citep{Hansen1974} with a fractional
variance $b$ = 0.1 for the deep cloud of large particles and 0.05 for
the other layers.  
  
Later analysis reveals that the spectral effects of
these layers are roughly as follows.  Layer 1 controls the
short-wavelength absorption and  the depth of methane
absorption bands. Layer 2 mainly controls the pseudo-continuum I/F
values at wavelengths from about 700 nm to near-IR wavelengths. Layer
3 affects the I/F in the 2-\mum and 2.7-3.1 \mum region.  Layer 4 has
its main influence in limiting the I/F in the 4.7--5.1 \mum emission
region.

\subsection{Parameterization of the chromophore absorption}\label{Sec:chrompar}

\begin{figure}[!htb]\centering
\includegraphics[width=3.1in]{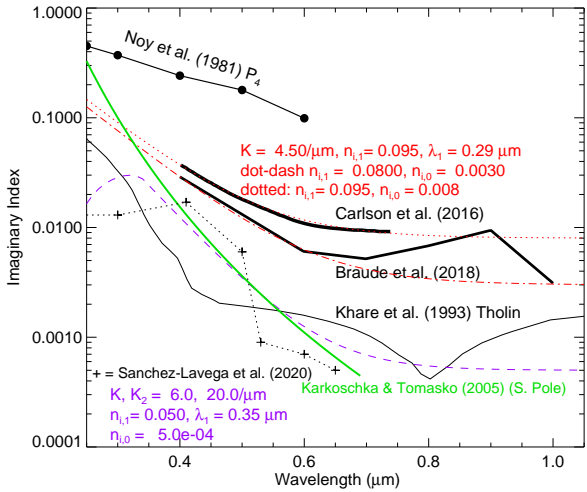}
\caption{Comparison of imaginary index spectra for
candidate chromophores with parameterized model fits,
using Eq,\ \ref{Eq:chromod}. For the red curves $K_1 = K_2 = K$. The purple dashed curve is one example
of a best fit model to a Saturn polar spectrum. Index spectra from
\cite{Noy1981}, \cite{Khare1993}, and \cite{Carlson2016} are based
on laboratory measurements.}
\label{Fig:chromod}
\end{figure}

Saturn requires some population of particles in the upper troposphere
that absorbs strongly at short wavelengths in order to produce its
generally tan color.  It seems plausible that these
chromophores consist of complex hydrocarbons created by the
photochemical reactions driven by solar UV radiation.  Candidate
chromophores produced in laboratory experiments, such as the so called
tholin material produced by \cite{Khare1993}, the phosphorus compound
produced by \cite{Noy1981}, or the potential Jovian chromophore
produced by \cite{Carlson2016}, have all yielded materials with a
strong UV absorption declining roughly log-linearly with wavelength to
much lower baseline values at long wavelengths. This is also roughly
the characteristic of the imaginary index for a Jovian chromophore
inferred by \cite{Braude2018} from an analysis of center-to-limb
spectra obtained by VLT/MUSE observations of Jupiter.  These are
compared with our model parameterizations in Fig.\ \ref{Fig:chromod},
which are based on the functional form
\begin{eqnarray}
n_i^*(\lambda) = & \qquad \notag \\
 \begin{cases}
  n_{i,0} + ( n_{i,1} -  n_{i,0})\times 10^{-K_1 (\lambda - \lambda_1)} &
  \text{$\lambda > \lambda_1$}  \\ 
  n_{i,1} \times 10^{-K_2 (\lambda_1 - \lambda)} &
 \text{$\lambda \le \lambda_1$}\\
0 &  \text{$\lambda \le \lambda_c$}
 \end{cases}\label{Eq:chromod}\\
n_i(\lambda) = \frac{1}{\Delta}\int_{\lambda-\Delta/2}^{\lambda+\Delta/2}n_i^*(\lambda')d\lambda' \qquad \label{Eq:boxav}
\end{eqnarray}

\noindent which defines a peak at wavelength $\lambda_1$ and falloff
rates away from the peak defined by log slope $K_1$ on the
long-wavelength side and $K_2$ on the shortwave side. The zero value
below wavelength $\lambda_c = 0.15$ \mum ensures that the integral in
Eq.\ \ref{Eq:nr} will remain finite.  The box-car average defined by
Eq. \ref{Eq:boxav} produces a more rounded peak that better fits the
I/F spectra for $\Delta \approx$ 0.11 \mumx.  The chromophore
characteristics so defined are compared with other chromophore models
in Fig.\ \ref{Fig:chromod}.  As evident in the figure, this
parameterization can provide a decent fit to the suggested chromophore
materials with appropriate choices of the model parameters. Since we
know so little about the spectral character of the Saturnian
chromophore it will be constrained by fitting the VIMS
observations. The key parameters we adjust to fit model spectra are
$n_{i,1}$, $\lambda_1$ and the log slope $K_1$. The other parameters
$K_2$ (which controls the rate of decline on the shortwave side of the
peak) and $n_{i,0}$ (which is the baseline absorption at long
wavelengths) are not well constrained by the observations and usually
set to somewhat arbitrary values of 20 \mum$^{-1}$ and 5$\times
10^{-4}$ respectively. The latter value is in approximate agreement
with the long-wavelength value inferred from modeling HST observations
of Saturn by \cite{Sanchez-Lavega2020hazes}. It also agrees with the
minimum value inferred by \cite{Kark2005} for wavelengths below 700
nm, although it exceeds their estimated upper limit of $10^{-4}$ for
wavelengths near 1 \mumx.

Because
the imaginary index can reach significant peak values, we use a real
index consistent with the Kramers-Kronig relation:
\begin{eqnarray}
n_r(\lambda) =  n_{r,0} + \frac{2}{\pi} P \int_0^\infty\frac{\nu' n_i(\nu')}{\nu'^2-\nu^2}d\nu'
\label{Eq:nr}
\end{eqnarray}
\noindent where $\nu = 1/\lambda$ and $P$ indicates the principal value of the 
integral that follows it, which we computed using Maclaurin's formula \citep{Ohta1988}.
 For $n_{r,0}$, the real index at infinite
wavelength (zero wavenumber), we ultimately adopted a value of 1.4, comparable to
values expected for stratospheric hydrocarbons.  Although none of the measured
imaginary index spectra plotted in Fig.\ \ref{Fig:chromod} captured a short-wavelength
peak, a peak was observed near 400 nm in recent modeling of HST observations of 
Saturn \citep{Sanchez-Lavega2020hazes}.

\subsection{Constraining the composition of the deep cloud layer}\label{Sec:deep}

\begin{figure*}[!ht]\centering
\includegraphics[width=5in]{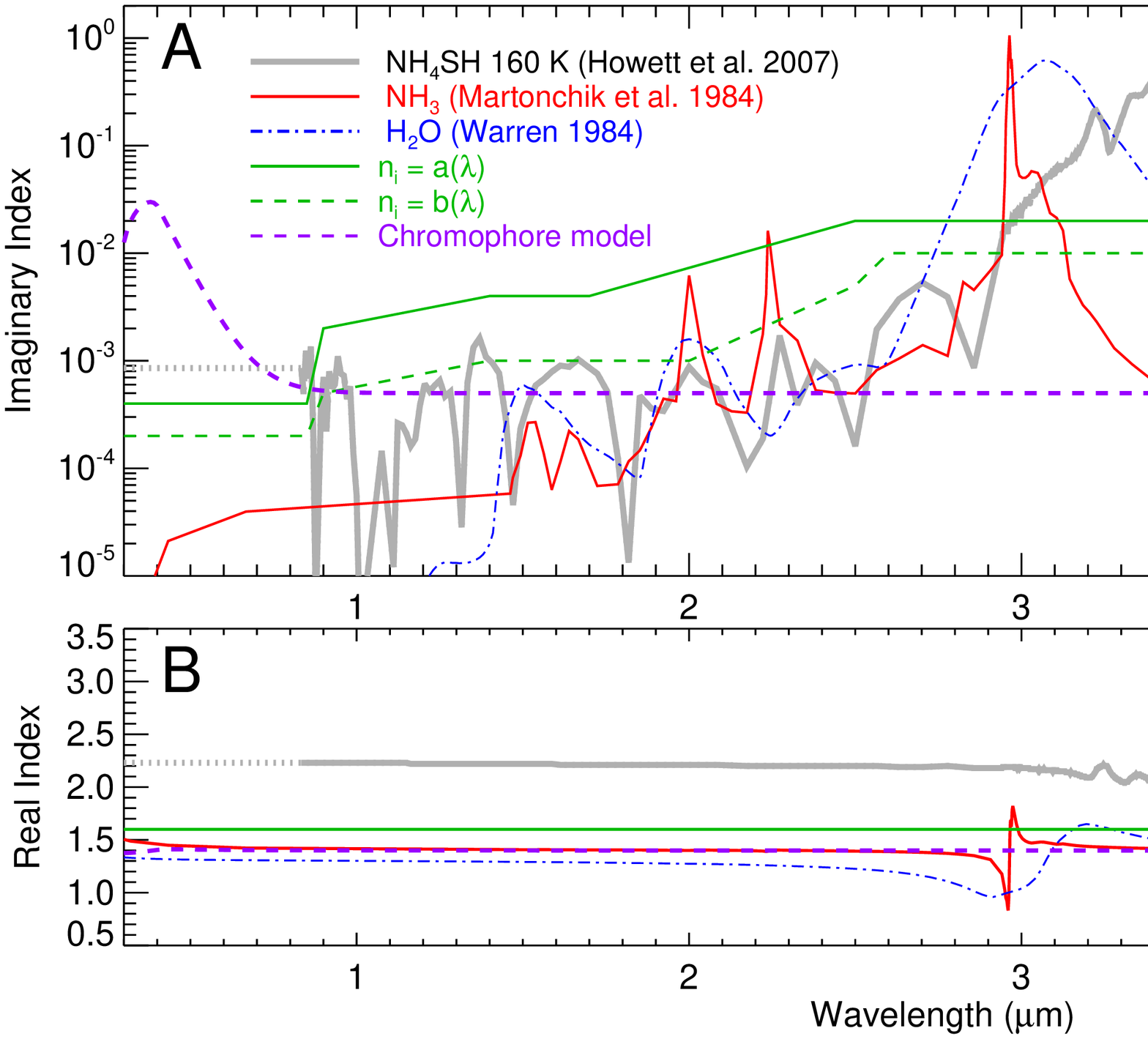}
\caption{Imaginary refractive index (A) and real index (B)
  spectra for candidate cloud particle compositions of \nhfsh
  \citep{Howett2007}, \nht \citep{Martonchik1984}, and \hto
  \citep{Warren1984}, compared to empirical deep cloud models (green curves). The dashed purple line
  displays a typical imaginary index model for the stratospheric haze layer
  that is the assumed source of Saturn's tan color (its chromophore
  layer), computed using Eqns. 2-4 using n$_{i,0}$ = 5$\times10^{-4}$,
n$_{i,1}$ = 0.05, $\lambda_1$ = 0.35 \mumx, K$_1$= 6.0, K$_2$= 20, and $\Delta$ = 0.11 \mumx.
 }\label{Fig:varic}
\end{figure*}

The combination of visual and near-IR spectra provides an opportunity to
better constrain the composition of the deep cloud that from an ECCM perspective
might be dominated by \nhfsh or H$_2$O ice or a mixture of the two, as the deep
layer has a significant effect on both short wavelength continuum I/F values
as well as the region of thermal emission.  Because these different compositions
have different spectral absorption signatures (Fig.\ \ref{Fig:varic}), we hoped to find which
provided the best spectral fits.   Using 2013 spectra inside the hexagon,
which provided a better view of deeper layers because of reduced optical depth
of the other layers, we found that water ice particles provided the worst
fit and \nhfsh particles a better fit, although neither provided as good a
fit as the simple wavelength-independent refractive index model in which
the index of the deep layer is given by $n_4$ = 2 + 0.01$i$. To minimize
residual errors in the 0.8 -- 1.7 \mum region, we further improved
that model by setting the real index to 1.6 (between that of \hto and \nhfshx),
and varying the imaginary index with wavelength as shown in Fig.\ \ref{Fig:varic}.
We adopted the imaginary index model labeled $a(\lambda)$ for the deep layer in
carrying out the latitude dependent fits for both 2013 and 2016.

\begin{figure*}[!ht]\centering
\hspace{-0.15in}\includegraphics[width=3.25in]{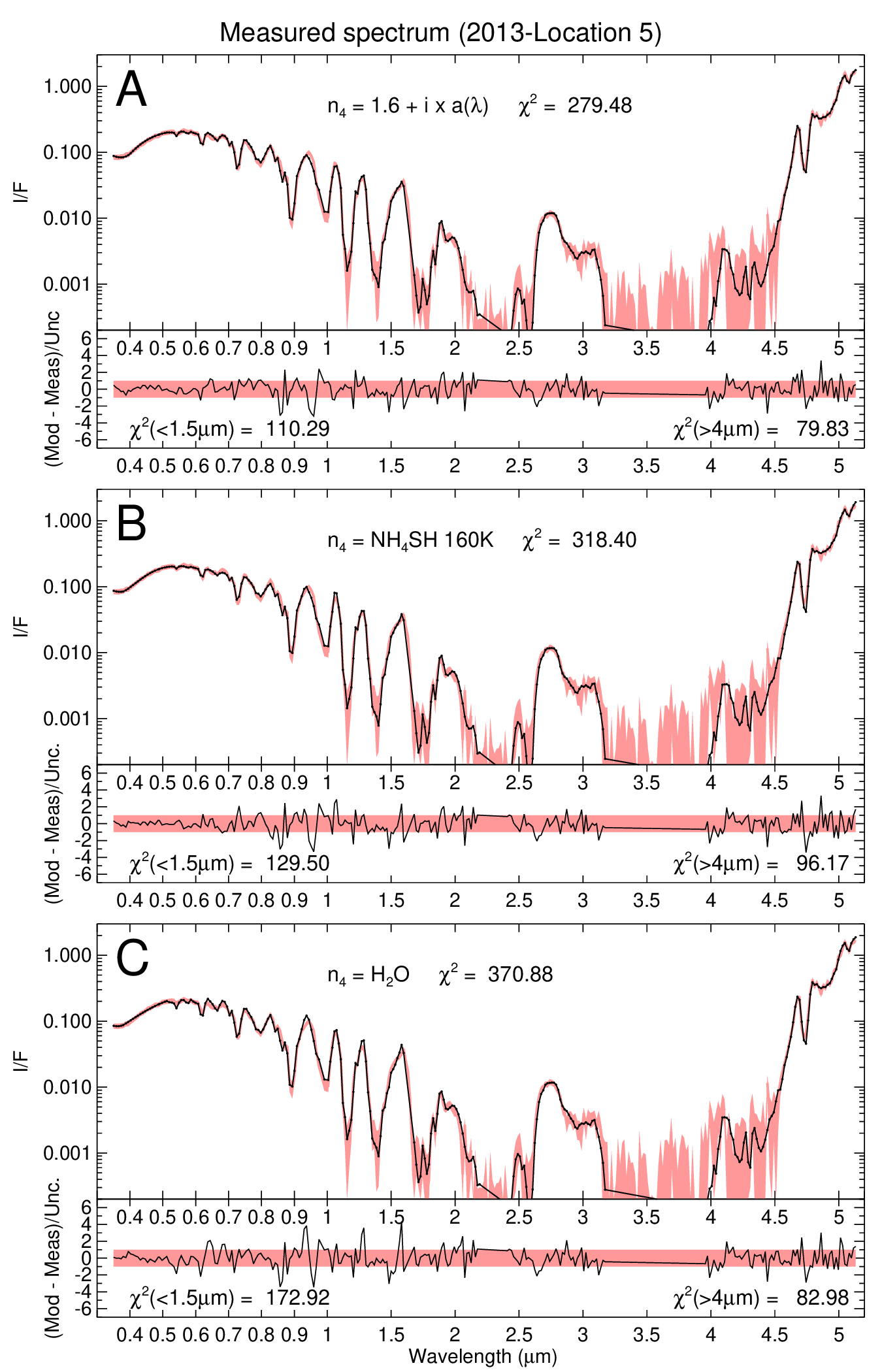}\hspace{-0.2in}
\includegraphics[width=3.25in]{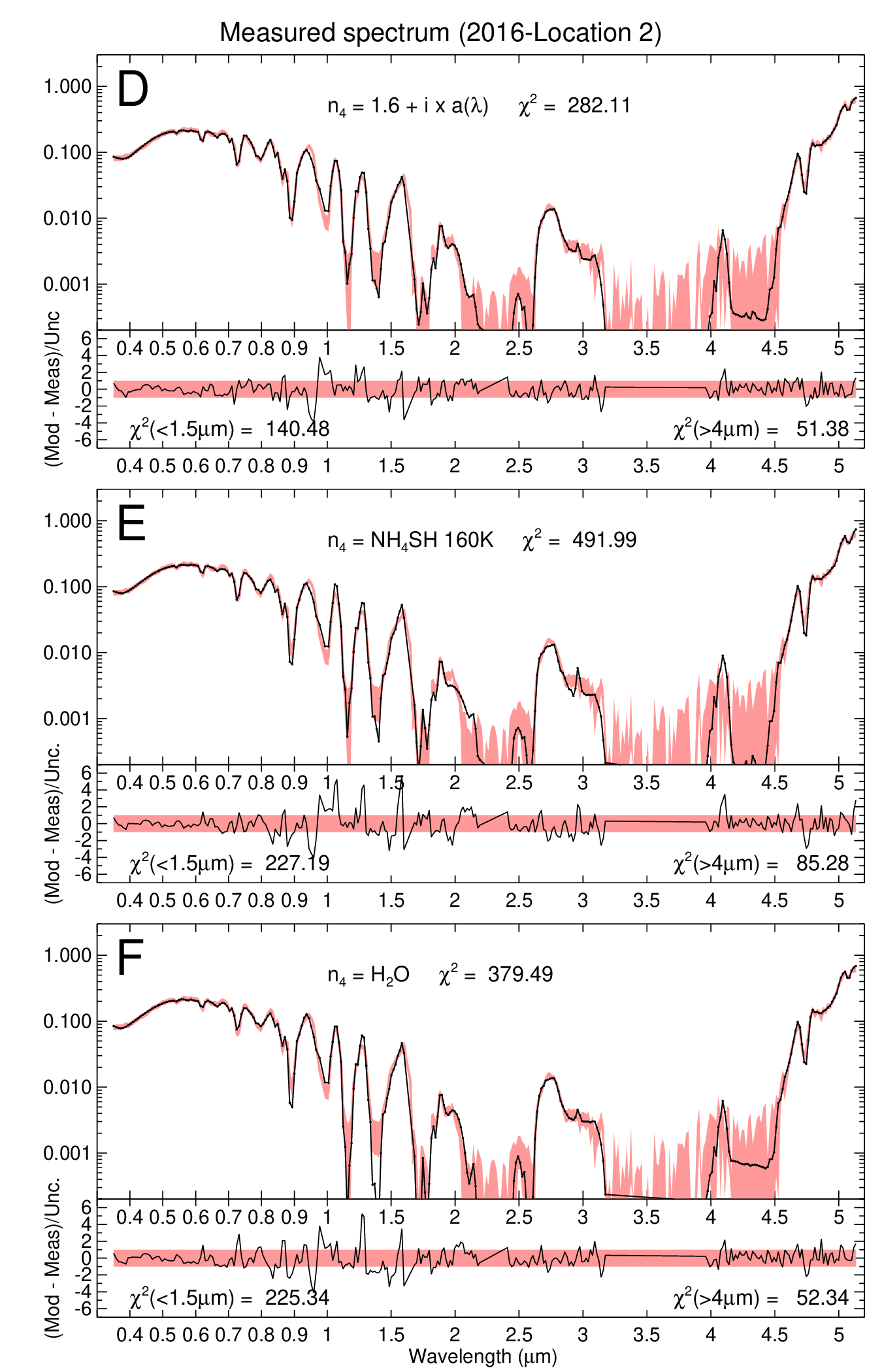}
\caption{Spectral fits to the 2013-5 (A-C) mid-hexagon spectrum and the 2016-2 eye spectrum (D-F) and
 using alternative composition models for deep cloud layers: Synthetic models (A and D),
 \nhfsh (B and E), and H$_2$O ice (C and F). The top figure in each panel displays the measured
spectrum as colored with shading width equal to the combined estimated uncertainty of measurement 
and modeling, and the model spectrum as a solid line.  In the bottom sub-figure of each panel
the model-measured difference ratio to uncertainty is plotted, with the shaded region indicating
the $\pm \sigma$ range.  These fits all used \nhtx$v1$ = 12 ppm. Note the \chisq computations for spectral
subregions of $\lambda <$ 1.5 \mum and $\lambda >$ 4.5 \mum shown at lower left and
lower right corners of the ratio subplots. }\label{Fig:deepfits} 
\end{figure*}

\begin{table*}[!htb]\centering
\caption{Best-fit parameters for alternate deep cloud (layer 4) composition models.}\label{Tbl:deepfits}
\begin{footnotesize}
\setlength\tabcolsep{2pt}

\begin{tabular}{|l|c c c c|c c c c|}
\hline\\[-0.08in]
Spectrum:  & \multicolumn{4}{|c|}{2013-Location 5 in Fig. 14 (81.6\degx N)} &\multicolumn{4}{|c|}{2016-Location 2 in Fig. 15 (89\degx N)} \\[0.05in]
5.12-\mum I/F:  & \multicolumn{4}{|c|}{1.4} &\multicolumn{4}{|c|}{0.7} \\[0.05in]
\hline\\[-0.05in]
                          &   {\bf   1.6 + }                &      1.6 +          &         \nhfsh       &             &   {\bf    1.6 +   }   &      1.6 +   &      \nhfsh   &   \\[0.05in]
Layer 4 index             &   {\bf  i $\times$ a($\lambda)$}  &  i  $\times$ b($\lambda$)  &      (160 K)  &      \hto  &  {\bf i $\times$ a($\lambda$)}  &   i   $\times$ b($\lambda$)   &   (160 K)    &       \hto\\[0.05in]
\hline\\[-0.05in]
 $p_1$ (bar)$\times 10^2$ &   3.7$^{+  1.5}_{-  1.1}$ &   3.2$^{+ 1.3}_{- 0.9}$ &   4.2$^{+ 0.9}_{- 0.7}$ &   3.8$^{+ 1.2}_{- 0.9}$ &   4.7$^{+ 1.6}_{- 1.2}$ &   5.0$^{+ 1.7}_{- 1.3}$ &   7.9$^{+ 1.5}_{- 1.3}$ &  14.9$^{+ 1.1}_{- 1.1}$\\[0.05in]
 $p_2$ (bar)$\times 10$ &   2.7$^{+  0.2}_{-  0.2}$ &   2.5$^{+ 0.2}_{- 0.2}$ &   2.6$^{+ 0.2}_{- 0.2}$ &   2.4$^{+ 0.2}_{- 0.2}$ &   2.7$^{+ 0.3}_{- 0.3}$ &   2.5$^{+ 0.3}_{- 0.3}$ &   2.6$^{+ 0.8}_{- 0.5}$ &   2.8$^{+ 0.5}_{- 0.4}$\\[0.05in]
 $p_3$ (bar)$\times 10$ &   9.6$^{+  0.3}_{-  0.2}$ &   9.4$^{+ 0.2}_{- 0.2}$ &   9.7$^{+ 0.3}_{- 0.3}$ &   9.4$^{+ 0.3}_{- 0.2}$ &   8.6$^{+ 0.4}_{- 0.4}$ &   8.2$^{+ 0.4}_{- 0.4}$ &   7.1$^{+ 0.4}_{- 0.3}$ &   8.8$^{+ 0.6}_{- 0.6}$\\[0.05in]
              $p_4$ (bar) &   3.2$^{+  0.3}_{-  0.2}$ &   3.1$^{+ 0.2}_{- 0.2}$ &   4.2$^{+ 0.0}_{- 0.0}$ &   3.2$^{+ 0.2}_{- 0.2}$ &   3.1$^{+ 0.0}_{- 0.0}$ &   3.1$^{+ 0.0}_{- 0.0}$ &   3.3$^{+ 0.1}_{- 0.1}$ &   2.8$^{+ 0.1}_{- 0.1}$\\[0.05in]
\hline\\[-0.1in]
$r_1$ (\mum)$\times 10^2$ &   9.3$^{+  1.6}_{-  1.4}$ &   6.9$^{+ 2.3}_{- 1.8}$ &   5.0$^{+ 2.0}_{- 1.5}$ &   7.2$^{+ 1.3}_{- 1.1}$ &   5.8$^{+ 3.6}_{- 2.2}$ &   5.8$^{+ 3.0}_{- 2.0}$ &  18.3$^{+ 1.3}_{- 1.3}$ &   3.6$^{+ 0.1}_{- 0.1}$\\[0.05in]
$r_2$ (\mum)$\times 10$ &   2.1$^{+  0.1}_{-  0.1}$ &   2.0$^{+ 0.1}_{- 0.1}$ &   2.2$^{+ 0.1}_{- 0.1}$ &   2.2$^{+ 0.1}_{- 0.1}$ &   2.4$^{+ 0.2}_{- 0.2}$ &   2.2$^{+ 0.2}_{- 0.2}$ &   0.6$^{+ 3.5}_{- 0.5}$ &   5.0$^{+ 0.5}_{- 0.5}$\\[0.05in]
             $r_3$ (\mum) &   3.0$^{+  0.4}_{-  0.3}$ &   3.1$^{+ 0.4}_{- 0.4}$ &   2.9$^{+ 0.3}_{- 0.3}$ &   3.4$^{+ 0.5}_{- 0.5}$ &   9.0$^{+ 0.6}_{- 0.6}$ &   8.6$^{+ 0.8}_{- 0.7}$ &  17.0$^{+ 0.5}_{- 0.5}$ &  12.4$^{+ 0.7}_{- 0.7}$\\[0.05in]
             $r_4$ (\mum) &  12.8$^{+  1.8}_{-  1.7}$ &  33.2$^{+ 0.8}_{- 1.3}$ &  33.0$^{+ 1.0}_{- 1.8}$ &  29.7$^{+ 0.8}_{- 1.0}$ &  11.8$^{+ 1.7}_{- 1.6}$ &  33.3$^{+ 0.7}_{- 1.1}$ &  44.5$^{+ 0.4}_{- 1.1}$ &  30.9$^{+ 0.9}_{- 1.1}$\\[0.05in]
\hline\\[-0.1in]
    $\tau_1$$\times 10^2$ &   5.1$^{+  1.3}_{-  1.1}$ &   4.3$^{+ 0.8}_{- 0.7}$ &   5.6$^{+ 1.0}_{- 0.9}$ &   5.3$^{+ 0.9}_{- 0.8}$ &   6.7$^{+ 1.5}_{- 1.2}$ &   6.6$^{+ 1.7}_{- 1.4}$ &  11.0$^{+ 2.4}_{- 2.0}$ &  30.2$^{+ 2.2}_{- 2.1}$\\[0.05in]
    $\tau_2$$\times 10$ &   3.6$^{+  0.3}_{-  0.3}$ &   3.4$^{+ 0.3}_{- 0.3}$ &   3.5$^{+ 0.3}_{- 0.3}$ &   3.2$^{+ 0.3}_{- 0.2}$ &   2.4$^{+ 0.2}_{- 0.2}$ &   2.3$^{+ 0.2}_{- 0.2}$ &   0.9$^{+ 0.3}_{- 0.2}$ &   0.6$^{+ 0.2}_{- 0.1}$\\[0.05in]
    $\tau_3$$\times 10$ &   1.7$^{+  0.1}_{-  0.1}$ &   1.7$^{+ 0.1}_{- 0.1}$ &   1.8$^{+ 0.2}_{- 0.2}$ &   1.6$^{+ 0.1}_{- 0.1}$ &   3.4$^{+ 0.3}_{- 0.2}$ &   3.3$^{+ 0.2}_{- 0.2}$ &   5.0$^{+ 0.3}_{- 0.3}$ &   4.2$^{+ 0.3}_{- 0.3}$\\[0.05in]
                 $\tau_4$ &   3.1$^{+  0.5}_{-  0.4}$ &   3.2$^{+ 0.4}_{- 0.3}$ &  50.0                  &   3.2$^{+ 0.4}_{- 0.3}$ &  50.0                &  50.0                 &  35.8$^{+14.2}_{-34.1}$ &   5.2$^{+ 0.7}_{- 0.6}$\\[0.05in]
\hline\\[-0.1in]
  $\lambda_1$ (\mumx) $ \times 10$&   3.5$^{+  0.1}_{-  0.1}$ &   3.5$^{+ 0.1}_{- 0.1}$ &   3.3$^{+ 0.1}_{- 0.1}$ &   3.3$^{+ 0.1}_{- 0.1}$ &   3.4$^{+ 0.1}_{- 0.1}$ &   3.5$^{+ 0.1}_{- 0.1}$ &   3.3$^{+ 0.2}_{- 0.2}$ &   3.2$^{+ 0.1}_{- 0.1}$\\[0.05in]
       $n_{i,1} \times 10^2$ &   6.4$^{+  1.2}_{-  1.0}$ &   5.1$^{+ 1.0}_{- 0.8}$ &   4.4$^{+ 1.8}_{- 1.3}$ &   4.2$^{+ 0.4}_{- 0.4}$ &   6.2$^{+ 4.4}_{- 2.6}$ &   6.4$^{+ 3.8}_{- 2.4}$ &  31.3$^{+35.1}_{-16.5}$ &   6.4$^{+ 1.7}_{- 1.3}$\\[0.05in]
      $K_1$ (\mumx$^{-1}$) &   5.8$^{+  0.8}_{-  0.7}$ &   6.2$^{+ 0.7}_{- 0.7}$ &   7.7$^{+ 1.6}_{- 1.4}$ &   3.2$^{+ 0.4}_{- 0.4}$ &   8.0$^{+ 2.3}_{- 1.8}$ &   8.3$^{+ 2.1}_{- 1.7}$ &   6.0$^{+ 1.8}_{- 1.4}$ &  12.2$^{+ 0.6}_{- 0.6}$\\[0.05in]
\hline\\[-0.1in]
           \pht $p_b$ (bar) &  1.12$^{+ 0.09}_{- 0.07}$ &  1.09$^{+0.08}_{-0.07}$ &  1.21$^{+0.09}_{-0.08}$ &  1.14$^{+0.07}_{-0.08}$ &  0.77$^{+0.12}_{-0.11}$ &  0.68$^{+0.11}_{-0.09}$ &  0.62$^{+0.12}_{-0.10}$ &  0.68$^{+0.15}_{-0.12}$\\[0.05in]
     \pht $\alpha_0$  (ppm) &  5.37$^{+ 0.51}_{- 0.47}$ &  5.23$^{+0.51}_{-0.47}$ &  7.04$^{+0.38}_{-0.36}$ &  5.96$^{+0.43}_{-0.40}$ &  6.35$^{+0.61}_{-0.56}$ &  6.16$^{+0.56}_{-0.51}$ &  7.12$^{+0.45}_{-0.42}$ &  5.50$^{+0.68}_{-0.62}$\\[0.05in]
 \asht VMR (ppb) &  2.27$^{+ 0.50}_{- 0.44}$ &  2.29$^{+0.48}_{-0.42}$ &  1.95$^{+0.35}_{-0.31}$ &  1.70$^{+0.33}_{-0.28}$ &  1.75$^{+0.45}_{-0.37}$ &  1.65$^{+0.47}_{-0.38}$ &  2.26$^{+0.53}_{-0.46}$ &  1.55$^{+0.39}_{-0.32}$\\[0.05in]
 \hline\\[-0.1in]
$\chi^2$ &   {\bf279.48} & 277.08 & 318.40 & 370.88 & {\bf  282.11} & 297.91 & 491.99 & 379.49\\[0.05in]
$\chi^2/N_F$ &   {\bf 1.09} &   1.08 &   1.24 &   1.45 &   {\bf 1.10} &   1.16 &   1.92 &   1.48\\[0.05in]
\hline
\end{tabular}
\parbox[t]{5.7in}{NOTE: The chromophore parameters are defined in Eqs. \ref{Eq:chromod} and \ref{Eq:boxav}, 
among which we assumed fixed values of $K_2$ = 20/\mumx, and $n_{i,0} = 5\times 10^{-4}$. Bold is used in
first and fifth columns to indicate our preferred solution. These fits all used \nht$v1$ = 12 ppm.
 Here $N_F$ is the number of degrees of freedom (number of
constraints less the number of fitted parameters). }
\end{footnotesize}
\end{table*}

To investigate this issue further we considered four different
composition models for the bottom cloud layer, all of which are
plotted in Fig.\ \ref{Fig:varic}: (1) our default synthetic model
characterized by $n_4 = 1.6 +i\times a(\lambda)$, (2) a variant
synthetic model for which\ $n_4 = 1.6 +i\times b(\lambda)$, (3)
\nhfshx, and (4) \htox.  We tried to constrain these models by fitting
observed spectra from regions with low upper aerosol opacity so that
the properties of the deep layer would be less attenuated.  We chose a
mid hexagon spectrum from 2013 location 5 (shown in Figs. \ref{Fig:vimsvis2013},
 \ref{Fig:vimsnir2013} and \ref{Fig:2013combined}), which has a relatively
high 5-\mum emission (and presumable low deep cloud opacity) and
another spectrum from location 2 in 2016, which has about half the
5-\mum emission of the first case.  The observed and fitted spectra
are displayed in Fig.\ \ref{Fig:deepfits} for 3 of the four models.
The best-fit parameter values for all of the models are listed in
Table\ \ref{Tbl:deepfits}.  The table shows that models (1) and (2)
have closely similar parameter values and fit qualities, which is why
model (2) was omitted from the spectral plots of
Fig.\ \ref{Fig:deepfits}.


The first column of Fig.\ \ref{Fig:deepfits} displays fits to
the spectrum from 2013 location 5, which is in the middle of the hexagon, in a region
with higher thermal emission and thus likely reduced deep cloud opacity.  In this case
\nhfsh (\chisq = 318.40) provides a better fit than water ice  (\chisq = 370.88), with
our default synthetic index providing significantly better overall fit (\chisq = 279.48), as well
as beating them both in short wavelength and long wavelength subregions.  We also fit
these models to a 2016 spectrum from the middle of the hexagon (location 6) in a
region with higher upper level aerosol opacity. In this case (not shown if 
Fig.\ \ref{Fig:deepfits}) the three models all
achieved comparable fit qualities, with our synthetic index providing the best
fit, but only slightly better than \nhfshx, which was only slightly better than \htox.

The second column of Fig.\ \ref{Fig:deepfits} displays results for location 2 in the eye region in 2016,
which has a low burden of aerosols above the deep layer, but provides a significant
attenuation of Saturn's thermal emission, presumably due to the deep cloud.
In this case models assuming a deep cloud made of \hto provided
a somewhat better fit (\chisq = 379.49) than a cloud made of \nhfsh (\chisq = 491.99),
 and both of these compositions
were soundly beaten by our standard model (1) with \chisq = 282.11. Here the main advantage
of \hto ice over \nhfsh is in the thermal emission region, with \chisq contributions
for $\lambda > 4.5$ \mum of only 52.34 vs. 85.28 for \nhfshx. 


Our synthetic index models seem to beat the single-component compositions for the deep cloud
because they contain important short-wavelength absorption lacking in water ice, as well
as significant and spectrally flat long-wavelength absorption lacking in \nhfshx.  Our synthetic
models seem to simulate the effects of combining both water ice and \nhfshx, which suggests that
the real deep cloud is some combination of these two materials, possibly a coating of \nhfsh
on top of water ice, or possibly a water cloud layer underneath a layer of \nhfsh particles.
In the left column of Fig.\ \ref{Fig:deepfits} it appears that the deep layer might have \nhfsh as the
dominant component, while in the right column perhaps \hto dominates.
However, our attempts to fit the spectra with these two layers in close proximity did
not even reach the fit quality achieved with single-layer models.  The problem
seems to be mainly in the wavelength region between 1 and 1.5 \mumx.  More work is needed
in finding combinations of these two substances. Coated particles are prohibitively
costly in computation time for the large particle sizes that seem to be required in this layer.

\begin{figure*}[!ht]\centering
\hspace{-0.15in}\includegraphics[width=5.in]{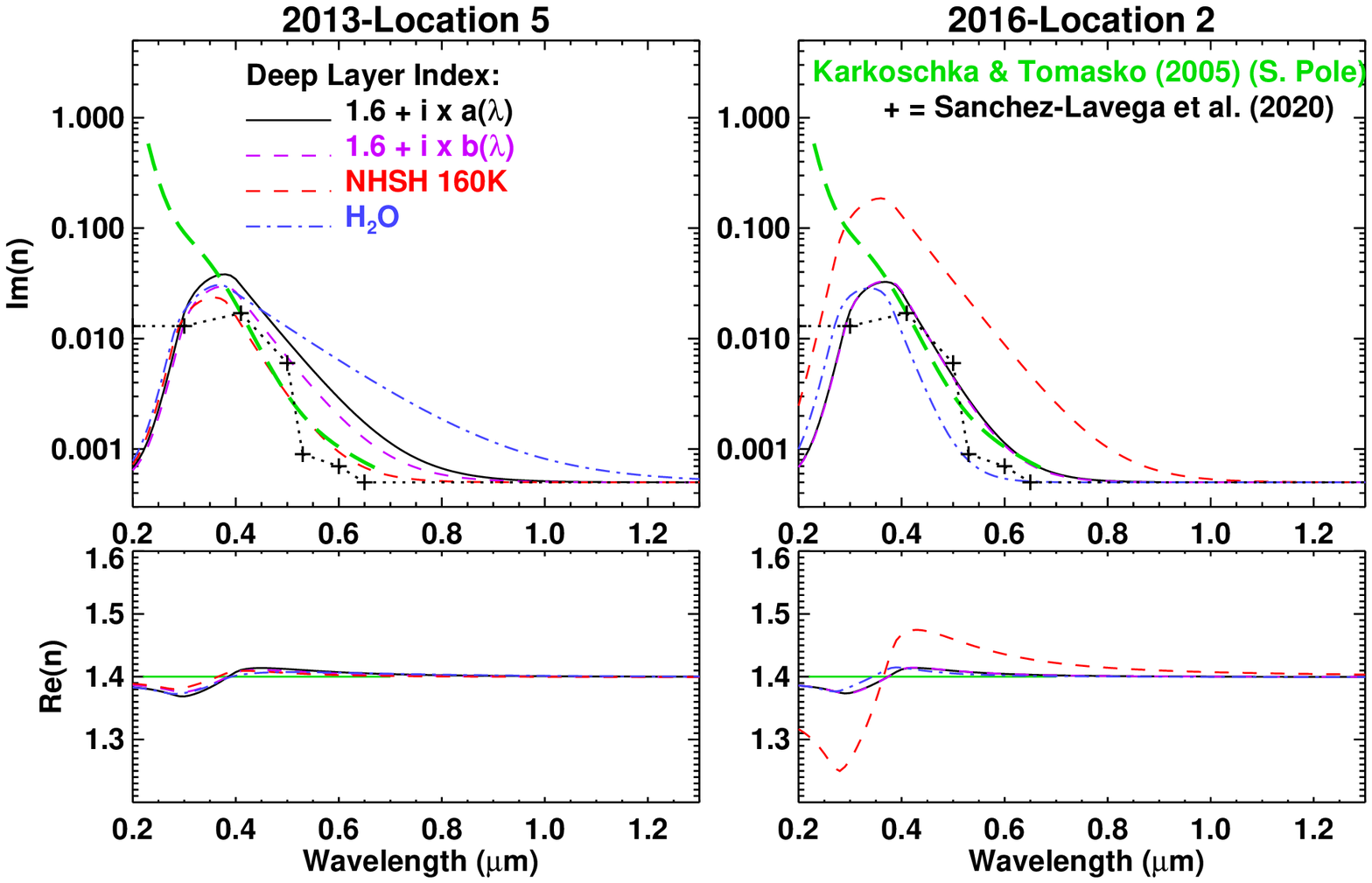}
\caption{Chromophore spectral model fits from Table 4 for four different assumed deep cloud compositions.}\label{Fig:chromdeep} 
\end{figure*}

Changing the deep cloud model composition does have an effect on other parameters in the fit, as displayed in
Table\ \ref{Tbl:deepfits}.  For a deep cloud composed of only \nhfshx, blocking thermal radiation from
the deep atmosphere requires a much larger optical depth of layer 4 than is needed by a deep cloud of
water vapor because of the latter's more significant imaginary index.  The pressures of most
layers do not vary much for different models, with the exception of the pressure of the ammonia
layer (layer 3) for fits to the eye spectrum with a deep layer composition of \nhfshx, which is
a low 0.71 bar compared to 0.82 and 0.88 bars for the other models. That model also has 
deviant values for particle radii.   The fits to the 2013-Location 5 spectrum (left 4 columns in Fig.\ \ref{Fig:deepfits})
show generally much less variability due to differences in deep layer models (with the exception
of the optical depth of the deep layer for \nhfshx). There are also significant differences
in the retrieved chromophore parameters. The resultant differences in wavelength dependence
of refractive index components are shown in Fig.\ \ref{Fig:chromdeep}. The outlier for the 2013 fits
is the model in which the bottom layer is pure water ice, in imaginary index (blue dot-dash curve)
as well as in \chisqx.  The outlier for the 2016 fits is the model in which the bottom layer is made of
pure \nhfshx, also both in imaginary index (red dot-dash curve) and in \chisqx.



\subsection{Vertical location of Saturn's chromophore.}\label{Sec:vertloc}

 \cite{Kark2005} used HST observations to make a
strong case for absorption of shortwave light in the stratosphere, and were
successful in modeling their observations with spherical particles of
 the same composition
 in both the stratosphere and upper troposphere. \cite{Perez-Hoyos2005}
also placed short wave absorbers in both layers.  The analysis closest in
temporal and spatial coverage to our observations, is that of \cite{Sanz-Requena2018},
which is based on Cassini ISS observations of the north polar region in 2013. That 
analysis found extremely low shortwave absorption in the upper troposphere,
and stronger absorption in the stratosphere, but with a much flatter
spectral variation than found by \cite{Kark2005}.  Stratospheric
absorption in the north polar stratosphere was also confirmed
by the detection of a UV dark polar hexagon at 180 nm by
the Cassini Ultraviolet Imaging Spectrometer \citep{Pryor2019JGRE}.

At early stages of the modeling, we placed the chromophore in the
stratosphere, with the top (stratospheric) layer of our model
comprised entirely of chromophore particles. That worked reasonably
well in fitting the observed VIMS spectra, but given the variability
in prior results and the nature of our \pht retrievals, we also
considered other options.  The fact that our fits generally found a
relatively rapid decline of \pht with altitude above the putative
diphosphine cloud suggested that the stratospheric haze might not be
providing much protection against the destruction of \pht by solar UV light.
Another finding, i.e. that the putative diphosphine cloud did seem to
provide that protection, suggested that diphosphine itself might
be the chromophore, placing it well below the stratosphere.  That
model would reduce the attenuation of longwave UV light above that
cloud, promoting the destruction of \pht above it, which seemed more
compatible with our \pht results, although this ignores the important
role of eddy mixing in controlling the \pht profile.  More evidence on
this topic is provided by the \cite{Braude2019EPSC} analysis of
VLT/MUSE 480-930 nm spectra, which placed the chromophore within or
just above the tropospheric haze, although their rather long
short-wavelength limit of 480 nm might make them less able to
constrain stratospheric contributions.  If our layer 2 is actually
composed of diphosphine and does not act as the chromophore at
wavelengths longer than 0.35 \mumx, it would still absorb significant
amounts of UV radiation up to $\sim$260 nm \citep{Ferris1981}, and
thus would still provide considerable protection against \pht destruction.
  Although the VIMS observations do not extend to wavelengths
below 350 nm, and thus cannot directly constrain the degree of UV
protection provided by either stratospheric or upper tropospheric
layers, they do have the potential to constrain the location of the
color-producing absorber. In the following subsection we describe
our attempt to constrain these possibilities using VIMS observation
to  evaluate four alternative vertical distributions of the chromophore.

\subsubsection{Fit results for alternative models of chromophore vertical distribution}
\begin{figure*}[!ht]\centering
\hspace{-0.15in}\includegraphics[width=6.3in]{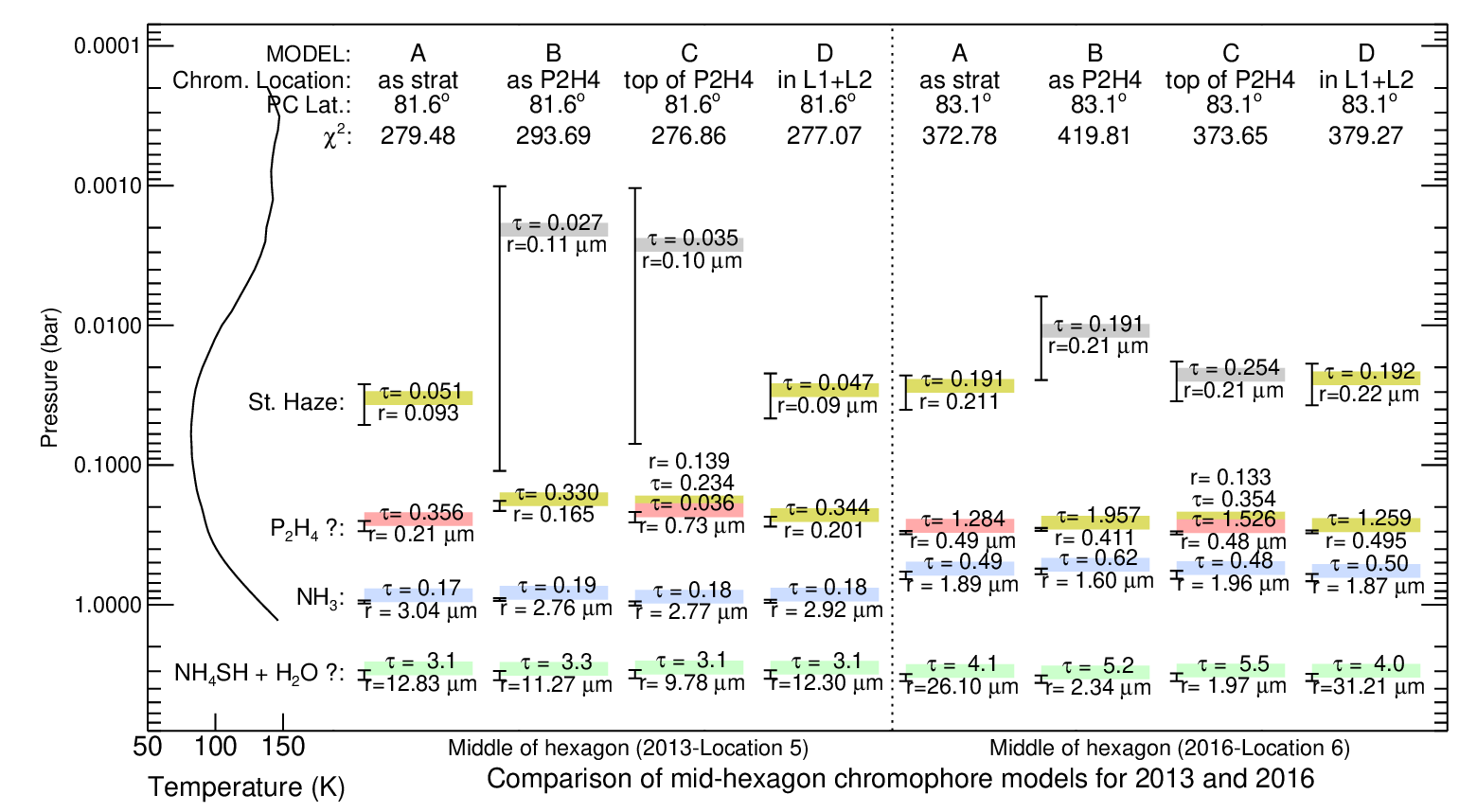}
\caption{Alternative best-fit vertical structure model fits to mid hexagon
spectra for 2013 location 5 (left quartet) and 2016 location 6 (right quartet). A gold
color indicates the location of the chromophore: in the stratosphere (first of each
quartet), in the putative diphosphine layer (2nd of each quartet), attached
to the top of the putative diphosphine layer (3rd of each quartet, or
in both layers 1 and 2 (4th of each quartet). These fits all used \nht$v1$ = 12 ppm. 
}\label{Fig:chromcartoon} 
\end{figure*}

To try to constrain the vertical location of the chromophore, we
 focused on four model structures:
(A) a four-layer model with the chromophore in the stratosphere; (B) a
four-layer model with the chromophore consisting of the
particles in the putative diphosphine layer (our lack of knowledge of
the optical properties of diphosphine makes it impossible to rule
out this possibility); (C) a
5-layer model with a high stratospheric haze providing some shortwave
absorption, but most of the color provided by a separate chromophore
layer just above the putative diphosphine layer, and (D) a 4-layer model in
which the compositions of Layer 1 and Layer 2 were identical, an assumption
made by both \cite{Kark2005} and \cite{Sanchez-Lavega2020hazes}. These structures are
illustrated in Fig.\ \ref{Fig:chromcartoon}, with best-fit parameter values given in
Table\ \ref{Tbl:chromcomp}. (Spectral plots are not shown because the different model
cases are difficult to distinguish visually.)

We did one set of fits for location 5 in 2013 (left four columns in
Fig.\ \ref{Fig:chromcartoon} and Table\ \ref{Tbl:chromcomp}) and
another for location 6 in 2016 (right four columns in
Fig.\ \ref{Fig:chromcartoon} and Table\ \ref{Tbl:chromcomp}).  For
both spectra, the model with the worst fit quality is the 4-layer
model that assumes that the putative diphosphine layer is entirely
made of chromophore particles (Model B).  The other three options,
i.e. a 4-layer model with the chromophore in the stratosphere (Model A) 
or a 5-layer model with a chromophore layer just above the putative
diphosphine layer (Model C), and a 4-layer model with chromophore in
both Layer 1 and Layer 2 (Model D), provide comparable fit qualities,
but very different chromophore properties, as listed in
Table\ \ref{Tbl:chromcomp} and displayed in the spectral plot in
Fig.\ \ref{Fig:imagchrom}.

\subsubsection{Evaluation of alternative chromophore models}
Placing the chromophore either within the putative diphosphine layer
(Model B in Fig.\ \ref{Fig:chromcartoon} ) or as a separate layer just
above the diphosphine layer (Model C), has the potential virtue of
being more compatible with our findings regarding the vertical
distribution of phosphine gas.  That distribution is characterized by
a deep mixing ratio, a pressure breakpoint, and above that breakpoint
a rapid decline with altitude.  Because UV radiation dissociates
\phtx, putting the UV blocking chromophore close to the \pht
breakpoint is plausible. That allows dissociation of \pht above the
chromophore, and protects \pht below that layer.  In fact, it seemed
plausible that the chromophore is being created by the very UV
photolysis that is breaking down the phosphine gas above that layer.
However, the fit quality is worst for the case in which the
chromophore is distributed throughout layer 2 (Model B).  And, as noted earlier,
\pthf has significant UV absorption even without the longer wavelength
absorption needed to serve as a chromophore to shape the visible
spectrum.  The fit qualities for the other two options (Models A and D) are
significantly better and comparable to each other.

\begin{table*}[!htb]\centering
\caption{Best-fit parameters for alternate chromophore vertical structure models.}\label{Tbl:chromcomp}
\renewcommand{\baselinestretch}{0.8}
\begin{footnotesize}
\setlength\tabcolsep{2pt}
\begin{tabular}{|l | c c c c | c c c c|}
\hline\\[-0.05in]
Spectrum:  & \multicolumn{4}{c|}{2013-Location 5 (81.6\degx N)} &\multicolumn{4}{c|}{2016-Location 6 (83.1\degx N)} \\[0.05in]
\hline\\[-0.05in]
MODEL:   & A & B & C & D  & A & B & C & D   \\[0.05in]
\hline\\[-0.05in]
   Chrom. Layer:    &          1            &        2      &    2C$^*$      & 1 + 2  &     1          & 2  &  2C   & 1 + 2     \\[0.05in]
       $\lambda_1$$\times 10^1$ &   3.5$^{+  0.1}_{-  0.1}$ &   2.6$^{+ 0.9}_{- 0.4}$ &   2.2$^{+ 2.8}_{- 0.2}$ &   3.4$^{+ 0.1}_{- 0.1}$ &   3.5$^{+ 0.1}_{- 0.1}$ &   3.4$^{+ 1.9}_{- 1.2}$ &   2.6$^{+ 2.9}_{- 0.5}$ &   3.2$^{+ 0.5}_{- 0.5}$\\[0.05in]
                    $n_{i,1}$ &   0.064$^{+  0.01}_{-  0.01}$ &   2.5$^{+ 3.0}_{- 1.9}$ &   6.0$^{+ 8.9}_{- 4.2}$ &   0.12$^{+ 0.02}_{- 0.02}$ &   0.2$^{+ 0.1}_{- 0.0}$ &   0.4$^{+ 7.5}_{- 0.4}$ &   4.5$^{+23.0}_{- 4.4}$ &   0.4$^{+ 0.9}_{- 0.3}$\\[0.05in]
                      $K_1$ &   5.8$^{+  0.8}_{-  0.7}$ &  10.9$^{+ 1.4}_{- 1.3}$ &  10.5$^{+ 1.5}_{- 1.4}$ &   7.3$^{+ 1.2}_{- 1.0}$ &   5.2$^{+ 0.6}_{- 0.6}$ &  13.1$^{+ 3.1}_{- 2.7}$ &  11.6$^{+ 3.7}_{- 3.0}$ &   8.5$^{+ 0.9}_{- 0.9}$\\[0.05in]
\hline\\[-0.1in]
 $p_1$ (bar)$\times 10^2$ &   3.7$^{+  1.5}_{-  1.1}$ &  0.23$^{+10}_{-0.13}$&  0.30$^{+6.8}_{-0.19}$ & 3.26$^{+1.4}_{-1.1}$ &  3.0$^{+ 1.0}_{- 0.8}$ &  1.22$^{+1.24}_{-0.60}$  &  2.53$^{+1.1}_{-0.7}$ & 2.68$^{+1.06}_{-0.80}$ \\[0.05in]
 $p_{2C}$ (bar) &                         &                         &  0.9$\times p_2$ & &                       &                     &  0.9$\times p_2$ &\\[0.05in]
 $p_2$ (bar) &   0.27$^{+  0.02}_{-  0.02}$ &   0.197$^{+ 0.17}_{- 0.16}$   & 0.24$^{+ 0.02}_{- 0.02}$ & 0.255$^{+0.02}_{-0.02}$  & 0.30$^{+ 0.01}_{- 0.01}$ &   0.29$^{+ 0.07}_{- 0.07}$ &   0.31$^{+ 0.01}_{- 0.01}$&  0.30$^{+ 0.01}_{- 0.01}$ \\[0.05in]
 $p_3$ (bar)$\times 10^1$ &   9.6$^{+  0.3}_{-  0.2}$ &   9.2$^{+ 0.2}_{- 0.2}$ &   9.8$^{+ 0.3}_{- 0.3}$ &   9.5$^{+ 0.2}_{- 0.2}$ &   6.2$^{+ 0.4}_{- 0.4}$ &   5.8$^{+ 0.3}_{- 0.3}$ &   6.1$^{+ 0.4}_{- 0.4}$ &   6.4$^{+ 0.4}_{- 0.4}$\\[0.05in]
              $p_4$ (bar) &   3.2$^{+  0.3}_{-  0.2}$ &   3.2$^{+ 0.3}_{- 0.2}$ &   3.2$^{+ 0.2}_{- 0.2}$ &   3.2$^{+ 0.2}_{- 0.2}$ &   3.3$^{+ 0.2}_{- 0.2}$ &   3.4$^{+ 0.2}_{- 0.2}$ &   3.3$^{+ 0.2}_{- 0.2}$ &   3.3$^{+ 0.2}_{- 0.2}$\\[0.05in]
\hline\\[-0.1in]
$r_1$ (\mum)$\times 10^2$ &   9.3$^{+  1.6}_{-  1.4}$ & 10.5$^{+4.4}_{-3.2}$  &  9.8$^{+3.1}_{-2}$ &  9.4$^{+1.2}_{-1}$ &  21.1$^{+ 1.6}_{- 1.6}$ & 20.6$^{+1.7}_{-1.6}$ & 20.6$^{+1.6}_{-1.5}$ & 21.7$^{+1.5}_{-1.4}$\ \\[0.05in]
$r_{2C}$ (\mum)$\times 10^1$ &  & &  1.39$^{+0.12}_{-0.11}$ && &  &  1.33$^{+0.5}_{-0.4}$ &\\[0.05in]
$r_2$ (\mum)$\times 10^1$ &   2.1$^{+  0.1}_{-  0.1}$ &   1.65$^{+ 0.07}_{- 0.07}$    &   7.3$^{+ 1.4}_{- 2.0}$ &  2.0$^{+ 0.1}_{-0.1}$  &  4.9$^{+ 0.1}_{- 0.1}$ &  4.11$^{+ 0.1}_{-0.1}$    &   4.8$^{+ 0.2}_{- 0.2}$ &  4.95$^{+0.1}_{-0.1}$\\[0.05in]
             $r_3$ (\mum) &   3.0$^{+  0.4}_{-  0.3}$ &   2.8$^{+ 0.3}_{- 0.2}$ &   2.8$^{+ 0.4}_{- 0.4}$ &   2.9$^{+ 0.3}_{- 0.3}$ &   1.9$^{+ 0.3}_{- 0.3}$ &   1.6$^{+ 0.2}_{- 0.2}$ &   2.0$^{+ 0.4}_{- 0.3}$ &   1.9$^{+ 0.3}_{- 0.3}$\\[0.05in]
             $r_4$ (\mum) &  12.8$^{+  1.8}_{-  1.7}$ &  11.3$^{+ 1.7}_{- 1.6}$ &   9.8$^{+ 2.2}_{- 2.0}$ &  12.3$^{+ 2.0}_{- 1.8}$ &  26.1$^{+ 3.5}_{- 8.6}$ &   2.3$^{+ 0.6}_{- 0.5}$ &   2.0$^{+ 0.6}_{- 0.5}$ &  31.2$^{+ 7.1}_{- 9.9}$\\[0.05in]
 \hline\\[-0.1in]
   $\tau_1$$\times 10^2$ &   5.1$^{+  1.3}_{-  1.1}$ & 2.7$^{+1.4}_{-0.9}$  &  3.5$^{+1.5}_{-1.1}$ &  4.72$^{+1.37}_{-1.06}$ & 19.1$^{+ 2.8}_{- 2.4}$ & 19.2$^{+2.6}_{-2.4}$ & 25.4$^{+3.3}_{-3.0}$ & 19.19$^{+2.66}_{-2.34}$\\[0.05in]
    $\tau_2$$\times 10^1$ &   3.6$^{+  0.3}_{-  0.3}$ &    3.3$^{+ 0.3}_{-0.3}$ &   0.4$^{+ 0.5}_{- 0.2}$ &   0.34$^{+0.03}_{- 0.03}$ &  12.8$^{+ 0.8}_{- 0.7}$ &  19.6$^{+1.4}_{-1.3}$  &  15.3$^{+ 1.8}_{- 1.7}$ &   12.6$^{+ 0.7}_{- 0.8}$\\[0.05in]
    $\tau_3$$\times 10^1$ &   1.7$^{+  0.1}_{-  0.1}$ &   1.9$^{+ 0.1}_{- 0.1}$ &   1.8$^{+ 0.3}_{- 0.2}$ &   1.8$^{+ 0.1}_{- 0.1}$ &   4.9$^{+ 0.6}_{- 0.5}$ &   6.2$^{+ 0.8}_{- 0.7}$ &   4.8$^{+ 0.8}_{- 0.7}$ &   5.0$^{+ 0.6}_{- 0.5}$\\[0.05in]
                 $\tau_4$ &   3.1$^{+  0.5}_{-  0.4}$ &   3.3$^{+ 0.5}_{- 0.4}$ &   3.1$^{+ 0.4}_{- 0.3}$ &   3.1$^{+ 0.4}_{- 0.4}$ &   4.1$^{+ 0.8}_{- 0.7}$ &   5.2$^{+ 1.4}_{- 1.0}$ &   5.5$^{+ 1.9}_{- 1.3}$ &   4.0$^{+ 0.9}_{- 0.7}$\\[0.05in]
\hline\\[-0.1in]
   Im($n_1$)$\times 10^2$ & & 1.03$^{+0.59}_{-0.39}$ & 1.29$^{+0.44}_{-0.34}$ & & & 3.1$^{+0.3}_{-0.3}$ &  3.29$^{+0.38}_{-0.34}$&\\[0.05in]
    \pht $p_b$ (bar) &  1.12$^{+ 0.09}_{- 0.07}$ &  1.08$^{+0.07}_{-0.07}$ &  1.11$^{+0.10}_{-0.09}$ &  1.12$^{+0.08}_{-0.08}$ &  0.28$^{+0.02}_{-0.02}$ &  0.30$^{+0.02}_{-0.02}$ &  0.27$^{+0.02}_{-0.02}$ &  0.28$^{+0.02}_{-0.02}$\\[0.05in]
  \pht $\alpha_0$    (ppm) &  5.37$^{+ 0.51}_{- 0.47}$ &  5.35$^{+0.53}_{-0.47}$ &  5.34$^{+0.56}_{-0.52}$ &  5.39$^{+0.53}_{-0.49}$ &  4.95$^{+0.34}_{-0.32}$ &  4.95$^{+0.38}_{-0.35}$ &  4.75$^{+0.38}_{-0.35}$ &  4.94$^{+0.34}_{-0.31}$\\[0.05in]
    \asht VMR (ppb) &  2.27$^{+ 0.50}_{- 0.44}$ &  2.14$^{+0.50}_{-0.42}$ &  2.19$^{+0.48}_{-0.41}$ &  2.18$^{+0.48}_{-0.41}$ &  1.67$^{+0.39}_{-0.33}$ &  1.63$^{+0.34}_{-0.29}$ &  1.59$^{+0.35}_{-0.29}$ &  1.63$^{+0.37}_{-0.31}$\\[0.05in]
\hline\\[-0.1in]
$\chi^2$ & 279.48 & 293.69 & 276.86 & 277.07 & 372.78 & 419.81 & 373.65 & 379.27 \\[0.05in]
$\chi^2/N_F$ &   1.09 &   1.15 &   1.09 &   1.08 &   1.46 &   1.65 &   1.48 &   1.48\\[0.05in]
\hline
\end{tabular}
\end{footnotesize}
\parbox[c]{6in}{$^*$2C denotes a chromophore layer placed at the top of layer 2, with base pressure $p_{C}$ = 0.9$\times p_2$ and top pressure at 90\% of base pressure.  These fits all used \nht$v1$ = 12 ppm.}
\end{table*}

In the case of Model B  (the chromophore being the constituent
that makes up the putative diphosphine layer), it is also necessary to have
a stratospheric layer that performs two spectral shaping functions: filling
in the deep methane absorption bands as needed and absorbing some
of the light reflected by the putative diphosphine layer.  For this option
the stratospheric layer is treated as a pseudo-gray layer, which
means using a wavelength-independent imaginary index, which produces a useful
amount of absorption at short wavelengths, while not absorbing too much
light at longer wavelengths.  

Model D above (the stratospheric
and upper tropospheric layers composed of the
same material, with the same wavelength-dependent refractive index) is
not attractive if diphosphine is the material, because it is hard to conceive of a diphosphine
layer created in the stratosphere, where the phosphine abundance is expected to be vanishingly small.

\begin{figure*}[!ht]\centering
\includegraphics[width=5in]{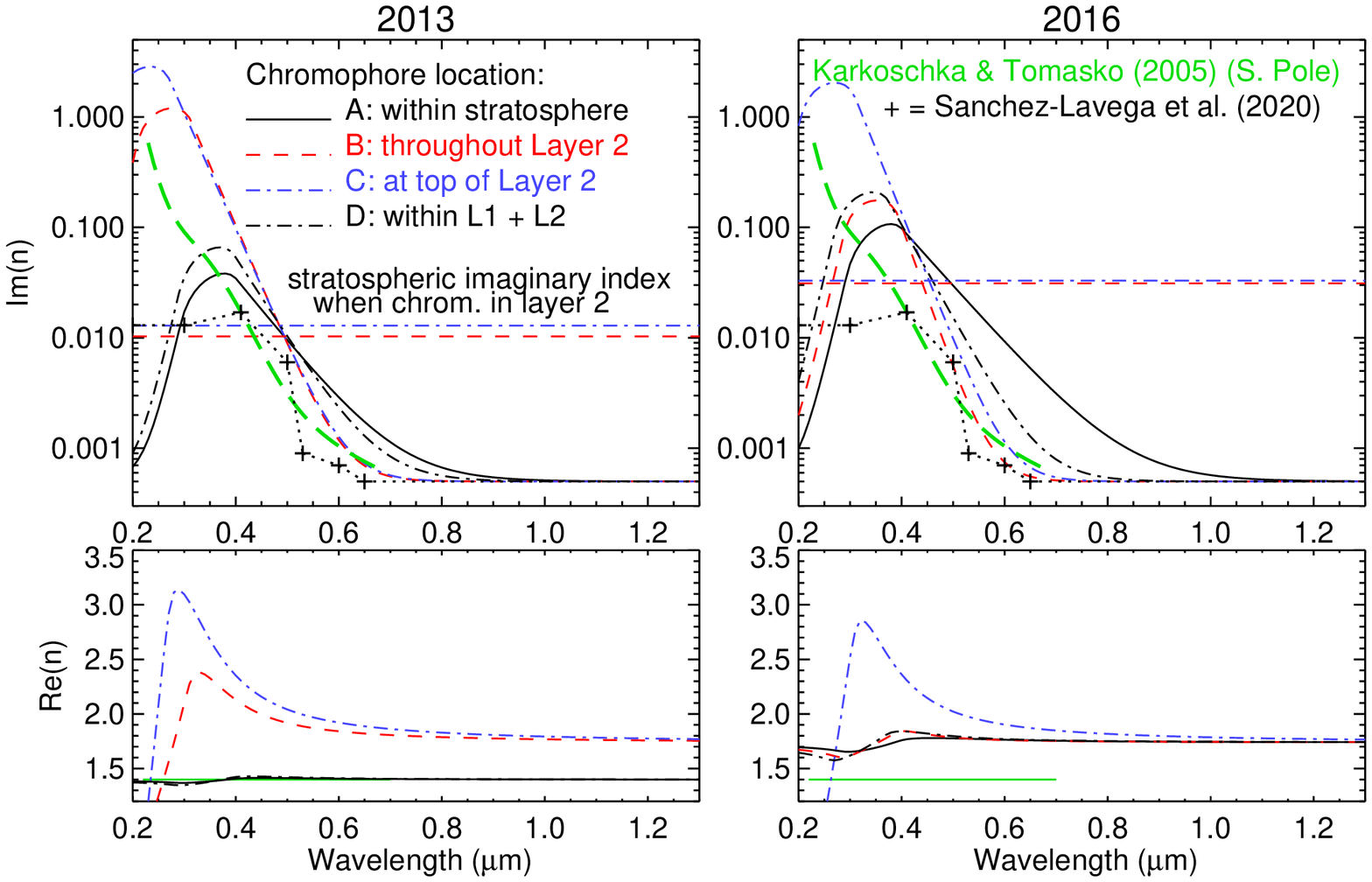}
\caption{Imaginary refractive index (upper) and real index (lower) for
  best-fit 2013 (left) and 2016 (right) chromophore models in which
  the chromophore is in the stratosphere (A), throughout layer 2
  (B), immediately above layer 2 (C), or within top two layers (D).
 Parameters and uncertainties for these mid-hexagon fits
  are given in Table\ \ref{Tbl:chromcomp}. Chromophore index models of
  \cite{Kark2005} (green, long dash, for south pole) and
  \cite{Sanchez-Lavega2020hazes} (+, based on HST nadir observations)
  are also shown.}\label{Fig:imagchrom}
\end{figure*}

A problem with placing the chromophore entirely within layer 2 (Model B) or at the top of
layer 2 (Model C) is that the imaginary index required for the chromophore then becomes extremely large,
reaching peak values exceeding 1.0, with accompanying large variations in the real
index, as illustrated in Fig.\ \ref{Fig:imagchrom}. Even more extreme values are
needed outside the hexagon for the 5-layer model.  The other problem is that \cite{Kark2005}, using
HST imaging observations covering a wide range of filter peak wavelengths from 0.23 \mum to 2.37 \mumx, 
established that the 
stratospheric haze contained aerosols with significant UV absorption. Their HST-based
stratospheric haze index model, also shown in Fig.\ \ref{Fig:imagchrom}, has a
spectral slope similar to our stratospheric model, but does not show a peak below 0.4 \mumx,
instead continuing upward to rather large values at their shortest wavelength.
However, the imaginary index variation they inferred
is for an assumed constant real index, which violates the Kramers-Kronig relation.
Another difference from our modeling is that they assumed the same refractive
index model for tropospheric particles as for the stratosphere and adopted
a fixed radius of 0.08 \mumx, while we allowed that parameter to adjust as needed
to improve the fit.  Inside the hexagon we found aerosol radii from 0.06 \mum to 0.12 \mum in
2013, when optical depths were very low, and a nearly uniform radius of 0.2 \mum
in 2016, when optical depths were much larger.  It is not clear to what degree
these differences in analysis can account for differences in the inferred
stratospheric refractive index.  However, it is worth noting that the refractive index
model provided in supplemental material associated with the \cite{Sanchez-Lavega2020hazes}
paper, which is based on an analysis of nadir HST observations, and also assumed
the same index for stratospheric and tropospheric particles, did find a peak in
the imaginary index near 0.4 \mumx, also shown in Fig.\ \ref{Fig:imagchrom}.

An additional problem with placing the main chromophore absorber in or at the top
of Layer 2 is that there are more parameters to constrain, including the assumed wavelength-independent
 imaginary index of the stratosphere and the additional optical depth of the
overlying layer (2C in the table).  The result is increased uncertainty in
the chromophore parameters, without providing an overall improvement of fit
quality.  Although we cannot rule out a vertically distributed chromophore, the
least extreme and most convenient choice for modeling was to put the chromophore into
the stratosphere.

\subsection{Sensitivity of spectra to model parameters}

\begin{figure*}[!b]
\hspace{-0.2in}
\includegraphics[width=6.4in]{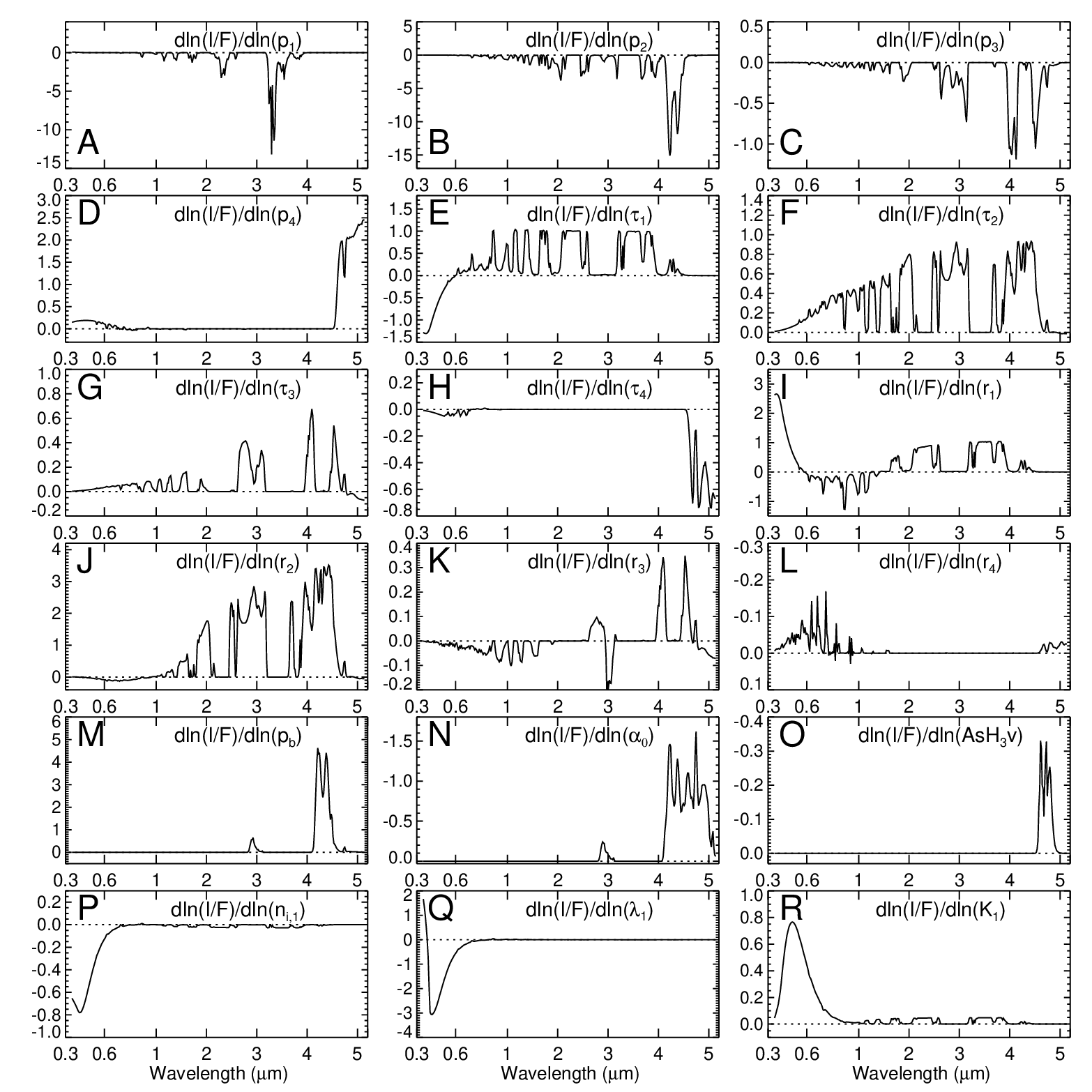}\\
\caption{Logarithmic derivative spectra for pressures (A-D), optical depths (E-H),
particle radii (I-L), minor gas parameters (M-O), and chromophore refractive
index parameters (P-R). See Table\ \ref{Tbl:paramlist} and Eqns. 2-4 for definitions of variable names.}
\label{Fig:deriv}
\end{figure*}

The degree to which model parameters can be constrained by model spectra
is illustrated by logarithmic derivative spectra in Fig.\ \ref{Fig:deriv}.
These spectra were computed by perturbing each model parameter one at a time,
then computing the fractional change in the model spectrum divided by
the fractional change in the model parameter. The reference model for
these calculations is for location 6 in 2016. The derivative spectra  
take the form 
\begin{eqnarray} \frac{\partial \log (I(\lambda; x_1, \cdots,
    x_N))}{\partial \log (x_i)} = \frac{(1/I)\partial I(\lambda; x_1,
    \cdots, x_N)}{(1/x_i)\partial x_i}, \quad \label{Eq:lnder}
\end{eqnarray}
where $x_i$ is any of the model parameters $x_1, \cdots, x_N$, and $I$ is the
model spectral radiance for the given set of parameters.  The relation also holds
if radiance $I$ is replaced by reflectivity $I/F$.
The utility of these derivatives is that they have a simple interpretation.
For example, panel E shows that a 1\% increase in the effective stratospheric haze
pressure will produce a 10\% decrease in I/F at 3.25 \mumx, a 3\% decrease
at 2.3 \mumx, but only a 1\% decrease at 0.89 \mumx.  Thus, it is apparent that
the near-IR I/F spectrum provides much stronger constraints on stratospheric
haze vertical location ($p_1$) than do visible or CCD spectra. The vertical location
of the second layer (the putative diphosphine layer at base pressure P$_2$) is also best constrained
by the near-IR spectra, but different key wavelengths are involved. The
parameters $p_1$ and $p_2$ have different spectral derivative signatures and are relatively
uncorrelated.
  
\begin{figure*}[!htb]\centering
\includegraphics[width=6in]{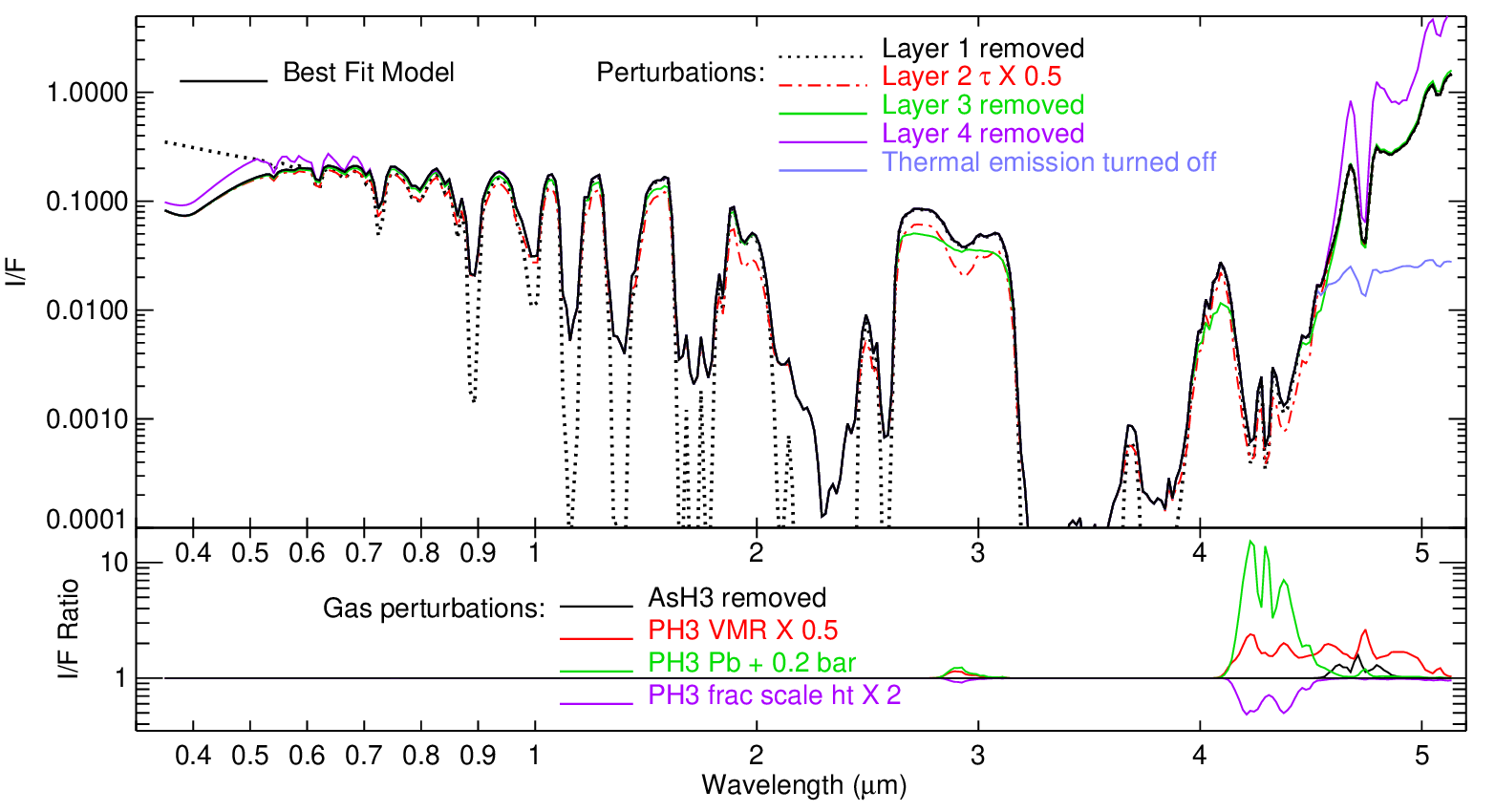}
\caption{{\bf Top:} Best-fit 2016 model spectrum for location 6 in Fig.\ \ref{Fig:vimsvis2016},
with perturbed spectra for one at a time perturbations of each model layer as indicated in the legend.
Also shown is model spectrum with Saturn's thermal emission turned off (blue curve). {\bf Bottom:} Ratio
of spectra with perturbed gas parameters to the unperturbed spectrum.}
\label{Fig:sens2}
\end{figure*}

A second way to assess the sensitivity of spectra to model structural
parameters is illustrated in Fig.\ \ref{Fig:sens2}, which displays in
the top panel a best fit model spectrum (solid black curve) along with
spectra for models with layers removed (1, 3, 4) or halved in optical
depth (2), all changes taken one at a time. Note that the dip near 2.9
\mum is removed when the \nht ice cloud is removed (layer 3).  Further
note that the deep layer (4) only has significant spectral effects at
short wavelengths, where \chf absorption is negligible, and at thermal
emission wavelengths, where \chf absorption is also
negligible. However, if the deep layer is replaced by particles with
the same scattering parameters, but with a unit single scattering
albedo, discrepancies then appear in the pseudo continuum regions with
low methane absorption between 0.9 and 1.5 \mumx.

Also shown in the top panel is a model spectrum with the emitted contribution turned off (blue curve),
showing the dominance of Saturn's thermal emission to the spectrum at wavelengths beyond 4.5 \mumx.
In the bottom panel are ratio plots of perturbed spectra to the best-fit spectrum, where
the perturbations are to the main variable gases \pht and \ashtx.

\subsection{Sensitivity of fits to initial guesses}

\begin{figure*}[!t]\centering
\includegraphics[width=5in]{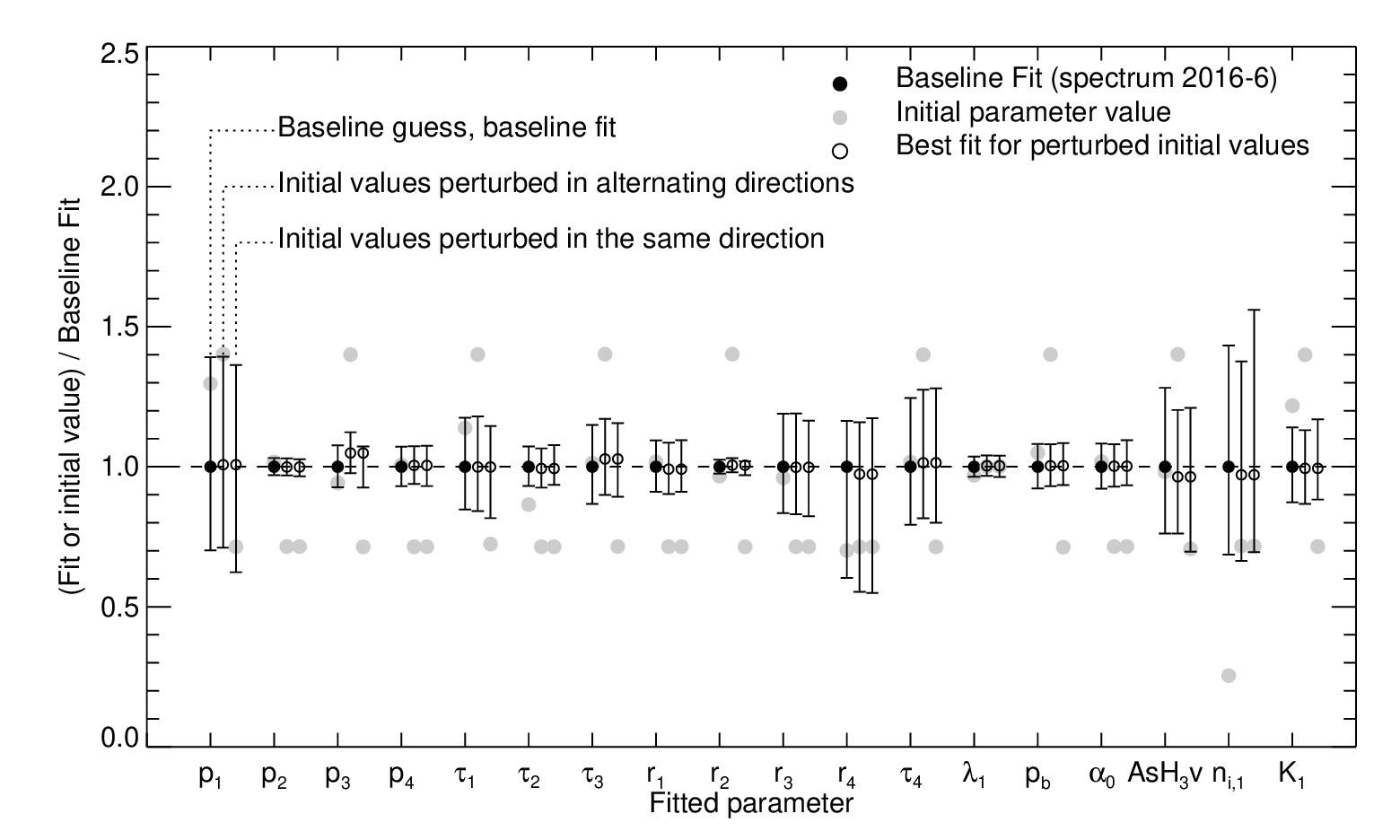}\par
\caption{Effects of different initial values on best fit parameters. The gray
dots show the initial parameter values, the filled circles the initial fit to spectrum 2016-6,
and the open circles indicate fit results for initial values perturbed
from best fit values.  Each parameter is plotted as a ratio to that parameter for the
initial fit (that is why the first fit results are all on the unit line).}\label{Fig:perturb}
\end{figure*}

Physical insight, guided
by the logarithmic derivative spectra, was used to formulate initial
guesses that provide at least crude fits to the observed spectra,
which were then refined by our L-M algorithm.  To see how sensitive the
final results were to the initial crude estimates, we did some trial perturbed
calculations, samples of which are illustrated in Fig.\ \ref{Fig:perturb}. 
 Three fits are shown here, with
initial guess and best-fit parameter values plotted for each case,
shown as ratios to the parameter values obtained from the first best fit.
The initial guess values for each case are plotted using gray filled
circles.  The best fit values for the first case are plotted as black dots with
error bars.  These all have unit central values because they have a
unit ratio to themselves.  The other fit results used initial values
that were perturbed by either multiplying or dividing by 1.4.
In the first alternative fit, the perturbations were in alternating
directions, while they were all in the same direction in the
second alternate.  The resulting best fits using that guess, shown by open circles,
were very close to the initial fit values, all well within the
fitting uncertainties.  We conclude that the fits are not very easily
perturbed, indicating that there is no nearby fit that is better than
the one we obtained.

\section{Fit results}

\subsection{Overview of latitude-dependent vertical structure fits}

\begin{figure*}[!t]\centering
\hspace{-0.15in}\includegraphics[width=3.15in]{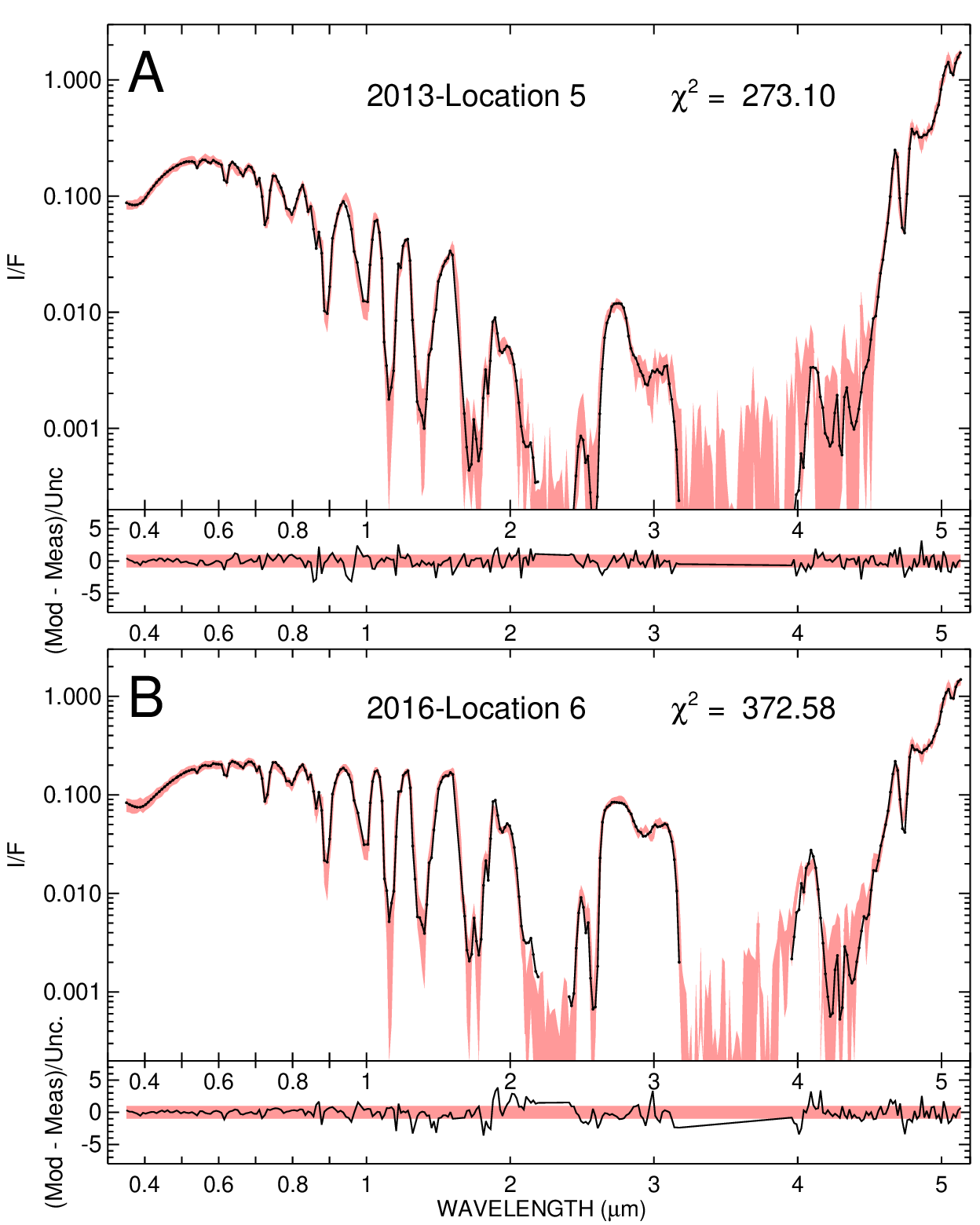}\hspace{-0.05in}
\includegraphics[width=3.15in]{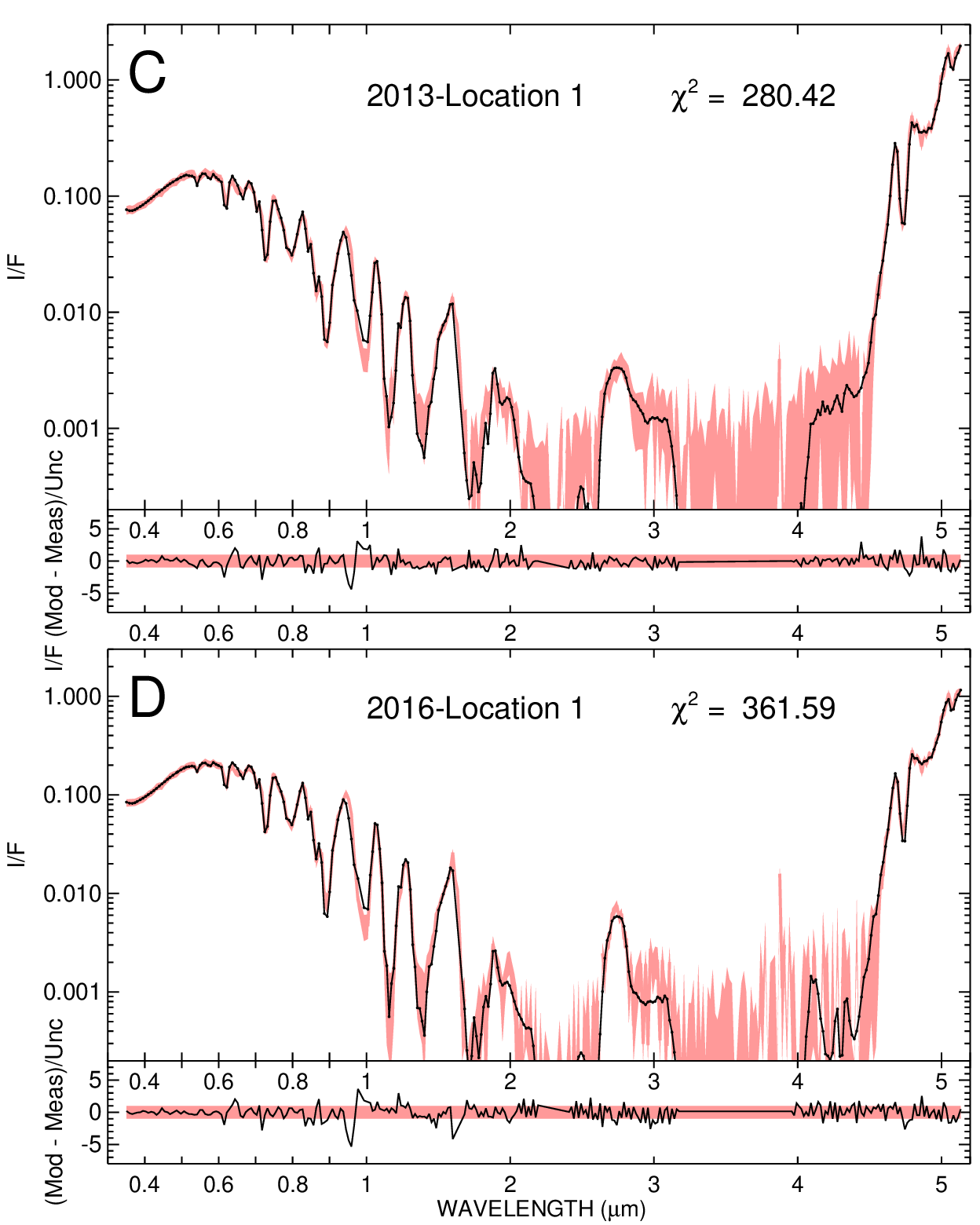}
\caption{Sample fits (black) compared to VIMS observations (pink, with vertical
spread indicating combined uncertainties) inside the hexagon (A, B)
and in the eye (C, D) for 2013 (A, C) and 2016 (B, D).  For each fit panel
the subpanel displays the measured - model difference divided by the
estimated combined uncertainty, with the shaded red bar indicating the
expected spread of the ratio (between -1 and +1).}
\label{Fig:specfits}
\end{figure*}

\begin{figure*}[!ht]\centering
\includegraphics[width=3.1in]{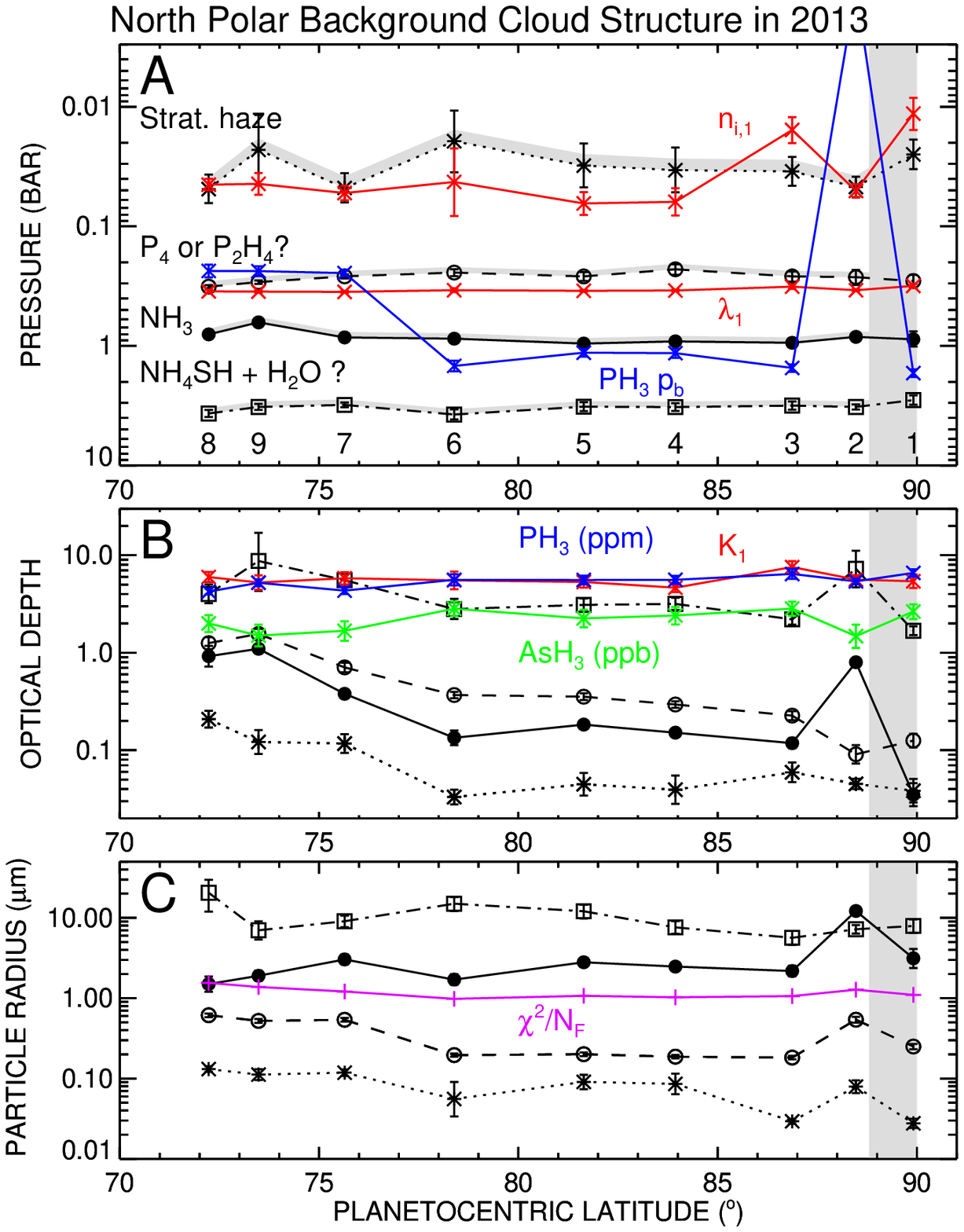}
\includegraphics[width=3.1in]{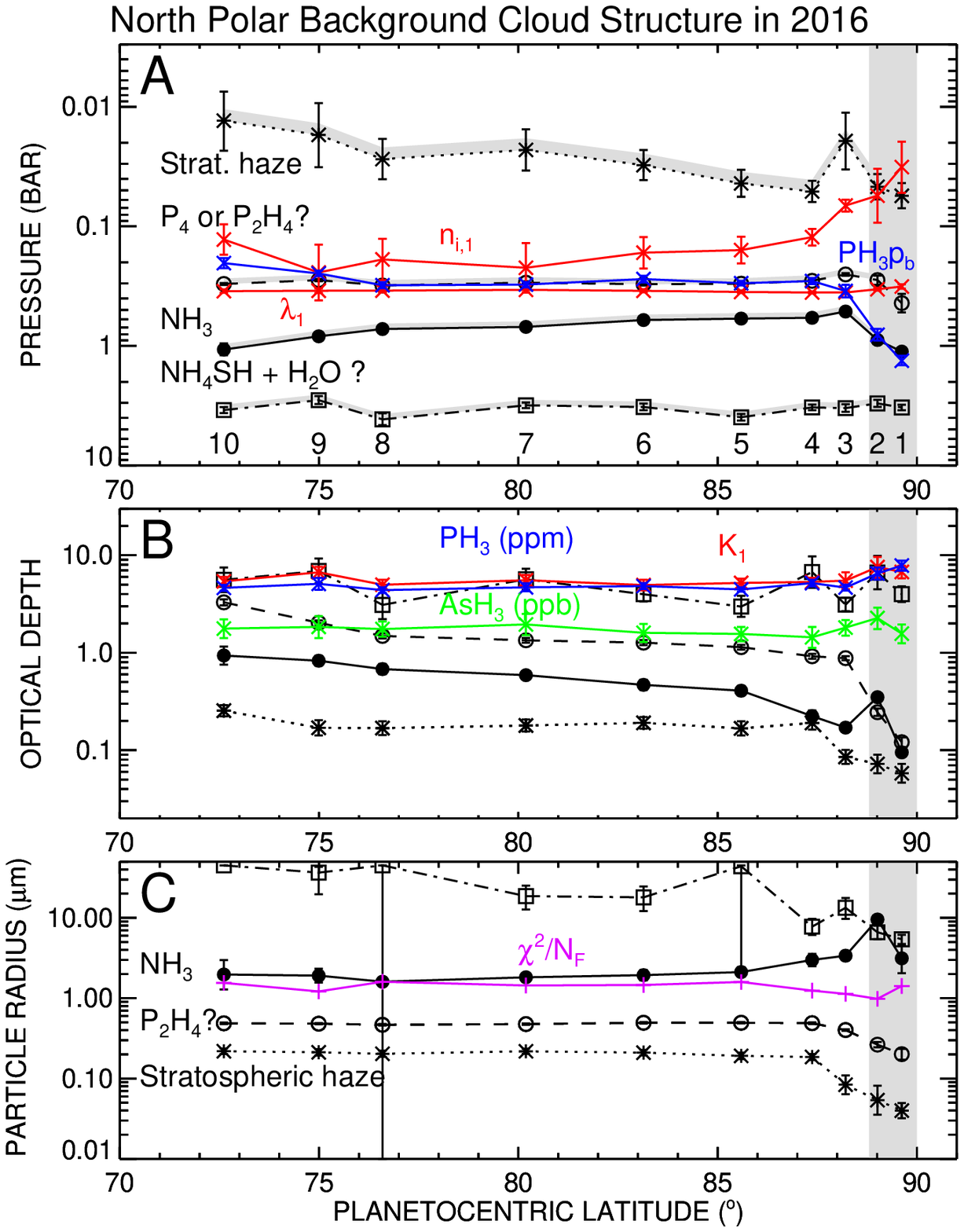}
\caption{Model fit parameters and uncertainties versus latitude for 2013 ({\bf Left}) and
2016 ({\bf Right}). For each year pressures are displayed in panel A, optical depths in panel B, and
particle radii in panel C. Chromophore parameters are overlaid in red, with $n_{i,1}$ and
$\lambda_1$ shown in panel A, and $K_1$ in panel B. Phosphine parameters are overlaid in blue,
with the breakpoint pressure $p_b$ displayed in panel A and the deep mixing ratio in ppm is
displayed in panel B, where the arsine mixing ratio in ppb is displayed in green. The vertical
gray bars mark the eye region. The very small value of $p_b$ for location 2 is
an anomaly discussed in Sec.\ \ref{Sec:nh3ice}.}
\label{Fig:2013+2016}
\end{figure*}

The results of fitting VIMS combined visual and near-IR spectra of the background clouds in the
north polar region are summarized in Table\ \ref{Tbl:bestfit2013} for
2013 and Table\ \ref{Tbl:bestfit2016} for 2016. Both tables contain results
from ten locations that avoid bright discrete cloud features. 
  A sampling of the spectral fit quality for mid hexagon
and eye regions for both years is provided in Fig.\ \ref{Fig:specfits}.  The best-fit
parameters and their uncertainties are
compared in side-by-side plots versus latitude in Fig.\ \ref{Fig:2013+2016}.
Fig.\ \ref{Fig:5-panel} provides an alternative set of plots that
makes it easier to appreciate the evolutionary changes between 2013
and 2016.  Similar comparison plots versus latitude of the column mass densities and their estimated
uncertainties are presented in Fig.\ \ref{Fig:massden}.

In general the pressures of the four layers are reasonably
well constrained by the observations and do not vary dramatically with
latitude except near the eye boundary in 2016.  This stability of
pressure levels is consistent with the aerosols being formed from
condensates, but do not compel that conclusion.  The biggest variation
with latitude is in the optical depths of model layers in 2013, for
which optical depths inside the hexagon were substantially lower,
decreasing by a factor of 4 to 5 relative to those outside the hexagon,
with small additional decreases inside the eye region.  In 2016 there
was gradient of descent near the eye in the two middle layers, as
their base pressures increased and optical depths dropped sharply inside
the eye, to levels that are similar to those present over most of the
region interior to the hexagon in 2013. One could roughly characterize the
entire region inside the hexagon in 2013 as having an aerosol structure
similar to that of the eye, but brightening substantially by 2016,
leaving only the eye as a remnant.  The increased optical depths of
aerosol layers inside the hexagon by 2016 also contributed to the color
change, which was further aided by increased absorption of blue light
by the stratospheric haze as its imaginary index at short wavelengths
increased as well as its optical depth.

\subsection{Aerosol structure results by layer}

\begin{table*}[!htb]\centering
\caption{Best-fit model parameters as constrained by 2013 observations.}
\begin{scriptsize}
\setlength\tabcolsep{2pt}
\begin{tabular}{l c c c c c c c c c c c}
\hline\\[-0.05in]
Locations:&                   8&                   9&                   7&                  10&                   6&                   5&                   4&                   3&                   2&                   1\\[0.05in]
Parameter $\setminus$ Lat. &           72.2\degx  &          73.5\degx  &          75.6\degx  &          76.2\degx  &          78.4\degx  &          81.6\degx  &          83.9\degx  &          86.9\degx  &          88.5\degx &          89.9\degx \\[0.05in]
\hline\\[-0.05in]
 $p_1$ (bar)$\times 10^2$ &   4.9$^{+  1.5}_{-  1.2}$ &   2.3$^{+  2.2}_{-  1.1}$ &   4.8$^{+  1.5}_{-  1.2}$ &   3.4$^{+  1.6}_{-  1.1}$ &   1.9$^{+  1.6}_{-  0.9}$ &   3.1$^{+  1.6}_{-  1.1}$ &   3.4$^{+  1.8}_{-  1.2}$ &   3.5$^{+  1.1}_{-  0.8}$ &   4.7$^{+  1.0}_{-  0.8}$ &   2.5$^{+  0.8}_{-  0.6}$\\[0.05in]
 $p_2$ (bar)$\times 10^1$ &   3.2$^{+  0.1}_{-  0.1}$ &   2.9$^{+  0.1}_{-  0.1}$ &   2.6$^{+  0.1}_{-  0.1}$ &   2.5$^{+  0.1}_{-  0.1}$ &   2.4$^{+  0.2}_{-  0.2}$ &   2.6$^{+  0.2}_{-  0.2}$ &   2.3$^{+  0.3}_{-  0.2}$ &   2.6$^{+  0.3}_{-  0.2}$ &   2.7$^{+  0.4}_{-  0.3}$ &   2.9$^{+  0.4}_{-  0.3}$\\[0.05in]
 $p_3$ (bar)$\times 10^1$ &   8.0$^{+  0.7}_{-  0.6}$ &   6.4$^{+  0.4}_{-  0.3}$ &   8.5$^{+  0.5}_{-  0.5}$ &   9.8$^{+  0.4}_{-  0.4}$ &   8.7$^{+  0.5}_{-  0.5}$ &   9.6$^{+  0.2}_{-  0.2}$ &   9.2$^{+  0.3}_{-  0.3}$ &   9.4$^{+  0.3}_{-  0.3}$ &   8.4$^{+  0.3}_{-  0.3}$ &   8.8$^{+  1.3}_{-  1.2}$\\[0.05in]
              $p_4$ (bar) &   3.7$^{+  0.3}_{-  0.2}$ &   3.2$^{+  0.2}_{-  0.2}$ &   3.1$^{+  0.1}_{-  0.1}$ &   2.8$^{+  0.4}_{-  0.4}$ &   3.8$^{+  0.4}_{-  0.4}$ &   3.2$^{+  0.3}_{-  0.3}$ &   3.3$^{+  0.2}_{-  0.2}$ &   3.2$^{+  0.3}_{-  0.2}$ &   3.2$^{+  0.2}_{-  0.2}$ &   2.8$^{+  0.3}_{-  0.3}$\\[0.05in]
\hline\\[-0.1in]
$r_1$ (\mum)$\times 10^2$ &  13.1$^{+  1.1}_{-  1.0}$ &  11.2$^{+  1.9}_{-  1.6}$ &  11.8$^{+  1.2}_{-  1.1}$ &  10.5$^{+  1.7}_{-  1.5}$ &   5.6$^{+  3.5}_{-  2.2}$ &   9.1$^{+  2.1}_{-  1.7}$ &   8.6$^{+  2.9}_{-  2.2}$ &   2.9$^{+  0.2}_{-  0.2}$ &   7.9$^{+  1.6}_{-  1.3}$ &   2.8$^{+  0.4}_{-  0.3}$\\[0.05in]
$r_2$ (\mum)$\times 10^1$ &   6.1$^{+  0.3}_{-  0.3}$ &   5.2$^{+  0.3}_{-  0.3}$ &   5.4$^{+  0.3}_{-  0.3}$ &   5.0$^{+  0.3}_{-  0.2}$ &   2.0$^{+  0.1}_{-  0.1}$ &   2.0$^{+  0.1}_{-  0.1}$ &   1.9$^{+  0.1}_{-  0.1}$ &   1.8$^{+  0.1}_{-  0.1}$ &   5.4$^{+  0.5}_{-  0.5}$ &   2.5$^{+  0.2}_{-  0.2}$\\[0.05in]
             $r_3$ (\mum) &   1.5$^{+  0.4}_{-  0.3}$ &   1.9$^{+  0.3}_{-  0.3}$ &   3.0$^{+  0.5}_{-  0.4}$ &   1.5$^{+  0.3}_{-  0.2}$ &   1.7$^{+  0.3}_{-  0.2}$ &   2.8$^{+  0.3}_{-  0.3}$ &   2.5$^{+  0.3}_{-  0.3}$ &   2.2$^{+  0.4}_{-  0.3}$ &  12.2$^{+  0.6}_{-  0.6}$ &   3.1$^{+  1.0}_{-  0.8}$\\[0.05in]
             $r_4$ (\mum) &  20.5$^{+  9.2}_{-  8.6}$ &   7.0$^{+  2.0}_{-  1.6}$ &   9.1$^{+  1.6}_{-  1.4}$ &   5.5$^{+  0.8}_{-  0.7}$ &  15.0$^{+  3.1}_{-  2.8}$ &  12.1$^{+  2.5}_{-  2.2}$ &   7.6$^{+  1.5}_{-  1.3}$ &   5.6$^{+  1.2}_{-  1.0}$ &   7.2$^{+  0.9}_{-  0.8}$ &   7.9$^{+  1.6}_{-  1.4}$\\[0.05in]
\hline\\[-0.1in]
    $\tau_1$$\times 10^2$ &  20.8$^{+  4.5}_{-  3.7}$ &  12.1$^{+  3.9}_{-  3.0}$ &  11.7$^{+  2.9}_{-  2.3}$ &   9.1$^{+  2.4}_{-  1.9}$ &   3.3$^{+  0.6}_{-  0.5}$ &   4.5$^{+  1.4}_{-  1.0}$ &   3.9$^{+  1.6}_{-  1.1}$ &   5.9$^{+  1.5}_{-  1.2}$ &   4.5$^{+  0.6}_{-  0.5}$ &   3.8$^{+  1.2}_{-  0.9}$\\[0.05in]
    $\tau_2$$\times 10^1$ &  12.5$^{+  1.9}_{-  1.7}$ &  15.6$^{+  2.0}_{-  1.8}$ &   7.1$^{+  0.8}_{-  0.8}$ &   7.4$^{+  0.8}_{-  0.7}$ &   3.7$^{+  0.3}_{-  0.3}$ &   3.5$^{+  0.3}_{-  0.3}$ &   2.9$^{+  0.2}_{-  0.2}$ &   2.3$^{+  0.3}_{-  0.2}$ &   0.9$^{+  0.2}_{-  0.2}$ &   1.3$^{+  0.2}_{-  0.2}$\\[0.05in]
    $\tau_3$$\times 10^1$ &   9.2$^{+  2.5}_{-  2.0}$ &  11.0$^{+  1.3}_{-  1.1}$ &   3.8$^{+  0.3}_{-  0.3}$ &   4.5$^{+  1.0}_{-  0.8}$ &   1.3$^{+  0.3}_{-  0.2}$ &   1.8$^{+  0.2}_{-  0.2}$ &   1.5$^{+  0.1}_{-  0.1}$ &   1.2$^{+  0.1}_{-  0.1}$ &   8.0$^{+  0.5}_{-  0.5}$ &   0.3$^{+  0.1}_{-  0.1}$\\[0.05in]
                 $\tau_4$ &   4.0$^{+  1.0}_{-  0.8}$ &   8.7$^{+  8.3}_{-  4.4}$ &   5.5$^{+  0.9}_{-  0.8}$ &  47.0$^{+ 23.0}_{- 46.5}$ &   2.8$^{+  0.8}_{-  0.6}$ &   3.1$^{+  0.5}_{-  0.4}$ &   3.2$^{+  0.5}_{-  0.4}$ &   2.2$^{+  0.3}_{-  0.3}$ &   7.3$^{+  3.9}_{-  2.6}$ &   1.7$^{+  0.2}_{-  0.2}$\\[0.05in]
\hline\\[-0.1in]
       $\lambda_1\times 10^1$ &   3.5$^{+  0.1}_{-  0.1}$ &   3.5$^{+  0.1}_{-  0.1}$ &   3.5$^{+  0.1}_{-  0.1}$ &   3.6$^{+  0.1}_{-  0.1}$ &   3.4$^{+  0.1}_{-  0.1}$ &   3.5$^{+  0.1}_{-  0.1}$ &   3.4$^{+  0.1}_{-  0.1}$ &   3.2$^{+  0.1}_{-  0.1}$ &   3.4$^{+  0.1}_{-  0.1}$ &   3.1$^{+  0.1}_{-  0.1}$\\[0.05in]
       $n_{i,1}\times 10^2$ &   4.5$^{+  0.5}_{-  0.5}$ &   4.4$^{+  1.0}_{-  0.8}$ &   5.3$^{+  0.8}_{-  0.7}$ &   4.9$^{+  0.9}_{-  0.8}$ &   4.2$^{+  3.9}_{-  2.0}$ &   6.4$^{+  1.6}_{-  1.3}$ &   6.2$^{+  1.9}_{-  1.4}$ &   1.6$^{+  0.5}_{-  0.3}$ &   5.1$^{+  0.7}_{-  0.6}$ &   1.1$^{+  0.4}_{-  0.3}$\\[0.05in]
                      $K_1$ &   6.0$^{+  0.8}_{-  0.7}$ &   5.2$^{+  1.0}_{-  0.8}$ &   5.8$^{+  0.9}_{-  0.8}$ &   5.7$^{+  0.9}_{-  0.8}$ &   5.5$^{+  1.3}_{-  1.1}$ &   5.3$^{+  0.8}_{-  0.7}$ &   4.7$^{+  0.5}_{-  0.5}$ &   7.6$^{+  1.1}_{-  1.0}$ &   5.6$^{+  0.6}_{-  0.6}$ &   5.4$^{+  1.0}_{-  0.8}$\\[0.05in]
\hline\\[-0.1in]
       \nht$v1$ (ppm) &   1.6$^{+16}_{-  1.6}$ &   0.3$^{+32}_{-  0.3}$ &   2.6$^{+ 11.3}_{-  2.1}$ & 103.0$^{+ 85}_{- 48}$ &  44.2$^{+ 45}_{- 23}$ &  34.8$^{+ 32}_{- 17}$ &  46.8$^{+ 23}_{- 16}$ &  54.8$^{+ 23}_{- 17}$ &  42.7$^{+ 26}_{- 16}$ &  79.4$^{+ 25}_{- 19}$\\[0.05in]
         \pht $p_b$ (bar) &  0.24$^{+ 0.03}_{- 0.03}$ &  0.24$^{+ 0.02}_{- 0.02}$ &  0.25$^{+ 0.02}_{- 0.02}$ &  0.25$^{+ 0.02}_{- 0.01}$ &  1.47$^{+ 0.16}_{- 0.14}$ &  1.14$^{+ 0.09}_{- 0.08}$ &  1.15$^{+ 0.11}_{- 0.09}$ &  1.53$^{+ 0.12}_{- 0.10}$ &  see note &  1.69$^{+ 0.13}_{- 0.12}$\\[0.05in]
     \pht $\alpha_0$ (ppm) &  4.24$^{+ 0.3}_{- 0.3}$ &  5.22$^{+ 0.4}_{- 0.4}$ &  4.34$^{+ 0.4}_{- 0.3}$ &  4.49$^{+ 0.5}_{- 0.4}$ &  5.60$^{+ 0.8}_{- 0.7}$ &  5.57$^{+ 0.6}_{- 0.5}$ &  5.60$^{+ 0.6}_{- 0.5}$ &  6.43$^{+ 0.8}_{- 0.7}$ &  5.40$^{+ 0.4}_{- 0.4}$ &  6.59$^{+ 0.6}_{- 0.6}$\\[0.05in]
    \asht VMR (ppb) &  2.00$^{+ 0.4}_{- 0.4}$ &  1.50$^{+ 0.4}_{- 0.3}$ &  1.68$^{+ 0.4}_{- 0.4}$ &  2.15$^{+ 0.7}_{- 0.6}$ &  2.83$^{+ 0.5}_{- 0.5}$ &  2.25$^{+ 0.5}_{- 0.4}$ &  2.41$^{+ 0.5}_{- 0.5}$ &  2.84$^{+ 0.5}_{- 0.5}$ &  1.48$^{+ 0.5}_{- 0.4}$ &  2.65$^{+ 0.5}_{- 0.4}$\\[0.05in]
\hline\\[-0.1in]
M1 ($\mu$g/cm$^2$)&     37$^{+ 11}_{-  10}$&     31$^{+ 17}_{- 15}$&     26$^{+  9}_{-  8}$&     28$^{+ 14}_{- 13}$&     56$^{+107}_{-56}$&     19.8$^{+ 14}_{- 13}$&     20$^{+ 21}_{- 20}$&    454$^{+132}_{-111}$&     29$^{+ 17}_{- 17}$&    325$^{+130}_{-111}$\\[0.05in]
M2 ($\mu$g/cm$^2$)&     31.2$^{+  5.4}_{-  4.9}$&     30.3$^{+  4.5}_{-  4.2}$&     14.4$^{+  2.0}_{-  1.9}$&     13.5$^{+  1.7}_{-  1.6}$&      8.1$^{+  0.9}_{-  0.9}$&      7.5$^{+  0.9}_{-  0.9}$&      7.0$^{+  0.9}_{-  0.9}$&      5.7$^{+  0.8}_{-  0.8}$&      1.9$^{+  0.5}_{-  0.4}$&      2.0$^{+  0.4}_{-  0.3}$\\[0.05in]
M3 ($\mu$g/cm$^2$)&     72$^{+ 31}_{- 28}$&    115$^{+ 26}_{- 25}$&     66$^{+ 14}_{- 14}$&     37$^{+ 12}_{- 11}$&     12.4$^{+  3.4}_{-  3.1}$&     29.5$^{+  4.6}_{-  4.3}$&     21.2$^{+  3.8}_{-  3.7}$&     14.4$^{+  3.2}_{-  3.1}$&    613$^{+ 48}_{- 47}$&      6.4$^{+  2.9}_{-  2.6}$\\[0.05in]
M4 (mg/cm$^2$)&      5.3$^{+  2.7}_{-  2.6}$&      3.8$^{+  3.7}_{-  2.2}$&      3.1$^{+  0.8}_{-  0.7}$&     15.7$^{+  8.0}_{- 15.7}$&      2.7$^{+  0.9}_{-  0.8}$&      2.4$^{+  0.6}_{-  0.6}$&      1.5$^{+  0.4}_{-  0.4}$&      0.8$^{+  0.2}_{-  0.2}$&      3.2$^{+  1.8}_{-  1.2}$&      0.8$^{+  0.2}_{-  0.2}$\\[0.05in]
\hline\\[-0.1in]
$\chi^2$ & 395.46 & 351.42 & 308.77 & 260.40 & 250.47 & 273.10 & 261.58 & 270.87 & 325.39 & 280.42\\[0.05in]
$\chi^2/N_F$ &   1.54 &   1.37 &   1.21 &   1.02 &   0.98 &   1.07 &   1.02 &   1.06 &   1.27 &   1.10\\[0.05in]
\hline\\[-0.1in]
\end{tabular}
\label{Tbl:bestfit2013}
\end{scriptsize}
\begin{small}
\parbox[c]{6in}{NOTE: Here location numbers refer to encircled pixels
  shown in Figs.\ \ref{Fig:vimsvis2013}, \ref{Fig:vimsnir2013}, and
  \ref{Fig:2013combined}A. Optical depths are given at a wavelength of
  1 \mumx. Refractive index values for each aerosol layer are
  discussed in the text.  Scattering is computed for homogeneous
  spheres with a gamma size distribution. Pressures are for the bottom
  of physically thin clouds with top pressures equal to 90\% of bottom
  pressures. M$i$ denotes column mass density for layer $i$, which
  was computed as $\tau_i r_i\frac{4}{3}\rho/Q_\mathrm{ext}$ assuming
a particle mass density $\rho$ of 1 g/cm$^3$.  The column mass densities are
particularly uncertain for small particles because of the strong dependence
of extinction efficiency $Q_\mathrm{ext}$ on particle size.  Fixed
  values were $n_{i,0} = 5\times10^{-4}$ and $K_2=20$. See Sec.\ \ref{Sec:nh3ice} for
discussion of the anomalous $p_b$ result of $p_b < 0.15$ bar for location 2.}
\end{small}
\end{table*}

\begin{figure*}[!ht]\centering
\hspace{-0.12in}\includegraphics[width=3.1in]{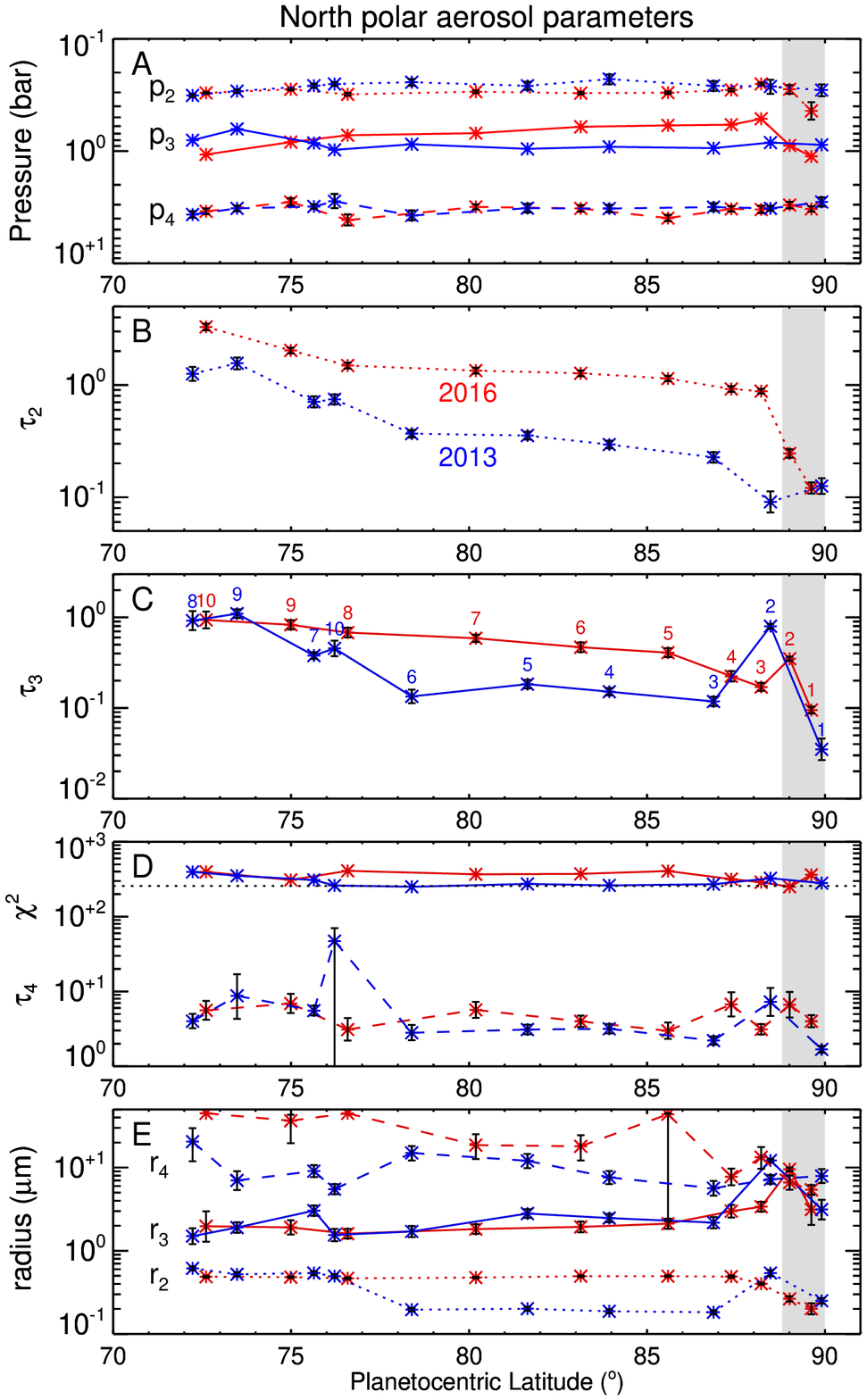}\hspace{-0.04in}
\includegraphics[width=3.1in]{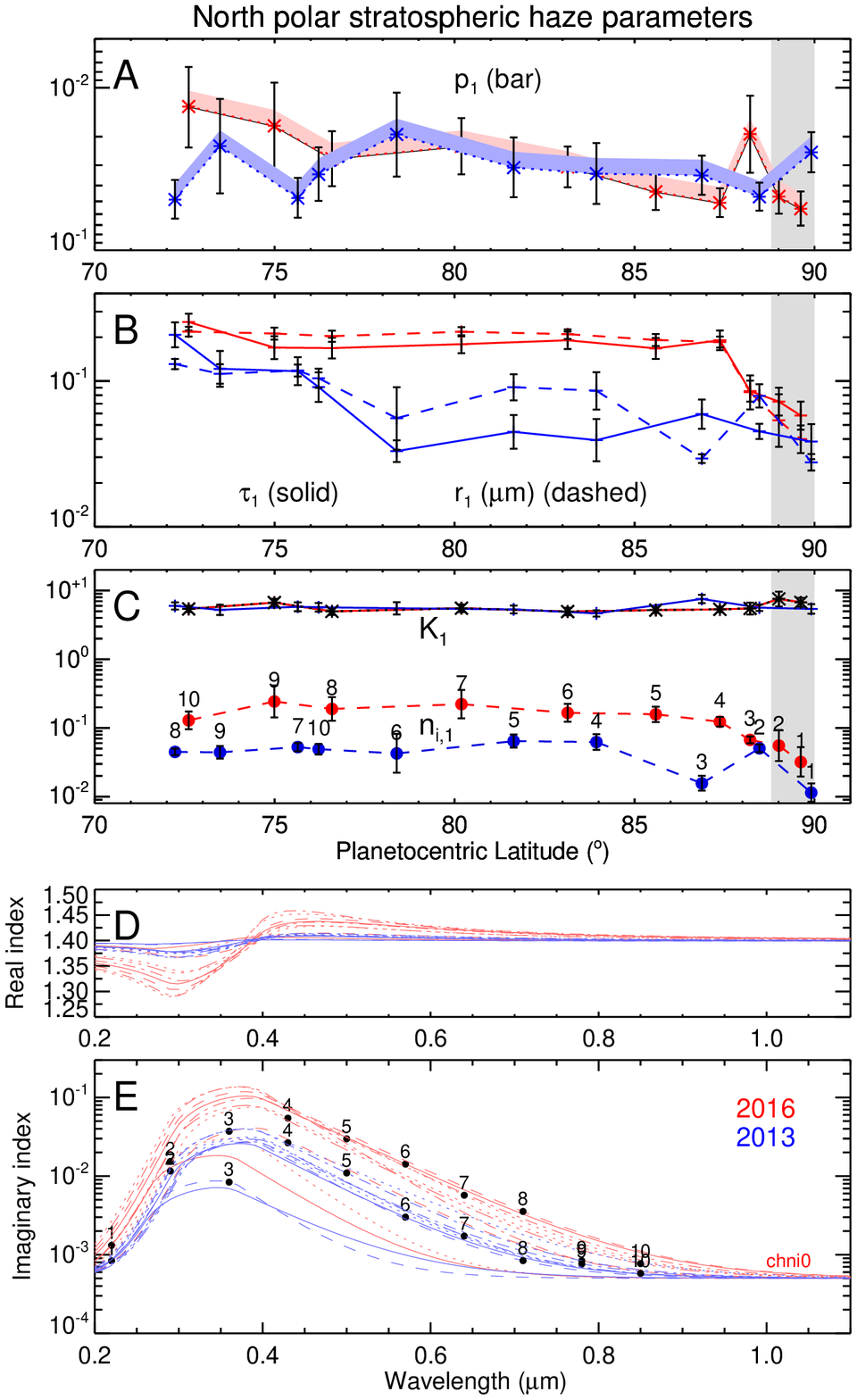}
\caption{Model fit tropospheric aerosol parameters and uncertainties
  versus latitude ({\bf Left}) and stratospheric haze parameters ({\bf
    Right}). In both columns blue is used for 2013 and red for 2016.
 On the left pressures are displayed in panel A, optical
  depths for layer 2 in panel B, for layer 3 in panel C, and layer 4
  in panel D, with 2013 results displayed in blue and 2016 results in
  red.  Panel D also displays \chisqx, and particle radii are displayed in panel
  E. On the right, panel A displays the effective pressure of the stratospheric
haze, panel B its optical depth and particle radius, panel C the
chromophore parameters $K_1$ and $n_{i,1}$, and panels D and E display
the real and imaginary refractive indexes for the stratospheric haze particles.
Note that the three 2016 curves (red) that overlap the 2013 curves (blue)
are from the eye region.}\label{Fig:5-panel}
\end{figure*}

\paragraph{Layer 1 (stratosphere).} The stratospheric haze parameters are
best summarized in the right column of Fig.\ \ref{Fig:5-panel},
which displays results for 2013 in blue and for 2016 in red. In 2013,
the effective pressure of this layer is roughly in the 15 mbar to 50
mbar range.  Uncertainties in effective pressure are so large that a
clear trend with latitude is hard to discern in the 2013 results.
There seems to be a descending trend towards the pole in 2016, but
that is also uncertain.  The optical depth trends are much better
established.  In 2016 the 1-\mum optical depth of the stratospheric
haze is near 0.2 from just south of the hexagon to just outside the eye, at
which point it begins a rapid drop by a factor of five, to a value of
0.04, just twice the level seen in 2013 at the same location. But
outside the eye, out to the edge of the hexagon, the 2013 optical depth
of this layer remains near 0.04, about five times below the value found in 2016. The
large increase in optical depth of the stratospheric haze between 2013
and 2016 is a major factor in changing the color of the region
interior to the hexagon because that layer is a strong absorber of blue
light, as evident from the fitted imaginary index model displayed in
panel E of the right column of Fig.\ \ref{Fig:5-panel}. Increased
absorption in 2016 was further aided by an increase in the imaginary
index by about a factor of 3 between 2013 and 2016.  This indicates
that the stratospheric haze particles are not of uniform composition.
One possibility is that they are a mixture of two components, and the
fraction of the short-wave absorber component is increased over time
by UV photolysis.  A more complex changing composition is also
possible. Two parameters of the imaginary index that did not change
significantly between 2013 and 2016 are the log slope parameter $K_1$,
which averaged 5 to 6 at both times, and the wavelength of the peak, which
remained near 0.35 \mum during both
periods. It is also noteworthy that the best fit chromophore index
models in the eye region for the two years are much more similar
than the mid-hexagon values.
In 2013 the  particle radius for this layer was roughly 0.05 \mum
inside the hexagon with perhaps a decrease inside the eye,
though uncertainties are too large to confirm the eye gradient.
In 2016 the particle radius outside the eye was close to
a latitude independent value near 0.2 \mumx, but dropped to
2013 values inside the eye.

\begin{table*}[!htb]\centering
\caption{Best-fit model parameters as constrained by 2016 observations.}
\begin{scriptsize}
\setlength\tabcolsep{2pt}
\begin{tabular}{l l l l l l l l l l l l}
\hline\\[-0.05in]
Locations:&                  10&                   9&                   8&                   7&                   6&                   5&                   4&                   3&                   2&                   1\\[0.05in]
Parameter $\setminus$ Lat. &           72.6\degx  &          75.0\degx  &          76.6\degx  &          80.2\degx  &          83.1\degx  &          85.6\degx  &          87.4\degx  &          88.2\degx  &          89.0\degx &          89.6\degx \\[0.05in]
\hline\\[-0.05in]
 $p_1$ (bar)$\times 10^2$ &   1.3$^{+  1.0}_{-  0.6}$ &   1.7$^{+  1.5}_{-  0.8}$ &   2.7$^{+  1.3}_{-  0.9}$ &   2.3$^{+  1.1}_{-  0.8}$ &   3.1$^{+  1.0}_{-  0.8}$ &   4.4$^{+  1.3}_{-  1.0}$ &   5.1$^{+  1.1}_{-  0.9}$ &   1.9$^{+  1.4}_{-  0.8}$ &   4.7$^{+  1.3}_{-  1.0}$ &   5.6$^{+  1.5}_{-  1.2}$\\[0.05in]
 $p_2$ (bar)$\times 10^1$ &   3.0$^{+  0.1}_{-  0.1}$ &   2.8$^{+  0.1}_{-  0.1}$ &   3.1$^{+  0.1}_{-  0.1}$ &   3.0$^{+  0.1}_{-  0.1}$ &   3.0$^{+  0.1}_{-  0.1}$ &   3.0$^{+  0.1}_{-  0.1}$ &   2.9$^{+  0.1}_{-  0.1}$ &   2.5$^{+  0.1}_{-  0.1}$ &   2.8$^{+  0.3}_{-  0.2}$ &   4.4$^{+  0.8}_{-  0.7}$\\[0.05in]
 $p_3$ (bar)$\times 10^1$ &  10.7$^{+  1.2}_{-  1.2}$ &   8.3$^{+  0.5}_{-  0.4}$ &   7.2$^{+  0.4}_{-  0.4}$ &   6.9$^{+  0.4}_{-  0.3}$ &   6.1$^{+  0.4}_{-  0.4}$ &   5.9$^{+  0.4}_{-  0.3}$ &   5.8$^{+  0.5}_{-  0.5}$ &   5.2$^{+  0.4}_{-  0.4}$ &   8.9$^{+  0.4}_{-  0.4}$ &  11.2$^{+  0.4}_{-  0.4}$\\[0.05in]
              $p_4$ (bar) &   3.4$^{+  0.2}_{-  0.2}$ &   2.8$^{+  0.2}_{-  0.2}$ &   4.1$^{+  0.5}_{-  0.5}$ &   3.1$^{+  0.2}_{-  0.2}$ &   3.2$^{+  0.2}_{-  0.2}$ &   4.0$^{+  0.3}_{-  0.3}$ &   3.3$^{+  0.2}_{-  0.2}$ &   3.3$^{+  0.3}_{-  0.2}$ &   3.0$^{+  0.2}_{-  0.2}$ &   3.3$^{+  0.2}_{-  0.2}$\\[0.05in]
\hline\\[-0.1in]
$r_1$ (\mum)$\times 10^2$ &  21.8$^{+  1.7}_{-  1.6}$ &  21.2$^{+  2.1}_{-  1.9}$ &  20.3$^{+  1.9}_{-  1.7}$ &  21.8$^{+  1.7}_{-  1.6}$ &  21.0$^{+  1.7}_{-  1.6}$ &  19.2$^{+  1.9}_{-  1.8}$ &  18.5$^{+  1.7}_{-  1.6}$ &   8.4$^{+  2.6}_{-  2.0}$ &   5.4$^{+  2.7}_{-  1.8}$ &   4.0$^{+  1.0}_{-  0.8}$\\[0.05in]
$r_2$ (\mum)$\times 10^1$ &   4.9$^{+  0.1}_{-  0.1}$ &   4.8$^{+  0.1}_{-  0.1}$ &   4.7$^{+  0.1}_{-  0.1}$ &   4.8$^{+  0.1}_{-  0.1}$ &   4.9$^{+  0.1}_{-  0.1}$ &   4.9$^{+  0.1}_{-  0.1}$ &   4.9$^{+  0.1}_{-  0.1}$ &   4.0$^{+  0.1}_{-  0.1}$ &   2.6$^{+  0.2}_{-  0.2}$ &   2.0$^{+  0.3}_{-  0.3}$\\[0.05in]
             $r_3$ (\mum) &   2.0$^{+  1.0}_{-  0.7}$ &   1.9$^{+  0.4}_{-  0.3}$ &   1.6$^{+  0.2}_{-  0.2}$ &   1.8$^{+  0.3}_{-  0.2}$ &   1.9$^{+  0.3}_{-  0.3}$ &   2.1$^{+  0.3}_{-  0.3}$ &   3.0$^{+  0.6}_{-  0.5}$ &   3.4$^{+  0.5}_{-  0.4}$ &   9.5$^{+  0.5}_{-  0.5}$ &   3.1$^{+  1.6}_{-  1.1}$\\[0.05in]
             $r_4$ (\mum) &  45 &  36.5$^{+  6.7}_{- 16.9}$ &  45.0$^{+  0.0}_{- 45.0}$ &  18.6$^{+  6.5}_{-  5.9}$ &  18.0$^{+  6.5}_{-  5.8}$ &  43.8$^{+  1.2}_{- 41.5}$ &   7.8$^{+  2.0}_{-  1.6}$ &  13.3$^{+  4.3}_{-  3.6}$ &   6.6$^{+  1.4}_{-  1.2}$ &   5.4$^{+  0.8}_{-  0.7}$\\[0.05in]
\hline\\[-0.1in]
    $\tau_1$$\times 10^2$ &  25.4$^{+  3.6}_{-  3.2}$ &  17.0$^{+  3.4}_{-  2.8}$ &  16.9$^{+  3.0}_{-  2.5}$ &  17.9$^{+  2.7}_{-  2.4}$ &  19.1$^{+  2.8}_{-  2.5}$ &  16.8$^{+  3.0}_{-  2.6}$ &  19.1$^{+  3.2}_{-  2.7}$ &   8.5$^{+  1.5}_{-  1.3}$ &   7.2$^{+  1.8}_{-  1.4}$ &   5.8$^{+  1.4}_{-  1.2}$\\[0.05in]
    $\tau_2$$\times 10^1$ &  32.9$^{+  2.3}_{-  2.2}$ &  20.3$^{+  1.2}_{-  1.1}$ &  14.9$^{+  0.9}_{-  0.9}$ &  13.4$^{+  0.8}_{-  0.7}$ &  12.7$^{+  0.8}_{-  0.7}$ &  11.4$^{+  0.7}_{-  0.6}$ &   9.2$^{+  0.5}_{-  0.5}$ &   8.8$^{+  0.4}_{-  0.4}$ &   2.5$^{+  0.2}_{-  0.2}$ &   1.2$^{+  0.1}_{-  0.1}$\\[0.05in]
    $\tau_3$$\times 10^1$ &   9.4$^{+  2.2}_{-  1.8}$ &   8.3$^{+  1.0}_{-  0.9}$ &   6.8$^{+  0.9}_{-  0.8}$ &   5.9$^{+  0.6}_{-  0.5}$ &   4.7$^{+  0.6}_{-  0.5}$ &   4.1$^{+  0.5}_{-  0.4}$ &   2.2$^{+  0.3}_{-  0.3}$ &   1.7$^{+  0.2}_{-  0.2}$ &   3.5$^{+  0.2}_{-  0.2}$ &   1.0$^{+  0.1}_{-  0.1}$\\[0.05in]
                 $\tau_4$ &   5.6$^{+  1.9}_{-  1.4}$ &   6.9$^{+  2.4}_{-  1.8}$ &   3.1$^{+  1.3}_{-  0.9}$ &   5.7$^{+  1.6}_{-  1.2}$ &   4.0$^{+  0.8}_{-  0.6}$ &   3.0$^{+  0.9}_{-  0.6}$ &   6.7$^{+  3.0}_{-  2.1}$ &   3.1$^{+  0.5}_{-  0.5}$ &   6.7$^{+  3.2}_{-  2.2}$ &   4.0$^{+  0.8}_{-  0.7}$\\[0.05in]
\hline\\[-0.1in]
       $\lambda_1 \times 10^1$ &   3.5$^{+  0.1}_{-  0.1}$ &   3.5$^{+  0.1}_{-  0.1}$ &   3.4$^{+  0.1}_{-  0.1}$ &   3.4$^{+  0.2}_{-  0.2}$ &   3.5$^{+  0.1}_{-  0.1}$ &   3.5$^{+  0.1}_{-  0.1}$ &   3.6$^{+  0.1}_{-  0.1}$ &   3.6$^{+  0.1}_{-  0.1}$ &   3.4$^{+  0.1}_{-  0.1}$ &   3.2$^{+  0.1}_{-  0.1}$\\[0.05in]
       $n_{i,1}\times 10^2$ &  12.9$^{+  4}_{-  3}$ &  24.3$^{+ 17}_{- 10}$ &  18.9$^{+  9}_{-  6}$ &  22.2$^{+ 14}_{-  8}$ &  16.6$^{+  6}_{-  4}$ &  15.8$^{+  5}_{-  4}$ &  12.3$^{+  2}_{-  2}$ &   6.7$^{+  0.8}_{-  0.7}$ &   5.5$^{+  4}_{-  2}$ &   3.2$^{+  2}_{-  1}$\\[0.05in]
                 $K_1$ &   5.4$^{+  0.6}_{-  0.6}$ &   6.7$^{+  0.9}_{-  0.8}$ &   5.0$^{+  0.6}_{-  0.6}$ &   5.5$^{+  0.7}_{-  0.7}$ &   4.9$^{+  0.7}_{-  0.6}$ &   5.2$^{+  0.6}_{-  0.5}$ &   5.3$^{+  0.5}_{-  0.5}$ &   5.5$^{+  1.2}_{-  1.0}$ &   7.6$^{+  2.1}_{-  1.7}$ &   6.7$^{+  1.1}_{-  0.9}$\\[0.05in]
\hline\\[-0.1in]
       \nht$v1$ (ppm) &   0.3$^{+  6}_{-  0.2}$ &   5.8$^{+94}_{-  5.8}$ &   1.1$^{+9}_{-  1.1}$ &   0.8$^{+ 34}_{-  0.7}$ &   0.0$^{+  28}_{-  0.0}$ &   0.0$^{+16}_{-0}$ &  17.3$^{+ 45}_{- 13}$ &   4.7$^{+ 42}_{-  4.2}$ &  88$^{+ 30}_{- 20}$ & 108$^{+ 23}_{- 18}$\\[0.05in]

     \pht $p_b$ (bar) &  0.20$^{+ 0.02}_{- 0.02}$ &  0.25$^{+ 0.01}_{- 0.01}$ &  0.31$^{+ 0.02}_{- 0.02}$ &  0.31$^{+ 0.02}_{- 0.02}$ &  0.28$^{+ 0.02}_{- 0.02}$ &  0.30$^{+ 0.01}_{- 0.01}$ &  0.29$^{+ 0.02}_{- 0.02}$ &  0.35$^{+ 0.04}_{- 0.04}$ &  0.81$^{+ 0.1}_{- 0.1}$ &  1.33$^{+ 0.1}_{- 0.1}$\\[0.05in]
    \pht $\alpha_0$ (ppm) &  4.63$^{+ 0.34}_{- 0.32}$ &  5.09$^{+ 0.76}_{- 0.67}$ &  4.39$^{+ 0.30}_{- 0.29}$ &  4.69$^{+ 0.47}_{- 0.42}$ &  4.83$^{+ 0.35}_{- 0.33}$ &  4.46$^{+ 0.22}_{- 0.21}$ &  5.18$^{+ 0.43}_{- 0.40}$ &  4.66$^{+ 0.33}_{- 0.31}$ &  6.50$^{+ 1.0}_{- 0.9}$ &  7.84$^{+ 1.0}_{- 0.9}$\\[0.05in]
     \asht VMR (ppb) &  1.77$^{+ 0.43}_{- 0.36}$ &  1.84$^{+ 0.53}_{- 0.43}$ &  1.75$^{+ 0.34}_{- 0.29}$ &  1.95$^{+ 0.55}_{- 0.44}$ &  1.60$^{+ 0.37}_{- 0.31}$ &  1.56$^{+ 0.28}_{- 0.25}$ &  1.44$^{+ 0.40}_{- 0.32}$ &  1.81$^{+ 0.35}_{- 0.31}$ &  2.27$^{+ 0.6}_{- 0.5}$ &  1.57$^{+ 0.4}_{- 0.3}$\\[0.05in]
\hline\\[-0.1in]
M1 ($\mu$g/cm$^2$)&     15.9$^{+  3.1}_{-  2.9}$&     11.1$^{+  2.9}_{-  2.7}$&     11.8$^{+  2.9}_{-  2.6}$&     11.1$^{+  2.2}_{-  2.1}$&     12.6$^{+  2.6}_{-  2.4}$&     13.2$^{+  3.4}_{-  3.2}$&     16.0$^{+  3.9}_{-  3.6}$&     46.7$^{+ 42}_{- 42}$&    134$^{+204}_{-203}$&    231$^{+151}_{-147}$\\[0.05in]
M2 ($\mu$g/cm$^2$)&     58.5$^{+  4.4}_{-  4.3}$&     35.6$^{+  2.2}_{-  2.1}$&     25.2$^{+  1.6}_{-  1.6}$&     23.2$^{+  1.5}_{-  1.4}$&     23.0$^{+  1.5}_{-  1.5}$&     20.7$^{+  1.3}_{-  1.2}$&     16.4$^{+  1.1}_{-  1.1}$&     13.2$^{+  0.7}_{-  0.6}$&      3.8$^{+  0.4}_{-  0.4}$&      2.6$^{+  0.7}_{-  0.7}$\\[0.05in]
M3 ($\mu$g/cm$^2$)&    103$^{+ 61}_{- 60}$&     87$^{+ 24}_{- 24}$&     58$^{+ 13}_{- 12}$&     59$^{+ 11}_{- 11}$&     50$^{+ 10}_{- 10}$&     48$^{+ 10}_{- 10}$&     39$^{+ 10}_{-  10}$&     34$^{+  4}_{-  4}$&    209$^{+ 13}_{- 12}$&     17$^{+  9}_{-  9}$\\[0.05in]
M4 (mg/cm$^2$)&     16.4&     16.4$^{+  6.4}_{-  5.2}$&      9.1$^{+  3.9}_{-  2.6}$&      6.7$^{+  3.1}_{-  2.8}$&      4.6$^{+  1.9}_{-  1.8}$&      8.5$^{+  2.5}_{-  1.9}$&      3.2$^{+  1.7}_{-  1.3}$&      2.6$^{+  1.0}_{-  0.9}$&      2.7$^{+  1.4}_{-  1.1}$&      1.3$^{+  0.3}_{-  0.3}$\\[0.05in]
\hline\\[-0.1in]
$\chi^2$ & 395.45 & 309.16 & 408.63 & 367.39 & 372.58 & 406.32 & 316.83 & 288.69 & 250.42 & 361.59\\[0.05in]
$\chi^2/N_F$ &   1.54 &   1.21 &   1.60 &   1.44 &   1.46 &   1.59 &   1.24 &   1.13 &   0.98 &   1.41\\[0.05in]
\hline\\[-0.1in]
\label{Tbl:bestfit2016}
\end{tabular}
\end{scriptsize}
\begin{small}
\parbox[c]{6in}{NOTE: Here location numbers refer to encircled pixels
  shown in Figs.\ \ref{Fig:vimsvis2016}, \ref{Fig:vimsnir2016}, and
  \ref{Fig:2016combined}A. Optical depths are given at a wavelength of
  1 \mumx. Refractive index values for each aerosol layer are
  discussed in the text.  Fixed values and computational details are
  as described for Table\ \ref{Tbl:bestfit2013}. Meaningful uncertainties
for $r_4$ and $M_4$ for location 10 were not possible because the best fit value of
$r_4$ reached the upper limit of our allowed range.}
\end{small}
\end{table*}

\paragraph{Layer 2 (putative diphosphine layer).}  This layer remained
very close to a latitude independent pressure near 300 mbar for both
2013 and 2016, with one exception, that being in the eye region in
2016 for which it descended to a pressure near 590 mbar.  A different
story is told by the optical depth of this layer.  In 2013 it
 declined from 1-2 outside the hexagon to one tenth that value
 inside the eye, with a decline that was distributed over latitude
rather than marked by sharp changes.  In 2016, however, the optical
depth of this layer reached 3 outside the hexagon and remained
above 1.0 out to the edge of the eye, at which point it
sharply dropped to the level it had in 2013. The particle 
radius for particles in this layer in 2016 was essentially independent
of latitude at about 0.5 \mum from just south of the hexagon
all the way to the edge of the eye, at which point it
dropped to just above 0.1 \mum inside the eye.  Between
2013 and 2016 the mid-hexagon column mass density for this
layer actually did increase by a factor of two, as might be
expected from the large increase in optical depth of the layer.  In 2013
its particle radius was $\sim$0.5 \mum from outside the hexagon to 76\degx N,
but stayed near 2 \mum from 78\degx N almost to the edge of the eye
at 88.8\degx N (except for the increase at Location 2, which has
an unusually high optical depth in the ammonia layer).

\paragraph{Layer 3 (\nht ice).}  This layer in 2013 was at a nearly constant
pressure near 900 mbar inside the hexagon, rising just slightly (to a lower
pressure near 630 mbar) at the edge of the hexagon.  In 2016 however, there
was a rise in altitude with latitude, moving from 1 bar just outside
the hexagon to 520 mbar at the edge of the eye, but inside the eye
the layer dropped in altitude to a pressure near 1 bar again.
In 2013 the optical depth inside the hexagon was fairly flat at about 0.1,
rising at the edge of the hexagon and at the edge of the eye to values
near 1.0, a factor of 10 increase, but dropped sharply inside the eye,
to about 0.025.  In 2016 the optical depths inside the hexagon increased
by about a factor of 3 with a decreasing slope towards the eye.  Not 
surprisingly, there was also a large (factor of two) increase in
column mass density of this layer.
The particle radius versus latitude for this layer was fairly flat for both 2013 and 2016,
staying near 2 \mumx, except near the eye in 2013 where an increase to
about 12 \mum was followed by a decline with increasing latitude to about 4 \mumx.
This large layer-3 particle size for location 2 is somewhat of an anomaly,
which is discussed in Sec.\ \ref{Sec:nh3ice}.

\begin{figure}[!t]\centering
\hspace{-0.1in}\includegraphics[width=3.0in]{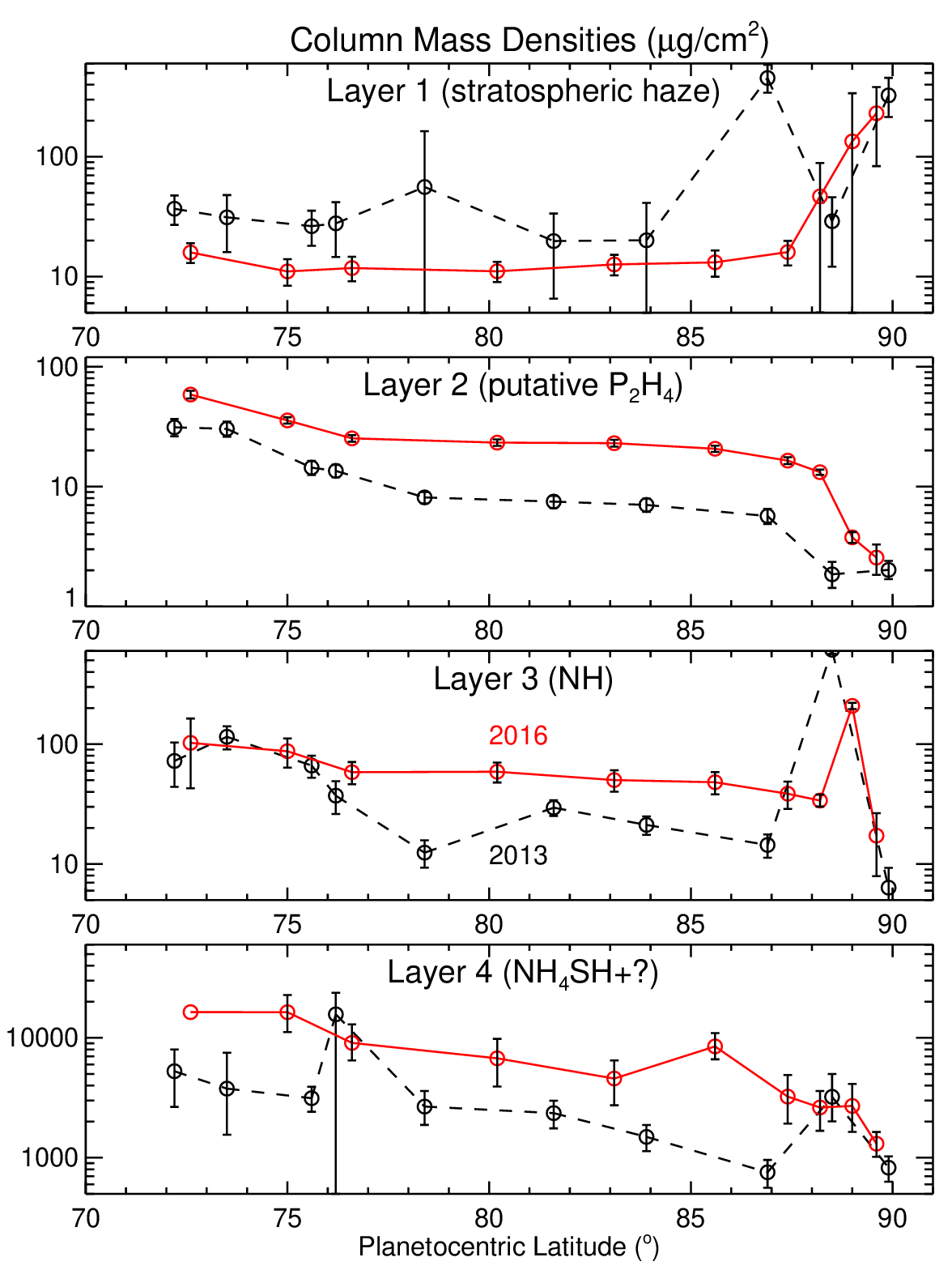}\hspace{-0.15in}
\caption{Column mass densities versus latitude by layer for 2013 (dashed)
and 2016 (solid red).}
\label{Fig:massden}
\end{figure}

\paragraph{Layer 4 (\nhfsh + \hto + ?).} The base pressure of this layer ranged
 from 2.7 to 4.5 bars, with most of the
fits close to 3.3 bars.  The pressures are well defined because of their
strong effect on emitted radiance.  The optical depths retrieved for this
layer have been between 2 and 7, with particle sizes ranging from
about 20 to 30 \mumx, although the higher numbers are not well
constrained.  Exceptions are seen at the eye, where optical depths
seem to be too large to allow defining an upper limit.  These results
are for an assumed real index of 1.6, which is between that of water
ice and \nhfshx, and for an imaginary index profile displayed in
Fig.\ \ref{Fig:varic}, which has in a very crude sense the wavelength
dependence of a mix of water ice and \nhfshx, but of a higher
magnitude.  The need for extra absorption in this layer, beyond that
provided by pure \nhfsh or water ice, was also
noted by \cite{Barstow2016} and \cite{Sro2020spole}.

\begin{figure}[!ht]\centering
\includegraphics[width=3.1in]{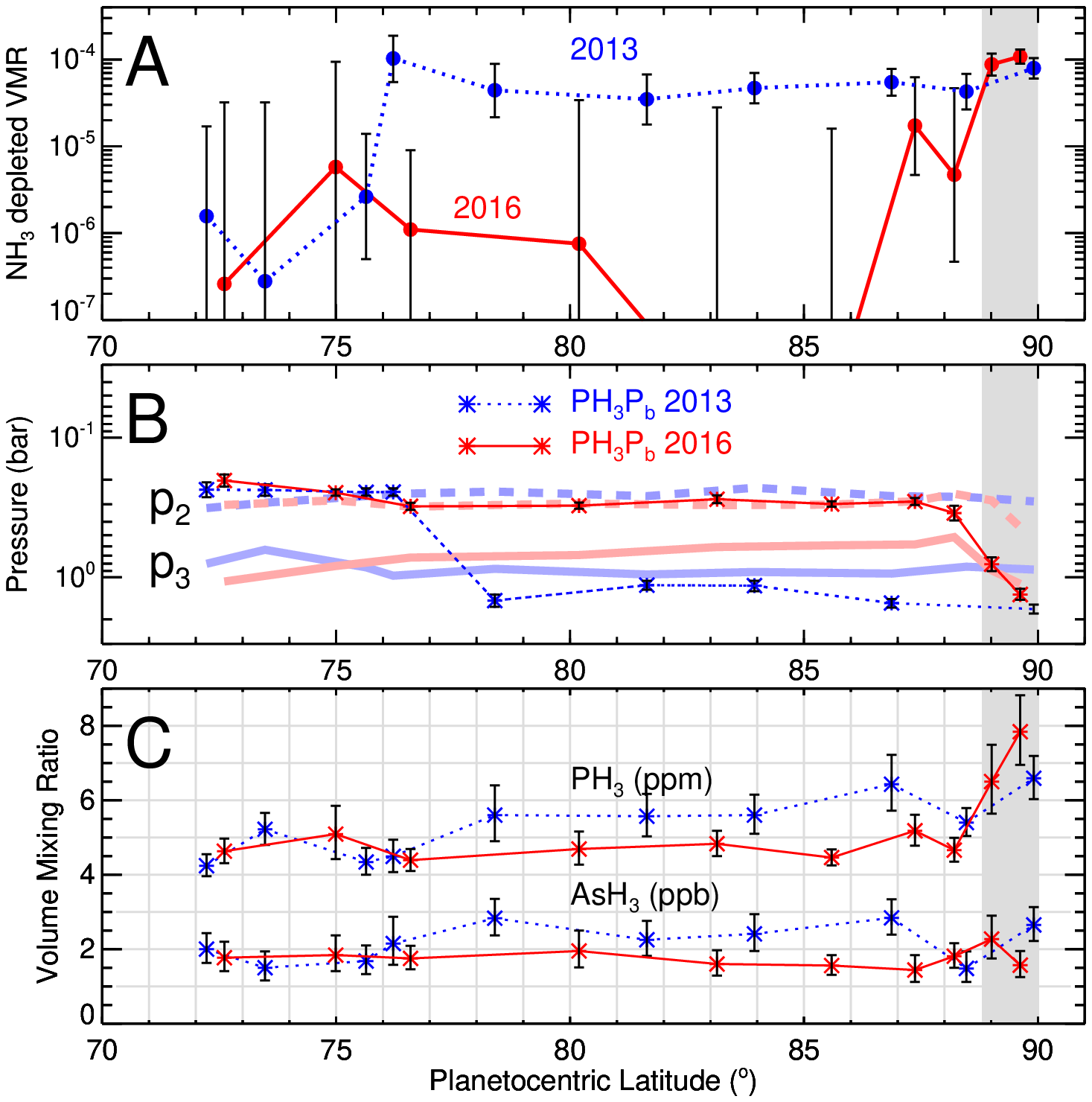}
\caption{Best-fit minor gas parameters for 2013 (blue) and  2016 (red). {\bf A:} Ammonia VMR in
the depleted region ($\sim$ to 4 bars).  {\bf B:} Phosphine
pressure break point fits (thin solid lines) compared to effective cloud pressures for layer 2 (thick dashed)
and layer 3 (thick solid). The large deviation at 88\degx N away from the 2013 trend (see Fig.\ \ref{Fig:2013+2016})
was omitted from this plot, as it  appears to be an anomalous case (see Section \ref{Sec:nh3ice}).
 {\bf C:} Volume mixing ratios of \asht and \pht versus latitude, the latter for pressures exceeding the breakpoint 
pressure.}\label{Fig:gasfits}
\end{figure}

\paragraph{Aerosol column mass densities. }

The column mass densities from Tables\ \ref{Tbl:bestfit2013}
 and \ref{Tbl:bestfit2016} are compared in Fig.\ \ref{Fig:massden} as a function
of latitude.    
The stratospheric haze layer is surprising in
two respects: first, its column mass density is comparable to that of layer 2, even
though its optical depth is 4-10 times smaller, and second,
its increase in optical depth from 2013 to 2016 is accompanied by a
decrease in column mass density rather than the expected increase.
 A possible cause is that increased
photochemical production of stratospheric condensables in 2016 results
in more condensation on existing nuclei, which increase in size and
fall faster to a level where they evaporate, taking more mass out of
the layer, but still serving to increase the effective particle size in the
stratosphere enough to produce an increase in optical depth because of
greatly increased scattering efficiency.  This scenario would suggest
that the bottom of the stratospheric haze would need to be at
pressures exceeding 100 mbar.  Particles above that level and falling to higher pressures
would experience decreasing temperatures as they fell and thus would
not evaporate before that level. The actual situation appears to be
much more complicated, given the evidence of multiple compact haze
layers in limb observations \citep{Sanchez-Lavega2020hazes}.
The column mass density of the putative diphosphine layer (layer 2) did increase
between 2013 and 2016, while its particle size changed very little.  The
ammonia ice layer column mass density generally increased over time but
displayed some large deviations, most notably for location 2 in 2013, which
was roughly two orders of magnitude more massive than its neighboring locations, 
along with its much larger particle size and optical depth. The cloud structure
in Location 2 is dominated by the ammonia ice layer and further discussed
in Section\ \ref{Sec:nh3ice}.  Layer 4 is the
most massive (note that M$_4$ is given in mg/cm$^2$ instead of $\mu$g/cm$^2$) but more variable
with latitude.

\subsection{Minor gas fit results}

\paragraph{Ammonia.} The VIMS north polar spectra are somewhat
sensitive to the volume mixing ratio of ammonia in the depleted region
above the 4-bar level. The greatest sensitivity was obtained
for observations in 2013, when
overlying aerosol opacity was lower than in 2016.
Fig.\ \ref{Fig:gasfits}A indicates that the depleted-region VMR in
2013 (blue dotted line)) was relatively independent of latitude from
76\degx N (just inside the hex boundary) all the way to the eye, with
a mean value over that region of 50.5$\pm$6 ppm, with a standard
deviation of 16 ppm.  This is comparable to the lowest value (at
10\degx S) inferred by \cite{Fletcher2011vims} for their 2-cloud
scattering model of aerosols. Their more typical values in the
northern hemisphere, up to 65\degx N, were about double that value,
with higher values near the equator, indicative of upwelling motions
there. (The \cite{Barstow2016} reanalysis of the nightside VIMS
observations used by Fletcher et. al, found about twice their levels
of \nhtx.)  Our lower 2013 values in the north polar region, even
lower than the 70-110 ppm depleted values of \cite{Briggs1989}, are
indicative of downwelling motions. However, our results for 2016 
favor much lower values, although these are very uncertain
because of the obscuring effects of increased aerosol optical
depths. The upper limits of the uncertainty range for these values can
reach almost to the 2013 levels in some cases.  Just outside the
hexagon both 2013 and 2016 values are low and comparable, but also
with greater uncertainty because of high aerosol optical depths in
that location for both years. The suggested downwelling in regions of
increased aerosol opacity is hard to understand.  The alternative
interpretation, i.e. that the retrievals are biased by the aerosol
obscuration, cannot be entirely ruled out.  One feature of our 2013
retrievals that seems at least qualitatively consistent with 2015
microwave observations at the Very Large Array is a local depletion of
the \nht abundance at the outer boundary of the hexagon
\citep{Li2020DPS}.  However, the VIMS results closest in time (for
2016) do not agree with relative variation with latitude
indicated by the VLA results, casting further doubt on the
large depletions found near the hexagon boundary in our 2013 retrievals.

\paragraph{Arsine. } Our results for \asht are relatively simple.
  Its VMR is about 2 ppb and displays only small variations with latitude
  and time (see
  Fig.\ \ref{Fig:gasfits}C).  Between latitudes 72\degx N and 90\deg N
  we found mean values of 2.18$\pm$0.19 ppb in 2013 and 1.76$\pm$0.08
  ppb in 2016, with standard deviations of 0.54 ppb and 0.24 ppb
  respectively. Individual measurement errors were typically about 0.5
  ppb.  The increased mean value for 2013 was mainly due to increased
  values derived in regions of low aerosol opacity, suggesting that
  the arsine mixing ratio declines with altitude above the cloud
  level, although this is not very dramatic and does not provide much
  support for the results of \cite{Bezard1989} who derived \asht mixing
  ratios of 2.4$^{+1.4}_{-1.2}$ ppb for what they
  call the thermal component and 0.39$^{+0.21}_{-0.13}$ ppb for what
  they refer to as the reflected solar component. The latter is
  probably representative of their effective value above the 200 -- 400
  mbar range where they inferred a haze layer, while the former
  applies to the deep mixing ratio.  Our result for 2016 is comparable
  to the 1.8$\pm$0.1 ppb found for the south polar region between
  71\degx S and 86\degx S by \cite{Sro2020spole}, which is derived
  from VIMS day-side near-IR spectra.  Our 2013 result is a
  close match to the \cite{Fletcher2011vims} global value of
  2.2$\pm$0.3 ppb, obtained from nighttime VIMS observations, but our
  2016 value is in slight disagreement. Both results are lower than the
  less accurate global estimate of 3$\pm$1 ppb by \cite{Noll1990}.

\paragraph{Phosphine.} Our results for \pht are displayed in Fig.\ \ref{Fig:gasfits}B and C.
The deep mixing ratio seems to have very little latitudinal variation
outside the eye. Between 72\degx N and 88\deg N, we found mean values
of 5.30$\pm$0.3 ppm in 2013 and 4.74$\pm$0.1 ppm in 2016, with
standard deviations of 0.71 ppm and 0.28 ppm respectively. These are
somewhat larger than the 4.4 ppm (std. dev.  0.2 ppm) found for
background clouds in the south polar region \citep{Sro2020spole}, but
well below the 6 ppm we estimated by averaging the 2007 CIRS-based results of
\cite{Fletcher2008} over a similar northern latitude range.  We also found no evidence of their
inferred decrease in VMR from 7.4 ppm at 88\deg N to 4.8 ppm at 90\deg
N.  Instead we see almost the exact opposite in our 2016 results: an
increase from 4.8$\pm$0.5 ppm at 88 N to about 7.5$\pm$0.5 ppm near
the pole. This discrepancy might be due to differences in how the \pht
profile was parameterized, or possibly a temporal variation.  A much
smaller latitudinal variation is indicated in our 2013 results: just a
1 ppm increase over a similar latitude interval, which is almost
consistent with no increase.  Global average results of
\cite{Orton2000ph3}, corrected by \cite{Orton2001ph3}, are consistent
with a deep \pht VMR of 7.4 ppm, which is also consistent with the
7$^{+3}_{-2}$ ppm deep value derived by \cite{Noll1991} from analysis
of Saturn's 4.5-5 \mum spectrum measured in 1981.  The two analyses
of the same VIMS nightside observations disagreed about the \pht VMR
values. Except for a local minimum near the equator, \cite{Fletcher2011vims} 
retrieved values near 3 ppm, almost independent of latitude for their
scattering cloud model, while \cite{Barstow2016} retrieved values about twice
as large, and ranged between 4 ppm and 6 ppm in the northern
hemisphere.  Our results would seem to be in better agreement
with \cite{Barstow2016}, although neither of the two nightside
analyses covered the polar region that we observed.

 What we found to be most variable in the \pht profile is the
 breakpoint pressure, which largely followed the effective pressure of
 the main cloud layer except where its optical depth decreased to just
 a few tenths, at which point the breakpoint pressure increased dramatically. This is well illustrated
 by the difference between 2013 and 2016 results displayed in
 Fig.\ \ref{Fig:gasfits}B. This shows that the \pht breakpoint
 pressure for 2016 is almost exactly equal to the base pressure of
 aerosol layer 2, which outside the eye has an optical depth greater
 than 1.0. But inside the eye that layer's optical depth declines
 dramatically, which is accompanied by a sharp increase in the
 breakpoint pressure.  In 2013, layer 2 reaches optical depths greater
 than one only outside the hexagon, where the breakpoint pressure
 again matches that layer's base pressure.  But inside the hexagon in
 2013 the optical depth of layer 2 declines by a factor of 10,
 resulting in the breakpoint pressure increasing
beyond the base pressure of the ammonia cloud (layer 3).  This behavior is consistent
 with the idea that the upper tropospheric cloud layer is shielding
 the vertically mixed \pht from destruction by incident solar UV
 light, as discussed in review chapters by \cite{Fouchet2009} and
 \cite{Fletcher2019book}.  

Our models are generally consistent with a steep
 decline in the \pht VMR above the relatively thick cloud layers. However,
an exception occurs in regions with high optical depths in the
ammonia layer, such as the 2013 location 2 and location 11, where the
breakpoint seems to occur at much lower pressures (see discussion in Sec.\ \ref{Sec:nh3ice}).  It
 should also be noted that the chromophore, which we assume to be in the
 stratosphere, should also provide protection, but has such a low
 optical depth that it only absorbs a modest fraction of the UV
 light. Furthermore, we cannot be sure that the chromophore particles
 remain strong absorbers at wavelengths much below the 350-nm lower
 limit of the VIMS observations, which might further reduce protection
 by the stratospheric layer.  Although we assumed no UV absorption in
 our modeling of the putative diphosphine layer, if that layer is
 actually composed of diphosphine, it would block UV light at
 wavelengths below the lower limit of VIMS observations, as already
 noted, in which case there indeed would be protection provided by
 that layer.

\subsection{Changes in cloud structure between 2013 and 2016}

\begin{figure*}[!ht]\centering
\includegraphics[width=6.2in]{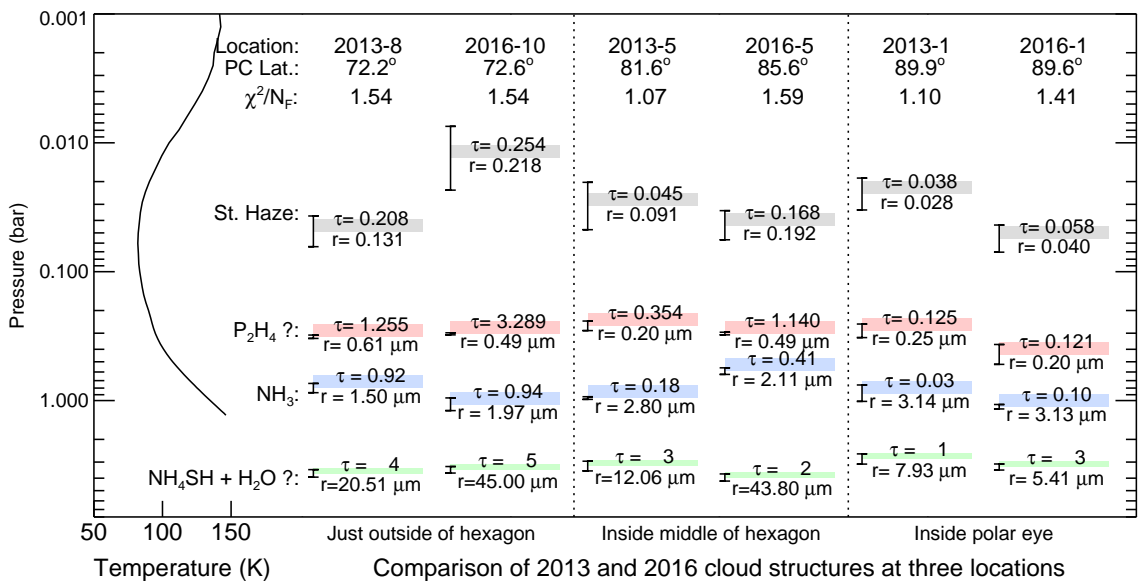}
\caption{Best fit model aerosol structures for 2013 and 2016 for three regions: just outside the
hexagon (left pair), in the middle of the hexagon (middle pair) and in the eye (right pair).
Locations and latitudes are indicated at the top of each column.  Error bars indicate the
uncertainties of the base pressure for each layer, which are small for all but the stratospheric layer.
}
\label{Fig:cartoon3}
\end{figure*}

Another way to look at changes in aerosol structure is provided in
Fig.\ \ref{Fig:cartoon3}, which displays model structures for three
typical regions for 2013 and 2016 side by side.  Just outside the
hexagon the most significant change is in the optical depth of the
putative diphosphine layer, which increases from 1.25 to 3.31. The
main change in the stratospheric haze is the increase in particle
radius from 0.13 \mum to 0.22 \mumx.  Inside the hexagon there are
significant changes in both the stratosphere and the deeper layers.
The stratospheric haze particle size doubled and the optical depth
increased by a factor of four.  The putative diphosphine layer
experienced similar changes, while the main change in the ammonia
layer was a nearly tripling of the optical depth and a nearly 300 mb
decrease in pressure.  Inside the eye region there was a doubling of
the optical depth of the stratospheric haze up to a still small value
of 0.05, with little change in particle size, and a small 30 mbar
increase in pressure.  There were also substantial changes in the deep
cloud layer, but these are likely due to different sampling of
significant deep spatial variations, temporal changes in that layer
seem to occur with a speed not characteristic of slow seasonal
changes.

\subsection{The nature of the north polar transformation}

\begin{figure*}[!b]\centering
\includegraphics[width=4in]{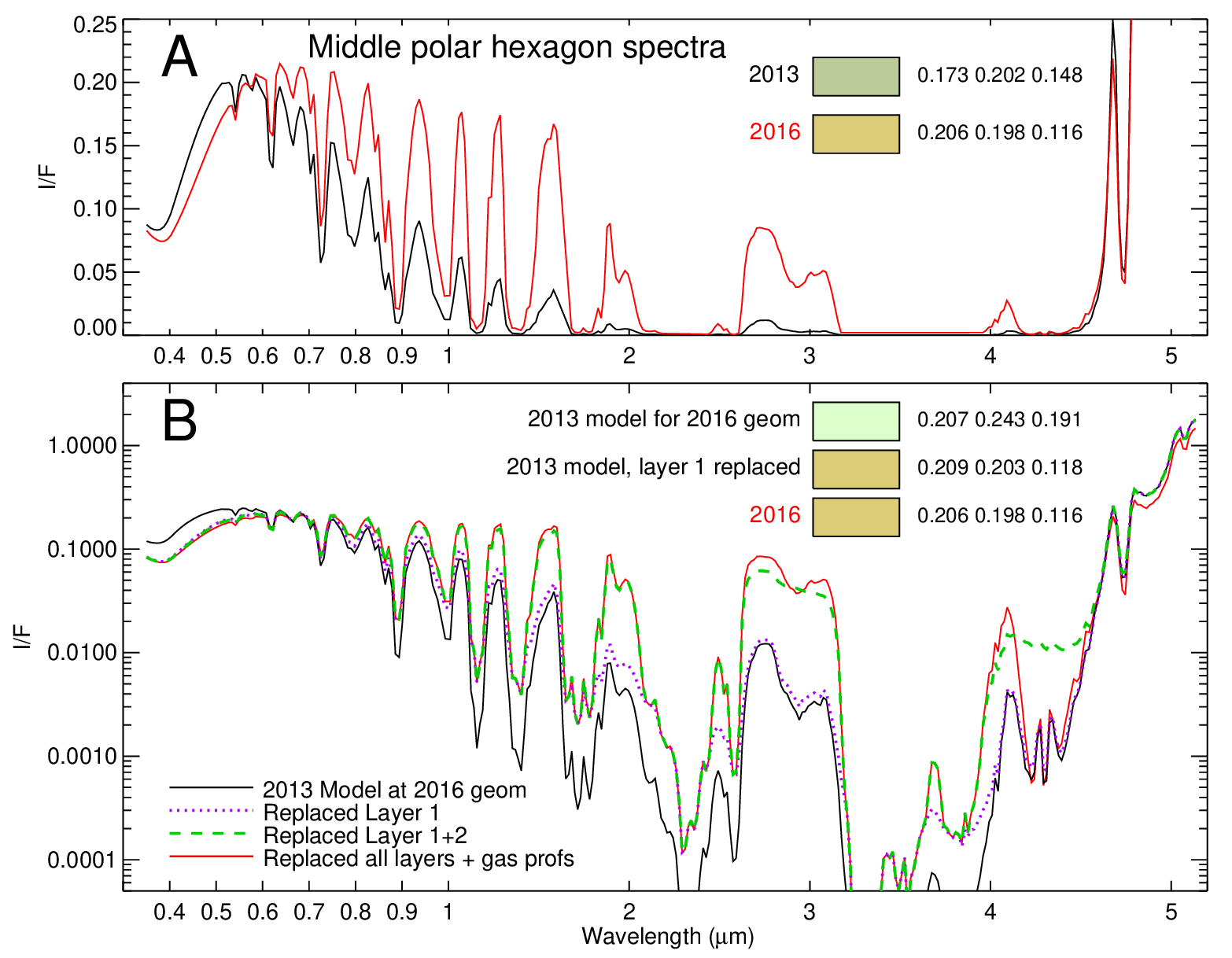}
\caption{{\bf A:} Model spectra in the middle of hexagon region in 2013
  (black, from location 5 in Fig.\ \ref{Fig:2013combined} and
  Table\ \ref{Tbl:bestfit2013}) and 2016 (red, from location 5 in
  Fig.\ \ref{Fig:2016combined} and Table\ \ref{Tbl:bestfit2016}), with
  corresponding colors computed for each. The spectra are from
  locations 5 and 6 respectively from Figs.\ \ref{Fig:2013combined}
  and \ref{Fig:2016combined}, at latitudes of 81.6\degx N and
  83.1\degx N respectively. {\bf B:} Transformation of 2013 model
  spectrum computed at 2016 observing geometry (black) by replacing
  layer 1 (dotted), by replacing layers 1 and 2 (dashed) and by
  replacing layers 1 and 2 and minor gas profiles (red), where
  replacement layers come from the 2016 model. This shows that most of
  the visual color change is due to the top layer, while most of the
  near-IR spectral change is due to layer 2. The numbers at the right
  of each color bar are RGB tristimulus values normalized by the white
  reference values.}
\label{Fig:transform}
\end{figure*}

Spectra from the middle of the hexagon region are shown in
Fig.\ \ref{Fig:transform}A for 2013 and 2016, along with their
computed colors.  Although the spectra are dramatically different, the
color differences are more subtle, with 2013 being somewhat greenish
blue and 2016 being somewhat gold.  At longer wavelengths, where
Rayleigh scattering contributes much less and methane absorption
becomes more significant, the 2016 I/F becomes many times as large as
seen in 2013.  But these large differences don't affect visual color
very much.

To better understand how aerosol changes created these spectral
differences, we show in Fig.\ \ref{Fig:transform}B the effect of
replacing 2013 model components with 2016 model components.  We begin
with the 2013 model structure with its corresponding spectrum
 recomputed for the observing geometry of
2016, which is shown as the black curve.  The blue-green color is
brightened at this observing geometry due to more favorable
illumination, which especially brightens the shortest wavelengths.  By
replacing layer 1 of the 2013 model with layer 1 of the 2016 model,
this excess short-wavelength brightness is almost entirely eliminated
and the color is transformed to what is essentially identical
to the color of the complete 2016 model. Thus the visual color change from
2013 to 2016 can be entirely attributed to changes in the stratospheric layer.
According to Tables\ \ref{Tbl:bestfit2013} and \ref{Tbl:bestfit2016}, the main changes
in this layer included particle radius increase from 0.09$\pm$0.015 \mum to 0.21$\pm$0.02 \mumx,
an optical depth increase from 0.05$\pm$0.01 to 0.19$\pm$0.03, and an important increase
in the imaginary index peak from 0.06$\pm$0.01 to 0.17$\pm$0.06. The wavelength peak
and log slope of the imaginary index did not change significantly. The change in
the peak amplitude suggests a compositional change, or a change in the fraction of
haze particles of fixed composition relative to a non-absorbing component that
was present in 2013 but not increased by photolysis over the three-year interval.  
The increase in particle radius by a factor of 2.3 and the increase in optical
depth by a factor of 3.8, strangely does not seem to imply a column mass
increase.  Although highly uncertain for the small particles in the stratospheric
layer, the calculated mass densities given in Tables \ref{Tbl:bestfit2013} and \ref{Tbl:bestfit2016}
actually indicate a column mass decrease as the optical depth of the stratospheric
layer increased.  Perhaps this is an indication that the extinction efficiency
of stratospheric particles has a size dependence different from that of spherical
particles.  The other effect
produced by the changes in the stratosphere was the increase in I/F in the deep methane
absorption bands. This was mainly due to increased scattering rather than the very slight
decrease in effective pressure. 

\begin{figure*}[!ht]\centering
\includegraphics[width=4in]{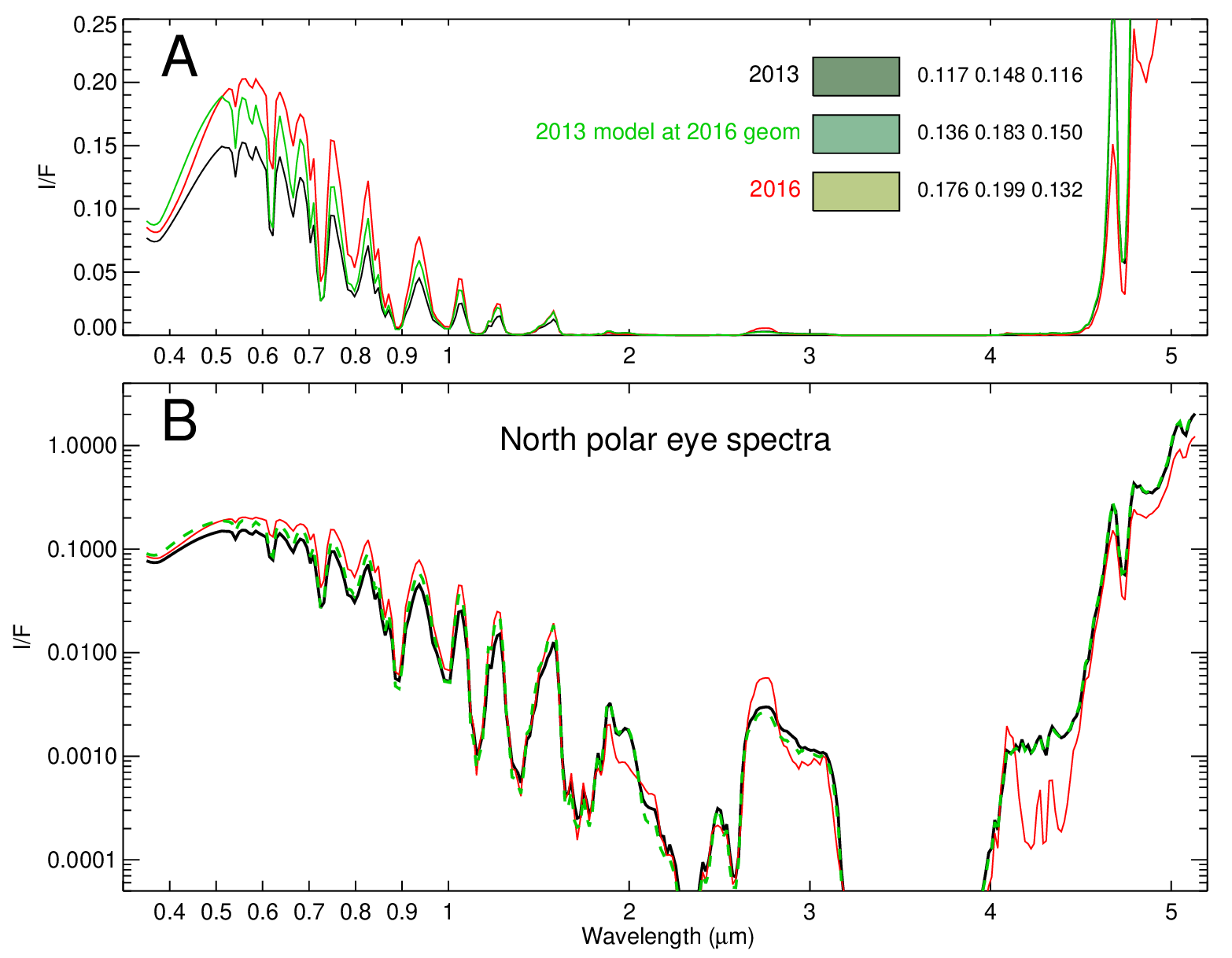}
\caption{{\bf A:} Spectra inside the north polar eye in 2013 (black,
  from location 1 in Fig.\ \ref{Fig:2013combined} and
  Table\ \ref{Tbl:bestfit2013}), in 2016 (red, from location 1 in
  Fig.\ \ref{Fig:2016combined} and Table\ \ref{Tbl:bestfit2016}), and
  the 2013 model spectrum computed for the 2016 observing geometry
  (green) {\bf B:} a logarithmic plot of results in A.}
\label{Fig:eyediff}
\end{figure*}

\begin{figure*}[!ht]\centering
\includegraphics[width=4in]{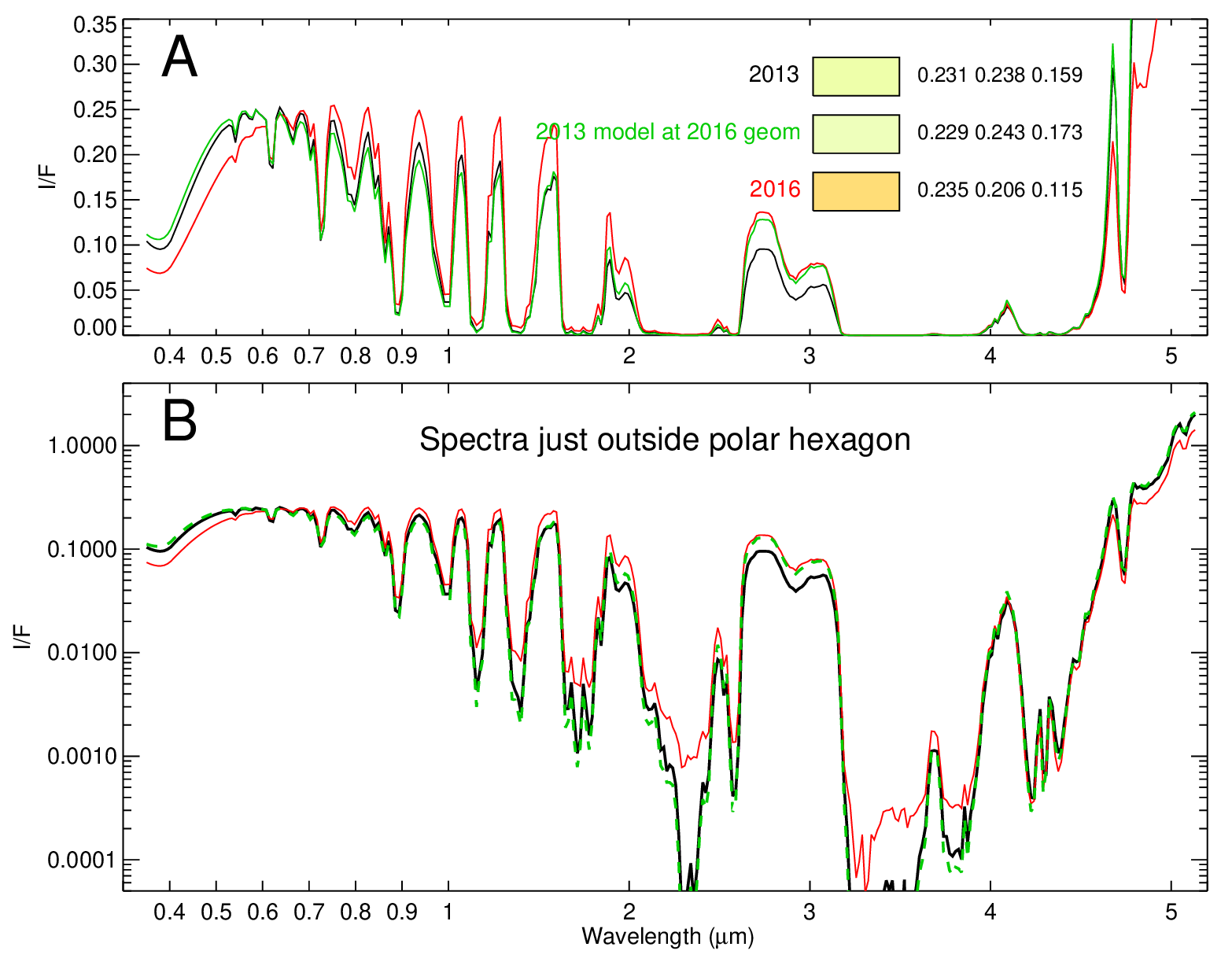}
\caption{{\bf A:} Spectra just outside the hexagon in 2013 (black, from
  location 8 in Fig.\ \ref{Fig:2013combined} and
  Table\ \ref{Tbl:bestfit2013} ), in 2016 (red, from location 10 in
  Fig.\ \ref{Fig:2016combined} and Table\ \ref{Tbl:bestfit2016}), and
  the 2013 model spectrum computed for the 2016 observing geometry
  (green) {\bf B:} a logarithmic plot of results in A. Note that the color
change here is not simply a result of a difference in observing geometry.}
\label{Fig:outdiff}
\end{figure*}

Most of the change in near-IR I/F continuum values between 2013 and 2016 can be explained by the changes that
happened in layer 2 (the putative diphosphine layer). Its pressure did not change much, but its particle
radius increased from 0.21$\pm$0.01 \mum to 0.49$\pm$0.01 \mum and its optical depth increased from 0.36$\pm$0.03 to
1.28$\pm$0.08, a factor of 3.6 increase. Replacing the 2013 layer with the 2016 layer (see the dashed
curve in Fig.\ \ref{Fig:transform}B) also had the
effect of filling in the \pht absorption bands, most notably the 4.3-\mum band.  That means
that the \pht pressure breakpoint had to move from 1.12$\pm$0.09 bars to 0.28$\pm$0.02 bars.
The change in the ammonia layer (layer 3) provided the remaining boost needed in the pseudo-continuum
I/F values (as evident from the red curve in Fig.\ \ref{Fig:transform}B). That layer's particle
size decreased slightly, from 3.0$\pm$0.4 \mum to 1.9$\pm$0.3 \mum, while its optical depth
increased from 1.7$\pm$0.1 to 4.9$\pm$0.6. The changes in the deep layer had little effect
on the external spectrum.  And the change in the \asht amount was inside the error bars.

The changes in the structure of the eye region between 2013 and 2016 are much more subtle, with
very little difference in the optical depths of the top two layers that are most responsible for the
visual color of the region, as evident from Fig.\ \ref{Fig:2013+2016}.  The color change that
is observed is mostly due to changes in the observing geometry, as evident from the model spectra
displayed in Fig.\ \ref{Fig:eyediff}. 

Outside the hexagon, model spectra in Fig.\ \ref{Fig:outdiff} show that the change in viewing
geometry has a negligible effect on the visual spectrum, with the substantial color shift
from yellow-green to gold, produced by increases in stratospheric haze absorption, as shown
in Fig.\ \ref{Fig:2013+2016}E.

\section{Discussion}

\subsection{Cloud features with strong \nht ice signatures}\label{Sec:nh3ice}

Although we intended to avoid discrete bright clouds in this study, at location 2 in
the 2013 VIMS observations we happened to sample a circular band of clouds with
an unusual spectrum having a large ratio of I/F at 2.7 \mum to that at
3.05 \mumx.  Our fit to that spectrum produced an ammonia ice layer
with unusually large particles (12 \mum instead of the more typical
2-3 \mumx) and an unusually large optical depth (0.8 vs 0.1-0.2 for mid
hexagon clouds).  To illustrate how well the composition of that layer
is constrained by the observations, we tried fitting that layer with
three different compositions, all with the $p_b$ set to 0.05 bar:
ammonia ice, \nhfshx, and a fixed real refractive index of 1.4 and zero
imaginary index.  The best fits with those compositions yielded
\chisq values of 322.94, 371.48, and 410.10, respectively, clearly
indicating a strong model preference for ammonia ice as the
composition of layer 3.  This is so obvious because in this case the ammonia ice
layer is not obscured very much by the overlying aerosols that here have
only about 1/9 the optical depth.  For most of the background cloud structures,
the overlying layers have about twice the optical depth of the ammonia ice layer,
which makes its composition far less detectable.  

Fitting the \pht
pressure breakpoint for location 2 also produced an anomalous result.
Instead of finding a value near 1.5 bars, as found just south of the
feature, the fitted value was driven to the lowest value in our fit
range.  By doing additional fits at fixed values of $p_b$ from 0.01
bar to 0.5 bars, we discovered that \chisq is low and varies
erratically up to about $p_b = 1.5$ bars, at which point it rises
dramatically.  Our initial automated fit apparently got stuck
at the lower boundary of our fit range, because of the bumpy
\chisq terrain in that region.  Our subsequent series of fits
shows that $p_b$ cannot be greater 0.15 bars,
which is at least ten times lower than for samples just south of
location 2. This indicates that within the ring of clouds sampled by
location 2 in 2013, there is an upwelling that lifts \pht to fairly
low pressures.  This upwelling might also be the origin of the
enhanced optical depth of the ammonia cloud layer.  

Another spectrum
with a strong \nht ice signature is that for location 11 in 2013,
as previously noted. Fitting our model structure to that spectrum,
we find an even larger optical depth in the ammonia ice layer (2.5 vs 1)
with a similarly low optical depth (0.23 vs 0.1) in the putative diphosphine layer.
We also found a high value for the \pht VMR (7 ppm versus the 5 ppm more generally
retrieved). 
 A more extensive analysis of \nhtx-signature
cloud features in the south polar region by \cite{Sro2020spole} also found an
enhancement of the deep \pht VMR compared to values derived from
background cloud spectra.   Our assumed composition for layer 3 was also tested
by fitting the same three alternatives used for the location 2 spectrum,  with the following
\chisq results: 394.47 (\nht ice), 1081.99 (\nhfshx), and 1468.82
(real index 1.4 and zero imaginary). In this case \nht ice is the overwhelmingly preferred composition
for layer 3. In the two very poorly fitting cases, the main fit discrepancies appear near
2 \mumx, 3 \mumx, and 4 \mumx.

\subsection{Is there enough \pht to make a cloud of \pthf particles?}

The plausibility of \pthf as the composition of layer-2 cloud particles
and photolysis of \pht as the source of those particles, depends
in part on the available supply of \pht. The idea is more plausible if the mass of the
cloud does not exceed the mass of the \pht vapor within the
cloud.  The column density of \pht over a pressure interval of
$\Delta P$ is given by (M$_{PH_3}$/M)$\alpha_{PH_3} \Delta P/g$, where
$g$ is gravity (12.13 m/s$^2$ at 88\degx N), $\alpha_{PH_3}$ is the
volume mixing ratio of \pht (about 5 ppm), and the ratio of molecular
weights of \pht to the total is given by 34/2.2 = 15.45.  This
evaluates to a column mass density of 6.4$\times 10^4$
$\mu$g/cm$^2$/bar, and for a layer thickness of 30 mbar would provide 1900
$\mu$g/cm$^2$ of \pht within the assumed thickness of the
layer 2 cloud, which is about 300 times the column mass density of the
cloud itself, and thus could provide about 150 times the number of
phosphorous atoms needed to create the cloud particles. Thus a net
upwelling flow of just 1/150th of the particle fall velocity could
sustain the cloud mass density, making \pht a very plausible source of
cloud material for layer 2. According to Fig. 11 of \cite{Roman2013}, the
fall speed of these particles (for $r$ = 0.2--0.5 \mum and $\rho \approx$ 0.8 g/cm$^3$)
should be about 0.02 -- 0.1 mm/s, which converts to 1.8 -- 9 km/year.

\subsection{Is there enough local \nht to supply the inferred \nht cloud?}

The small ammonia vapor mixing ratio we often find for the region above
the breakpoint pressure raises questions about the viability of
ammonia condensation as the source of particles in layer 3.  Applying
the analysis of the previous subsection to the ammonia cloud, assuming
a local \nht mixing ratio of about 10 ppm, and a typical assumed layer
thickness of about 90 mb we find an ammonia vapor mass density of 6386
$\mu$g/cm$^2$/bar, and thus about 576 $\mu$g/cm$^2$ within the cloud
layer, which is 25-50 times the column mass density of particles, and
thus could replenish that mass by an upwelling that was just 1/50th to
1/25th of the particle fall speed, which also makes ammonia a plausible
constituent for layer 3, even at significantly lower
mixing ratios, which is already plausible on other
grounds, namely that where its optical thickness is significant,
it actually displays obvious ammonia spectral
features.

\subsection{Could the chromophore be universal?}

Although the spectral shape of our chromophore fits for 2013 and 2016
are very similar, the absolute levels of absorption in the chromophore
particles as measured by the fitted imaginary index, is not constant.
Although the imaginary index is fairly constant as a function of
latitude for 2013, its absolute level outside the eye for 2016 is
roughly a factor of three higher than the 2013 value. \cite{Kark2005}
used a fixed spectral shape to fit all latitudes and years in their
data set, but also had to significantly adjust the absolute level of
absorption by a factor of ten between low and high latitudes.  Thus,
the inherent particle composition seems to vary, although this might
be only an apparent characteristic arising from a mix of two component
particle populations, one with a fixed composition that has a fixed
absorption level mixed with a second component that is
conservative. In this case it would be a varying ratio of two
components rather than a variation in the composition of either
component.  This situation is unlike that observed for the chromophore
on Jupiter, where \cite{Sro2017red} showed that a wide range of color
variations can be well fit with a nearly universal chromophore with a
single composition.  The lack of a universal chromophore on Saturn is
also suggested by the work of \cite{Braude2019EPSC}, who concluded
that two or more color-producing mechanisms were at work. We also made
some trial calculations to further explore this issue.  We tried to
fit a 2016 spectrum with a chromophore derived from 2013 and found
that the best-fit \chisq minimum increased from 317.02 to 742.93, a
huge worsening of the fit.  Going in the opposite direction, we tried
fitting a 2013 spectrum with a chromophore derived from a 2016
spectrum (which has a much larger imaginary index). In that case
\chisq increased from 281.46 to 312.13, still a significant worsening of the
fit, but not as dramatically as for the previous case.

\subsection{Can fit results apply to models with vertically extended layers?}\label{Sec:convert}

\begin{figure*}[!htb]\centering
\includegraphics[width=3in]{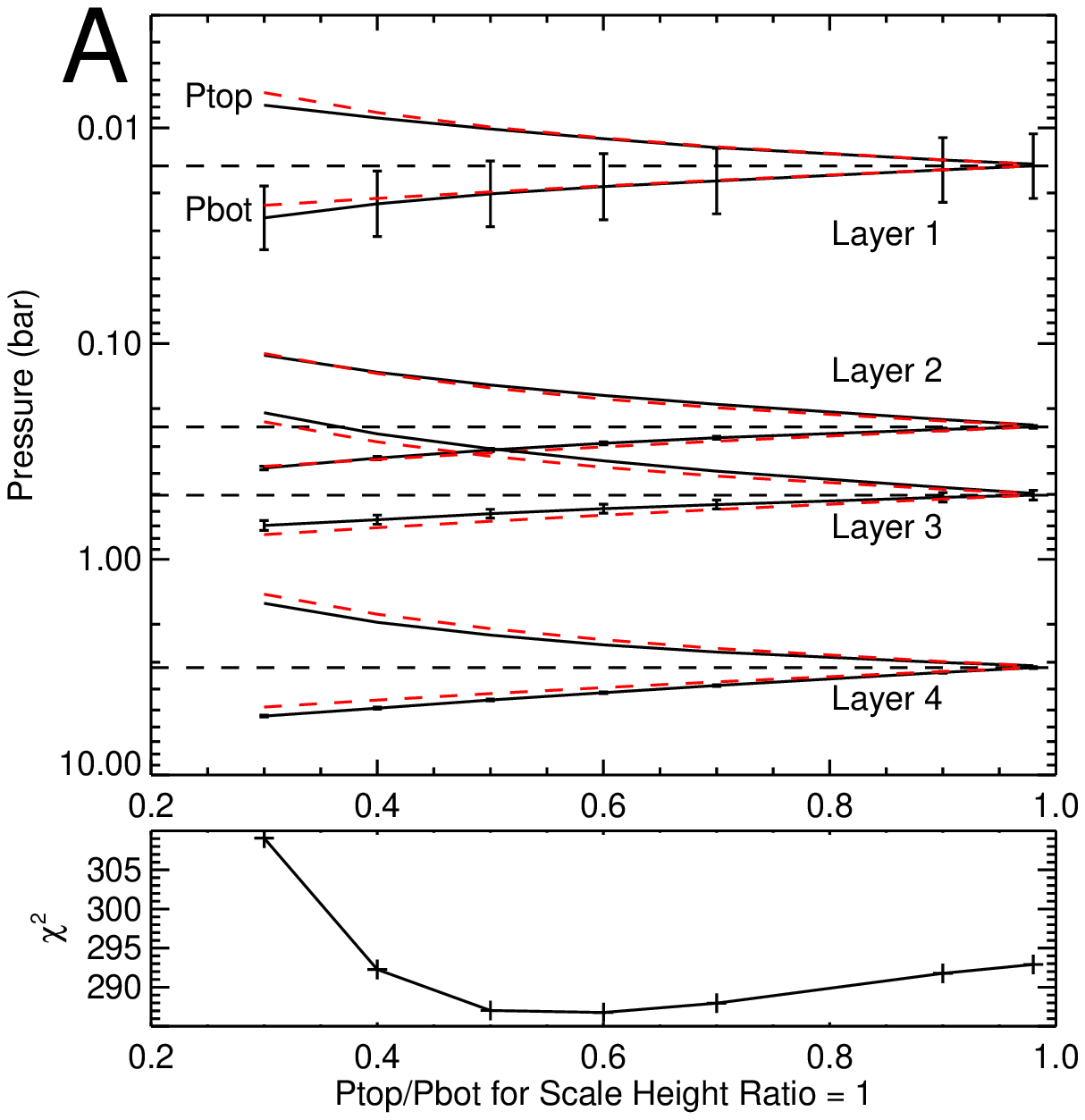}
\includegraphics[width=3in]{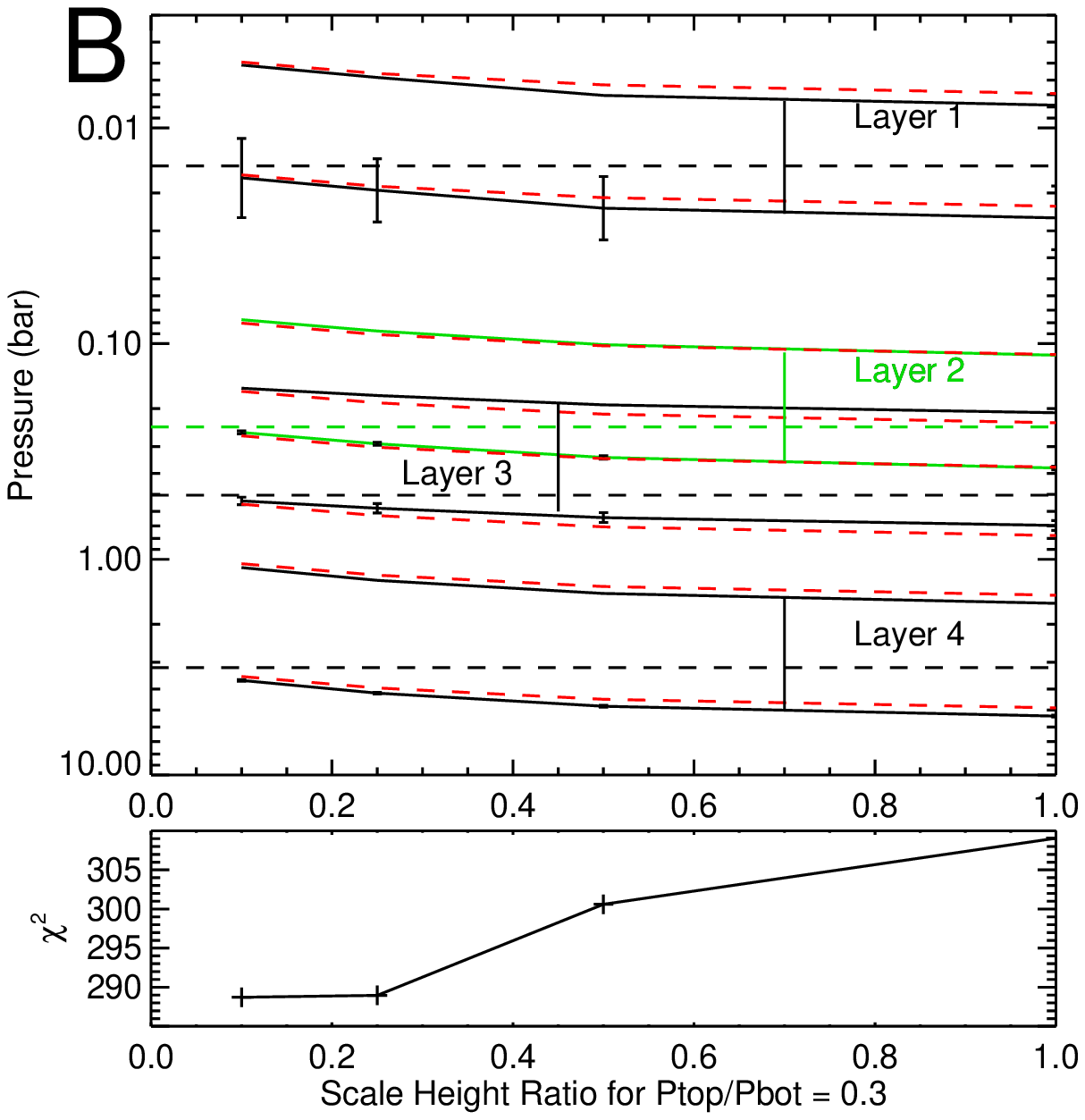}
\caption{{\bf A:} Best fit cloud pressure boundaries retrieved
from the location-3 2016 spectrum as a function of
  the assumed ratio of top to bottom pressures (solid curves) compared
  to approximations (red dashed curves) based on the assumption that
  the mean cloud pressure is fixed to the sheet cloud value (dashed
  curve).  In this case the particle scale height is equal to the pressure
scale height. {\bf B:} Best fit pressure boundaries as a function of
  particle to gas scale height ratio (solid curves) for diffuse particle layers
with a top pressure boundary that is just 30\% of the bottom pressure, compared to
  approximations (red dashed curves) for those boundaries based on assuming that the mean
  pressure of the cloud remains fixed at all scale height ratios.
  Layer 2 is plotted in green at the right to help follow the
  boundaries that overlap with those of layer 3.  The bottom panels
  display the \chisq values for each fit.}
\label{Fig:cloudthick}
\end{figure*}

Because the VIMS polar spectra are not strongly sensitive to the vertical
extent of the cloud particles in each layer (as long as the base of
each layer is not fixed), it is hard to constrain the eight additional
parameters needed to define the extra boundary and the fractional scale height
for the four layers in our standard model. We thus chose vertically thin
particle layers to simplify the model so that fewer parameters would
need to be constrained and the non-linear regression we used to
retrieve parameter values would be more
stable.  Although we did not find a need for vertically extended layers to
fit the spectra accurately, the spectra do allow such layers to be present, and
in some cases such layers might be more appropriate.  Thus it is worth
considering how we might translate our thin layer results to models with
more vertically extended layers.  From trial calculations shown in
Fig.\ \ref{Fig:cloudthick}, we found that very nearly the same fit
quality can be obtained just by refitting the top and bottom cloud
boundaries.  We also found that fairly accurate approximations for
the locations of the shifted cloud boundaries can be obtained by 
assuming that the average pressure of the cloud remains the same
independent of the vertical extent of the cloud boundaries and
independent of the particle scale height. Equations for that approximation
are provided in the following.

The most general equation for the mean pressure depends on the top and bottom
pressures and how the particles are distributed between them.
 Suppose an aerosol layer between pressures $p_\mathrm{t}$ and $p_\mathrm{b}$ has an
optical depth $\tau_\mathrm{tot}$. Further suppose that the aerosol optical depth is vertically
distributed with scale height $H_a$ and that $H_g$ is the gas (pressure) scale height.
That distribution can be written as \begin{eqnarray}
d\tau = \frac{\tau_\mathrm{tot}/f}{(p_\mathrm{b}^{1/f} - p_\mathrm{t}^{1/f})}p^{1/f -1} dp
\end{eqnarray}
where $f=H_a/H_g$ is the aerosol to gas scale height ratio.  The mean pressure within that layer can then be written as
\begin{eqnarray}
<p> = \int_{p\mathrm{t}}^{p\mathrm{b}}{p d\tau(p)}/\tau_\mathrm{tot} = \frac{p_\mathrm{b}}{(1+f)}
     \frac{(1-t^{1/f +1})}{(1-t^{1/f})} \qquad
\end{eqnarray}
where $t = p_t/p_b$.  This results in the following approximation for relating pressure boundaries for one set of $f$ and $t$ values, to those with a different set of values:
\begin{eqnarray}
p_{b2} = p_{b1} \frac{(1+t_1)}{(1+t_2)}\frac{(1-t_1^{1/f_1+1})}{(1-t_2^{1/f_2+1})}
     \frac{(1-t_2^{1/f_2})}{(1-t_1^{1/f_2})}, \\ \nonumber
p_{t2} = p_{b2}\times t_2 \label{Eq:thick}
\end{eqnarray}
where subscripts 1 and 2 refer to initial and final values respectively.  These approximations
are shown as red dashed curves in Fig.\ \ref{Fig:cloudthick}.  In panel A of that figure
all layers have $f_1=f_2$, which reduces Eq.\ \ref{Eq:thick} to the much simpler form
\begin{eqnarray}
p_{b2} = p_{b1}  \frac{(1+t_1)}{(1+t_2)},\\ \nonumber 
p_{t2} = p_{b2}\times t_2.
\end{eqnarray}

In that panel the plot of \chisq vs $t$ indicates that somewhat better
fits would have been obtained for this spectrum if we had used $t$ =
0.6, instead of $t=0.9$.  Our results for the bottom pressures would
then be increased by the ratio of 1.9/1.6 = 1.19 and our top pressures
decreased by the factor 0.6$\times$1.19 = 0.71.  That would change the
layer 2 boundaries from 252-280 mbar to 179-333 mbar, with the same
mean of 266 mbar.  Another conversion example is provided in
Fig.\ \ref{Fig:cloudthick}B. In this case we assumed a fixed value of
$t=0.3$ and varied the scale height ratio from 0.1 to 1.0.  The \chisq
plot indicates that aerosol layers vertically extended to this degree would have a
small scale height ratio, probably less than 0.3.  The boundary
model approximation for this vertical distribution also matches pretty
well the best fit values.  Thus, Eq.\ \ref{Eq:thick} provides a
general and reasonably accurate way to relate our results to models
with more vertically extended aerosols.  However, because the spectral
reflectivity effects of a layer are dominated by the upper few optical
depths these relations should not be used for conversion of optically
thick layers.

\subsection{Comparison with prior vertical structure models}

\begin{figure*}[!hbt]\centering
\includegraphics[width=6.24in]{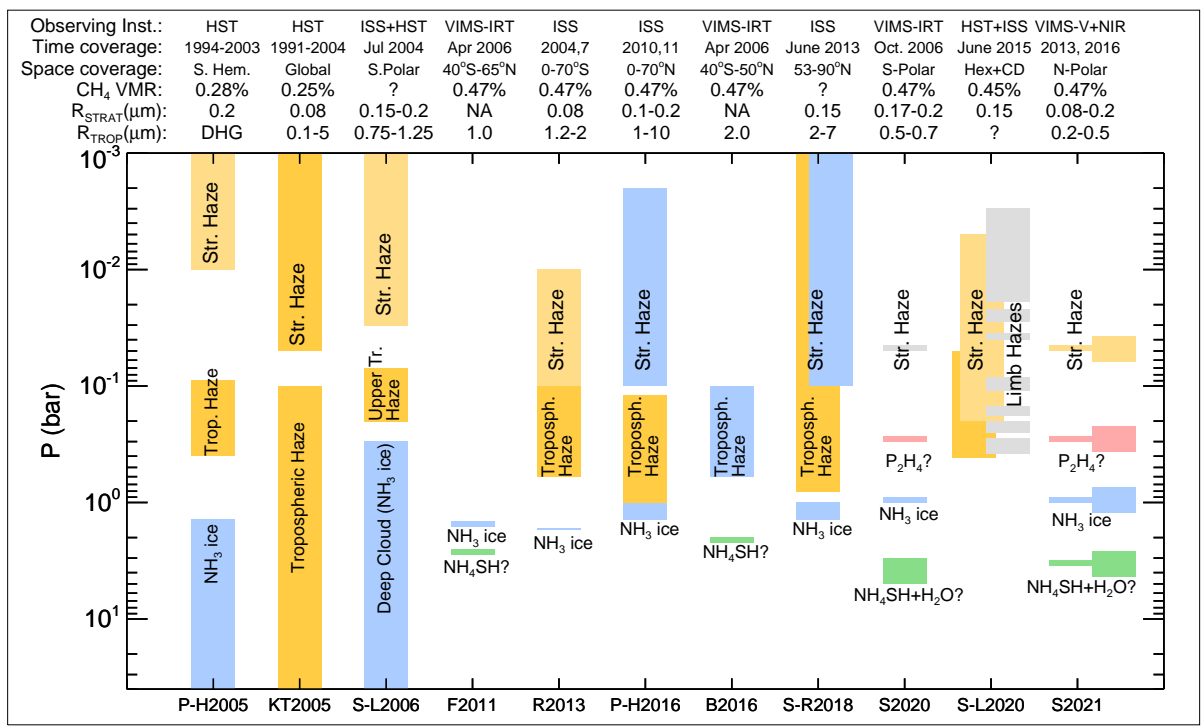}
\caption{Schematic representation of models of vertical aerosol structures used to fit
  observations characterized in the legend table above each reference
  column, where observing instrument, temporal
coverage, spatial coverage, assumed methane VMR, approximate stratospheric
and tropospheric particle radii are shown. The instrument label VIMS-IRT refers to
near IR observations including thermal emission near 5 \mumx.
 Note that successful models vary greatly in assumed vertical
  distributions. Reference labels on the x axis  are the same as defined in
the  Fig.\ \ref{Fig:seasons} caption.  Note the multiple compact layers detected
  by limb observations (gray) analyzed by S-L2020.  The tan colors mark
  layers with assumed or derived short wavelength absorption (the
  chromophore layers).  Blue denotes \nht ice
particle layers, pink possible \pthfx, and green possible \nhfsh or H$_2$O
ice particles.}\label{Fig:priormodels}
\end{figure*}

A summary of the time and space sampling of selected published works
on Saturn's cloud structure covering the last Saturnian year (29.5
earth years) was previously displayed in Fig.\ \ref{Fig:seasons}.  A
rough depiction of the vertical structure models of these various
works is presented in Fig.\ \ref{Fig:priormodels}, in order of
publication date.  The temporal sampling and spatial coverage are
indicated in tabular format at the top of the figure.  Most of these
are based on analysis of reflected sunlight in the CCD spectral range,
which have very limited sensitivity to the deeper aerosol properties.
Two (F2011 and B2016) are based on night-side VIMS observations of
thermal emission in the 5-\mum region, which have sensitivity to deep
clouds, but very limited sensitivity to the stratospheric and upper
tropospheric aerosols, and do not cover either polar region.  One
study (S2020) made use of VIMS near-IR dayside spectra of the south
polar region, and used both reflected sunlight and thermal emission
together to provide a wider range of vertical sensitivity.  At the
time of the previous overview of our understanding of Saturn's
aerosols by \cite{Fletcher2019book}, there was a consensus that
seasonal insolation changes induce hemispheric asymmetries in the
tropospheric (and likely stratospheric) hazes, with higher opacity in
the summer hemisphere and lower opacity and a bluer visual color in
the winter hemisphere.  Consistent asymmetries were also observed at
near-IR wavelengths, with increased upper tropospheric opacity
producing greater attenuation of Saturn's 5-\mum emission in the
summer hemisphere \citep{Baines2006DPS,Fletcher2011vims}.  There has
also been a consensus that that upper tropospheric aerosols reach to
higher altitudes at the equator and stratospheric aerosols reach their
greatest opacity in the polar regions.  The work we present here
(designated as S2021 in the aforementioned figures) is the only one
based on simultaneous constraints of VIMS visual and near-IR spectra
covering the 0.35-5.12 \mum spectral range and also uses both
reflected sunlight and thermal emission. In the following we first compare
our north polar results with other  models of the same region,
and then we compare our results with south polar results.

\subsubsection{North polar comparisons}
 
\paragraph{Comparison to 1991 results.}  \cite{Kark2005}
derived pole to pole stratospheric and tropospheric aerosol properties
from HST observations sampled from 1991 to 2004
(Fig.\ \ref{Fig:seasons}). Their model contains two uniformly mixed
particle layers of the same composition but they allowed that
composition to vary with time and latitude.  Their top (stratospheric)
layer extends upward from a fixed bottom pressure of 50 mbar and their
bottom (tropospheric) layer extends downward below a variable top
pressure. For the stratospheric layer, they chose a particle radius of
0.08 \mumx, independent of latitude and time. While it is a good match
to our 2013 results, it strongly disagrees with our 2016 value of 0.2
\mumx. Their inferred stratospheric optical depth in the 80\deg --
88\degx N latitude region is $\sim$0.35$\times$ the extinction
efficiency $Q_\mathrm{ext}(\lambda)$.  That evaluates to $\sim$0.008 at
1 \mum (4 years after solstice), which is about 1/5 of our 2013 value
(4 years before solstice) and about 1/20 of our 2016 value (1 year
before solstice).  Their low optical depth might be due to a rapid
decline in haze optical depth after solstice, but other factors may
also be at play.  Perhaps most significant is that our
assumed \chf VRM of 0.47\%
\citep{Fletcher2009ch4saturn} is nearly double their assumed value of
0.25\%. Their lower
value would make the stratospheric haze
much more visible in the 890-nm methane band and thus would require a
reduction in haze opacity above their fixed base pressure to fit the
I/F seen at that wavelength. Correcting for that inappropriate reduction
would bring their values closer to ours.

\begin{figure}[!hbt]\centering
\includegraphics[width=3.1in]{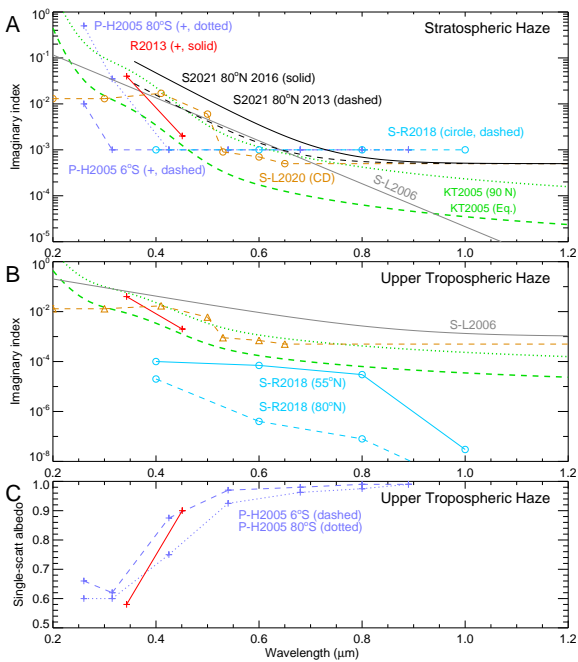}
\caption{Absorption properties of model chromophores for Saturn's
  stratosphere (A) and upper troposphere (B, C). Reference codes are
  the same as used in Figs.\ \ref{Fig:seasons} and
  \ref{Fig:priormodels}.  For those references that provided appropriate values,
  the imaginary index of refraction is plotted in A and B, with
  single-scattering albedo plotted in C.}\label{Fig:priorchrom}
\end{figure}

We do find approximate agreement in the wavelength dependence of the
imaginary refractive index, at least over the 0.35 \mum to 0.8 \mum
range.  As evident in Fig.\ \ref{Fig:priorchrom}, their imaginary
index spectral shape and absolute values for the north polar region in
1991 are in rough agreement with our results for 2013, but below the
increased values we found for 2016. We also differ of course on the
existence of a semi-infinite troposheric cloud, which is more of a
modeling convenience than a prediction, as \cite{Kark2005}, do not
claim sensitivity to particles deeper than $\sim$600 mbar.  From their
north polar observations in 1991, they inferred a tropospheric
particle radius of 0.15 \mumx, a top pressure of 80 mbar, and a
$\tau/Q_{ext}$ value of about 2/bar, which is equivalent to an optical
depth of 0.7/bar at 0.8 \mum or 0.37/bar at 1 \mumx.  However, they
found the best fit had no aerosols below the 400 mbar level, which
translates to a mean pressure of 240 mbar for their tropospheric layer
and an optical depth of 0.32 bar $\times$ 0.37/bar = 0.12 for the
total optical depth of that layer.  For comparison, our putative
diphosphine layer between 78\degx N and 87\degx N has a particle
radius of 0.2 \mumx, a mean pressure of 240 mbar, and an optical depth
of $\sim$0.3. This is relatively close agreement, except for optical
depth, which might easily a be due to a combination of their low \chf
VMR choice and temporal changes.

\paragraph{Comparison to 2013 results. }
The results of \cite{Sanz-Requena2018} are based on Cassini ISS
imagery of the north polar region taken in June of 2013, and thus
provide a close spatial and temporal match to the VIMS observations we
used for our 2013 analysis. Their model structure
(Fig.\ \ref{Fig:priormodels}) has a stratospheric haze, about which
they assumed a pressure range from 1 mbar to 100 mbar, a particle
number density independent of height, a very low spectrally flat
absorption (Fig.\ \ref{Fig:priorchrom}) and a fixed particle radius of
0.15 \mumx.  Our radius
range for 2013 is 0.04--0.08 \mum inside the eye to about 0.05--0.09
\mum inside the hexagon, and about 0.12 \mum outside the eye.   
From their assumptions and the ISS observations, they derived
an optical depth of about 0.2 at 0.4 \mum and 0.03 at 0.8 \mumx. These
convert to 0.02 and 0.015 respectively at 1 \mumx, which are 2 to 3
times smaller than our 2013 values inside the hexagon. The lower
values are not surprising because their assumed particle
size is so much larger than our fitted values. Their particles
thus have a much larger extinction efficiency, requiring
less optical depth to achieve the same observed I/F.

The upper tropospheric layer of \cite{Sanz-Requena2018} is modeled
with a fitted base pressure, a peak particle number density, and a
scale height, with a real index of 1.4 and fitted imaginary indexes at
0.4 \mumx, 0.6 \mumx, and 0.8 \mumx, which are plotted in
Fig.\ \ref{Fig:priorchrom} for their 80\degx N results.  The imaginary
values seem to be about a factor of 1000 too low to serve as Saturn's
chromophore, and since there is no chromophore in their stratospheric
model, it is hard to understand how they could fit their observations
with such a model. They found an effective particle radius for this
layer of 4 to 6 \mumx, and 1 to 2 optical depths respectively, with a
base at 800 mbar and a scale height of 32 km to 40 km, which are
70\%-90\% of the pressure scale height.  This is quite different from
our model which has a base pressure near 250 mbar, or an effective
pressure of 256 mb, a particle radius of about 0.2 \mumx, and an
optical depth of 0.3-0.4 for our 2013 mid-hexagon results.  Our values are in much better
agreement with those of \cite{Kark2005} than with those of \cite{Sanz-Requena2018},

The bottom aerosol layer of \cite{Sanz-Requena2018} is constrained by
the assumption of fixed top and bottom boundaries of 1.0 bar and 1.4
bars with a fixed refractive index the same as assumed for the
stratosphere, and a fixed particle radius of 10 \mumx.  The only
fitted parameter in this case is the optical depth, which they found
to be between 10 and 16, with uncertainties of 2 and 9 optical depths
respectively.  We did not find an aerosol layer in this pressure
range. And our closest layer is between their second and third layer,
with particle radii of 2-3 \mumx, and an optical depth of only
0.11--0.17.  Apparently our having many more spectral constraints over
a wide range of wavelengths (including thermal emission) and a model
with layers of distinct compositions and fitted particle sizes and
pressures instead of assumed values, can lead to very different
vertical distributions, particle sizes, and optical depths.

\subsubsection{Comparison to south polar results}

\paragraph{Comparison to a model with similar structure. }  From a modeling similarity standpoint, 
the results most easily compared to our north polar results are the south
polar results of \cite{Sro2020spole}, who used three compact upper
layers including an \nht ice layer, and also used as constraints both
reflected sunlight and thermal emission provided by VIMS near-IR
spectra, although they did not include visual bands.  Their observations
are from October 2006 (4 years after southern summer solstice and thus
at least 5 years in seasonal phase after our 2016 observations).  For the
stratospheric haze, they found an effective pressure of 40--60 mbar,
particle radii of about 0.18 \mum (in agreement with our 2016 radii)
with 1-\mum stratospheric optical depths about 0.025 outside the eye
(about 2/3 of our 2013 mid-hexagon value and 1/7th of our 2016 values)
and 0.01 inside the eye (about 1/4 of our value).  For the putative
\pthf layer, they found base pressures at 300--400 mbar, particle
radii of 0.3--0.6 \mumx, and optical depths of 0.3 inside the eye to
0.5--1 outside. Their results for this layer are in good agreement
with our north polar results for 2016, except that we find lower optical depths
inside the eye. The particle sizes and optical depths
are both somewhat larger than our north polar results for 2013.
 For the ammonia layer they found base pressures near 1
bar outside the eye, decreasing to 800 mb inside the eye, particle
radii of 1.5--2 \mumx, decreasing to 1--1.3 \mum inside the eye, and
optical depths of about 0.6--1.  Our 2016 analysis for this layer
yielded somewhat lower base pressures (0.6--1 bar), somewhat larger
particle radii, typically 2--3 \mumx, and similar optical depths
0.2--0.9 outside the eye, but 0.1 in the eye.  The biggest difference
between our north polar and the 2006 south polar results is that the
latter infer a stratospheric haze with larger particles and a smaller
optical depth, which might be related to the difference in seasonal
phase, considering that their south polar observations were obtained 4
years after solstice, while our north polar observations were obtained
4 years before solstice in 2013, and one year before in 2016.  If
north-south symmetry applies to the stratospheric haze behavior, these
results suggest that its optical depth increases as solstice is
approached, reaches a maximum near solstice, and declines after
solstice, but particle size, while increasing during the approach to
solstice, is slower to decline afterwards. On the other hand, the
south polar results may have found somewhat larger particles because
they did not use the VIMS visual spectral constraints, or north-south
symmetry may not be valid (a plausible possibility, given the absence
of a south polar hexagon).

\paragraph{Comparison closest in seasonal phase. } Prior studies of the
 south polar region that provide the best seasonal phase match to our
 north polar results are those of \cite{Perez-Hoyos2005} for 2003.7,
 \cite{Kark2005} for 2004.3, and \cite{Sanchez-Lavega2006} for 2004.3
 and 2004.5, all less than 2 years past the southern summer solstice
 (October 2002) as shown Fig.\ \ref{Fig:seasons}, and thus less than
 three years apart in seasonal phase from our north polar 2016
 observations.  Turning first to \cite{Kark2005} results, we find the
 same disagreement on stratospheric particle size as for the north
 pole because they assumed the same radius of 0.08 \mum for all
 latitudes and years.  Their optical depth for the 80\degx S--90\degx
 S in March 2004 is $\sim$0.5$\times Q_{ext}$, which converts to 0.012
 at 1 \mumx.  That is about 1/4 of our 2013 mid-hexagon value and less
 than 1/10th of our 2016 mid-hexagon value, suggesting that the post
 solstice decline in haze optical depth is even more rapid than
 suggested by the comparison with \cite{Sro2020spole}.  However, as
 discussed for the north polar comparison, differences in the style of
 vertical structure models and nearly a factor two difference in
 assumed methane mixing ratios adds uncertainty to the meaning of
 these differences.

The comparison with \cite{Perez-Hoyos2005} is also clouded by their
assumption of a low \chf VMR of 0.28\%. Their model contained three
layers as illustrated in Fig.\ \ref{Fig:priormodels}.  They confined
their stratospheric haze to lie between 1 mbar and 10 mbar, which, combined
with their low \chf VMR, we would expect to lead to artificially low
optical depths.  However, their 80\degx S value at 889-nm is
equivalent to 0.02 to 0.04 at 1 \mumx, which is not far above our 
2013 mid-hexagon north polar results and close to the 80\degx S results
of \cite{Sro2020spole}, even though closer to solstice, and thus
suggesting an intermediate rate of decline. 
The inferred stratospheric absorption results of
\cite{Perez-Hoyos2005} are very different from our results and from
those of \cite{Kark2005}, displaying a very steep gradient from 260 nm
to 450 nm, but flat at longer wavelengths (see
Fig.\ \ref{Fig:priorchrom}).

It appears that much of the shortwave
absorption that is missing from their stratospheric haze is provided
by their upper tropospheric layer (see panel C of
Fig.\ \ref{Fig:priorchrom}), which is modeled using a wavelength-dependent 
single-scattering albedo and phase function, but a wavelength-independent
optical depth.  Their upper tropospheric layer in the polar region
extends from 90 mbar to 400 mbar, with about 3 optical depths for the entire
layer. Our optical thickness of this layer for
the mid-hexagon region of the north pole is about 0.3 optical depths
in 2013 and but close to 1 optical depth in 2016, both at 1 \mumx.
\cite{Sro2020spole} found particles in the 0.5 \mum to 0.6 \mum range
for this layer in the south polar region, at an effective pressure
near 300 mbar, which is somewhat deeper than the 245 mbar mean
pressure of \cite{Perez-Hoyos2005}, but with an optical depth of 0.6
to 0.8 outside the eye region, which is within the lower limit of
the optical depth range inferred by \cite{Perez-Hoyos2005}.

\cite{Sanchez-Lavega2006} used a model structure similar to that of \cite{Perez-Hoyos2005}
with the main exceptions being the use of spherical particles for the upper tropospheric
layer and the increased stratospheric absorption they inferred in the
stratosphere between 350 nm and 650 nm (see Fig.\ \ref{Fig:priorchrom}).  
Although their assumed value for the \chf VMR was not given, their paper predates
by three years the \cite{Fletcher2009ch4saturn} paper that established the current 
accepted value, and thus probably used the same value as  \cite{Perez-Hoyos2005}.  
They constrained a stratospheric haze extending from 1 mbar to a fitted bottom
of 30  mbar, with best fit particle radii of 0.15--0.2 \mumx, and an optical depth
of $\sim$0.05, which is about double that of \cite{Sro2020spole} and comparable to
our 2013 mid-hexagon results but about 1/4 of our 2016 mid-hexagon values. 
They placed their tropospheric haze layer between 70--80 mbar and 150--300 mbar,
with mean pressures from 110 mbar to 190 mbar, substantially lower in pressure
 than most other results.  They inferred particle radii of 0.75--1.0 \mum (comparable
to \cite{Sro2020spole} values outside the eye), and
optical depths of 0.9--1.2 \mumx, which are just slightly above the \cite{Sro2020spole} values
outside the eye, but both are above the levels we found in the north
polar region. The model of \cite{Sanchez-Lavega2006} also contains a putative ammonia
ice cloud that is taken to be dense and semi-infinite, with a top at 500$\pm$100 mb
at 80\degx S and 300$\pm$100 mbar at 87\degx S and 89.5\degx S.  This strongly conflicts
with the \cite{Sro2020spole} results, which put the effective pressure closer
to one bar and the optical depth near 1 as well.  A dense semi-infinite cloud
of ammonia ice is ruled out by the thermal emission constraints near 5-\mumx.

Unfortunately, many of the differences we find among various results cannot be
confidently interpreted as temporal (or seasonal phase) variations because of different styles of modeling,
different assumptions about the methane mixing ratio and pressure boundaries, and different observational
constraints, some covering just the CCD spectral range with a few bandpass filters,
and others covering a wide spectral range with complete spectra.  Clearly the deeper
aerosol layers benefit from near-IR observations and the combined use of reflected
sunlight and thermal emission.

\section{Summary and conclusions}

Using the latest VIMS calibrations, we combined co-located visual and infrared VIMS spectral
imaging observations of Saturn's north polar region from just outside the hexagon region to
inside the eye to create spectra spanning the range from 0.35 \mum to 5.12 \mumx, acquired in
daylight so that reflected sunlight and Saturn's thermal emission could both be used
to constrain Saturn's vertical aerosol structure and composition.  We used
a model containing four compact layers of spherical particles with different compositions in each layer to create
synthetic spectra for comparison with the observed VIMS
spectra. Models were constrained by the observations using a Levenberg-Marquardt non-linear
fitting code. The models fit the spectra well and most model parameters were
well constrained by the spectra.  The main results of our analysis, as constrained by
our assumed model structure as well as the observations, are described in the following.

\begin{enumerate}
\setlength{\itemsep}{0pt}
\item Layer 1, at an effective pressure generally in the 30 mbar to 50
  mbar range contains sub-micron particles with significant
  short-wavelength absorption that plays a major role in defining
  Saturn's color (this is the layer we assumed to be Saturn's chromophore). 
 From the hexagon boundary at about 75\degx N to the
  eye boundary at 88.8 \degx N the layer changed significantly between
 2013 and 2016: its particle sizes increased from below
  0.1 \mum to 0.2 \mumx, its optical depth increased by about a
  factor of 5, its column mass density decreased by a about
a factor of two,  and its imaginary index increased by a factor of three.
  In the eye (from 88.8\degx N to the pole) there was little change
in any of the parameters. The optical properties of the chromophore
did not vary much with latitude over this period.
 While a chromophore might be present in other layers as well,
  the best defined and simplest fits were obtained by placing the
  chromophore entirely in the stratosphere.

\item Layer 2, with base pressures found generally between 250 mbar and 300 mbar, lacks
  near-IR absorption features that might identify its composition. We
  cannot rule out the photochemically preferred \pthf as its main
  component because we do not know enough about the imaginary index of
  \pthfx.  Assuming the real index of \pthf for this layer (1.74)
  leads to somewhat more consistent fits than assuming an index of
  1.4. This layer has a column mass density that, if
  composed of \pthfx, could be replenished by an upwelling of \pht gas
  that was more than 150 times smaller than sedimentation
  speed of particles.  Between 2013 and 2016 this layer did not change much in base pressure,
 but between the hexagon outer boundary and the eye
its optical depth increased by a factor of 3--5, and its column mass
density by a similar amount, with little change inside the eye. 

\item Layer 3 has a base pressure located between 950 mbar and 1.05 bars, in a pressure
range where ammonia ice condensation is expected, and displays the
3-\mum absorption expected of ammonia ice, which is visible when upper layers
are optically thin or when the layer 3 optical depth is unusually
high. A small \nht VMR of only 10 ppm in the condensation pressure to
4 bar region is sufficient to resupply the particle
mass with an upwelling of gas only 1/25th of the sedimentation speed. Between
2013 and 2016 this layer also increased its optical depth and column mass
density by factors of 2-3, with little change in particle radius. Generally
much smaller changes were seen in the eye.

\item Layer 4, residing between 2.9 and 4.5 bars, mainly constrained
  by its effect on thermal emission, is plausibly partly composed in part of
  a mix of \nhfsh and water ice but seems to need additional absorption
  to better match the thermal emission spectrum near 5 \mum and to
  limit scattering contributions at short wavelengths. Our
empirical model provide better fits than either \nhfsh or \htox. Our retrievals
find very large particles, with radii of 10's of \mumx. Its column
mass density far exceeds all other layers, and outside the eye seems to have
increased by roughly a factor of three between 2013 and 2016.  This
layer also exhibits essentially no time dependence inside the eye.

\item An accurate measure of the color of the polar region, using full
  spectra instead of bandpass filter samples, shows that the color
  change within the hexagon region was from blue/green to gold and
  could be entirely explained by changes in a stratospheric
  chromophore layer: a doubling of its optical depth, a doubling of its
  particle radius, and a tripling of its imaginary index.  The large
  spectral changes in the near-IR region are mainly due to a factor of
  3--10 increase in the optical depth of the putative diphosphine
  layer (Layer 2 of our model).

\item VIMS observations are not sensitive to
the deep value of the \nht VMR, but somewhat sensitive to
its VMR in the depleted region between about 4 bars
and the \nht condensation level at generally less than 1 bar in our
results.  In 2013, aerosol optical depths were low enough that
within the hexagon, including the eye region,
 we were able to retrieve reasonably well constrained
values of about 50 ppm.  But where aerosol optical depths were much
greater, as at all latitudes in 2016 (except in the eye),
and just outside the hexagon boundary in both years, the retrieved 
best-fit VMR values were very low and very uncertain, with upper
limits of the uncertainty range approaching 2013 values in some cases.

\item We found average arsine VMR values of 2.18$\pm$0.19 ppb in 2013
  and 1.76$\pm$0.08 in 2016, with standard deviations of 0.54 ppb and
  0.24 ppb respectively.  The slightly higher values in 2013 are due
  mainly to higher values inside the hexagon, where aerosol opacity
  was generally low at that time, suggesting that arsine declines with
  altitude above the cloud level.

\item With an assumed small fractional scale height above an
  adjustable breakpoint pressure, below which a deep mixing ratio was
  derived, we found a virtually constant deep \pht VMR of 5 ppm, and a
  variable breakpoint pressure almost always near the top of the main
  aerosol layer (Layer 2).  But where that layer had very small
  optical depth, the best-fit pressure breakpoint moved just to beyond the next
  deeper layer (ammonia ice cloud), suggesting that the cloud layers
  provide protection against UV photolysis that efficiently destroys
  \pht above them, as suggested in previous publications.

\item While our compact layers were assumed have top pressures that
  were 90\% of their bottom pressures, we found that high quality fits
  could also be obtained with thicker layers, and provided equations
  to transform layer parameters to values appropriate to thicker
  layers.  In one case we found that setting top pressures of all
  layers to 60\% of their bottom pressures resulted in somewhat improved
  fit quality. Another case found thinner layers were preferred.
  This is an area that merits further investigation.

\item Location 2 from 2013 happened to sample a
ring of clouds at 88.5\degx N that displayed unusual spectral features that
are associated with enhanced optical depths in the ammonia ice layer.  Retrievals
confirming this also found  unusually low  \pht breakpoint 
pressures, and enhanced deep \pht VMR values, suggesting local upwelling motions.
The enhanced optical depth of layer 3 allowed us to confirm its identity by 
comparing fit quality assuming the \nht refractive index with 
the qualities obtained assuming \nhfsh (\chisq increased by 50),
 and a simple real index of 1.4 (\chisq increased by almost 90), providing
strong evidence in favor of our chosen composition. Applying
the same test to a bright discrete feature from 2013 location 11, the \chisq
increases were by huge values of 690 and 1070 respectively, providing overwhelming
preference for \nht ice as the composition of layer 3 particles. 

\item Comparisons with prior published models of southern and northern polar regions
are roughly consistent with the idea that stratospheric hazes in both polar
regions rise in optical depth as their summer solstice approaches and decline
following solstice, but differences in model styles, constraints, and assumed
values of the \chf VMR, as well as differences in chromophore models, add
considerable uncertainty to the comparisons with prior models.

\end{enumerate}

All of these results must be understood within the context of our assumed model
structure. There are likely other models that could be constructed to fit the spectra
just as well. It should also be remembered that we have not thoroughly explored the
enormous 18-20 dimensional space of parameter variation, so that there might also
be other solutions even within the constraints of our assumed model structure. Further
improvements might be made by using a variety of methods to effectively improve
the signal to noise ratios of the observations by spatial averaging or simultaneously
fitting observations at different view angles within a given latitude band. It would
also be valuable to have laboratory measurements of the optical properties of diphosphine
from UV to the thermal infrared. 

\section{Acknowledgments}

We are grateful to the two reviewers who provided prompt,  detailed,
and constructive review comments that led to significant improvements in the paper.  This
research was supported by NASA's Cassini Data Analysis Program via
award 80NSSC18K0966.  The archived data associated with this paper can be obtained
from the Planetary Data System (PDS) Atmospheres Node, including calibrated Cassini VIMS
and ISS datasets, as well as tabular data used in figures, and a public domain version
of the paper.



\begin{thebibliography}{85}
\expandafter\ifx\csname natexlab\endcsname\relax\def\natexlab#1{#1}\fi
\expandafter\ifx\csname url\endcsname\relax
  \def\url#1{\texttt{#1}}\fi
\expandafter\ifx\csname urlprefix\endcsname\relax\def\urlprefix{URL }\fi

\bibitem[{{Achterberg} and {Flasar}(2020)}]{Achterberg2020}
{Achterberg}, R.~K., {Flasar}, F.~M., 2020. {Saturn's Atmospheric Helium
  Abundance from Cassini Composite Infrared Spectrometer Data}. The Plan. Sci.
  J. 1:30, 14.

\bibitem[{{Achterberg} et~al.(2018){Achterberg}, {Flasar}, {Bjoraker},
  {Hesman}, {Gorius}, {Mamoutkine}, {Fletcher}, {Segura}, {Edgington}, and
  {Brooks}}]{Achterberg2018}
{Achterberg}, R.~K., {Flasar}, F.~M., {Bjoraker}, G.~L., {Hesman}, B.~E.,
  {Gorius}, N.~J.~P., {Mamoutkine}, A.~A., {Fletcher}, L.~N., {Segura}, M.~E.,
  {Edgington}, S.~G., {Brooks}, S.~M., 2018. {Thermal Emission From Saturn's
  Polar Cyclones}. \grl 45~(11), 5312--5319.

\bibitem[{{Acton}(1996)}]{Acton1996}
{Acton}, C.~H., 1996. {Ancillary data services of NASA's Navigation and
  Ancillary Information Facility}. Planet. and Space Sci. 44, 65--70.

\bibitem[{{Adriani} et~al.(2007){Adriani}, {Moriconi}, {Filacchione}, {Tosi},
  {Coradini}, {Brown}, {Baines}, {Buratti}, {Matson}, {Nelson}, {Momary},
  {Bellucci}, {Formisano}, {Capaccioni}, {Cerroni}, {Clark}, {Cruikshank},
  {Showalter}, {Drossart}, {Sicardy}, {Jaumann}, {Baugh}, {Griffith},
  {Langevin}, {McCord}, {Hansen}, {Hibbitts}, {Mennella}, {Nicholson}, and
  {Sotin}}]{Adriani2007}
{Adriani}, A., {Moriconi}, M.~L., {Filacchione}, G., {Tosi}, F., {Coradini},
  A., {Brown}, R.~H., {Baines}, K.~H., {Buratti}, B.~J., {Matson}, D.~L.,
  {Nelson}, R.~M., {Momary}, T.~W., {Bellucci}, G., {Formisano}, V.,
  {Capaccioni}, F., {Cerroni}, P., {Clark}, R.~N., {Cruikshank}, D.~P.,
 {et al.},
2007.

\bibitem[{{Anderson} et~al.(2004){Anderson}, {Sides}, {Soltesz}, {Sucharski},
  and {Becker}}]{Anderson2004}
{Anderson}, J.~A., {Sides}, S.~C., {Soltesz}, D.~L., {Sucharski}, T.~L.,
  {Becker}, K.~J., 2004. {Modernization of the Integrated Software for Imagers
  and Spectrometers}. In: {Mackwell}, S., {Stansbery}, E. (Eds.), Lunar and
  Planetary Science Conference. Vol.~35 of Lunar and Planetary Science
  Conference. p. 2039.

\bibitem[{{Antu{\~n}ano} et~al.(2018){Antu{\~n}ano}, {del
  R{\'\i}o-Gaztelurrutia}, {S{\'a}nchez-Lavega}, and
  {Rodr{\'\i}guez-Aseguinolaza}}]{Antunano2018}
{Antu{\~n}ano}, A., {del R{\'\i}o-Gaztelurrutia}, T., {S{\'a}nchez-Lavega}, A.,
  {Rodr{\'\i}guez-Aseguinolaza}, J., 2018. {Cloud morphology and dynamics in
  Saturn's northern polar region}. \icarus 299, 117--132.

\bibitem[{{Antu{\~n}ano} et~al.(2015){Antu{\~n}ano},
  {R{\'{\i}}o-Gaztelurrutia}, {S{\'a}nchez-Lavega}, and {Hueso}}]{Antunano2015}
{Antu{\~n}ano}, A., {R{\'{\i}}o-Gaztelurrutia}, T., {S{\'a}nchez-Lavega}, A.,
  {Hueso}, R., 2015. {Dynamics of Saturn's polar regions}. J. of Geophys. Res.
  (Planets) 120, 155--176.

\bibitem[{{Atreya} and {Wong}(2005)}]{Atreya2005SSR}
{Atreya}, S.~K., {Wong}, A., 2005. {Coupled Clouds and Chemistry of the Giant
  Planets - A Case for Multiprobes}. Space Sci. Rev. 116, 121--136.

\bibitem[{{Baines} et~al.(2009{\natexlab{a}}){Baines}, {Delitsky}, {Momary},
  {Brown}, {Buratti}, {Clark}, and {Nicholson}}]{Baines2009stormclouds}
{Baines}, K.~H., {Delitsky}, M.~L., {Momary}, T.~W., {Brown}, R.~H., {Buratti},
  B.~J., {Clark}, R.~N., {Nicholson}, P.~D., 2009{\natexlab{a}}. {Storm clouds
  on Saturn: Lightning-induced chemistry and associated materials consistent
  with Cassini/VIMS spectra}. \planss 57, 1650--1658.

\bibitem[{{Baines} et~al.(2005){Baines}, {Drossart}, {Momary}, {Formisano},
  {Griffith}, {Bellucci}, {Bibring}, {Brown}, {Buratti}, {Capaccioni},
  {Cerroni}, {Clark}, {Coradini}, {Combes}, {Cruikshank}, {Jaumann},
  {Langevin}, {Matson}, {McCord}, {Mennella}, {Nelson}, {Nicholson}, {Sicardy},
  and {Sotin}}]{Baines2005}
{Baines}, K.~H., {Drossart}, P., {Momary}, T.~W., {Formisano}, V., {Griffith},
  C., {Bellucci}, G., {Bibring}, J.~P., {Brown}, R.~H., {Buratti}, B.~J.,
  {Capaccioni}, F., {Cerroni}, P., {Clark}, R.~N., {Coradini}, A., {Combes},
  M., {Cruikshank}, D.~P., {Jaumann}, R., {Langevin}, Y., {Matson}, D.~L.,
  {McCord}, T.~B., {Mennella}, V., {Nelson}, R.~M., {Nicholson}, P.~D.,
  {Sicardy}, B., {Sotin}, C., 2005. {The Atmospheres of Saturn and Titan in the
  Near-Infrared First Results of Cassini/VIMS}. Earth Moon and Planets 96,
  119--147.

\bibitem[{{Baines} et~al.(2009{\natexlab{b}}){Baines}, {Momary}, {Fletcher},
  {Showman}, {Roos-Serote}, {Brown}, {Buratti}, {Clark}, and
  {Nicholson}}]{Baines2009cyclone}
{Baines}, K.~H., {Momary}, T.~W., {Fletcher}, L.~N., {Showman}, A.~P.,
  {Roos-Serote}, M., {Brown}, R.~H., {Buratti}, B.~J., {Clark}, R.~N.,
  {Nicholson}, P.~D., 2009{\natexlab{b}}. {Saturn's north polar cyclone and
  hexagon at depth revealed by Cassini/VIMS}. \planss 57, 1671--1681.

\bibitem[{{Baines} et~al.(2006){Baines}, {Momary}, {Roos-Serote}, and {the
  Cassini/VIMS Science Team}}]{Baines2006DPS}
{Baines}, K.~H., {Momary}, T.~W., {Roos-Serote}, M., {the Cassini/VIMS Science
  Team}, 2006. {North vs South on Saturn: Discovery of a Pronounced
  Hemispherical Asymmetry in Saturn's 5-Micron Emission and Associated Deep
  Cloud Structure by Cassini/VIMS}. In: AAS/Division for Planetary Sciences
  Meeting Abstracts. Vol.~38 of Bulletin of the American Astronomical Society.
  p. 488.

\bibitem[{{Baines} et~al.(2018){Baines}, {Sromovsky}, {Fry}, {Momary}, {Brown},
  {Buratti}, {Clark}, {Nicholson}, and {Sotin}}]{Baines2018GeoRL}
{Baines}, K.~H., {Sromovsky}, L.~A., {Fry}, P.~M., {Momary}, T.~W., {Brown},
  R.~H., {Buratti}, B.~J., {Clark}, R.~N., {Nicholson}, P.~D., {Sotin}, C.,
  2018. {The Eye of Saturn's North Polar Vortex: Unexpected Cloud Structures
  Observed at High Spatial Resolution by Cassini/VIMS}. \grl 45, 5867--5875.

\bibitem[{{Barstow} et~al.(2016){Barstow}, {Irwin}, {Fletcher}, {Giles}, and
  {Merlet}}]{Barstow2016}
{Barstow}, J.~K., {Irwin}, P.~G.~J., {Fletcher}, L.~N., {Giles}, R.~S.,
  {Merlet}, C., 2016. {Probing Saturn's tropospheric cloud with Cassini/VIMS}.
  Icarus 271, 400--417.

\bibitem[{{B\'ezard} et~al.(1989){B\'ezard}, {Drossart}, {Lellouch}, {Tarrago},
  and {Maillard}}]{Bezard1989}
{B\'ezard}, B., {Drossart}, P., {Lellouch}, E., {Tarrago}, G., {Maillard},
  J.~P., 1989. {Detection of arsine in Saturn}. \apj 346, 509--513.

\bibitem[{{Bjoraker} et~al.(1986){Bjoraker}, {Larson}, and
  {Kunde}}]{Bjoraker1986}
{Bjoraker}, G.~L., {Larson}, H.~P., {Kunde}, V.~G., 1986. {The gas composition
  of Jupiter derived from 5 micron airborne spectroscopic observations}. Icarus
  66, 579--609.

\bibitem[{{Borysow}(1991)}]{Borysow1991h2h2f}
{Borysow}, A., 1991. {Modeling of collision-induced infrared absorption spectra
  of H$_2$-H$_2$ pairs in the fundamental band at temperatures from 20 to 300
  K}. Icarus 92, 273--279.

\bibitem[{{Borysow}(1992)}]{Borysow1992h2he}
{Borysow}, A., 1992. {New model of collision-induced infrared absorption
  spectra of H$_2$-He pairs in the 2-2.5 micron range at temperatures from 20
  to 300 K - an update}. Icarus 96, 169--175.

\bibitem[{{Borysow}(1993)}]{Borysow1993errat}
{Borysow}, A., 1993. {Erratum}. Icarus 106, 614.

\bibitem[{{Bowles} et~al.(2008){Bowles}, {Calcutt}, {Irwin}, and
  {Temple}}]{Bowles2008}
{Bowles}, N., {Calcutt}, S., {Irwin}, P., {Temple}, J., 2008. {Band parameters
  for self-broadened ammonia gas in the range 0.74 to 5.24 {$\mu$}m to support
  measurements of the atmosphere of the planet Jupiter}. Icarus 196, 612--624.

\bibitem[{{Braude} et~al.(2019){Braude}, {Irwin}, {Fletcher}, and
  {Pall{\'e}}}]{Braude2019EPSC}
{Braude}, A., {Irwin}, P., {Fletcher}, L., {Pall{\'e}}, E., 2019.
  {Characterising the colour and vertical structure of Saturn's haze using
  observations from the VLT/MUSE instrument}. In: EPSC-DPS Joint Meeting 2019.
  Vol. 2019. pp. EPSC--DPS2019--652.

\bibitem[{{Braude} et~al.(2018){Braude}, {Irwin}, {Orton}, and
  {Fletcher}}]{Braude2018}
{Braude}, A.~S., {Irwin}, P., {Orton}, G., {Fletcher}, L., 2018. {Retrieving a
  universal chromophore to constrain visible changes in Jupiter's appearance
  between 2014-2018}. In: AAS/Division for Planetary Sciences Meeting
  Abstracts. p. 214.15.

\bibitem[{{Briggs} and {Sackett}(1989)}]{Briggs1989}
{Briggs}, F.~H., {Sackett}, P.~D., 1989. {Radio observations of Saturn as a
  probe of its atmosphere and cloud structure}. Icarus 80, 77--103.

\bibitem[{{Brown} et~al.(2004){Brown}, {Baines}, {Bellucci}, {Bibring},
  {Buratti}, {Capaccioni}, {Cerroni}, {Clark}, {Coradini}, {Cruikshank},
  {Drossart}, {Formisano}, {Jaumann}, {Langevin}, {Matson}, {McCord},
  {Mennella}, {Miller}, {Nelson}, {Nicholson}, {Sicardy}, and
  {Sotin}}]{Brown2004}
{Brown}, R.~H., {Baines}, K.~H., {Bellucci}, G., {Bibring}, J.-P., {Buratti},
  B.~J., {Capaccioni}, F., {Cerroni}, P., {Clark}, R.~N., {Coradini}, A.,
  {Cruikshank}, D.~P., {Drossart}, P., {Formisano}, V., {Jaumann}, R.,
  {Langevin}, Y., {Matson}, D.~L., {McCord}, T.~B., {Mennella}, V., {Miller},
  E., {Nelson}, R.~M., {Nicholson}, P.~D., {Sicardy}, B., {Sotin}, C., 2004.
  {The Cassini Visual and Infrared Mapping Spectrometer (VIMS) Investigation}.
  Space Sci. Rev. 115, 111--168.

\bibitem[{{Carlson} et~al.(2016){Carlson}, {Baines}, {Anderson}, {Filacchione},
  and {Simon}}]{Carlson2016}
{Carlson}, R.~W., {Baines}, K.~H., {Anderson}, M.~S., {Filacchione}, G.,
  {Simon}, A.~A., 2016. {Chromophores from photolyzed ammonia reacting with
  acetylene: Application to Jupiter's Great Red Spot}. Icarus 274, 106--115.

\bibitem[{{CIE}(1932)}]{CIE1932}
{CIE}, 1932. {Commission Internationale de l$'$$\acute{E}$clairage Proceedings,
  1931.} Cambridge: Cambridge University Press.

\bibitem[{{Clark} et~al.(2018){Clark}, {Brown}, {Lytle}, and
  {Hemdman}}]{Clark2018cal}
{Clark}, R.~N., {Brown}, R.~H., {Lytle}, D.~N., {Hemdman}, M., 2018. The vims
  wavelength and radiometric calibration 19, final report. Calibration Report:
  30 pages, NASA Planetary Data System, Atmospheres Node, Las Cruces, NM.

\bibitem[{{Dyudina} et~al.(2009){Dyudina}, {Ingersoll}, {Ewald}, {Vasavada},
  {West}, {Baines}, {Momary}, {Del Genio}, {Barbara}, {Porco}, {Achterberg},
  {Flasar}, {Simon-Miller}, and {Fletcher}}]{Dyudina2009}
{Dyudina}, U.~A., {Ingersoll}, A.~P., {Ewald}, S.~P., {Vasavada}, A.~R.,
  {West}, R.~A., {Baines}, K.~H., {Momary}, T.~W., {Del Genio}, A.~D.,
  {Barbara}, J.~M., {Porco}, C.~C., {Achterberg}, R.~K., {Flasar}, F.~M.,
  {Simon-Miller}, A.~A., {Fletcher}, L.~N., 2009. {Saturn's south polar vortex
  compared to other large vortices in the Solar System}. Icarus 202, 240--248.

\bibitem[{{Dyudina} et~al.(2008){Dyudina}, {Ingersoll}, {Ewald}, {Vasavada},
  {West}, {Del Genio}, {Barbara}, {Porco}, {Achterberg}, {Flasar},
  {Simon-Miller}, and {Fletcher}}]{Dyudina2008Sci}
{Dyudina}, U.~A., {Ingersoll}, A.~P., {Ewald}, S.~P., {Vasavada}, A.~R.,
  {West}, R.~A., {Del Genio}, A.~D., {Barbara}, J.~M., {Porco}, C.~C.,
  {Achterberg}, R.~K., {Flasar}, F.~M., {Simon-Miller}, A.~A., {Fletcher},
  L.~N., 2008. {Dynamics of Saturn's South Polar Vortex}. Science 319, 1801.

\bibitem[{{Ferris} and {Benson}(1981)}]{Ferris1981}
{Ferris}, J.~P., {Benson}, R., 1981. {An Investigation of the Mechanism of
  Phosphine Photolysis}. J. Am. Chem. Soc. 103, 1922--1927.

\bibitem[{{Fletcher} et~al.(2011{\natexlab{a}}){Fletcher}, {Baines}, {Momary},
  {Showman}, {Irwin}, {Orton}, M., and {Merlit}}]{Fletcher2011vims}
{Fletcher}, L.~N., {Baines}, K.~H., {Momary}, T.~M., {Showman}, A.~S., {Irwin},
  P.~G.~J., {Orton}, G.~S., M., R., {Merlit}, C., 2011{\natexlab{a}}. {Saturn's
  tropospheric composition and clouds from Cassini/VIMS 4.6 -- 5.1 $\mu$m
  nightside spectroscopy}. Icarus 214, 510--533.

\bibitem[{{Fletcher} et~al.(2019){Fletcher}, {Greathouse}, {Guerlet}, {Moses},
  and {West}}]{Fletcher2019book}
{Fletcher}, L.~N., {Greathouse}, T.~K., {Guerlet}, S., {Moses}, J.~I., {West},
  R.~A., 2019. {Saturn's Seasonally Changing Atmosphere}. {Cambridge Univ.
  Press}, p. 448.

\bibitem[{{Fletcher} et~al.(2011{\natexlab{b}}){Fletcher}, {Hesman}, {Irwin},
  {Baines}, {Momary}, {Sanchez-Lavega}, {Flasar}, {Read}, {Orton},
  {Simon-Miller}, {Hueso}, {Bjoraker}, {Mamoutkine}, {del Rio-Gaztelurrutia},
  {Gomez}, {Buratti}, {Clark}, {Nicholson}, and {Sotin}}]{Fletcher2011Sci}
{Fletcher}, L.~N., {Hesman}, B.~E., {Irwin}, P.~G.~J., {Baines}, K.~H.,
  {Momary}, T.~W., {Sanchez-Lavega}, A., {Flasar}, F.~M., {Read}, P.~L.,
  {Orton}, G.~S., {Simon-Miller}, A., {Hueso}, R., {Bjoraker}, G.~L.,
  {Mamoutkine}, A., {del Rio-Gaztelurrutia}, T., {Gomez}, J.~M., {Buratti}, B.,
  {Clark}, R.~N., {Nicholson}, P.~D., {Sotin}, C., 2011{\natexlab{b}}. {Thermal
  Structure and Dynamics of Saturn's Northern Springtime Disturbance}. Science
  332, 1413--1417.

\bibitem[{{Fletcher} et~al.(2008){Fletcher}, {Irwin}, {Orton}, {Teanby},
  {Achterberg}, {Bjoraker}, {Read}, {Simon-Miller}, {Howett}, {de Kok},
  {Bowles}, {Calcutt}, {Hesman}, and {Flasar}}]{Fletcher2008}
{Fletcher}, L.~N., {Irwin}, P.~G.~J., {Orton}, G.~S., {Teanby}, N.~A.,
  {Achterberg}, R.~K., {Bjoraker}, G.~L., {Read}, P.~L., {Simon-Miller}, A.~A.,
  {Howett}, C., {de Kok}, R., {Bowles}, N., {Calcutt}, S.~B., {Hesman}, B.,
  {Flasar}, F.~M., 2008. {Temperature and Composition of Saturn's Polar Hot
  Spots and Hexagon}. Science 319, 79.

\bibitem[{{Fletcher} et~al.(2018){Fletcher}, {Orton}, {Sinclair}, {Guerlet},
  {Read}, {Antu{\~n}ano}, {Achterberg}, {Flasar}, {Irwin}, {Bjoraker},
  {Hurley}, {Hesman}, {Segura}, {Gorius}, {Mamoutkine}, and
  {Calcutt}}]{Fletcher2018NatCo}
{Fletcher}, L.~N., {Orton}, G.~S., {Sinclair}, J.~A., {Guerlet}, S., {Read},
  P.~L., {Antu{\~n}ano}, A., {Achterberg}, R.~K., {Flasar}, F.~M., {Irwin},
  P.~G.~J., {Bjoraker}, G.~L., {Hurley}, J., {Hesman}, B.~E., {Segura}, M.,
  {Gorius}, N., {Mamoutkine}, A., {Calcutt}, S.~B., 2018. {A hexagon in
  Saturn's northern stratosphere surrounding the emerging summertime polar
  vortex}. Nature Communications 9, 3564.

\bibitem[{{Fletcher} et~al.(2009{\natexlab{a}}){Fletcher}, {Orton}, {Teanby},
  and {Irwin}}]{Fletcher2009ph3}
{Fletcher}, L.~N., {Orton}, G.~S., {Teanby}, N.~A., {Irwin}, P.~G.~J.,
  2009{\natexlab{a}}. {Phosphine on Jupiter and Saturn from Cassini/CIRS}.
  Icarus 202, 543--564.

\bibitem[{{Fletcher} et~al.(2009{\natexlab{b}}){Fletcher}, {Orton}, {Teanby},
  {Irwin}, and {Bjoraker}}]{Fletcher2009ch4saturn}
{Fletcher}, L.~N., {Orton}, G.~S., {Teanby}, N.~A., {Irwin}, P.~G.~J.,
  {Bjoraker}, G.~L., 2009{\natexlab{b}}. {Methane and its isotopologues on
  Saturn from Cassini/CIRS observations}. Icarus 199, 351--367.

\bibitem[{{Fouchet} et~al.(2009){Fouchet}, {Moses}, and
  {Conrath}}]{Fouchet2009}
{Fouchet}, T., {Moses}, J.~I., {Conrath}, B.~J., 2009. {Saturn: Composition and
  Chemistry}. In: {Dougherty}, M.~K., {Esposito}, L.~W., {Krimigis}, S.~M.
  (Eds.), Saturn from Cassini-Huygens. Springer Dordrecht Heidelberg London New
  York, pp. 83--112.

\bibitem[{{Hansen} and {Travis}(1974)}]{Hansen1974}
{Hansen}, J.~E., {Travis}, L.~D., 1974. {Light scattering in planetary
  atmospheres}. Space Sci. Rev. 16, 527--610.

\bibitem[{{Howett} et~al.(2007){Howett}, {Carlson}, {Irwin}, and
  {Calcutt}}]{Howett2007}
{Howett}, C.~J.~A., {Carlson}, R.~W., {Irwin}, P.~G.~J., {Calcutt}, S.~B.,
  2007. {Optical constants of ammonium hydrosulfide ice and ammonia ice}.
  Journal of the Optical Society of America B Optical Physics 24, 126--136.

\bibitem[{{IEC}(1999)}]{IEC99}
{IEC}, 1999. {International Standard 61966-2-1: Part 2-1: Colour management --
  Default RGB colour space -- sRGB}. Geneva: IEC Webstore.

\bibitem[{{Karkoschka} and {Tomasko}(2005)}]{Kark2005}
{Karkoschka}, E., {Tomasko}, M., 2005. {Saturn's vertical and latitudinal cloud
  structure 1991-2004 from HST imaging in 30 filters}. Icarus 179, 195--221.

\bibitem[{{Karkoschka} and {Tomasko}(2010)}]{Kark2010ch4}
{Karkoschka}, E., {Tomasko}, M.~G., 2010. {Methane absorption coefficients for
  the jovian planets from laboratory, Huygens, and HST data}. Icarus 205,
  674--694.

\bibitem[{{Kerola} et~al.(1997){Kerola}, {Larson}, and {Tomasko}}]{Kerola1997}
{Kerola}, D.~X., {Larson}, H.~P., {Tomasko}, M.~G., 1997. {Analysis of the
  Near-IR Spectrum of Saturn: A Comprehensive Radiative Transfer Model of Its
  Middle and Upper Troposphere}. Icarus 127, 190--212.

\bibitem[{{Khare} et~al.(1993){Khare}, {Thompson}, {Cheng}, {Chyba}, {Sagan},
  {Arakawa}, {Meisse}, and {Tuminello}}]{Khare1993}
{Khare}, B.~N., {Thompson}, W.~R., {Cheng}, L., {Chyba}, C., {Sagan}, C.,
  {Arakawa}, E.~T., {Meisse}, C., {Tuminello}, P.~S., 1993. {Production and
  optical constraints of ice tholin from charged particle irradiation of (1:6)
  C2H6/H2O at 77 K}. Icarus 103, 290--300.

\bibitem[{{Laraia} et~al.(2013){Laraia}, {Ingersoll}, {Janssen}, {Gulkis},
  {Oyafuso}, and {Allison}}]{Laraia2013}
{Laraia}, A.~L., {Ingersoll}, A.~P., {Janssen}, M.~A., {Gulkis}, S., {Oyafuso},
  F., {Allison}, M., 2013. {Analysis of Saturn's thermal emission at 2.2-cm
  wavelength: Spatial distribution of ammonia vapor}. Icarus 226, 641--654.

\bibitem[{{Li} et~al.(2020){Li}, {de Pater}, {Sault}, {Butler}, {Hayes}, and
  {Zhang}}]{Li2020DPS}
{Li}, C., {de Pater}, I., {Sault}, R., {Butler}, B., {Hayes}, A., {Zhang}, Z.,
  2020. {New insights into Saturn's Polar Hexagon}. In: AAS/Division for
  Planetary Sciences Meeting Abstracts. Vol.~52 of AAS/Division for Planetary
  Sciences Meeting Abstracts. p. 204.04.

\bibitem[{{Lindal} et~al.(1985){Lindal}, {Sweetnam}, and
  {Eshleman}}]{Lindal1985}
{Lindal}, G.~F., {Sweetnam}, D.~N., {Eshleman}, V.~R., 1985. {The atmosphere of
  Saturn - an analysis of the Voyager radio occultation measurements}. \aj 90,
  1136--1146.

\bibitem[{{Martonchik} et~al.(1984){Martonchik}, {Orton}, and
  {Appleby}}]{Martonchik1984}
{Martonchik}, J.~V., {Orton}, G.~S., {Appleby}, J.~F., 1984. {Optical
  properties of NH$_3$ ice from the far infrared to the near ultraviolet}.
  Appl. Optics 23, 541--547.

\bibitem[{{Noll} and {Larson}(1991)}]{Noll1991}
{Noll}, K.~S., {Larson}, H.~P., 1991. {The spectrum of Saturn from 1990 to
  2230/cm - Abundances of AsH3, CH3D, CO, GeH4, NH3, and PH3}. Icarus 89,
  168--189.

\bibitem[{{Noll} et~al.(1990){Noll}, {Larson}, and {Geballe}}]{Noll1990}
{Noll}, K.~S., {Larson}, H.~P., {Geballe}, T.~R., 1990. {The abundance of
  AsH$_3$ in Jupiter}. Icarus 83, 494--499.

\bibitem[{{Noy} et~al.(1981){Noy}, {Podolak}, and {Bar-Nun}}]{Noy1981}
{Noy}, N., {Podolak}, M., {Bar-Nun}, A., 1981. {Photochemistry of phosphine and
  Jupiter's Great Red Spot}. \jgr 86, 11985--11988.

\bibitem[{{Ohta} and {Ishida}(1988)}]{Ohta1988}
{Ohta}, K., {Ishida}, H., 1988. {Comparison Among Several Numerical Integration
  Methods for Kramers-Kronig Transformation}. Applied Spectr. 42, 952--957.

\bibitem[{{Orton} et~al.(2000){Orton}, {Serabyn}, and {Lee}}]{Orton2000ph3}
{Orton}, G.~S., {Serabyn}, E., {Lee}, Y.~T., 2000. {Vertical distribution of PH
  $_{3}$ in Saturn from observations of its 1-0 and 3-2 rotational lines}.
  Icarus 146, 48--59.

\bibitem[{{Orton} et~al.(2001){Orton}, {Serabyn}, and {Lee}}]{Orton2001ph3}
{Orton}, G.~S., {Serabyn}, E., {Lee}, Y.~T., 2001. {Erratum, Volume 146, Number
  1, pages 48-59 (2000), in the article ``Vertical Distribution of PH$_{3}$ in
  Saturn from Observations of Its 1-0 and 3-2 Rotational Lines,''}. Icarus 149,
  489--490.

\bibitem[{{Orton} and {Yanamandra-Fisher}(2005)}]{Orton2005Sci}
{Orton}, G.~S., {Yanamandra-Fisher}, P.~A., 2005. {Saturn's Temperature Field
  from High-Resolution Middle-Infrared Imaging}. Science 307~(5710), 696--698.

\bibitem[{{P{\'e}rez-Hoyos} et~al.(2005){P{\'e}rez-Hoyos},
  {S{\'a}nchez-Lavega}, {French}, and {Rojas}}]{Perez-Hoyos2005}
{P{\'e}rez-Hoyos}, S., {S{\'a}nchez-Lavega}, A., {French}, R.~G., {Rojas},
  J.~F., 2005. {Saturn's cloud structure and temporal evolution from ten years
  of Hubble Space Telescope images (1994 -- 2003)}. Icarus 176, 155--174.

\bibitem[{{P{\'e}rez-Hoyos} et~al.(2016){P{\'e}rez-Hoyos}, {Sanz-Requena},
  {S{\'a}nchez-Lavega}, {Irwin}, and {Smith}}]{Perez-Hoyos2016}
{P{\'e}rez-Hoyos}, S., {Sanz-Requena}, J.~F., {S{\'a}nchez-Lavega}, A.,
  {Irwin}, P.~G.~J., {Smith}, A., 2016. {Saturn's tropospheric particles phase
  function and spatial distribution from Cassini ISS 2010-11 observations}.
  Icarus 277, 1--18.

\bibitem[{{Porco} et~al.(2004){Porco}, {West}, {Squyres}, {McEwen}, {Thomas},
  {Murray}, {Delgenio}, {Ingersoll}, {Johnson}, {Neukum}, {Veverka}, {Dones},
  {Brahic}, {Burns}, {Haemmerle}, {Knowles}, {Dawson}, {Roatsch}, {Beurle}, and
  {Owen}}]{Porco2004SSR}
{Porco}, C.~C., {West}, R.~A., {Squyres}, S., {McEwen}, A., {Thomas}, P.,
  {Murray}, C.~D., {Delgenio}, A., {Ingersoll}, A.~P., {Johnson}, T.~V.,
  {Neukum}, G., {Veverka}, J., {Dones}, L., {Brahic}, A., {Burns}, J.~A.,
  {Haemmerle}, V., {Knowles}, B., {Dawson}, D., {Roatsch}, T., {Beurle}, K.,
  {Owen}, W., 2004. {Cassini Imaging Science: Instrument Characteristics and
  Anticipated Scientific Investigations at Saturn}. Space Science Reviews 115,
  363--497.

\bibitem[{{Press} et~al.(1992){Press}, {Teukolsky}, {Vetterling}, and
  {Flannery}}]{Press1992}
{Press}, W.~H., {Teukolsky}, S.~A., {Vetterling}, W.~T., {Flannery}, B.~P.,
  1992. {Numerical recipes in FORTRAN. The art of scientific computing, 2nd
  ed.} Cambridge: University Press.

\bibitem[{{Pryor} et~al.(2019){Pryor}, {Esposito}, {Jouchoux}, {West},
  {Grodent}, {G{\'e}rard}, {Radioti}, {Lamy}, and {Koskinen}}]{Pryor2019JGRE}
{Pryor}, W.~R., {Esposito}, L.~W., {Jouchoux}, A., {West}, R.~A., {Grodent},
  D., {G{\'e}rard}, J.~C., {Radioti}, A., {Lamy}, L., {Koskinen}, T., 2019.
  {Cassini UVIS Detection of Saturn's North Polar Hexagon in the Grand Finale
  Orbits}. Journal of Geophysical Research (Planets) 124~(7), 1979--1988.

\bibitem[{{Roman} et~al.(2013){Roman}, {Banfield}, and {Gierasch}}]{Roman2013}
{Roman}, M.~T., {Banfield}, D., {Gierasch}, P.~J., 2013. {Saturn's cloud
  structure inferred from Cassini ISS}. Icarus 225~(1), 93--110.

\bibitem[{{S{\'a}nchez-Lavega} et~al.(2020){S{\'a}nchez-Lavega},
  {Garc{\'\i}a-Mu{\~n}oz}, {del R{\'\i}o-Gaztelurrutia}, {P{\'e}rez-Hoyos},
  {Sanz-Requena}, {Hueso}, {Guerlet}, and {Peralta}}]{Sanchez-Lavega2020hazes}
{S{\'a}nchez-Lavega}, A., {Garc{\'\i}a-Mu{\~n}oz}, A., {del
  R{\'\i}o-Gaztelurrutia}, T., {P{\'e}rez-Hoyos}, S., {Sanz-Requena}, J.~F.,
  {Hueso}, R., {Guerlet}, S., {Peralta}, J., 2020. {Multilayer hazes over
  Saturn's hexagon from Cassini ISS limb images}. Nature Communications 11,
  2281.

\bibitem[{{S{\'a}nchez-Lavega} et~al.(2006){S{\'a}nchez-Lavega}, {Hueso},
  {P{\'e}rez-Hoyos}, and {Rojas}}]{Sanchez-Lavega2006}
{S{\'a}nchez-Lavega}, A., {Hueso}, R., {P{\'e}rez-Hoyos}, S., {Rojas}, J.~F.,
  2006. {A strong vortex in Saturn's South Pole}. Icarus 184, 524--531.

\bibitem[{{Sanz-Requena} et~al.(2018){Sanz-Requena}, {P{\'e}rez-Hoyos},
  {S{\'a}nchez-Lavega}, {Antu{\~n}ano}, and {Irwin}}]{Sanz-Requena2018}
{Sanz-Requena}, J.~F., {P{\'e}rez-Hoyos}, S., {S{\'a}nchez-Lavega}, A.,
  {Antu{\~n}ano}, A., {Irwin}, P.~G.~J., 2018. {Haze and cloud structure of
  Saturn's North Pole and Hexagon Wave from Cassini/ISS imaging}. Icarus 305,
  284--300.

\bibitem[{{Sayanagi} et~al.(2019){Sayanagi}, {Baines}, {Dyudina}, {Fletcher},
  {S{\'a}nchez-Lavega}, and {West}}]{Sayanagi2019book}
{Sayanagi}, K.~M., {Baines}, K.~H., {Dyudina}, U.~A., {Fletcher}, L.~N.,
  {S{\'a}nchez-Lavega}, A., {West}, R.~A., 2019. Ch. {Saturn's Polar
  Atmosphere}, pp. 337--376.

\bibitem[{{Sayanagi} et~al.(2017){Sayanagi}, {Blalock}, {Dyudina}, {Ewald}, and
  {Ingersoll}}]{Sayanagi2017}
{Sayanagi}, K.~M., {Blalock}, J.~J., {Dyudina}, U.~A., {Ewald}, S.~P.,
  {Ingersoll}, A.~P., 2017. {Cassini ISS observation of Saturn's north polar
  vortex and comparison to the south polar vortex}. Icarus 285, 68--82.

\bibitem[{{Sayanagi} et~al.(2016){Sayanagi}, {Blalock}, {Ingersoll}, {Dyudina},
  and {Ewald}}]{Sayanagi2016DPS}
{Sayanagi}, K.~M., {Blalock}, J.~J., {Ingersoll}, A.~P., {Dyudina}, U.~A.,
  {Ewald}, S.~P., 2016. {Formation of a Bright Polar Hood over the Summer North
  Pole of Saturn in 2016}. In: AAS/Division for Planetary Sciences Meeting
  Abstracts \#48. Vol.~48 of AAS/Division for Planetary Sciences Meeting
  Abstracts. p. 501.03.

\bibitem[{{Sromovsky} et~al.(2013){Sromovsky}, {Baines}, and
  {Fry}}]{Sro2013gws}
{Sromovsky}, L.~A., {Baines}, K.~H., {Fry}, P.~M., 2013. {Saturn's Great Storm
  of 2010-2011: Evidence for ammonia and water ices from analysis of VIMS
  spectra}. Icarus 226, 402--418.

\bibitem[{{Sromovsky} et~al.(2018){Sromovsky}, {Baines}, and
  {Fry}}]{Sro2018dark}
{Sromovsky}, L.~A., {Baines}, K.~H., {Fry}, P.~M., 2018. {Models of bright
  storm clouds and related dark ovals in Saturn's Storm Alley as constrained by
  2008 Cassini/VIMS spectra}. Icarus 302, 360--385.

\bibitem[{{Sromovsky} et~al.(2020{\natexlab{a}}){Sromovsky}, {Baines}, and
  {Fry}}]{Sro2020spole}
{Sromovsky}, L.~A., {Baines}, K.~H., {Fry}, P.~M., 2020{\natexlab{a}}.
  {Saturn's south polar cloud structure inferred from 2006 Cassini VIMS
  spectra}. Icarus 344, 113398 (1--24).

\bibitem[{{Sromovsky} et~al.(2017){Sromovsky}, {Baines}, {Fry}, and
  {Carlson}}]{Sro2017red}
{Sromovsky}, L.~A., {Baines}, K.~H., {Fry}, P.~M., {Carlson}, R.~W., 2017. {A
  Possibly universal red chromophore for modeling color variations on Jupiter}.
  Icarus 291, 232--244.

\bibitem[{{Sromovsky} et~al.(2016){Sromovsky}, {Baines}, {Fry}, and
  {Momary}}]{Sro2016}
{Sromovsky}, L.~A., {Baines}, K.~H., {Fry}, P.~M., {Momary}, T.~W., 2016.
  {Cloud clearing in the wake of Saturn's Great Storm of 2010-2011 and
  suggested new constraints on Saturn's He/H$_{2}$ ratio}. Icarus 276,
  141--162.

\bibitem[{{Sromovsky} and {Fry}(2010{\natexlab{a}})}]{Sro2010iso}
{Sromovsky}, L.~A., {Fry}, P.~M., 2010{\natexlab{a}}. {The source of 3-{$\mu$}m
  absorption in Jupiter's clouds: Reanalysis of ISO observations using new
  NH$_{3}$ absorption models}. Icarus 210, 211--229.

\bibitem[{{Sromovsky} and {Fry}(2010{\natexlab{b}})}]{Sro2010vims}
{Sromovsky}, L.~A., {Fry}, P.~M., 2010{\natexlab{b}}. {The source of widespread
  3-{$\mu$}m absorption in Jupiter's clouds: Constraints from 2000 Cassini VIMS
  observations}. Icarus 210, 230--257.

\bibitem[{{Sromovsky} et~al.(2020{\natexlab{b}}){Sromovsky}, {Fry}, and
  {Baines}}]{Sro2020shadows}
{Sromovsky}, L.~A., {Fry}, P.~M., {Baines}, K.~H., 2020{\natexlab{b}}.
  {Interpretation of south polar cloud shadows and antishadows on Saturn and
  the evidence against south polar eyewalls}. Icarus 344, 113399 (1--13).

\bibitem[{{Sromovsky} et~al.(2012){Sromovsky}, {Fry}, {Boudon}, {Campargue},
  and {Nikitin}}]{Sro2012LBL}
{Sromovsky}, L.~A., {Fry}, P.~M., {Boudon}, V., {Campargue}, A., {Nikitin}, A.,
  2012. {Comparison of line-by-line and band models of near-IR methane
  absorption applied to outer planet atmospheres}. Icarus 218, 1--23.

\bibitem[{{Thekaekara}(1974)}]{Thekaekara1974}
{Thekaekara}, M.~P., 1974. {Extraterrestrial solar spectrum, 3000 - 6100 {\AA}
  at 1-{\AA} intervals.} Appl. Opt. 13, 518--522.

\bibitem[{{Thompson} et~al.(2015){Thompson}, {Seidel}, {Gao}, {Gierach},
  {Green}, {Kudela}, and {Mouroulis}}]{Thompson2015sun}
{Thompson}, D.~R., {Seidel}, F.~C., {Gao}, B.~C., {Gierach}, M.~M., {Green},
  R.~O., {Kudela}, R.~M., {Mouroulis}, P., 2015. {Optimizing irradiance
  estimates for coastal and inland water imaging spectroscopy}. \grl 42~(10),
  4116--4123.

\bibitem[{{Warren}(1984)}]{Warren1984}
{Warren}, S.~G., 1984. {Optical constants of ice from the ultraviolet to the
  microwave}. Appl. Optics 23, 1206--1225.

\bibitem[{{Weidenschilling} and {Lewis}(1973)}]{Weidenschilling1973}
{Weidenschilling}, S.~J., {Lewis}, J.~S., 1973. {Atmospheric and cloud
  structures of the jovian planets}. Icarus 20, 465--476.

\bibitem[{{West} et~al.(2009){West}, {Baines}, {Karkoschka}, and
  {S{\'a}nchez-Lavega}}]{West2009satbook}
{West}, R.~A., {Baines}, K.~H., {Karkoschka}, E., {S{\'a}nchez-Lavega}, A.,
  2009. {Clouds and Aerosols in Saturn's Atmosphere}. In: Dougherty, M.~K.,
  Esposito, L.~W., Krimigis, S.~M. (Eds.), {Saturn from Cassini-Huygens}.
  {Springer}, pp. 161--179.

\bibitem[{{West} et~al.(1986){West}, {Strobel}, and {Tomasko}}]{West1986}
{West}, R.~A., {Strobel}, D.~F., {Tomasko}, M.~G., 1986. {Clouds, aerosols, and
  photochemistry in the Jovian atmosphere}. Icarus 65, 161--217.

\bibitem[{Wohlfarth(2008)}]{Wohlfarth2008}
Wohlfarth, C., 2008. Refractive index of diphosphine: Datasheet from
  Landolt-B{\"o}rnstein - Group III Condensed Matter {\textperiodcentered}
  Volume 47: ``Refractive Indices of Pure Liquids and Binary Liquid Mixtures
  (Supplement to III/38)'' in SpringerMaterials). Springer-Verlag Berlin
  Heidelberg, copyright 2008 Springer-Verlag Berlin Heidelberg.

\bibitem[{{Zheng} and {Borysow}(1995)}]{Zheng1995h2h2o1}
{Zheng}, C., {Borysow}, A., 1995. {Modeling of collision-induced infrared
  absorption spectra of H$_2$ pairs in the first overtone band at temperatures
  from 20 to 500 K}. Icarus 113, 84--90.

\end{thebibliography}

\end{document}